\documentclass[12pt]{article}
\def\baselinestretch{1.2}

\usepackage{jheppub, bm, wrapfig,float,array}
\usepackage[utf8]{inputenc}
\numberwithin{equation}{section}
\renewcommand\afterTocRuleSpace{\bigskip}

\usepackage{amsfonts}
\usepackage{amsmath}
\usepackage{color}
\usepackage{graphicx}
\usepackage{float}
\usepackage[T1]{fontenc}
\usepackage[utf8]{inputenc}
\usepackage{alphabeta}
\usepackage{fancyvrb}
\usepackage[dvipsnames]{xcolor}
\usepackage{bold-extra}
\usepackage{lmodern}

\usepackage{tikz}
\usetikzlibrary{arrows}
\usetikzlibrary{shapes.geometric,calc,arrows, positioning,shapes.misc,decorations.markings}
\tikzset{
  big arrow/.style={
    decoration={markings,mark=at position 1 with {\arrow[scale=2,#1]{>}}},
    postaction={decorate},
    shorten >=0.4pt},
  big arrow/.default=black}

\newcommand{\bea}{\begin{eqnarray}}
\newcommand{\eea}{\end{eqnarray}}
\newcommand{\be}{\begin{equation}}
\newcommand{\ee}{\end{equation}}
\newcommand{\bit}{\begin{itemize}}
\newcommand{\eit}{\end{itemize}}
\newcommand{\ben}{\begin{enumerate}}
\newcommand{\een}{\end{enumerate}}
\newcommand{\nn}{\nonumber}
\renewcommand{\ni}{\noindent}

\newcommand{\wt}{\widetilde}
\newcommand{\wh}{\widehat}
\newcommand{\ot}{\otimes}
\newcommand{\half}{\frac{1}{2}}

\newcommand{\mbb}{\mathbb}
\newcommand{\Z}{{\mathbb Z}}
\newcommand{\R}{{\mathbb R}}
\newcommand{\C}{{\mathbb C}}
\newcommand{\Q}{{\mathbb Q}}
\renewcommand{\P}{{\mathbb P}}

\newcommand{\cA}{\mathcal{A}}
\newcommand{\cB}{\mathcal{B}}
\newcommand{\cC}{\mathcal{C}}
\newcommand{\cD}{\mathcal{D}}
\newcommand{\cE}{\mathcal{E}}
\newcommand{\cF}{\mathcal{F}}
\newcommand{\cG}{\mathcal{G}}
\newcommand{\cH}{\mathcal{H}}
\newcommand{\cI}{\mathcal{I}}
\newcommand{\cJ}{\mathcal{J}}
\newcommand{\cK}{\mathcal{K}}
\newcommand{\cL}{\mathcal{L}}
\newcommand{\cM}{\mathcal{M}}
\newcommand{\cN}{\mathcal{N}}
\newcommand{\cO}{\mathcal{O}}
\newcommand{\cP}{\mathcal{P}}
\newcommand{\cQ}{\mathcal{Q}}
\newcommand{\cR}{\mathcal{R}}
\newcommand{\cS}{\mathcal{S}}
\newcommand{\cT}{\mathcal{T}}
\newcommand{\cU}{\mathcal{U}}
\newcommand{\cV}{\mathcal{V}}
\newcommand{\cW}{\mathcal{W}}
\newcommand{\cX}{\mathcal{X}}
\newcommand{\cY}{\mathcal{Y}}
\newcommand{\cZ}{\mathcal{Z}}

\newcommand{\F}{\mathsf{F}}
\renewcommand{\S}{\mathsf{S}}
\newcommand{\Asym}{\mathsf{\Lambda}^2}
\newcommand{\bAsym}{\bar{\mathsf{\Lambda}}^2}
\newcommand{\tAsym}{\mathsf{\Lambda}^3}
\newcommand{\btAsym}{\bar{\mathsf{\Lambda}}^3}
\newcommand{\nAsym}{\mathsf{\Lambda}^n}
\newcommand{\bnAsym}{\bar{\mathsf{\Lambda}}^n}
\newcommand{\mAsym}{\mathsf{\Lambda}^m}
\newcommand{\Sym}{\mathsf{S}^2}
\newcommand{\bSym}{\bar{\mathsf{S}}^2}
\renewcommand{\C}{\mathsf{C}}
\newcommand{\A}{\mathsf{A}}
\newcommand{\B}{\mathsf{B}}
\renewcommand{\C}{\mathsf{C}}
\newcommand{\D}{\mathsf{D}}
\renewcommand{\L}{\mathsf{\Lambda}}

\newcommand{\mf}{\mathfrak}
\newcommand{\fT}{\mathfrak{T}}
\newcommand{\fB}{\mathfrak{B}}
\newcommand{\fe}{\mathfrak{e}}
\newcommand{\ff}{\mathfrak{f}}
\newcommand{\fg}{\mathfrak{g}}
\newcommand{\fh}{\mathfrak{h}}
\newcommand{\su}{\mathfrak{su}}
\renewcommand{\sp}{\mathfrak{sp}}
\newcommand{\so}{\mathfrak{so}}
\renewcommand{\u}{\mathfrak{u}}

\newcommand{\PP}{Pin$^+$}
\newcommand{\PM}{Pin$^-$}

\newcommand{\ubf}[1]{\underline{\bf #1}}
\newcommand{\res}[1]{\left(#1\right)|_\Sigma}
\def\tr{\mathop{\mathrm{tr}}\nolimits}

\newcommand{\lra}{\leftrightarrow}
\newcommand{\llra}{\longleftrightarrow}
\newcommand{\Lra}{\longrightarrow}

\newcommand{\bF}{{\mathbb F}}
\newcommand{\dP}{\mathbf{dP}}

\title{Flavor Symmetry of $5d$ SCFTs, Part 1: General Setup}

\author{Lakshya Bhardwaj}

\affiliation{Mathematical Institute, University of Oxford,\\Andrew Wiles Building, Woodstock Road, Oxford, OX2 6GG, UK}

\abstract{A large class of $5d$ superconformal field theories (SCFTs) can be constructed by integrating out BPS particles from $6d$ SCFTs compactified on a circle. We describe a general method for extracting the flavor symmetry of any $5d$ SCFT lying in this class. For this purpose, we utilize the geometric engineering of $5d$ $\cN=1$ theories in M-theory, where the flavor symmetry is encoded in a collection of non-compact surfaces.
}

\begin{document}

\maketitle

\section{Introduction} \label{I}
By now, the existence of a huge number of interacting superconformal field theories (SCFTs) in five dimensions has been proposed. These theories are typically defined as certain decoupling limits of string theory compactifications on singular geometries \cite{Morrison:1996xf,Intriligator:1997pq,Diaconescu:1998cn,DelZotto:2017pti,Xie:2017pfl,Closset:2018bjz,Jefferson:2018irk,Apruzzi:2018nre,Bhardwaj:2018yhy,Bhardwaj:2018vuu,Bhardwaj:2019fzv,Apruzzi:2019vpe,Apruzzi:2019opn,Apruzzi:2019enx,Bhardwaj:2019jtr,Saxena:2019wuy,Bhardwaj:2019xeg,Apruzzi:2019syw,Bhardwaj:2020gyu,Eckhard:2020jyr,Closset:2020scj,Hubner:2020uvb,Bhardwaj:2020kim} or of intersecting brane configurations in string theory \cite{Seiberg:1996bd,Aharony:1997ju,Aharony:1997bh,DeWolfe:1999hik,Brandhuber:1999yo,Bergman:2013aca,Zafrir:2014ywa,Zafrir:2015ftn,Hayashi:2015zka,Hayashi:2015fsa,Bergman:2015dpa,Hayashi:2018lyv,Hayashi:2018bkd,Hayashi:2019yxj}\footnote{See also \cite{Witten:1996qb,Kim:2012gu,Zafrir:2015rga,Hayashi:2015vhy,Kim:2015jba,Ohmori:2015pia,Yonekura:2015ksa,Zafrir:2015uaa,Tachikawa:2015mha,Hayashi:2016abm,Ohmori:2016shy,Jefferson:2017ahm,Mekareeya:2017jgc,Ashok:2017bld,Bastian:2018fba,Assel:2018rcw,Bhardwaj:2019ngx,Closset:2019mdz,Hayashi:2020sly,Morrison:2020ool,Bhardwaj:2020phs,BenettiGenolini:2020doj}, and see \cite{Bergman:2012rgz,DHoker:2016wak,DHoker:2017prl,DHoker:2017muf,DHoker:2017gcu,Chaney:2018gjc,Bah:2018lyv,Uhlemann:2019ypp,Uhlemann:2019ors} for study of $5d$ SCFTs using holography.}. A purely field theoretic handle on these theories can also be obtained in certain cases where the SCFT reduces in the IR, upon a mass deformation, to a $5d$ $\cN=1$ supersymmetric gauge theory.

In this series of papers (Part 1 and Part 2 \cite{Bhardwaj:2020avz}), we study the flavor symmetry algebras of $5d$ SCFTs. This topic has been explored using various approaches in the past. See \cite{Apruzzi:2019vpe,Apruzzi:2019opn,Apruzzi:2019enx,Apruzzi:2019syw,Eckhard:2020jyr,Hubner:2020uvb} for a recent approach based on Combined Fiber Diagrams (CFDs), \cite{Kim:2012gu,Bergman:2013aca,Zafrir:2014ywa,Zafrir:2015ftn,Hayashi:2015fsa,Bergman:2015dpa,Hayashi:2019yxj,Hayashi:2015vhy} for approaches based on brane-webs and superconformal index, \cite{Yonekura:2015ksa,Zafrir:2015uaa,Tachikawa:2015mha,Jefferson:2017ahm,Zafrir:2015rga} for approach based on instanton operators, and \cite{Hayashi:2020sly} for an approach based on complete prepotentials. 

We will approach this topic from the point of view of the construction of $5d$ SCFTs in terms of M-theory compactified on Calabi-Yau threefolds (CY3). The advantage of using this approach is that it applies to all known $5d$ SCFTs. Indeed, at the time of writing this paper, there is no known $5d$ SCFT which (or some discretely gauged\footnote{One can also gauge after stacking by an SPT phase. Here we include gaugings of higher-form symmetries and also more general gaugings, some of which were discussed in \cite{Bhardwaj:2017xup}.} version of it) does not admit such a construction\footnote{See also the recent paper \cite{Closset:2020scj} where some examples of exotic $5d$ SCFTs have been proposed which, despite admitting a CY3 construction in M-theory, cannot be directly studied using methods of this paper, since the CY3 cannot be completely resolved in those cases.}. Moreover, we will restrict ourselves to the sub-class of those $5d$ SCFTs which can be obtained (on their extended\footnote{Extended Coulomb branch of a $5d$ SCFT is defined to be the total space formed by fibering Coulomb branch of its mass deformations upon the space of supersymmetric mass deformations.} Coulomb branch) by integrating out BPS particles from the extended Coulomb branch of a known $6d$ SCFT compactified on a circle (possibly with a twist by a discrete flavor symmetry of the $6d$ SCFT). This sub-class is expected to capture ``almost all'' of the $5d$ SCFTs that admit a smooth M-theory construction. Evidence for this statement was provided in \cite{Bhardwaj:2020gyu,Bhardwaj:2019xeg} where it was shown that among the class of $5d$ SCFTs which can be deformed to some $5d$ $\cN=1$ gauge theory with a simple gauge algebra, the $5d$ SCFTs not arising by integrating out of BPS particles from circle compactified $6d$ SCFTs form an extremely tiny subset.

In Part 2 \cite{Bhardwaj:2020avz}, we apply the method discussed in this paper to determine the flavor symmetry of all $5d$ SCFTs which can be deformed to some $5d$ $\cN=1$ gauge theory with a simple gauge algebra and lying in the above class. For many of these theories, the flavor symmetry at the conformal point is larger than the flavor symmetry of the $\cN=1$ gauge theory, a phenomenon referred to as the \emph{enhancement} of the flavor symmetry at the conformal point. Thus, in this context, our method can be viewed as a tool to determine this enhancement. Some of our results in Part 2 \cite{Bhardwaj:2020avz} have been found using other methods discussed above, but many of the results are new.

Let us now provide an overview of our method, which is similar in spirit to the CFD based method \cite{Apruzzi:2019vpe,Apruzzi:2019opn,Apruzzi:2019enx} and bears many resemblances to it. The essential idea is that the flavor symmetry algebra of a $5d$ SCFT is captured in the spectrum of non-compact surfaces inside the resolved non-compact Calabi-Yau threefold describing the theory on the extended Coulomb branch of the $5d$ SCFT. The non-abelian part of this flavor symmetry admits an even nicer description in this context, which is captured in the spectrum of $\P^1$ fibered non-compact surfaces having a non-compact base curve. It is easy to see that this should be the case from the well-known statement that non-abelian part of the \emph{gauge} symmetry is captured by $\P^1$ fibered compact surfaces (possibly with additional blowups). M2 branes wrapping the fibers of such compact surfaces give rise to W-bosons for the gauge algebra, and M2 branes wrapping the base curves of these compact surfaces give rise to instantonic BPS particles whose mass is controlled by the inverse gauge coupling of the gauge algebra. Since the mass of a particle arising from an M2 brane wrapping a compact curve is controlled by the volume of that curve, the volumes of the base curves of the Hirzebruch surfaces are then identified with the inverse gauge coupling of the gauge algebra. Turning the gauge coupling off turns the gauge algebra into a flavor algebra, and correspondingly decompactifies the base curves of the compact $\P^1$ fibered surfaces. Thus, we expect that non-abelian part of the flavor symmetry algebra is captured by the $\P^1$ fibered non-compact surfaces.

We start by providing a recipe to explicitly identify the non-compact surfaces capturing the non-abelian part of the flavor symmetry algebra of a $6d$ SCFT compactified on a circle\footnote{From this point on, we will use the term ``$5d$ KK theory'' to denote a (twisted or untwisted) circle compactification of a $6d$ SCFT.} (possibly with a twist). We do not need to identify the non-compact surfaces corresponding to the abelian part of the flavor symmetry algebra. The reason for this is that the number of abelian factors in the flavor symmetry algebra can be easily tracked when we integrate out BPS particles, as we will explain in more detail below.

Now, the process of integrating out a BPS particle is achieved by performing a sequence of flops on the set of compact surfaces until a compact curve is taken out of the compact surfaces into the non-compact surfaces. The process of integrating out is fully completed when we send this compact curve to infinite volume after it has been taken out of the set of compact surfaces. This last step of decompactifying the compact curve ends up modifying the non-compact surfaces so that in general some of the $\P^1$ fibered non-compact surfaces do not remain $\P^1$ fibered anymore. Thus, after integrating out the BPS particle we obtain a CY3 describing a $5d$ SCFT with a different collection of non-compact surfaces admitting $\P^1$ fibration. We propose that this set of $\P^1$ fibered non-compact surfaces is the maximal set of $\P^1$ fibered non-compact surfaces for the CY3 obtained after the integrating out process is complete. In other words, the non-abelian part of the flavor symmetry of the corresponding $5d$ SCFT can be read from the data of the $\P^1$ fibered non-compact surfaces that we have deduced from the data of the $\P^1$ fibered non-compact surfaces of the parent $5d$ KK theory.

The abelian part of the flavor symmetry of the descendant $5d$ SCFT is obtained by noticing that the above integrating out procedure reduces the total rank of the flavor symmetry by one. Thus, the number of abelian factors in the flavor symmetry of the $5d$ SCFT can be deduced from the knowledge of the rank of the non-abelian part of the flavor symmetry of the $5d$ SCFT along with the number of abelian factors in flavor symmetry algebra of the parent $5d$ KK theory.

One can go on in this fashion to integrate out another BPS particle from the above descendant $5d$ SCFT, which again in general destroys the $\P^1$ fibrations of some of the $\P^1$ fibered non-compact surfaces, resulting in a reduction of the non-abelian part of the flavor symmetry. The number of abelian factors can again be computed by using the fact that the integrating out of each BPS particle reduces the total rank of the flavor symmetry algebra by one. Similarly, we can continue to integrate out even more BPS particles to obtain even more $5d$ SCFTs and determine their corresponding flavor symmetry algebras.

The paper is organized as follows:\\
In Section \ref{6dF}, we discuss the flavor symmetry of $6d$ SCFTs, making a key distinction between localized and delocalized flavor symmetries. This prepares the ground for a discussion of flavor symmetry of $5d$ KK theories in Section \ref{5dF}. In Section \ref{GR}, we discuss the coupling of non-compact surfaces corresponding to flavor symmetries of $5d$ KK theories to the compact surfaces describing the extended Coulomb branch of the $5d$ KK theories. In Section \ref{FB}, we illustrate using a simple example how flavor symmetry of $5d$ SCFTs obtained by integrating out BPS particles from $5d$ KK theories can be obtained by using the formalism set up in the previous sections of this paper. Further more complicated examples are discussed in Part 2 \cite{Bhardwaj:2020avz} of this series of papers. We finish this paper with a quick discussion in Section \ref{CFD} on how the method for computing flavor symmetry discussed in this paper compares with another method based on CFDs \cite{Apruzzi:2019vpe,Apruzzi:2019opn,Apruzzi:2019enx}.

Throughout this paper, we use notation and background about geometric constructions and $5d$ KK theories that can be found in Section 5 and Appendix A of \cite{Bhardwaj:2019fzv}. Background and notation about geometric construction of $5d$ $\cN=1$ gauge theories can be found in Section 2 of \cite{Bhardwaj:2019ngx} and Section 3.2 of \cite{Bhardwaj:2020gyu}. Background on flops can be found in \cite{Bhardwaj:2019jtr}.

\section{Flavor symmetry of $6d$ SCFTs}\label{6dF}
The starting point of our analysis of flavor symmetry of $5d$ SCFTs is the flavor symmetry of $6d$ SCFTs\footnote{See \cite{Bertolini:2015bwa,Merkx:2017jey,DelZotto:2018tcj} for prior work on this topic.}. In this paper, we describe $6d$ SCFTs\footnote{All the currently known $6d$ SCFTs are classified in \cite{Heckman:2015bfa,Bhardwaj:2019hhd} (upto discrete gaugings).} in terms of their tensor branch data which can be captured in terms of a graph. Briefly, the structure of the graph captures the Green-Schwarz matrix $\Omega^{ij}$ associated to the $6d$ theory, the algebras on the nodes describe the gauge algebras appearing on the tensor branch of the $6d$ theory, and the edges between the nodes encode the hypermultiplet matter content transforming under the gauge algebras. See Section 2 of \cite{Bhardwaj:2019fzv} for more details on this graphical notation which we will employ throughout this paper.

The flavor symmetry of $6d$ SCFTs can be classified into two types: localized and delocalized. Localized flavor symmetries are associated to a single node or edge in the graph associated to the $6d$ SCFT, and delocalized flavor symmetries are associated to multiple edges and nodes in the graph associated to $6d$ SCFTs. Moreover, as discussed in the introduction, we will only need to focus on the non-abelian part of the flavor symmetry for the purposes of this paper. In what follows, we will often use the term ``flavor symmetry'' even when we are talking about non-abelian part of the flavor symmetry. The precise meaning of the phrase ``flavor symmetry'' should be clear from the context of discussion.

\subsection{Localized flavor symmetries}
Let us consider a localized flavor symmetry associated to a node $i$ in the graph associated to the $6d$ SCFT. If the gauge algebra $\fg_i$ associated to the node is non-trivial, then this flavor symmetry acts on some hypers that are not gauged by any other gauge algebra $\fg_j$ for $j\neq i$. We can represent localized flavor symmetries by extending the graph of the $6d$ SCFT by some extra nodes that we refer to as `flavor nodes'. So, a localized flavor symmetry $\ff_i$ associated to node $i$ is represented by a flavor node $\hat i$ carrying label $[\ff_i]$ and joined to the node $i$ by an edge. The edge can be of different kinds (solid, dashed etc.) depending on the representation formed by the hypers under $\fg_i$ that $\ff_i$ rotates. Moreover, the edge between $\hat i$ and $i$ can carry a label which is specified by the positive integer $-\Omega^{i\hat i}$ appearing in the Green-Schwarz coupling
\be
\int\Omega^{i\hat i} B_i\wedge\text{tr}(F^2_{\hat i})
\ee
where $F_{\hat i}$ denotes the field strength of the background gauge field associated to $\ff_i$. When $-\Omega^{i\hat i}=1$, the edge between $i$ and $\hat i$ carries no label. But when $-\Omega^{i\hat i}>1$, we insert a label in the middle of the edge specifying the value of $-\Omega^{i\hat i}$. We can compute $\Omega^{i\hat i}$ concretely as
\be
\Omega^{i\hat i}=\frac14\text{ind}(R_i)\text{ind}(R_{\hat i})n_i n_{\hat i}
\ee
where $R_i\otimes R_{\hat i}$ is the irrep of $\fg_i\oplus\ff_i$ formed by the hypers; $R_i\otimes R_{\hat i}$ is assumed to be a non-pseudo-real irrep of $\fg_i\oplus\ff_i$; $\text{ind}(R_i),\text{ind}(R_{\hat i})$ denote the index of representation\footnote{This is the index normalized by the index of the fundamental representation of the corresponding algebra.} $R_i, R_{\hat i}$; and $n_i,n_{\hat i}$ are numbers associated to algebra $\fg_i,\ff_i$ according to the Table 1 of \cite{Bhardwaj:2015xxa}. If $R_i\otimes R_{\hat i}$ is a pseudo-real irrep of $\fg_i\oplus\ff_i$, then we have
\be
\Omega^{i\hat i}=\frac18\text{ind}(R_i)\text{ind}(R_{\hat i})n_i n_{\hat i}
\ee

All the possible localized flavor symmetries for non-trivial $\fg_i$ have been collected in Tables \ref{TR1} and \ref{TR2}. In these tables we show the data of the node $i$, the gauge algebra $fg_i$ and the attached localized flavor symmetries $\ff_i$. Let us explain these tables a little: 
\bit
\item For example the first entry
\be
\begin{tikzpicture}
\node (v1) at (-0.5,0.45) {1};
\node at (-0.45,0.9) {$\sp(n)$};
\begin{scope}[shift={(1.7,0)}]
\node (v2) at (-0.5,0.45) {$\left[\so(m)\right]$};
\end{scope}
\draw  (v1) edge (v2);
\end{tikzpicture}
\ee
arises because an $\sp(n)$ node with $\Omega^{ii}=1$ can have $m$ half-hypers not transforming under any other gauge algebra, which result in an $\so(m)$ flavor symmetry. Since the total number of half-hypers at such an $\sp(n)$ node are $4n+16$, we find that $m\le 4n+16$.
\item It is possible to have two different flavor algebras associated to the same gauge node as in the case of
\be
\begin{tikzpicture}
\node (v1) at (-0.5,0.45) {1};
\node at (-0.45,0.9) {$\su(4)$};
\begin{scope}[shift={(1.7,0)}]
\node (v2) at (-0.5,0.45) {$\left[\su(m)\right]$};
\end{scope}
\draw  (v1) edge (v2);
\begin{scope}[shift={(-1.8,0)}]
\node (v3) at (-0.5,0.45) {$\left[\su(2)\right]$};
\end{scope}
\draw [densely dashed] (v3) edge (v1);
\end{tikzpicture}
\ee
where the $\su(4)$ node with $\Omega^{ii}=1$ contains a total of $12$ hypers in fundamental and one hyper in 2-index antisymmetric. The hypers can be further gauged by other neighboring gauge algebras, but the antisymmetric cannot be gauged in a $6d$ SCFT. So, we obtain an $\su(m\le 12)$ flavor symmetry from the fundamentals and an $\su(2)$ flavor symmetry from the antisymmetric. The flavor symmetry arising from the fundamentals is joined by a solid edge and the flavor symmetry arising from the antisymmetric is joined by a dashed edge.
\item Consider a node $\fg_i=\su(2)$ and $\Omega^{ii}=2$. There are a total number of 8 half-hypers transforming in fundamental representation of $\su(2)$. It is known that the maximal flavor symmetry algebra is only $\so(7)$ under which the hypers transform in the spinor representation of $\so(7)$. The spinor representations are always denoted by a dashed edge, as in the case of
\be
\begin{tikzpicture}
\node (v1) at (-0.5,0.45) {2};
\node at (-0.45,0.9) {$\su(2)$};
\begin{scope}[shift={(1.7,0)}]
\node (v2) at (-0.5,0.45) {$\left[\so(7)\right]$};
\end{scope}
\draw [densely dashed] (v1) edge (v2);
\end{tikzpicture}
\ee
It is also possible for a $6d$ SCFT carrying such a node $i$ to have only 6 or 7 out of the 8 half-hypers not transforming under some other neighboring gauge algebra. When we have 7 such half-hypers we have a $\fg_2$ flavor symmetry, and when we have 6 such half-hypers we have an $\su(3)$ flavor symmetry.
\item For $\fg_i=\so(n)$, we join flavor symmetries associated to hypers transforming in spinor representation of $\so(n)$ by a dashed edge, and the flavor symmetries associated to hypers transforming in vector representation of $\so(n)$ by a solid edge. A dotted edge is also used in the case of $\so(n)=\so(8)$ to denote hypers transforming in cospinor representation of $\so(n)=\so(8)$.
\eit

\begin{table}[htbp]
\begin{center}
\begin{tabular}{|c|c|}
\hline
 \raisebox{-.4\height}{\begin{tikzpicture}
\node (v1) at (-0.5,0.45) {1};
\node at (-0.45,0.9) {$\sp(n)$};
\begin{scope}[shift={(1.7,0)}]
\node (v2) at (-0.5,0.45) {$\left[\so(m)\right]$};
\end{scope}
\draw  (v1) edge (v2);
\end{tikzpicture}}&$m\le 4n+16$
\\ \hline
 \raisebox{-.4\height}{\begin{tikzpicture}
\node (v1) at (-0.5,0.45) {1};
\node at (-0.45,0.9) {$\su(3)$};
\begin{scope}[shift={(1.7,0)}]
\node (v2) at (-0.5,0.45) {$\left[\su(m)\right]$};
\end{scope}
\draw  (v1) edge (v2);
\end{tikzpicture}}&$m\le 12$
\\ \hline
\raisebox{-.4\height}{\begin{tikzpicture}
\node (v1) at (-0.5,0.45) {1};
\node at (-0.45,0.9) {$\su(4)$};
\begin{scope}[shift={(1.7,0)}]
\node (v2) at (-0.5,0.45) {$\left[\su(m)\right]$};
\end{scope}
\draw  (v1) edge (v2);
\begin{scope}[shift={(-1.8,0)}]
\node (v3) at (-0.5,0.45) {$\left[\su(2)\right]$};
\end{scope}
\draw [densely dashed] (v3) edge (v1);
\end{tikzpicture}}&$m\le 12$
\\ \hline
\raisebox{-.4\height}{\begin{tikzpicture}
\node (v1) at (-0.5,0.45) {1};
\node at (-0.45,0.9) {$\su(n)$};
\begin{scope}[shift={(1.7,0)}]
\node (v2) at (-0.5,0.45) {$\left[\su(m)\right]$};
\end{scope}
\draw  (v1) edge (v2);
\end{tikzpicture}}&$n\ge 5$; $m\le n+8$
\\ \hline
\raisebox{-.4\height}{\begin{tikzpicture}
\node (v1) at (-0.5,0.45) {1};
\node at (-0.45,0.9) {$\su(\wh n)$};
\begin{scope}[shift={(1.7,0)}]
\node (v2) at (-0.5,0.45) {$\left[\su(m)\right]$};
\end{scope}
\draw  (v1) edge (v2);
\end{tikzpicture}}&$m\le n-8$
\\ \hline
\raisebox{-.4\height}{\begin{tikzpicture}
\node (v1) at (-0.5,0.45) {1};
\node at (-0.45,0.9) {$\su(\tilde 6)$};
\begin{scope}[shift={(1.7,0)}]
\node (v2) at (-0.5,0.45) {$\left[\su(m)\right]$};
\end{scope}
\draw  (v1) edge (v2);
\end{tikzpicture}}&$m\le 15$
\\ \hline
\raisebox{-.4\height}{\begin{tikzpicture}
\node (v1) at (-0.5,0.45) {2};
\node at (-0.45,0.9) {$\su(2)$};
\begin{scope}[shift={(1.7,0)}]
\node (v2) at (-0.5,0.45) {$\left[\so(7)\right]$};
\end{scope}
\draw [densely dashed] (v1) edge (v2);
\end{tikzpicture}}&
\\ \hline
\raisebox{-.4\height}{\begin{tikzpicture}
\node (v1) at (-0.5,0.45) {2};
\node at (-0.45,0.9) {$\su(2)$};
\begin{scope}[shift={(1.7,0)}]
\node (v2) at (-0.5,0.45) {$\left[\fg\right]$};
\end{scope}
\draw (v1) edge (v2);
\end{tikzpicture}}&$\fg=\fg_2,\su(3),\su(2)$
\\ \hline
\raisebox{-.4\height}{\begin{tikzpicture}
\node (v1) at (-0.5,0.45) {2};
\node at (-0.45,0.9) {$\su(n)$};
\begin{scope}[shift={(1.7,0)}]
\node (v2) at (-0.5,0.45) {$\left[\su(m)\right]$};
\end{scope}
\draw  (v1) edge (v2);
\end{tikzpicture}}&$m\le 2n$
\\ \hline
\raisebox{-.4\height}{\begin{tikzpicture}
\node (v1) at (-0.5,0.45) {4};
\node at (-0.45,0.9) {$\so(n)$};
\begin{scope}[shift={(1.7,0)}]
\node (v2) at (-0.5,0.45) {$\left[\sp(m)\right]$};
\end{scope}
\draw  (v1) edge (v2);
\end{tikzpicture}}&$m\le n-8$
\\ \hline
\raisebox{-.4\height}{\begin{tikzpicture}
\node (v1) at (-0.5,0.45) {$k$};
\node at (-0.45,0.9) {$\so(8)$};
\begin{scope}[shift={(1.7,0)}]
\node (v2) at (-0.5,0.45) {$\left[\sp(n)\right]$};
\end{scope}
\draw  (v1) edge (v2);
\begin{scope}[shift={(-1.7,0)}]
\node (v3) at (-0.5,0.45) {$\left[\sp(m)\right]$};
\end{scope}
\draw [densely dashed] (v3) edge (v1);
\begin{scope}[shift={(0,-1.1)}]
\node (v4) at (-0.5,0.45) {$\left[\sp(l)\right]$};
\end{scope}
\draw [densely dotted] (v1) edge (v4);
\end{tikzpicture}}&$1\le k\le 3$; $l,m,n\le 4-k$
\\ \hline
\raisebox{-.4\height}{\begin{tikzpicture}
\node (v1) at (-0.5,0.4) {$k$};
\node at (-0.45,0.85) {$\so(7)$};
\begin{scope}[shift={(1.7,0)}]
\node (v2) at (-0.5,0.4) {$\left[\sp(n)\right]$};
\end{scope}
\draw  (v1) edge (v2);
\begin{scope}[shift={(-1.9,0)}]
\node (v3) at (-0.3,0.4) {$\left[\sp(m)\right]$};
\end{scope}
\draw [densely dashed] (v3) edge (v1);
\end{tikzpicture}}&$1\le k\le 3$; $n\le 3-k$; $m\le 8-2k$
\\ \hline
\raisebox{-.4\height}{\begin{tikzpicture}
\node (v1) at (-0.5,0.4) {$k$};
\node at (-0.45,0.85) {$\so(9)$};
\begin{scope}[shift={(1.7,0)}]
\node (v2) at (-0.5,0.4) {$\left[\sp(n)\right]$};
\end{scope}
\draw  (v1) edge (v2);
\begin{scope}[shift={(-2.5,0)}]
\node (v3) at (-0.5,0.4) {$\left[\sp(4-k)\right]$};
\end{scope}
\node (v4) at (-1.4,0.4) {\tiny{2}};
\draw [densely dashed] (v3) edge (v4);
\draw [densely dashed] (v4) edge (v1);
\end{tikzpicture}}&$1\le k\le 3$; $n\le5-k$
\\ \hline
\raisebox{-.4\height}{\begin{tikzpicture}
\node (v1) at (-0.5,0.4) {$k$};
\node at (-0.45,0.85) {$\so(10)$};
\begin{scope}[shift={(1.7,0)}]
\node (v2) at (-0.5,0.4) {$\left[\sp(n)\right]$};
\end{scope}
\draw  (v1) edge (v2);
\begin{scope}[shift={(-2.5,0)}]
\node (v3) at (-0.5,0.4) {$\left[\su(4-k)\right]$};
\end{scope}
\node (v4) at (-1.4,0.4) {\tiny{4}};
\draw [densely dashed] (v3) edge (v4);
\draw [densely dashed] (v4) edge (v1);
\end{tikzpicture}}&$1\le k\le 3$; $n\le6-k$
\\ \hline
\raisebox{-.4\height}{\begin{tikzpicture}
\node (v1) at (-0.5,0.4) {$k$};
\node at (-0.45,0.85) {$\so(n)$};
\begin{scope}[shift={(1.7,0)}]
\node (v2) at (-0.5,0.4) {$\left[\sp(m)\right]$};
\end{scope}
\draw  (v1) edge (v2);
\begin{scope}[shift={(-2.5,0)}]
\node (v3) at (-0.5,0.4) {$\left[\so(4-k)\right]$};
\end{scope}
\node (v4) at (-1.4,0.4) {\tiny{8}};
\draw [densely dashed] (v3) edge (v4);
\draw [densely dashed] (v4) edge (v1);
\end{tikzpicture}}&$n=11,12$; $1\le k\le 3$; $m\le n-4-k$
\\ \hline
\end{tabular}
\end{center}
\caption{The possible localized flavor symmetries in the context of $6d$ SCFTs. See the text for more explanation. Continued in Table \ref{TR2}.}\label{TR1}
\end{table}

\begin{table}[htbp]
\begin{center}
\begin{tabular}{|c|c|}
\hline
\raisebox{-.4\height}{\begin{tikzpicture}
\node (v1) at (-0.5,0.4) {$k$};
\node at (-0.45,0.85) {$\so(\wh{12})$};
\begin{scope}[shift={(1.7,0)}]
\node (v2) at (-0.5,0.4) {$\left[\sp(m)\right]$};
\end{scope}
\draw  (v1) edge (v2);
\end{tikzpicture}}&$k=1,2$; $m\le 8-k$
\\ \hline
\raisebox{-.4\height}{\begin{tikzpicture}
\node (v1) at (-0.5,0.4) {2};
\node at (-0.45,0.85) {$\so(13)$};
\begin{scope}[shift={(1.7,0)}]
\node (v2) at (-0.5,0.4) {$\left[\sp(m)\right]$};
\end{scope}
\draw  (v1) edge (v2);
\end{tikzpicture}}&$m\le 7$
\\ \hline
\raisebox{-.4\height}{\begin{tikzpicture}
\node (v1) at (-0.5,0.4) {$k$};
\node at (-0.45,0.85) {$\fg_2$};
\begin{scope}[shift={(1.7,0)}]
\node (v2) at (-0.5,0.4) {$\left[\sp(m)\right]$};
\end{scope}
\draw  (v1) edge (v2);
\end{tikzpicture}}&$1\le k\le 3$; $m\le 10-3k$
\\ \hline
\raisebox{-.4\height}{\begin{tikzpicture}
\node (v1) at (-0.5,0.4) {$k$};
\node at (-0.45,0.85) {$\ff_4$};
\begin{scope}[shift={(2.5,0)}]
\node (v2) at (-0.5,0.4) {$\left[\sp(5-k)\right]$};
\end{scope}
\node (v3) at (0.4,0.4) {\tiny{3}};
\draw  (v1) edge (v3);
\draw  (v3) edge (v2);
\end{tikzpicture}}&$1\le k\le5$
\\ \hline
\raisebox{-.4\height}{\begin{tikzpicture}
\node (v1) at (-0.5,0.4) {$k$};
\node at (-0.45,0.85) {$\fe_6$};
\begin{scope}[shift={(2.5,0)}]
\node (v2) at (-0.5,0.4) {$\left[\su(6-k)\right]$};
\end{scope}
\node (v3) at (0.4,0.4) {\tiny{6}};
\draw  (v1) edge (v3);
\draw  (v3) edge (v2);
\end{tikzpicture}}&$1\le k\le6$
\\ \hline
\raisebox{-.4\height}{\begin{tikzpicture}
\node (v1) at (-0.5,0.4) {$k$};
\node at (-0.45,0.85) {$\fe_7$};
\begin{scope}[shift={(2.5,0)}]
\node (v2) at (-0.5,0.4) {$\left[\so(8-k)\right]$};
\end{scope}
\node (v3) at (0.4,0.4) {\tiny{12}};
\draw  (v1) edge (v3);
\draw  (v3) edge (v2);
\end{tikzpicture}}&$1\le k\le8$
\\ \hline
\end{tabular}
\end{center}
\caption{Continuation of Table \ref{TR1}.}\label{TR2}
\end{table}

Now we turn to a discussion of localized flavor symmetries for nodes $i$ with trivial $\fg_i$. First consider the case of $\Omega^{ii}=1$ for which we denote the trivial gauge algebra as $\fg_i=\sp(0)$. In the case of $6d$ SCFTs, all the neighboring nodes $j$ of such a node $i$ have $\Omega^{ij}=-1$ and $\oplus_j\fg_j\subseteq\fe_8$. Then, the flavor symmetry algebra associated to node $i$ is the commutant of $\oplus_j\fg_j$ inside $\fe_8$. If one of the neighbors has $\fg_j=\su(8)$, then there are two possible embeddings into $\fe_8$. One of them has commutant $\su(2)$ and the other has a $\u(1)$ commutant. We differentiate these two cases by attaching a theta angle $\theta$ to the $\sp(0)$, with $\theta=0$ if the commutant is $\su(2)$ and $\theta=\pi$ if the commutant is $\u(1)$. It is possible for some factors of the non-abelian flavor symmetry algebra to have embedding index larger than 1. In such cases, we insert a label denoting the embedding index in the middle of the edge connecting the node $\hat i$ associated to to this factor of flavor symmetry algebra and the node $i$. We do so because the value of $\Omega^{i\hat i}$ coincides with this embedding index when $\fg_i$ is trivial.

Second, consider the case of $\Omega^{ii}=2$ for which we denote the trivial gauge algebra as $\fg_i=\su(1)$. If this node has a neighbor with $\fg_j=\su(1)$ or $\fg_j=\su(2)$, then there is no non-trivial localized flavor symmetry. Only if the node has no neighbor or has a neighbor with $\fg_j=\sp(0)$ do we obtain an $\su(2)$ localized flavor symmetry. 

All the possible localized flavor symmetries for these two types of nodes are collected in Tables \ref{TR3}, \ref{TR4}, \ref{TR5} and \ref{TR6}, where we list all the neighbors of the node $i$. In these cases, we display all the abelian factors in the flavor symmetry explicitly (for which, unlike the non-abelian case, we do not need to define a notion of embedding index) since it is difficult to track the number of abelian factors for non-gauge-theoretic nodes as compared to the case of gauge-theoretic nodes discussed above.

\begin{table}[htbp]
\begin{center}
\begin{tabular}{|c|c|}
\hline
\raisebox{-.4\height}{\begin{tikzpicture}
\node (v1) at (-0.5,0.4) {1};
\node at (-0.45,0.85) {$\sp(0)_0$};
\begin{scope}[shift={(1.5,-0.05)}]
\node (v2) at (-0.5,0.45) {2};
\node at (-0.45,0.9) {$\su(8)$};
\end{scope}
\draw  (v1) edge (v2);
\node (v3) at (-2.3,0.4) {$\left[\su(2)\right]$};
\draw  (v3) edge (v1);
\end{tikzpicture}}&\raisebox{-.4\height}{\begin{tikzpicture}
\node (v1) at (-0.5,0.4) {1};
\node at (-0.45,0.85) {$\sp(0)_\pi$};
\begin{scope}[shift={(1.5,-0.05)}]
\node (v2) at (-0.5,0.45) {2};
\node at (-0.45,0.9) {$\su(8)$};
\end{scope}
\draw  (v1) edge (v2);
\node (v3) at (-2.3,0.4) {$\left[\u(1)\right]$};
\draw  (v3) edge (v1);
\end{tikzpicture}}
\\ \hline
\raisebox{-.4\height}{\begin{tikzpicture}
\node (v1) at (-0.5,0.4) {1};
\node at (-0.45,0.85) {$\sp(0)$};
\begin{scope}[shift={(1.5,-0.05)}]
\node (v2) at (-0.5,0.45) {2};
\node at (-0.45,0.9) {$\su(7)$};
\end{scope}
\draw  (v1) edge (v2);
\node (v3) at (-2.3,0.4) {$\left[\su(2)\right]$};
\draw  (v3) edge (v1);
\node (v4) at (-0.5,-0.9) {$\left[\u(1)\right]$};
\draw  (v1) edge (v4);
\end{tikzpicture}}&\raisebox{-.4\height}{\begin{tikzpicture}
\node (v1) at (-0.5,0.4) {1};
\node at (-0.45,0.85) {$\sp(0)$};
\begin{scope}[shift={(1.5,-0.05)}]
\node (v2) at (-0.5,0.45) {2};
\node at (-0.45,0.9) {$\su(6)$};
\end{scope}
\draw  (v1) edge (v2);
\node (v3) at (-2.3,0.4) {$\left[\su(3)\right]$};
\draw  (v3) edge (v1);
\node (v4) at (-0.5,-0.9) {$\left[\su(2)\right]$};
\draw  (v1) edge (v4);
\end{tikzpicture}}
\\ \hline
\raisebox{-.4\height}{\begin{tikzpicture}
\node (v1) at (-0.5,0.4) {1};
\node at (-0.45,0.85) {$\sp(0)$};
\begin{scope}[shift={(1.5,-0.05)}]
\node (v2) at (-0.5,0.45) {2};
\node at (-0.45,0.9) {$\su(5)$};
\end{scope}
\draw  (v1) edge (v2);
\node (v3) at (-2.3,0.4) {$\left[\su(5)\right]$};
\draw  (v3) edge (v1);
\end{tikzpicture}}&\raisebox{-.4\height}{\begin{tikzpicture}
\node (v1) at (-0.5,0.4) {1};
\node at (-0.45,0.85) {$\sp(0)$};
\begin{scope}[shift={(1.5,-0.05)}]
\node (v2) at (-0.5,0.45) {2};
\node at (-0.45,0.9) {$\su(4)$};
\end{scope}
\draw  (v1) edge (v2);
\node (v3) at (-2.3,0.4) {$\left[\so(10)\right]$};
\draw  (v3) edge (v1);
\end{tikzpicture}}
\\ \hline
\raisebox{-.4\height}{\begin{tikzpicture}
\node (v1) at (-0.5,0.4) {1};
\node at (-0.45,0.85) {$\sp(0)$};
\begin{scope}[shift={(1.5,-0.05)}]
\node (v2) at (-0.5,0.45) {$k$};
\node at (-0.45,0.9) {$\su(3)$};
\end{scope}
\draw  (v1) edge (v2);
\node (v3) at (-2.3,0.4) {$\left[\fe_6\right]$};
\draw  (v3) edge (v1);
\end{tikzpicture}}&\raisebox{-.4\height}{\begin{tikzpicture}
\node (v1) at (-0.5,0.4) {1};
\node at (-0.45,0.85) {$\sp(0)$};
\begin{scope}[shift={(1.5,-0.05)}]
\node (v2) at (-0.5,0.45) {2};
\node at (-0.45,0.9) {$\su(2)$};
\end{scope}
\draw  (v1) edge (v2);
\node (v3) at (-2.3,0.4) {$\left[\fe_7\right]$};
\draw  (v3) edge (v1);
\end{tikzpicture}}
\\ \hline
\raisebox{-.4\height}{\begin{tikzpicture}
\node (v1) at (-0.5,0.4) {1};
\node at (-0.45,0.85) {$\sp(0)$};
\begin{scope}[shift={(1.5,-0.05)}]
\node (v2) at (-0.5,0.45) {2};
\node at (-0.45,0.9) {$\su(1)$};
\end{scope}
\draw  (v1) edge (v2);
\node (v3) at (-2.3,0.4) {$\left[\fe_8\right]$};
\draw  (v3) edge (v1);
\end{tikzpicture}}&\raisebox{-.4\height}{\begin{tikzpicture}
\node (v1) at (-0.5,0.4) {1};
\node at (-0.45,0.85) {$\sp(0)$};
\begin{scope}[shift={(1.5,-0.05)}]
\node (v2) at (-0.5,0.45) {$k$};
\node at (-0.45,0.9) {$\so(14)$};
\end{scope}
\draw  (v1) edge (v2);
\node (v3) at (-2.3,0.4) {$\left[\u(1)\right]$};
\draw  (v3) edge (v1);
\end{tikzpicture}}
\\ \hline
\raisebox{-.4\height}{\begin{tikzpicture}
\node (v1) at (-0.5,0.4) {1};
\node at (-0.45,0.85) {$\sp(0)$};
\begin{scope}[shift={(1.5,-0.05)}]
\node (v2) at (-0.5,0.45) {$k$};
\node at (-0.45,0.9) {$\so(13)$};
\end{scope}
\draw  (v1) edge (v2);
\node (v3) at (-2.6,0.4) {$\left[\su(2)\right]$};
\node (v4) at (-1.3,0.4) {\tiny{2}};
\draw  (v3) edge (v4);
\draw  (v4) edge (v1);
\end{tikzpicture}}&\raisebox{-.4\height}{\begin{tikzpicture}
\node (v1) at (-0.5,0.4) {1};
\node at (-0.45,0.85) {$\sp(0)$};
\begin{scope}[shift={(1.5,-0.05)}]
\node (v2) at (-0.5,0.45) {$k$};
\node at (-0.45,0.9) {$\so(12)$};
\end{scope}
\draw  (v1) edge (v2);
\node (v3) at (-2.3,0.4) {$\left[\su(2)\right]$};
\draw  (v3) edge (v1);
\node (v4) at (-0.5,-0.9) {$\left[\su(2)\right]$};
\draw  (v1) edge (v4);
\end{tikzpicture}}
\\ \hline
\raisebox{-.4\height}{\begin{tikzpicture}
\node (v1) at (-0.5,0.4) {1};
\node at (-0.45,0.85) {$\sp(0)$};
\begin{scope}[shift={(1.5,-0.05)}]
\node (v2) at (-0.5,0.45) {$k$};
\node at (-0.45,0.9) {$\so(11)$};
\end{scope}
\draw  (v1) edge (v2);
\node (v3) at (-2.3,0.4) {$\left[\sp(2)\right]$};
\draw  (v3) edge (v1);
\end{tikzpicture}}&\raisebox{-.4\height}{\begin{tikzpicture}
\node (v1) at (-0.5,0.4) {1};
\node at (-0.45,0.85) {$\sp(0)$};
\begin{scope}[shift={(1.5,-0.05)}]
\node (v2) at (-0.5,0.45) {$k$};
\node at (-0.45,0.9) {$\so(10)$};
\end{scope}
\draw  (v1) edge (v2);
\node (v3) at (-2.3,0.4) {$\left[\su(4)\right]$};
\draw  (v3) edge (v1);
\end{tikzpicture}}
\\ \hline
\raisebox{-.4\height}{\begin{tikzpicture}
\node (v1) at (-0.5,0.4) {1};
\node at (-0.45,0.85) {$\sp(0)$};
\begin{scope}[shift={(1.5,-0.05)}]
\node (v2) at (-0.5,0.45) {$k$};
\node at (-0.45,0.9) {$\so(9)$};
\end{scope}
\draw  (v1) edge (v2);
\node (v3) at (-2.3,0.4) {$\left[\so(7)\right]$};
\draw  (v3) edge (v1);
\end{tikzpicture}}&\raisebox{-.4\height}{\begin{tikzpicture}
\node (v1) at (-0.5,0.4) {1};
\node at (-0.45,0.85) {$\sp(0)$};
\begin{scope}[shift={(1.5,-0.05)}]
\node (v2) at (-0.5,0.45) {$k$};
\node at (-0.45,0.9) {$\so(8)$};
\end{scope}
\draw  (v1) edge (v2);
\node (v3) at (-2.3,0.4) {$\left[\so(8)\right]$};
\draw  (v3) edge (v1);
\end{tikzpicture}}
\\ \hline
\raisebox{-.4\height}{\begin{tikzpicture}
\node (v1) at (-0.5,0.4) {1};
\node at (-0.45,0.85) {$\sp(0)$};
\begin{scope}[shift={(1.5,-0.05)}]
\node (v2) at (-0.5,0.45) {$k$};
\node at (-0.45,0.9) {$\so(7)$};
\end{scope}
\draw  (v1) edge (v2);
\node (v3) at (-2.3,0.4) {$\left[\so(9)\right]$};
\draw  (v3) edge (v1);
\end{tikzpicture}}&\raisebox{-.4\height}{\begin{tikzpicture}
\node (v1) at (-0.5,0.4) {1};
\node at (-0.45,0.85) {$\sp(0)$};
\begin{scope}[shift={(1.5,-0.05)}]
\node (v2) at (-0.5,0.45) {$k$};
\node at (-0.45,0.9) {$\fg_2$};
\end{scope}
\draw  (v1) edge (v2);
\node (v3) at (-2.3,0.4) {$\left[\ff_4\right]$};
\draw  (v3) edge (v1);
\end{tikzpicture}}
\\ \hline
\raisebox{-.4\height}{\begin{tikzpicture}
\node (v1) at (-0.5,0.4) {1};
\node at (-0.45,0.85) {$\sp(0)$};
\begin{scope}[shift={(1.5,-0.05)}]
\node (v2) at (-0.5,0.45) {$k$};
\node at (-0.45,0.9) {$\ff_4$};
\end{scope}
\draw  (v1) edge (v2);
\node (v3) at (-2.3,0.4) {$\left[\fg_2\right]$};
\draw  (v3) edge (v1);
\end{tikzpicture}}&\raisebox{-.4\height}{\begin{tikzpicture}
\node (v1) at (-0.5,0.4) {1};
\node at (-0.45,0.85) {$\sp(0)$};
\begin{scope}[shift={(1.5,-0.05)}]
\node (v2) at (-0.5,0.45) {$k$};
\node at (-0.45,0.9) {$\fe_6$};
\end{scope}
\draw  (v1) edge (v2);
\node (v3) at (-2.3,0.4) {$\left[\su(3)\right]$};
\draw  (v3) edge (v1);
\end{tikzpicture}}
\\ \hline
\raisebox{-.4\height}{\begin{tikzpicture}
\node (v1) at (-0.5,0.4) {1};
\node at (-0.45,0.85) {$\sp(0)$};
\begin{scope}[shift={(1.5,-0.05)}]
\node (v2) at (-0.5,0.45) {$k$};
\node at (-0.45,0.9) {$\fe_7$};
\end{scope}
\draw  (v1) edge (v2);
\node (v3) at (-2.3,0.4) {$\left[\su(2)\right]$};
\draw  (v3) edge (v1);
\end{tikzpicture}}&\raisebox{-.4\height}{\begin{tikzpicture}
\node (v1) at (-0.5,0.4) {2};
\node at (-0.5,0.85) {$\su(1)$};
\begin{scope}[shift={(1.5,-0.05)}]
\node (v2) at (-0.5,0.45) {1};
\node at (-0.45,0.9) {$\sp(0)$};
\end{scope}
\draw  (v1) edge (v2);
\node (v3) at (-2.3,0.4) {$\left[\su(2)\right]$};
\draw  (v3) edge (v1);
\end{tikzpicture}}
\\ \hline
\end{tabular}
\end{center}
\caption{Possible localized flavor symmetries for nodes with a trivial gauge algebra. Continued in Table \ref{TR4}.}\label{TR3}
\end{table}

\begin{table}[htbp]
\begin{center}
\begin{tabular}{|c|c|}
\hline
\raisebox{-.4\height}{\begin{tikzpicture}
\node (v1) at (-0.5,0.4) {1};
\node (v7) at (-0.5,0.85) {$\sp(0)$};
\begin{scope}[shift={(1.5,-0.05)}]
\node (v2) at (-0.5,0.45) {$2$};
\node at (-0.45,0.9) {$\su(6)$};
\end{scope}
\draw  (v1) edge (v2);
\node (v3) at (-2.1,0.4) {$3$};
\draw  (v3) edge (v1);
\node (v4) at (-0.5,-0.9) {$\left[\su(2)\right]$};
\draw  (v1) edge (v4);
\node at (-2.1,0.85) {$\su(3)$};
\end{tikzpicture}}&\raisebox{-.4\height}{\begin{tikzpicture}
\node (v1) at (-0.5,0.4) {1};
\node (v7) at (-0.5,0.85) {$\sp(0)$};
\begin{scope}[shift={(1.5,-0.05)}]
\node (v2) at (-0.5,0.45) {$2$};
\node at (-0.45,0.9) {$\su(5)$};
\end{scope}
\draw  (v1) edge (v2);
\node (v3) at (-2.1,0.4) {$3$};
\draw  (v3) edge (v1);
\node (v4) at (-0.5,-0.9) {$\left[\su(2)\right]$};
\draw  (v1) edge (v4);
\node at (-2.1,0.85) {$\su(3)$};
\node (v5) at (-0.5,2.2) {$\left[\u(1)\right]$};
\draw  (v5) edge (v7);
\end{tikzpicture}}
\\ \hline
\raisebox{-.4\height}{\begin{tikzpicture}
\node (v1) at (-0.5,0.4) {1};
\node (v7) at (-0.5,0.85) {$\sp(0)$};
\begin{scope}[shift={(1.5,-0.05)}]
\node (v2) at (-0.5,0.45) {$2$};
\node at (-0.45,0.9) {$\su(4)$};
\end{scope}
\draw  (v1) edge (v2);
\node (v3) at (-2.1,0.4) {$k$};
\draw  (v3) edge (v1);
\node (v4) at (-0.5,-0.9) {$\left[\u(1)\right]$};
\draw  (v1) edge (v4);
\node at (-2.1,0.85) {$\so(8)$};
\end{tikzpicture}}&\raisebox{-.4\height}{\begin{tikzpicture}
\node (v1) at (-0.5,0.4) {1};
\node (v7) at (-0.5,0.85) {$\sp(0)$};
\begin{scope}[shift={(1.5,-0.05)}]
\node (v2) at (-0.5,0.45) {$3$};
\node at (-0.45,0.9) {$\so(7)$};
\end{scope}
\draw  (v1) edge (v2);
\node (v3) at (-2.1,0.4) {$2$};
\draw  (v3) edge (v1);
\node (v4) at (-0.5,-1.1) {$\left[\su(2)\right]$};
\node at (-2.1,0.85) {$\su(4)$};
\node (v5) at (-0.5,-0.3) {\tiny{2}};
\draw  (v1) edge (v5);
\draw  (v5) edge (v4);
\end{tikzpicture}}
\\ \hline
\raisebox{-.4\height}{\begin{tikzpicture}
\node (v1) at (-0.5,0.4) {1};
\node (v7) at (-0.5,0.85) {$\sp(0)$};
\begin{scope}[shift={(1.5,-0.05)}]
\node (v2) at (-0.5,0.45) {$2$};
\node at (-0.45,0.9) {$\su(4)$};
\end{scope}
\draw  (v1) edge (v2);
\node (v3) at (-2.1,0.4) {3};
\draw  (v3) edge (v1);
\node (v4) at (-0.5,-0.9) {$\left[\u(1)\right]$};
\draw  (v1) edge (v4);
\node at (-2.1,0.85) {$\su(3)$};
\node (v5) at (-1.5,2) {$\left[\su(2)\right]$};
\node (v6) at (0.6,2) {$\left[\su(2)\right]$};
\draw  (v5) edge (v7);
\draw  (v6) edge (v7);
\end{tikzpicture}}&\raisebox{-.4\height}{\begin{tikzpicture}
\node (v1) at (-0.5,0.4) {1};
\node (v7) at (-0.5,0.85) {$\sp(0)$};
\begin{scope}[shift={(1.5,-0.05)}]
\node (v2) at (-0.5,0.45) {$l$};
\node at (-0.45,0.9) {$\fg_2$};
\end{scope}
\draw  (v1) edge (v2);
\node (v3) at (-2.1,0.4) {$k$};
\draw  (v3) edge (v1);
\node (v4) at (-0.5,-1.1) {$\left[\su(3)\right]$};
\node at (-2.1,0.85) {$\su(3)$};
\node (v5) at (-0.5,-0.3) {\tiny{2}};
\draw  (v1) edge (v5);
\draw  (v5) edge (v4);
\end{tikzpicture}}
\\ \hline
\raisebox{-.4\height}{\begin{tikzpicture}
\node (v1) at (-0.5,0.4) {1};
\node at (-0.45,0.85) {$\sp(0)$};
\begin{scope}[shift={(1.5,-0.05)}]
\node (v2) at (-0.5,0.45) {$k$};
\node at (-0.45,0.9) {$\so(10)$};
\end{scope}
\draw  (v1) edge (v2);
\node (v3) at (-2.1,0.4) {$l$};
\draw  (v3) edge (v1);
\node (v4) at (-0.5,-0.9) {$\left[\u(1)\right]$};
\draw  (v1) edge (v4);
\node at (-2.1,0.85) {$\su(3)$};
\end{tikzpicture}}&\raisebox{-.4\height}{\begin{tikzpicture}
\node (v1) at (-0.5,0.4) {1};
\node at (-0.45,0.85) {$\sp(0)$};
\begin{scope}[shift={(1.5,-0.05)}]
\node (v2) at (-0.5,0.45) {$k$};
\node at (-0.45,0.9) {$\so(9)$};
\end{scope}
\draw  (v1) edge (v2);
\node (v3) at (-2.1,0.4) {$l$};
\draw  (v3) edge (v1);
\node (v4) at (-0.5,-0.9) {$\left[\u(1)\right]$};
\draw  (v1) edge (v4);
\node at (-2.1,0.85) {$\su(3)$};
\end{tikzpicture}}
\\ \hline
\raisebox{-.4\height}{\begin{tikzpicture}
\node (v1) at (-0.5,0.4) {1};
\node (v7) at (-0.5,0.85) {$\sp(0)$};
\begin{scope}[shift={(1.5,-0.05)}]
\node (v2) at (-0.5,0.45) {$k$};
\node at (-0.45,0.9) {$\so(8)$};
\end{scope}
\draw  (v1) edge (v2);
\node (v3) at (-2.1,0.4) {$l$};
\draw  (v3) edge (v1);
\node (v4) at (-0.5,-0.9) {$\left[\u(1)\right]$};
\draw  (v1) edge (v4);
\node at (-2.1,0.85) {$\su(3)$};
\node (v5) at (-0.5,2.2) {$\left[\u(1)\right]$};
\draw  (v5) edge (v7);
\end{tikzpicture}}&\raisebox{-.4\height}{\begin{tikzpicture}
\node (v1) at (-0.5,0.4) {1};
\node (v7) at (-0.5,0.85) {$\sp(0)$};
\begin{scope}[shift={(1.5,-0.05)}]
\node (v2) at (-0.5,0.45) {$k$};
\node at (-0.45,0.9) {$\so(7)$};
\end{scope}
\draw  (v1) edge (v2);
\node (v3) at (-2.1,0.4) {$l$};
\draw  (v3) edge (v1);
\node (v4) at (-0.5,-0.9) {$\left[\u(1)\right]$};
\draw  (v1) edge (v4);
\node at (-2.1,0.85) {$\su(3)$};
\node (v5) at (-0.5,2.6) {$\left[\su(2)\right]$};
\node (v6) at (-0.5,1.7) {\tiny{2}};
\draw  (v5) edge (v6);
\draw  (v6) edge (v7);
\end{tikzpicture}}
\\ \hline
\end{tabular}
\end{center}
\caption{Possible localized flavor symmetries for nodes with a trivial gauge algebra continued from Table \ref{TR3}. Continued further in Table \ref{TR5}.}\label{TR4}
\end{table}

\begin{table}[htbp]
\begin{center}
\begin{tabular}{|c|c|}
\hline
\raisebox{-.4\height}{\begin{tikzpicture}
\node (v1) at (-0.5,0.4) {1};
\node (v7) at (-0.5,0.85) {$\sp(0)$};
\begin{scope}[shift={(1.5,-0.05)}]
\node (v2) at (-0.5,0.45) {$k$};
\node at (-0.45,0.9) {$\su(3)$};
\end{scope}
\draw  (v1) edge (v2);
\node (v3) at (-2.1,0.4) {$3$};
\draw  (v3) edge (v1);
\node (v4) at (-0.5,-0.9) {$\left[\su(3)\right]$};
\draw  (v1) edge (v4);
\node at (-2.1,0.85) {$\su(3)$};
\node (v5) at (-0.5,2.2) {$\left[\su(3)\right]$};
\draw  (v5) edge (v7);
\end{tikzpicture}}&\raisebox{-.4\height}{\begin{tikzpicture}
\node (v1) at (-0.5,0.4) {1};
\node (v7) at (-0.5,0.85) {$\sp(0)$};
\begin{scope}[shift={(1.5,-0.05)}]
\node (v2) at (-0.5,0.45) {$2$};
\node at (-0.45,0.9) {$\su(2)$};
\end{scope}
\draw  (v1) edge (v2);
\node (v3) at (-2.1,0.4) {$3$};
\draw  (v3) edge (v1);
\node (v4) at (-0.5,-0.9) {$\left[\su(6)\right]$};
\draw  (v1) edge (v4);
\node at (-2.1,0.85) {$\su(3)$};
\end{tikzpicture}}
\\ \hline
\raisebox{-.4\height}{\begin{tikzpicture}
\node (v1) at (-0.5,0.4) {1};
\node at (-0.45,0.85) {$\sp(0)$};
\begin{scope}[shift={(1.5,-0.05)}]
\node (v2) at (-0.5,0.45) {$k$};
\node at (-0.45,0.9) {$\fe_6$};
\end{scope}
\draw  (v1) edge (v2);
\node (v3) at (-2.1,0.4) {$2$};
\draw  (v3) edge (v1);
\node (v4) at (-0.5,-0.9) {$\left[\u(1)\right]$};
\draw  (v1) edge (v4);
\node at (-2.1,0.85) {$\su(2)$};
\end{tikzpicture}}&\raisebox{-.4\height}{\begin{tikzpicture}
\node (v1) at (-0.5,0.4) {1};
\node (v7) at (-0.5,0.85) {$\sp(0)$};
\begin{scope}[shift={(1.5,-0.05)}]
\node (v2) at (-0.5,0.45) {$k$};
\node at (-0.45,0.9) {$\ff_4$};
\end{scope}
\draw  (v1) edge (v2);
\node (v3) at (-2.1,0.4) {2};
\draw  (v3) edge (v1);
\node (v4) at (-0.5,-1.3) {$\left[\su(2)\right]$};
\node at (-2.1,0.85) {$\su(2)$};
\node (v5) at (-0.5,-0.4) {\tiny{3}};
\draw  (v1) edge (v5);
\draw  (v5) edge (v4);
\end{tikzpicture}}
\\ \hline
\raisebox{-.4\height}{\begin{tikzpicture}
\node (v1) at (-0.5,0.4) {1};
\node (v7) at (-0.5,0.85) {$\sp(0)$};
\begin{scope}[shift={(1.5,-0.05)}]
\node (v2) at (-0.5,0.45) {$k$};
\node at (-0.45,0.9) {$\fg_2$};
\end{scope}
\draw  (v1) edge (v2);
\node (v3) at (-2.1,0.4) {2};
\draw  (v3) edge (v1);
\node (v4) at (-0.5,-0.9) {$\left[\sp(3)\right]$};
\draw  (v1) edge (v4);
\node at (-2.1,0.85) {$\su(2)$};
\end{tikzpicture}}&\raisebox{-.4\height}{\begin{tikzpicture}
\node (v1) at (-0.5,0.4) {1};
\node at (-0.45,0.85) {$\sp(0)$};
\begin{scope}[shift={(1.5,-0.05)}]
\node (v2) at (-0.5,0.45) {$k$};
\node at (-0.45,0.9) {$\so(12)$};
\end{scope}
\draw  (v1) edge (v2);
\node (v3) at (-2.1,0.4) {2};
\draw  (v3) edge (v1);
\node (v4) at (-0.5,-0.9) {$\left[\su(2)\right]$};
\draw  (v1) edge (v4);
\node at (-2.1,0.85) {$\su(2)$};
\end{tikzpicture}}
\\ \hline
\raisebox{-.4\height}{\begin{tikzpicture}
\node (v1) at (-0.5,0.4) {1};
\node at (-0.45,0.85) {$\sp(0)$};
\begin{scope}[shift={(1.5,-0.05)}]
\node (v2) at (-0.5,0.45) {$k$};
\node at (-0.45,0.9) {$\so(11)$};
\end{scope}
\draw  (v1) edge (v2);
\node (v3) at (-2.1,0.4) {2};
\draw  (v3) edge (v1);
\node (v4) at (-0.5,-0.9) {$\left[\su(2)\right]$};
\draw  (v1) edge (v4);
\node at (-2.1,0.85) {$\su(2)$};
\end{tikzpicture}}&\raisebox{-.4\height}{\begin{tikzpicture}
\node (v1) at (-0.5,0.4) {1};
\node (v7) at (-0.5,0.85) {$\sp(0)$};
\begin{scope}[shift={(1.5,-0.05)}]
\node (v2) at (-0.5,0.45) {$k$};
\node at (-0.45,0.9) {$\so(10)$};
\end{scope}
\draw  (v1) edge (v2);
\node (v3) at (-2.1,0.4) {$2$};
\draw  (v3) edge (v1);
\node (v4) at (-0.5,-0.9) {$\left[\u(1)\right]$};
\draw  (v1) edge (v4);
\node at (-2.1,0.85) {$\su(2)$};
\node (v5) at (-0.5,2.2) {$\left[\su(2)\right]$};
\draw  (v5) edge (v7);
\end{tikzpicture}}
\\ \hline
\raisebox{-.4\height}{\begin{tikzpicture}
\node (v1) at (-0.5,0.4) {1};
\node (v7) at (-0.5,0.85) {$\sp(0)$};
\begin{scope}[shift={(1.5,-0.05)}]
\node (v2) at (-0.5,0.45) {$k$};
\node at (-0.45,0.9) {$\so(9)$};
\end{scope}
\draw  (v1) edge (v2);
\node (v3) at (-2.1,0.4) {2};
\draw  (v3) edge (v1);
\node (v4) at (-0.5,-0.9) {$\left[\su(2)\right]$};
\draw  (v1) edge (v4);
\node at (-2.1,0.85) {$\su(2)$};
\node (v5) at (-0.5,2.6) {$\left[\su(2)\right]$};
\node (v6) at (-0.5,1.7) {\tiny{2}};
\draw  (v5) edge (v6);
\draw  (v6) edge (v7);
\end{tikzpicture}}&\raisebox{-.4\height}{\begin{tikzpicture}
\node (v1) at (-0.5,0.4) {1};
\node (v7) at (-0.5,0.85) {$\sp(0)$};
\begin{scope}[shift={(1.5,-0.05)}]
\node (v2) at (-0.5,0.45) {$k$};
\node at (-0.45,0.9) {$\so(8)$};
\end{scope}
\draw  (v1) edge (v2);
\node (v3) at (-2.1,0.4) {2};
\draw  (v3) edge (v1);
\node (v4) at (-0.5,-0.9) {$\left[\su(2)\right]$};
\draw  (v1) edge (v4);
\node at (-2.1,0.85) {$\su(2)$};
\node (v5) at (-1.5,2) {$\left[\su(2)\right]$};
\node (v6) at (0.6,2) {$\left[\su(2)\right]$};
\draw  (v5) edge (v7);
\draw  (v6) edge (v7);
\end{tikzpicture}}
\\ \hline
\end{tabular}
\end{center}
\caption{Possible localized flavor symmetries for nodes with a trivial gauge algebra continued from Table \ref{TR4}. Continued in Table \ref{TR6}.}\label{TR5}
\end{table}

\begin{table}[htbp]
\begin{center}
\begin{tabular}{|c|c|}
\hline
\raisebox{-.4\height}{\begin{tikzpicture}
\node (v1) at (-0.5,0.4) {1};
\node (v7) at (-0.5,0.85) {$\sp(0)$};
\begin{scope}[shift={(1.5,-0.05)}]
\node (v2) at (-0.5,0.45) {$3$};
\node at (-0.45,0.9) {$\so(7)$};
\end{scope}
\draw  (v1) edge (v2);
\node (v3) at (-2.1,0.4) {2};
\draw  (v3) edge (v1);
\node (v4) at (-0.5,-0.9) {$\left[\su(2)\right]$};
\draw  (v1) edge (v4);
\node at (-2.1,0.85) {$\su(2)$};
\node (v5) at (-0.5,2.2) {$\left[\sp(2)\right]$};
\draw  (v5) edge (v7);
\end{tikzpicture}}&\raisebox{-.4\height}{\begin{tikzpicture}
\node (v1) at (-0.5,0.4) {1};
\node (v7) at (-0.5,0.85) {$\sp(0)$};
\begin{scope}[shift={(1.5,-0.05)}]
\node (v2) at (-0.5,0.45) {$k$};
\node at (-0.45,0.9) {$\so(7)$};
\end{scope}
\draw  (v1) edge (v2);
\node (v3) at (-2.1,0.4) {$3$};
\draw  (v3) edge (v1);
\node (v4) at (-0.5,-0.9) {$\left[\u(1)\right]$};
\draw  (v1) edge (v4);
\node at (-2.1,0.85) {$\so(7)$};
\end{tikzpicture}}
\\ \hline
\raisebox{-.4\height}{\begin{tikzpicture}
\node (v1) at (-0.5,0.4) {1};
\node (v7) at (-0.5,0.85) {$\sp(0)$};
\begin{scope}[shift={(1.5,-0.05)}]
\node (v2) at (-0.5,0.45) {$k$};
\node at (-0.45,0.9) {$\fg_2$};
\end{scope}
\draw  (v1) edge (v2);
\node (v3) at (-2.1,0.4) {$l$};
\draw  (v3) edge (v1);
\node (v4) at (-0.5,-0.9) {$\left[\u(1)\right]$};
\draw  (v1) edge (v4);
\node at (-2.1,0.85) {$\so(7)$};
\end{tikzpicture}}&\raisebox{-.4\height}{\begin{tikzpicture}
\node (v1) at (-0.5,0.4) {1};
\node (v7) at (-0.5,0.85) {$\sp(0)$};
\begin{scope}[shift={(1.5,-0.05)}]
\node (v2) at (-0.5,0.45) {$3$};
\node at (-0.45,0.9) {$\fg_2$};
\end{scope}
\draw  (v1) edge (v2);
\node (v3) at (-2.1,0.4) {$2$};
\draw  (v3) edge (v1);
\node (v4) at (-0.5,-1.1) {$\left[\su(2)\right]$};
\node at (-2.1,0.85) {$\su(4)$};
\node (v5) at (-0.5,-0.3) {\tiny{2}};
\draw  (v1) edge (v5);
\draw  (v5) edge (v4);
\end{tikzpicture}}
\\ \hline
\raisebox{-.4\height}{\begin{tikzpicture}
\node (v1) at (-0.5,0.4) {1};
\node (v7) at (-0.5,0.85) {$\sp(0)$};
\begin{scope}[shift={(1.5,-0.05)}]
\node (v2) at (-0.5,0.45) {$3$};
\node at (-0.45,0.9) {$\fg_2$};
\end{scope}
\draw  (v1) edge (v2);
\node (v3) at (-2.1,0.4) {$k$};
\draw  (v3) edge (v1);
\node (v4) at (-0.5,-1) {$\left[\u(1)\right]$};
\node at (-2.1,0.85) {$\fg_2$};
\draw  (v1) edge (v4);
\end{tikzpicture}}&\raisebox{-.4\height}{}
\\ \hline
\end{tabular}
\end{center}
\caption{Possible localized flavor symmetries for nodes with a trivial gauge algebra continued from Table \ref{TR5}.}\label{TR6}
\end{table}

\subsection{Delocalized flavor symmetries}\label{delocal6d}
The reason for the delocalization of flavor symmetries has to do with nodes of type
\be\label{dnode}
\begin{tikzpicture}
\node (v1) at (-0.5,0.45) {$k$};
\node at (-0.45,0.9) {$\su(n)$};
\end{tikzpicture}
\ee
except for the $n=2, k=1$ case. For example for $k=2$ and $n\ge3$, one would naively expect a $\u(2n)$ flavor symmetry associated to this node. However, the diagonal $\u(1)$ inside $\u(2n)$ is anomalous at the quantum level. Thus, the above node is associated to an $\su(2n)$ flavor symmetry only which has rank one less than the naive $\u(2n)$ flavor symmetry. For $n=2$ and $k=2$, something similar occurs. Since the fundamental representation is pseudo-real, naively one would expect an $\so(8)$ flavor symmetry. However, the true flavor symmetry is only $\so(7)$ \cite{Heckman:2015bfa,Ohmori:2015pia} which again has rank one less as compared to the rank of the naively expected flavor symmetry. It should be noted that for $n=2$ and $k=1$, one does not find such a reduction in the rank of flavor symmetry.

Now consider a chain of the form
\be
\begin{tikzpicture} [scale=1.9]
\node (v1) at (-1.5,0.8) {2};
\node at (-1.5,1.1) {$\su(n_k)$};
\node (v3) at (-2.9,0.8) {2};
\node at (-2.9,1.1) {$\su(n_2)$};
\node (v2) at (-2.2,0.8) {$\cdots$};
\draw  (v2) edge (v3);
\draw  (v2) edge (v1);
\node (v4) at (-3.8,0.8) {2};
\node at (-3.8,1.1) {$\su(n_1)$};
\draw  (v4) edge (v3);
\end{tikzpicture}
\ee
where all $n_i\ge3$. Then there are various $\u(1)$ factors in the naive flavor symmetry. Each node $i$ carries $m_i$ hypers in fundamental of $\su(n_i)$ which are not charged under any $\su(n_j)$ for $j\neq i$. If $m_i\ge1$, then there is a diagonal $u(1)_i$ rotating all the $m_i$ hypers for each $i$. Moreover, the edge between any two consecutive nodes $i$ and $i+1$ denotes the presence of a hyper transforming in bifundamental of $\su(n_i)\oplus\su(n_{i+1})$. Then there is $\u(1)$ symmetry rotating each such bifundamental. $k$ linear combinations of all these $\u(1)$s are anomalous and thus the true flavor symmetry has rank $k$ less than the naive flavor symmetry. Moreover, the remaining $\u(1)$s are linear combinations of all the above mentioned localized $\u(1)$s, thus the remaining $\u(1)$s are ``delocalized''. We can say that each $\su(n_i)$ is responsible for reducing the rank of flavor symmetry by 1, thus leading to a difference of $k$ in the ranks of the true and the naive flavor symmetries.

The case of all $n_i$ equal is helpful to understand the delocalization:
\be\label{n>3}
\begin{tikzpicture} [scale=1.9]
\node (v1) at (-1.5,0.8) {2};
\node at (-1.5,1.1) {$\su(n)$};
\node (v3) at (-2.9,0.8) {2};
\node at (-2.9,1.1) {$\su(n)$};
\node (v2) at (-2.2,0.8) {$\cdots$};
\draw  (v2) edge (v3);
\draw  (v2) edge (v1);
\node (v4) at (-3.8,0.8) {2};
\node at (-3.8,1.1) {$\su(n)$};
\draw  (v4) edge (v3);
\end{tikzpicture}
\ee
In this case, along with bifundamentals between adjacent nodes, we have $n$ fundamentals transforming under the leftmost $\su(n)$ and $n$ 
fundamentals transforming under the rightmost $\su(n)$. Thus, we have a total of $k+1$ $\u(1)$ symmetry naively. Removing $k$ of these leaves only a single $\u(1)$ symmetry which is a linear combination of all the above $k+1$ $\u(1)$s, thus the remaining $\u(1)$ is not localized around any subset of nodes in (\ref{n>3}).

Now, consider a limit $n=2$ of (\ref{n>3})
\be
\begin{tikzpicture} [scale=1.9]
\node (v1) at (-1.5,0.8) {2};
\node at (-1.5,1.1) {$\su(2)$};
\node (v3) at (-2.9,0.8) {2};
\node at (-2.9,1.1) {$\su(2)$};
\node (v2) at (-2.2,0.8) {$\cdots$};
\draw  (v2) edge (v3);
\draw  (v2) edge (v1);
\node (v4) at (-3.8,0.8) {2};
\node at (-3.8,1.1) {$\su(2)$};
\draw  (v4) edge (v3);
\end{tikzpicture}
\ee
In this case, each bifundamental is a strictly-real representation and hence is rotated by an $\su(2)$ symmetry. The leftmost node carries an extra 2 fundamental hypers rotated by an $\so(4)=\su(2)\oplus\su(2)$ symmetry. Similarly, the rightmost node also carries an extra 2 fundamental hypers rotated by an $\so(4)=\su(2)\oplus\su(2)$ symmetry. In total the naive flavor symmetry is comprised of $k+3$ $\su(2)$s. We claim that just as in the $n>2$ case, each $\su(2)$ gauge node is responsible for the reduction of rank by 1. Thus the true flavor symmetry should comprise of 3 delocalized $\su(2)$s. For a general $6d$ SCFT, we propose that each node of the form
\be
\begin{tikzpicture}
\node (v1) at (-0.5,0.45) {2};
\node at (-0.45,0.9) {$\su(2)$};
\end{tikzpicture}
\ee
is also responsible for the reduction of flavor symmetry rank by 1.

Let us test this proposal. Consider the $6d$ SCFT
\be
\begin{tikzpicture} [scale=1.9]
\node (v1) at (-1.5,0.8) {2};
\node at (-1.5,1.1) {$\su(2)$};
\node (v3) at (-2.9,0.8) {2};
\node at (-2.9,1.1) {$\su(2)$};
\node (v2) at (-2.2,0.8) {$\cdots$};
\draw  (v2) edge (v3);
\draw  (v2) edge (v1);
\node (v4) at (-3.8,0.8) {1};
\node at (-3.8,1.1) {$\sp(0)$};
\draw  (v4) edge (v3);
\begin{scope}[shift={(0,0.45)}]
\node at (-2.2,-0.15) {$k$};
\draw (-3.1,0.15) .. controls (-3.1,0.1) and (-3.1,0.05) .. (-3,0.05);
\draw (-3,0.05) -- (-2.3,0.05);
\draw (-2.2,0) .. controls (-2.2,0.05) and (-2.25,0.05) .. (-2.3,0.05);
\draw (-2.2,0) .. controls (-2.2,0.05) and (-2.15,0.05) .. (-2.1,0.05);
\draw (-2.1,0.05) -- (-1.4,0.05);
\draw (-1.3,0.15) .. controls (-1.3,0.1) and (-1.3,0.05) .. (-1.4,0.05);
\end{scope}
\end{tikzpicture}
\ee
We expect its flavor symmetry to be $\fe_7\oplus\su(2)^3$ where the $\fe_7$ factor is the localized symmetry associated to the $\sp(0)$ node. In particular we expect the flavor symmetry to have rank $10$. If we would compactify this $6d$ SCFT on a circle of finite non-zero size, we would expect the flavor symmetry of the resulting $5d$ KK theory to have rank $11=10+1$ where the extra rank is associated to the size of the circle. It is known that this $5d$ KK theory on special loci in its extended Coulomb branch is also described by the $5d$ gauge theory
\be
\su(2k+2)_0+2\L^2+8\F
\ee
which has a rank 11 flavor symmetry (accounting for the instanton current). Thus, our proposal for the rank of the $6d$ SCFT passes this test. Later, we will solidify our proposal for the $\fe_7\oplus\su(2)^3$ flavor symmetry by manifesting it in the geometry.

Finally, consider the $n=1$ case of (\ref{n>3})
\be
\begin{tikzpicture} [scale=1.9]
\node (v1) at (-1.5,0.8) {2};
\node at (-1.5,1.1) {$\su(1)$};
\node (v3) at (-2.9,0.8) {2};
\node at (-2.9,1.1) {$\su(1)$};
\node (v2) at (-2.2,0.8) {$\cdots$};
\draw  (v2) edge (v3);
\draw  (v2) edge (v1);
\node (v4) at (-3.8,0.8) {2};
\node at (-3.8,1.1) {$\su(1)$};
\draw  (v4) edge (v3);
\end{tikzpicture}
\ee
which is well-known to have an $\su(2)$ delocalized flavor symmetry.

From the above considerations, we can derive the following rule for the determination of flavor symmetry $\ff$ of a $6d$ SCFT $\fT$ that has nodes of type (\ref{dnode}):\\
First, we define a \emph{naive} flavor symmetry $\ff_N$ associated to the theory. First part of $\ff_N$ arises from the flavor symmetry of the $6d$ $\cN=(1,0)$ gauge theory associated to $\fT$. Second part of $\ff_N$ arises from the flavor symmetries localized at all the $\sp(0)$ nodes in the theory. And the final and third part of $\ff_N$ arises from chains of $\su(1)$ nodes. Let $c$ be the number of connected subset of nodes of $\fT$ such that the connected subset contains only $\su(1)$ nodes and none of the nodes in the connected subset has a neighboring node carrying $\su(2)$ gauge algebra. Then the third contribution to $\ff_N$ is $\su(2)^{c}$.\\
Now, let $m_1$ be the number of nodes of type (\ref{dnode}) in $\fT$ that have $n\ge3$ and let $m_2$ be the number of nodes of type (\ref{dnode}) in $\fT$ that have $k=n=2$.\\
Then, we have
\be
\text{rank}(\ff)=\text{rank}(\ff_N)-m_1-m_2
\ee
More precisely, each node of type (\ref{dnode}) for $n\ge3$ removes a $\u(1)$ factor from $\ff_N$. The removal of this $\u(1)$ causes some of the remaining $\u(1)$ factors to become delocalized. On the other hand, each node of type (\ref{dnode}) for $k=n=2$ typically removes an $\su(2)$ factor, but can also remove a $\u(1)$ factor. Then, it is possible for some of the remaining $\su(2)$ and $\u(1)$ factors to delocalize. Moreover, if a node of type (\ref{dnode}) for $k=n=2$ is not a part of chain of nodes of type (\ref{n>3}), then it can also reduce the rank of naive flavor symmetry associated to it in more complicated ways, like $\so(8)\rightarrow\so(7)$ or $\so(7)\rightarrow\fg_2$ or $\so(6)\to\su(3)$ or $\so(5)\to\su(2)$ (see section \ref{delocal5d} for more details). The remaining delocalized factors in $\ff$ are $\su(2)$ factors associated to connected subsets comprising of more than one $\su(1)$ nodes without $\su(2)$ neighbors.

\section{Flavor symmetry of $5d$ KK theories}\label{5dF}
$5d$ KK theories are produced by compactifying $6d$ SCFTs on a circle of finite, non-zero radius. One then has the option of introducing, around the circle, holonomies for discrete flavor symmetries of the $6d$ SCFT. If no such holonomy is present, we say that the compactification is untwisted. If such a holonomy is present, we say that the compactification is twisted. From the point of view of the tensor branch of $6d$ SCFTs, the possible twists involve discrete symmetries related to outer automorphisms of the gauge algebras appearing on the tensor branch, permutation of the tensor multiplets, and parity transformation in $O(2n)$ flavor symmetry groups. See \cite{Bhardwaj:2018yhy,Bhardwaj:2018vuu,Bhardwaj:2019fzv,Bhardwaj:2020kim} for a discussion of all the known $5d$ KK theories.

Upon an untwisted compactification, it is known that the KK mode affinizes each gauge algebra $\fg_i$ appearing on the tensor branch to its untwisted affine version $\fg_i^{(1)}$. The same is true for flavor symmetries. Each localized or delocalized flavor symmetry is affinized to its affine version. This does not mean that the $5d$ KK theory admits a flavor symmetry given by an affine Lie algebra. It simply means that the mass parameters of the $5d$ KK theory are valued in the Cartan subalgebra of the affine Lie algebra. A linear combination of these mass parameters can be identified with the inverse radius of the compactification circle. Thus, as long as the radius of the compactification circle remains finite, the flavor symmetry of the $5d$ KK theory is always provided by a finite Lie algebra whose Dynkin diagram is a proper subset of the Dynkin diagram of the affine Lie algebra. Different such subsets arise at different loci in the extended Coulomb branch of the $5d$ KK theory. We represent a $5d$ KK theory arising via untwisted compactification of a $6d$ SCFT by a graph which is exactly the same as the graph for the $6d$ SCFT but replace each gauge algebra $\fg_i$ by $\fg_i^{(1)}$ and each localized flavor algebra $\ff_{\hat i}$ by $\ff_{\hat i}^{(1)}$.

For a $5d$ KK theory arising via a twisted compactification of a $6d$ SCFT, the graph for the $5d$ KK theory is obtained in general by folding the graph of the $6d$ SCFT and replacing the gauge algebras $\fg_i$ by affine Lie algebras $\fg_i^{(q_i)}$ if the twist involves an outer automorphism of $\fg_i$ of order $q_i$. Similarly, we replace the flavor algebras $\ff_{\hat i}$ by affine Lie algebras $\ff_{\hat i}^{(q_{\hat i})}$. The possible building blocks for graphs of $5d$ KK theories arising by twisted compactifications were discussed in \cite{Bhardwaj:2019fzv,Bhardwaj:2020kim} where the effect of the twist on flavor symmetry was ignored. In this paper, we extend their analysis to account for flavor symmetries in these building blocks.

\begin{table}[htbp]
\begin{center}
\begin{tabular}{|c|c|}
\hline
 \raisebox{-.4\height}{\begin{tikzpicture}
\node (v1) at (-0.5,0.45) {$k$};
\node at (-0.45,0.9) {$\su(n)^{(2)}$};
\begin{scope}[shift={(2,0)}]
\node (v2) at (-0.5,0.45) {$\left[\su(m)^{(2)}\right]$};
\end{scope}
\draw  (v1) edge (v2);
\end{tikzpicture}}&\raisebox{-.4\height}{\begin{tikzpicture}
\node (v1) at (-0.5,0.45) {$k$};
\node at (-0.45,0.9) {$\su(n)^{(2)}$};
\begin{scope}[shift={(2,0)}]
\node (v2) at (-0.5,0.45) {$\left[\su(2)^{(1)}\right]$};
\end{scope}
\draw  (v1) edge (v2);
\end{tikzpicture}}
\\ \hline
\raisebox{-.4\height}{\begin{tikzpicture}
\node (v1) at (-0.5,0.45) {1};
\node at (-0.45,0.9) {$\su(4)^{(2)}$};
\begin{scope}[shift={(1.9,0)}]
\node (v2) at (-0.5,0.45) {$\left[\su(m)^{(2)}\right]$};
\end{scope}
\draw  (v1) edge (v2);
\begin{scope}[shift={(-1.9,0)}]
\node (v3) at (-0.5,0.45) {$\left[\su(2)^{(1)}\right]$};
\end{scope}
\draw [densely dashed] (v3) edge (v1);
\end{tikzpicture}}&\raisebox{-.4\height}{\begin{tikzpicture}
\node (v1) at (-0.5,0.45) {1};
\node at (-0.45,0.9) {$\su(4)^{(2)}$};
\begin{scope}[shift={(1.9,0)}]
\node (v2) at (-0.5,0.45) {$\left[\su(2)^{(1)}\right]$};
\end{scope}
\draw  (v1) edge (v2);
\begin{scope}[shift={(-1.9,0)}]
\node (v3) at (-0.5,0.45) {$\left[\su(2)^{(1)}\right]$};
\end{scope}
\draw [densely dashed] (v3) edge (v1);
\end{tikzpicture}}
\\ \hline
\raisebox{-.4\height}{\begin{tikzpicture}
\node (v1) at (-0.5,0.45) {1};
\node at (-0.45,0.9) {$\su(\wh n)^{(2)}$};
\begin{scope}[shift={(2,0)}]
\node (v2) at (-0.5,0.45) {$\left[\su(m)^{(2)}\right]$};
\end{scope}
\draw  (v1) edge (v2);
\end{tikzpicture}}&\raisebox{-.4\height}{\begin{tikzpicture}
\node (v1) at (-0.5,0.45) {1};
\node at (-0.45,0.9) {$\su(\wh n)^{(2)}$};
\begin{scope}[shift={(2,0)}]
\node (v2) at (-0.5,0.45) {$\left[\su(2)^{(1)}\right]$};
\end{scope}
\draw  (v1) edge (v2);
\end{tikzpicture}}
\\ \hline
\raisebox{-.4\height}{\begin{tikzpicture}
\node (v1) at (-0.5,0.45) {1};
\node at (-0.45,0.9) {$\su(\tilde 6)^{(2)}$};
\begin{scope}[shift={(2,0)}]
\node (v2) at (-0.5,0.45) {$\left[\su(m)^{(2)}\right]$};
\end{scope}
\draw  (v1) edge (v2);
\end{tikzpicture}}&\raisebox{-.4\height}{\begin{tikzpicture}
\node (v1) at (-0.5,0.45) {1};
\node at (-0.45,0.9) {$\su(\tilde 6)^{(2)}$};
\begin{scope}[shift={(2,0)}]
\node (v2) at (-0.5,0.45) {$\left[\su(2)^{(1)}\right]$};
\end{scope}
\draw  (v1) edge (v2);
\end{tikzpicture}}
\\ \hline
\raisebox{-.4\height}{\begin{tikzpicture}
\node (v1) at (-0.5,0.45) {4};
\node at (-0.45,0.9) {$\so(2n)^{(2)}$};
\begin{scope}[shift={(2,0)}]
\node (v2) at (-0.5,0.45) {$\left[\sp(m)^{(1)}\right]$};
\end{scope}
\draw  (v1) edge (v2);
\end{tikzpicture}}&\raisebox{-.4\height}{\begin{tikzpicture}
\node (v1) at (-0.5,0.45) {$k$};
\node at (-0.45,0.9) {$\so(8)^{(2)}$};
\begin{scope}[shift={(2.6,0)}]
\node (v2) at (-0.5,0.45) {$\left[\sp(n)^{(1)}\right]$};
\end{scope}
\begin{scope}[shift={(-1.9,0)}]
\node (v3) at (-0.5,0.45) {$\left[\sp(l)^{(1)}\right]$};
\end{scope}
\draw [densely dotted] (v3) edge (v1);
\node (v4) at (0.5,0.45) {\tiny{2}};
\draw [densely dashed]  (v1) edge (v4);
\draw [->] (v4) edge (v2);
\end{tikzpicture}}
\\ \hline
\raisebox{-.4\height}{\begin{tikzpicture}
\node (v1) at (-0.5,0.45) {$k$};
\node at (-0.45,0.9) {$\so(8)^{(3)}$};
\begin{scope}[shift={(3.1,0)}]
\node (v2) at (-0.5,0.45) {$\left[\sp(n)^{(1)}\right]$};
\end{scope}
\node (v4) at (0.7,0.45) {\tiny{3}};
\draw [densely dashed]  (v1) edge (0.2,0.45);
\draw [densely dotted]  (0.2,0.45) edge (v4);
\draw [densely dotted]  (v4) edge (1.2,0.45);
\draw [->] (1.2,0.45) -- (v2);
\end{tikzpicture}}&\raisebox{-.4\height}{\begin{tikzpicture}
\node (v1) at (-0.5,0.4) {$1$};
\node at (-0.45,0.85) {$\so(10)^{(2)}$};
\begin{scope}[shift={(1.9,0)}]
\node (v2) at (-0.5,0.4) {$\left[\sp(n)^{(1)}\right]$};
\end{scope}
\draw  (v1) edge (v2);
\begin{scope}[shift={(-2.4,0)}]
\node (v3) at (-0.5,0.4) {$\left[\su(3)^{(2)}\right]$};
\end{scope}
\node (v4) at (-1.4,0.4) {\tiny{4}};
\draw [densely dashed] (v3) edge (v4);
\draw [densely dashed] (v4) edge (v1);
\end{tikzpicture}}
\\ \hline
\raisebox{-.4\height}{\begin{tikzpicture}
\node (v1) at (-0.5,0.4) {$2$};
\node at (-0.45,0.85) {$\so(10)^{(2)}$};
\begin{scope}[shift={(1.9,0)}]
\node (v2) at (-0.5,0.4) {$\left[\sp(n)^{(1)}\right]$};
\end{scope}
\draw  (v1) edge (v2);
\begin{scope}[shift={(-2.4,0)}]
\node (v3) at (-0.5,0.4) {$\left[\su(2)^{(1)}\right]$};
\end{scope}
\node (v4) at (-1.4,0.4) {\tiny{4}};
\draw [densely dashed] (v3) edge (v4);
\draw [densely dashed] (v4) edge (v1);
\end{tikzpicture}}&\raisebox{-.4\height}{\begin{tikzpicture}
\node (v1) at (-0.5,0.4) {$3$};
\node at (-0.45,0.85) {$\so(10)^{(2)}$};
\begin{scope}[shift={(1.9,0)}]
\node (v2) at (-0.5,0.4) {$\left[\sp(n)^{(1)}\right]$};
\end{scope}
\draw  (v1) edge (v2);
\end{tikzpicture}}
\\ \hline
\raisebox{-.4\height}{\begin{tikzpicture}
\node (v1) at (-0.5,0.4) {$2$};
\node at (-0.45,0.85) {$\so(\wh{12})^{(2)}$};
\begin{scope}[shift={(1.9,0)}]
\node (v2) at (-0.5,0.4) {$\left[\sp(n)^{(1)}\right]$};
\end{scope}
\draw  (v1) edge (v2);
\end{tikzpicture}}&\raisebox{-.4\height}{\begin{tikzpicture}
\node (v1) at (-0.5,0.4) {$k$};
\node at (-0.45,0.85) {$\fe_6^{(2)}$};
\begin{scope}[shift={(2.7,0)}]
\node (v2) at (-0.5,0.4) {$\left[\su(6-k)^{(2)}\right]$};
\end{scope}
\node (v3) at (0.4,0.4) {\tiny{6}};
\draw  (v1) edge (v3);
\draw  (v3) edge (v2);
\end{tikzpicture}}
\\ \hline
\raisebox{-.4\height}{\begin{tikzpicture}
\node (v1) at (-0.5,0.4) {$4$};
\node at (-0.45,0.85) {$\fe_6^{(2)}$};
\begin{scope}[shift={(2.5,0)}]
\node (v2) at (-0.5,0.4) {$\left[\su(2)^{(1)}\right]$};
\end{scope}
\node (v3) at (0.4,0.4) {\tiny{6}};
\draw  (v1) edge (v3);
\draw  (v3) edge (v2);
\end{tikzpicture}}&\raisebox{-.4\height}{\begin{tikzpicture}
\node (v1) at (-0.5,0.4) {$4$};
\node at (-0.45,0.85) {$\fe_7^{(1)}$};
\begin{scope}[shift={(2.5,0)}]
\node (v2) at (-0.5,0.4) {$\left[\so(3)^{(1)}\right]$};
\end{scope}
\node (v3) at (0.4,0.4) {\tiny{12}};
\draw  (v1) edge (v3);
\draw  (v3) edge (v2);
\begin{scope}[shift={(-1.9,0)}]
\node (v4) at (-0.5,0.4) {$\left[\Z_2^{(2)}\right]$};
\end{scope}
\draw  (v4) edge (v1);
\end{tikzpicture}}
\\ \hline
\raisebox{-.4\height}{\begin{tikzpicture}
\node (v1) at (-0.5,0.4) {$2$};
\node at (-0.45,0.85) {$\fe_7^{(1)}$};
\begin{scope}[shift={(2.5,0)}]
\node (v2) at (-0.5,0.4) {$\left[\so(5)^{(1)}\right]$};
\end{scope}
\node (v3) at (0.4,0.4) {\tiny{12}};
\draw  (v1) edge (v3);
\draw  (v3) edge (v2);
\begin{scope}[shift={(-1.9,0)}]
\node (v4) at (-0.5,0.4) {$\left[\Z_2^{(2)}\right]$};
\end{scope}
\draw  (v4) edge (v1);
\end{tikzpicture}}&
\\ \hline
\end{tabular}
\end{center}
\caption{Possible localized flavor symmetries for gauge-theoretic nodes in $5d$ KK theories.}\label{TR7}
\end{table}

\subsection{Localized flavor symmetries}
For nodes carrying a non-trivial gauge algebra, the localized flavor symmetries are collected\footnote{As explained above, we can ignore the completely untwisted cases as they have already been discussed in the last section.} in Table \ref{TR7}. Let us explain some parts of this table:
\bit
\item Consider
\be\label{K1}
\begin{tikzpicture}
\node (v1) at (-0.5,0.45) {$k$};
\node at (-0.45,0.9) {$\su(n)$};
\begin{scope}[shift={(2,0)}]
\node (v2) at (-0.5,0.45) {$\left[\su(m)\right]$};
\end{scope}
\draw  (v1) edge (v2);
\end{tikzpicture}
\ee
when we have in a $6d$ SCFT an $\su(n)$ node with an associated $\su(m)$ localized flavor symmetry for $m\ge3$. The edge between $\su(n)$ and $\su(m)$ carries hypers transforming in bifundamental $\F\otimes\F$ of $\su(n)\oplus\su(m)$. The fields inside this hypermultiplet are left invariant by an action of order two outer-automorphism on $\su(n)$ only if we simultaneously act by an order two outer-automorphism on $\su(n)$. This can be understood by noting that the two outer automorphisms send a hyper in $\F\otimes\F$ of $\su(n)\oplus\su(m)$ to a hyper in $\bar\F\otimes\bar\F$ which is simply a hyper in $\F\otimes\F$ since $\bar\F\otimes\bar\F$ is a complex conjugate of $\F\otimes\F$. Thus, the $5d$ KK theory produced after the above outer automorphism twist carries
\be
\begin{tikzpicture}
\node (v1) at (-0.5,0.45) {$k$};
\node at (-0.45,0.9) {$\su(n)^{(2)}$};
\begin{scope}[shift={(2,0)}]
\node (v2) at (-0.5,0.45) {$\left[\su(m)^{(2)}\right]$};
\end{scope}
\draw  (v1) edge (v2);
\end{tikzpicture}
\ee
If we consider the case $m=2$ of (\ref{K1}), then the outer automorphism of $\su(n)$ leaves the bifundamental invariant since the $\F$ of $\su(2)$ is real.
\item Consider now a $6d$ SCFT carrying
\be
\begin{tikzpicture}
\node (v1) at (-0.5,0.45) {1};
\node at (-0.45,0.9) {$\su(4)$};
\begin{scope}[shift={(1.7,0)}]
\node (v2) at (-0.5,0.45) {$\left[\su(m)\right]$};
\end{scope}
\draw  (v1) edge (v2);
\begin{scope}[shift={(-1.8,0)}]
\node (v3) at (-0.5,0.45) {$\left[\su(2)\right]$};
\end{scope}
\draw [densely dashed] (v3) edge (v1);
\end{tikzpicture}
\ee
for $m\ge3$. The edge between $\su(4)$ and $\su(m)$ carries a hyper in bifundamental, and the edge between $\su(2)$ and $\su(4)$ carries a half-hyper in $\F\otimes\L^2$ of $\su(2)\oplus\su(4)$. The outer automorphism of $\su(4)$ sends $\F\to\bar\F$ but leaves $\L^2$ invariant. Thus, to convert this outer automorphism into a symmetry, we need to only combine it with the outer automorphism of $\su(m)$. The resulting KK theory then carries
\be
\begin{tikzpicture}
\node (v1) at (-0.5,0.45) {1};
\node at (-0.45,0.9) {$\su(4)^{(2)}$};
\begin{scope}[shift={(1.9,0)}]
\node (v2) at (-0.5,0.45) {$\left[\su(m)^{(2)}\right]$};
\end{scope}
\draw  (v1) edge (v2);
\begin{scope}[shift={(-1.9,0)}]
\node (v3) at (-0.5,0.45) {$\left[\su(2)^{(1)}\right]$};
\end{scope}
\draw [densely dashed] (v3) edge (v1);
\end{tikzpicture}
\ee
\item Consider a $6d$ SCFT carrying
\be
\begin{tikzpicture}
\node (v1) at (-0.5,0.45) {$k$};
\node at (-0.45,0.9) {$\so(8)$};
\begin{scope}[shift={(1.7,0)}]
\node (v2) at (-0.5,0.45) {$\left[\sp(n)\right]$};
\end{scope}
\draw  (v1) edge (v2);
\begin{scope}[shift={(-1.7,0)}]
\node (v3) at (-0.5,0.45) {$\left[\sp(m)\right]$};
\end{scope}
\draw [densely dashed] (v3) edge (v1);
\begin{scope}[shift={(0,-1.1)}]
\node (v4) at (-0.5,0.45) {$\left[\sp(l)\right]$};
\end{scope}
\draw [densely dotted] (v1) edge (v4);
\end{tikzpicture}
\ee
Consider a $\Z_2$ outer automorphism which exchanges the fundamental and the spinor reps of $\so(8)$. This can be a symmetry of the theory if $m=n$. The resulting $5d$ KK theory then contains
\be
\begin{tikzpicture}
\node (v1) at (-0.5,0.45) {$k$};
\node at (-0.45,0.9) {$\so(8)^{(2)}$};
\begin{scope}[shift={(2.6,0)}]
\node (v2) at (-0.5,0.45) {$\left[\sp(n)^{(1)}\right]$};
\end{scope}
\begin{scope}[shift={(-1.9,0)}]
\node (v3) at (-0.5,0.45) {$\left[\sp(l)^{(1)}\right]$};
\end{scope}
\draw [densely dotted] (v3) edge (v1);
\node (v4) at (0.5,0.45) {\tiny{2}};
\draw [densely dashed]  (v1) edge (v4);
\draw [->] (v4) edge (v2);
\end{tikzpicture}
\ee
where we have folded the above graph to identify the flavor symmetries associated to $\F$ and $\S$ of $\so(8)$. The partially dashed and partially solid edge between $\so(8)^{(1)}$ and $\sp(n)^{(1)}$ edge is supposed to denote that this edge comes by combining a solid and a dashed edge of the $6d$ theory. The direction of the edge denotes the direction of folding and the label 2 in the middle of the edge can be thought of as denoting the fact that this directed edge arises from combining 2 undirected edges via a folding process. More precisely, the label 2 captures an off-diagonal entry in a matrix of Chern-Simons couplings arising in the low-energy abelian gauge theory on the extended Coulomb branch of the $5d$ KK theory \cite{Bhardwaj:2019fzv}.
\item Similarly,
\be

\ee
is obtained by twisting by a $\Z_3$ outer-automorphism of $\so(8)$, which forces $m=n=l$. The edge is denoted as partially solid, partially dashed and partially dotted to denote that it arises by combining solid, dashed and dotted edges, with the label 3 denoting that 3 edges have been identified by folding to produce the edge.
\item The $5d$ KK theory configuration
\be
%
\ee
is produced by compactifying the $6d$ SCFT configuration
\be
%
\ee
where the $\fe_7$ carries 6 half-hypers in $\mathbf{56}$ and hence the localized flavor symmetry of the $6d$ SCFT configuration is $\so(6)$. In fact, the flavor symmetry group is $O(6)$. The compactification resulting in the above $5d$ KK theory configuration contains a twist by a $\Z_2$ element lying in the disconnected component of $O(6)$. Thus one of the half-hypers is projected out, and we expect the $5d$ KK theory to have an $\so(5)^{(1)}$ flavor symmetry.
\eit

\begin{table}[htbp]
\begin{center}
\begin{tabular}{|c|c|}
\hline
 \raisebox{-.4\height}{\begin{tikzpicture}
\node (v1) at (-0.5,0.4) {1};
\node at (-0.45,0.85) {$\sp(0)^{(1)}_0$};
\begin{scope}[shift={(1.7,-0.05)}]
\node (v2) at (-0.5,0.45) {2};
\node at (-0.45,0.9) {$\su(8)^{(2)}$};
\end{scope}
\draw  (v1) edge (v2);
\node (v3) at (-2.7,0.4) {$\left[\su(2)^{(1)}\right]$};
\draw  (v3) edge (v1);
\end{tikzpicture}}&\raisebox{-.4\height}{\begin{tikzpicture}
\node (v1) at (-0.5,0.4) {1};
\node at (-0.45,0.85) {$\sp(0)^{(1)}$};
\begin{scope}[shift={(1.7,-0.05)}]
\node (v2) at (-0.5,0.45) {2};
\node at (-0.45,0.9) {$\su(7)^{(2)}$};
\end{scope}
\draw  (v1) edge (v2);
\node (v3) at (-2.7,0.4) {$\left[\su(2)^{(1)}\right]$};
\draw  (v3) edge (v1);
\end{tikzpicture}}
\\ \hline
\raisebox{-.4\height}{\begin{tikzpicture}
\node (v1) at (-0.5,0.4) {1};
\node at (-0.45,0.85) {$\sp(0)^{(1)}$};
\begin{scope}[shift={(1.7,-0.05)}]
\node (v2) at (-0.5,0.45) {2};
\node at (-0.45,0.9) {$\su(6)^{(2)}$};
\end{scope}
\draw  (v1) edge (v2);
\node (v3) at (-2.7,0.4) {$\left[\su(3)^{(2)}\right]$};
\draw  (v3) edge (v1);
\node (v4) at (-0.5,-0.9) {$\left[\su(2)^{(1)}\right]$};
\draw  (v1) edge (v4);
\end{tikzpicture}}&\raisebox{-.4\height}{\begin{tikzpicture}
\node (v1) at (-0.5,0.4) {1};
\node at (-0.45,0.85) {$\sp(0)^{(1)}$};
\begin{scope}[shift={(1.7,-0.05)}]
\node (v2) at (-0.5,0.45) {2};
\node at (-0.45,0.9) {$\su(5)^{(2)}$};
\end{scope}
\draw  (v1) edge (v2);
\node (v3) at (-2.7,0.4) {$\left[\su(5)^{(2)}\right]$};
\draw  (v3) edge (v1);
\end{tikzpicture}}
\\
\hline
\end{tabular}
\end{center}
\caption{Possible localized flavor symmetries in $5d$ KK theories for nodes with a trivial gauge algebra having no incoming directed edges. Continued in Table \ref{TR9}.}\label{TR8}
\end{table}

\begin{table}[htbp]
\begin{center}
\begin{tabular}{|c|c|}
\hline
\raisebox{-.4\height}{\begin{tikzpicture}
\node (v1) at (-0.5,0.4) {1};
\node (v7) at (-0.5,0.85) {$\sp(0)^{(1)}$};
\begin{scope}[shift={(1.7,-0.05)}]
\node (v2) at (-0.5,0.45) {$2$};
\node at (-0.45,0.9) {$\su(6)^{(2)}$};
\end{scope}
\draw  (v1) edge (v2);
\node (v3) at (-2.3,0.4) {$3$};
\draw  (v3) edge (v1);
\node (v4) at (-0.5,-0.9) {$\left[\su(2)^{(1)}\right]$};
\draw  (v1) edge (v4);
\node at (-2.3,0.85) {$\su(3)^{(2)}$};
\end{tikzpicture}}&\raisebox{-.4\height}{\begin{tikzpicture}
\node (v1) at (-0.5,0.4) {1};
\node (v7) at (-0.5,0.85) {$\sp(0)^{(1)}$};
\begin{scope}[shift={(1.7,-0.05)}]
\node (v2) at (-0.5,0.45) {$2$};
\node at (-0.45,0.9) {$\su(5)^{(2)}$};
\end{scope}
\draw  (v1) edge (v2);
\node (v3) at (-2.3,0.4) {$3$};
\draw  (v3) edge (v1);
\node (v4) at (-0.5,-0.9) {$\left[\su(2)^{(1)}\right]$};
\draw  (v1) edge (v4);
\node at (-2.3,0.85) {$\su(3)^{(2)}$};
\end{tikzpicture}}
\\ \hline
\raisebox{-.4\height}{\begin{tikzpicture}
\node (v1) at (-0.5,0.4) {1};
\node (v7) at (-0.5,0.85) {$\sp(0)^{(1)}$};
\begin{scope}[shift={(1.7,-0.05)}]
\node (v2) at (-0.5,0.45) {$k$};
\node at (-0.45,0.9) {$\so(8)^{(2)}$};
\end{scope}
\draw  (v1) edge (v2);
\node (v3) at (-2.3,0.4) {$2$};
\draw  (v3) edge (v1);
\node (v4) at (-0.5,-0.9) {$\left[\u(1)^{(1)}\right]$};
\draw  (v1) edge (v4);
\node at (-2.3,0.85) {$\su(4)^{(2)}$};
\end{tikzpicture}}&\raisebox{-.4\height}{\begin{tikzpicture}
\node (v1) at (-0.5,0.4) {1};
\node (v7) at (-0.5,0.85) {$\sp(0)^{(1)}$};
\begin{scope}[shift={(1.7,-0.05)}]
\node (v2) at (-0.5,0.45) {$3$};
\node at (-0.45,0.9) {$\so(7)^{(1)}$};
\end{scope}
\draw  (v1) edge (v2);
\node (v3) at (-2.3,0.4) {$2$};
\draw  (v3) edge (v1);
\node at (-2.3,0.85) {$\su(4)^{(2)}$};
\node (v5) at (-0.5,2.6) {$\left[\su(2)^{(1)}\right]$};
\node (v6) at (-0.5,1.7) {\tiny{2}};
\draw  (v7) edge (v6);
\draw (v6) edge (v5);
\end{tikzpicture}}
\\
\hline
\end{tabular}
\end{center}
\caption{Continued from \ref{TR8}: Possible localized flavor symmetries in $5d$ KK theories for nodes with a trivial gauge algebra having no incoming directed edges. Continued in Table \ref{TR10}.}\label{TR9}
\end{table}

\begin{table}[htbp]
\begin{center}
\begin{tabular}{|c|c|}
\hline
\raisebox{-.4\height}{\begin{tikzpicture}
\node (v1) at (-0.5,0.4) {1};
\node (v7) at (-0.5,0.85) {$\sp(0)^{(1)}$};
\begin{scope}[shift={(1.7,-0.05)}]
\node (v2) at (-0.5,0.45) {$k$};
\node at (-0.45,0.9) {$\su(3)^{(2)}$};
\end{scope}
\draw  (v1) edge (v2);
\node (v3) at (-2.3,0.4) {$3$};
\draw  (v3) edge (v1);
\node (v4) at (0.8,2.2) {$\left[\su(3)^{(2)}\right]$};
\node at (-2.3,0.85) {$\su(3)^{(2)}$};
\node (v5) at (-1.8,2.2) {$\left[\su(3)^{(2)}\right]$};
\draw  (v5) edge (v7);
\draw  (v7) edge (v4);
\end{tikzpicture}}&\raisebox{-.4\height}{\begin{tikzpicture}
\node (v1) at (-0.5,0.4) {1};
\node (v7) at (-0.5,0.85) {$\sp(0)^{(1)}$};
\begin{scope}[shift={(1.7,-0.05)}]
\node (v2) at (-0.5,0.45) {$l$};
\node at (-0.45,0.9) {$\su(3)^{(1)}$};
\end{scope}
\draw  (v1) edge (v2);
\node (v3) at (-2.3,0.4) {$k$};
\draw  (v3) edge (v1);
\node at (-2.3,0.85) {$\su(3)^{(2)}$};
\node (v5) at (-0.5,2.6) {$\left[\su(3)^{(1)}\right]$};
\node (v6) at (-0.5,1.7) {\tiny{2}};
\draw  (v7) edge (v6);
\draw [->] (v6) edge (v5);
\end{tikzpicture}}
\\
\hline
\end{tabular}
\end{center}
\caption{Continued from Table \ref{TR9}: Possible localized flavor symmetries in $5d$ KK theories for nodes with a trivial gauge algebra having no incoming directed edges.}\label{TR10}
\end{table}

\begin{table}[htbp]
\begin{center}
\begin{tabular}{|c|c|}
\hline
\raisebox{-.4\height}{\begin{tikzpicture}
\node (v1) at (-0.5,0.4) {1};
\node (v7) at (-0.5,0.85) {$\sp(0)^{(1)}$};
\begin{scope}[shift={(1.8,-0.05)}]
\node (v2) at (-0.5,0.45) {$k$};
\node at (-0.45,0.9) {$\fg^{(1)}$};
\end{scope}
\node (v3) at (-2.7,0.4) {$\left[\fh_2^{(1)}\right]$};
\node (v5) at (-0.5,2.7) {$\left[\fh_1^{(1)}\right]$};
\node (v6) at (-0.5,1.7) {\tiny{$f_1$}};
\draw  (v7) edge (v6);
\draw (v6) edge (v5);
\node (v6) at (0.4,0.4) {\tiny{$e$}};
\draw [<-] (v1) edge (v6);
\draw  (v6) edge (v2);
\node (v4) at (-1.4,0.4) {\tiny{$f_2$}};
\draw  (v4) edge (v1);
\draw  (v3) edge (v4);
\end{tikzpicture}}&\raisebox{-.4\height}{\begin{tikzpicture}
\node (v1) at (-0.5,0.4) {1};
\node (v7) at (-0.5,0.85) {$\sp(0)^{(1)}$};
\begin{scope}[shift={(1.8,-0.05)}]
\node (v2) at (-0.5,0.45) {$3$};
\node at (-0.45,0.9) {$\su(3)^{(2)}$};
\end{scope}
\node (v4) at (-2.4,0.4) {$\left[\fe_6^{(1)}\right]$};
\draw  (v1) edge (v4);
\node (v6) at (0.4,0.4) {\tiny{2}};
\draw [<-] (v1) edge (v6);
\draw (v6) edge (v2);
\end{tikzpicture}}
\\ \hline
\raisebox{-.4\height}{\begin{tikzpicture}
\node (v1) at (-0.5,0.4) {1};
\node (v7) at (-0.5,0.85) {$\sp(0)^{(1)}$};
\begin{scope}[shift={(1.8,-0.05)}]
\node (v2) at (-0.5,0.45) {$k$};
\node at (-0.45,0.9) {$\so(14)^{(2)}$};
\end{scope}
\node (v4) at (-2.4,0.4) {$\left[\u(1)^{(1)}\right]$};
\draw  (v1) edge (v4);
\node (v6) at (0.4,0.4) {\tiny{2}};
\draw [<-] (v1) edge (v6);
\draw (v6) edge (v2);
\end{tikzpicture}}&\raisebox{-.4\height}{\begin{tikzpicture}
\node (v1) at (-0.5,0.4) {1};
\node (v7) at (-0.5,0.85) {$\sp(0)^{(1)}$};
\begin{scope}[shift={(1.7,-0.05)}]
\node (v2) at (-0.5,0.45) {$k$};
\node at (-0.45,0.9) {$\so(12)^{(2)}$};
\end{scope}
\node (v3) at (-2.6,0.4) {$\left[\su(2)^{(1)}\right]$};
\draw  (v3) edge (v1);
\node (v5) at (-0.5,2.2) {$\left[\su(2)^{(1)}\right]$};
\node (v6) at (0.4,0.4) {\tiny{2}};
\draw [<-] (v1) edge (v6);
\draw  (v6) edge (v2);
\draw  (v5) edge (v7);
\end{tikzpicture}}
\\
\hline
\end{tabular}
\end{center}
\caption{Possible localized flavor symmetries in $5d$ KK theories for nodes with a trivial gauge algebra having incoming directed edges. See text for explanation of first entry in the table. Continued in \ref{TR12}.}\label{TR11}
\end{table}

\begin{table}[htbp]
\begin{center}
\begin{tabular}{|c|c|}
\hline
\raisebox{-.4\height}{\begin{tikzpicture}
\node (v1) at (-0.5,0.4) {1};
\node (v7) at (-0.5,0.85) {$\sp(0)^{(1)}$};
\begin{scope}[shift={(1.8,-0.05)}]
\node (v2) at (-0.5,0.45) {$k_1$};
\node at (-0.4,1) {$\fg_1^{(1)}$};
\end{scope}
\node (v3) at (-0.5,-1.4) {$\left[\fh_2^{(1)}\right]$};
\node (v5) at (-0.5,2.7) {$\left[\fh_1^{(1)}\right]$};
\node (v6) at (-0.5,1.7) {\tiny{$f_1$}};
\draw  (v7) edge (v6);
\draw (v6) edge (v5);
\node (v6) at (0.4,0.4) {\tiny{$e$}};
\draw [<-] (v1) edge (v6);
\draw  (v6) edge (v2);
\node (v4) at (-0.5,-0.4) {\tiny{$f_2$}};
\draw  (v4) edge (v1);
\draw  (v3) edge (v4);
\begin{scope}[shift={(-1.8,-0.05)}]
\node (v8) at (-0.5,0.45) {$k_2$};
\node at (-0.4,1) {$\fg_2^{(1)}$};
\end{scope}
\draw  (v8) edge (v1);
\end{tikzpicture}}&\raisebox{-.4\height}{\begin{tikzpicture}
\node (v1) at (-0.5,0.4) {1};
\node (v7) at (-0.5,0.85) {$\sp(0)^{(1)}$};
\begin{scope}[shift={(1.8,-0.05)}]
\node (v2) at (-0.5,0.45) {$4$};
\node at (-0.4,0.9) {$\so(10)^{(2)}$};
\end{scope}
\node (v5) at (-0.5,2.3) {$\left[\u(1)^{(1)}\right]$};
\node (v6) at (0.4,0.4) {\tiny{$2$}};
\draw [<-] (v1) edge (v6);
\draw  (v6) edge (v2);
\begin{scope}[shift={(-1.8,-0.05)}]
\node (v8) at (-0.5,0.45) {$3$};
\node at (-0.4,0.9) {$\su(3)^{(1)}$};
\end{scope}
\draw  (v8) edge (v1);
\draw  (v5) edge (v7);
\end{tikzpicture}}
\\
\hline
\end{tabular}
\end{center}
\caption{Continued from \ref{TR11}: Possible localized flavor symmetries for nodes with a trivial gauge algebra having incoming directed edges. See text for explanation of first entry in the table.}\label{TR12}
\end{table}

For nodes carrying a trivial gauge algebra, the flavor symmetries are collected in Tables \ref{TR8}--\ref{TR12}. Let us explain some parts of these tables below:
\bit
\item Consider the $6d$ SCFT
\be
%
\ee
The BPS string arising from the $\sp(0)$ node transforms as adjoint under the total $\fe_8$ symmetry of the $\sp(0)$. Under $\fe_8\to\su(5)\oplus\su(5)$, the adjoint of $\fe_8$ decomposes as
\be
\A\to(\A,1)+(1,\A)+(\L^2,\F)+(\bar\L^2,\bar\F)+(\F,\bar\L^2)+(\bar\F,\L^2)
\ee
The above decomposition is not preserved if we perform outer-automorphism on only one of the $\su(5)$ factors, but it is preserved if we perform outer-automorphism on both $\su(5)$ factors. Thus the configuration
\be
%
\ee
has an $\su(5)^{(2)}$ flavor symmetry localized at the $\sp(0)^{(1)}$ node, and we denote the full configuration as
\be
%
\ee
\item Consider $\fe_8\to\so(12)\oplus\su(2)\oplus\su(2)$ under which the adjoint of $\fe_8$ decomposes as
\be
\A\to (\A,1,1)+(1,\A,1)+(1,1,\A)+(\S,1,\F)+(\F,\F,\F)+(\C,\F,1)
\ee
From this we see that outer-automorphism of $\so(12)$ is a symmetry only if we exchange the two $\su(2)$. Thus, compactifying the configuration
\be
%
\ee
along with an outer-automorphism twist of $\so(12)$ induces a permutation twist on the two flavor $\su(2)$ nodes, and the resulting $5d$ KK theory configuration is
\be
%
\ee
which we want to compactify with an outer-automorphism twist of $\su(7)$. The adjoint of $\fe_8$ decomposes under $\fe_8\to\su(2)\oplus\su(7)\oplus\u(1)$ as
\begin{align}
\A\to&(\A,1)_0+(1,\A)_0+(1,\F)_8+(1,\bar\F)_{-8}+(1,\L^3)_{-4}+(1,\bar\L^3)_4+(\F,\F)_{-6}+\nn\\
&(\F,\L^2)_2+(\F,\bar\F)_6+(\F,\bar\L^2)_{-2}
\end{align}
where the subscripts denote $\u(1)$ charges. Thus, the outer-automorphism of $\su(7)$ is a symmetry only if we also perform outer-automorphism of $\u(1)$ which takes a charge $q$ representation of $\u(1)$ to charge $-q$ representation. Due to the presence of this outer-automorphism, the $5d$ KK theory does not obtain any continuous flavor symmetry from the $\u(1)$ factor present in the $6d$ SCFT, and hence we omit it and write the resulting $5d$ KK theory configuration simply as
\be
\begin{tikzpicture}
\node (v1) at (-0.5,0.4) {1};
\node at (-0.45,0.85) {$\sp(0)^{(1)}$};
\begin{scope}[shift={(1.7,-0.05)}]
\node (v2) at (-0.5,0.45) {2};
\node at (-0.45,0.9) {$\su(7)^{(2)}$};
\end{scope}
\draw  (v1) edge (v2);
\node (v3) at (-2.7,0.4) {$\left[\su(2)^{(1)}\right]$};
\draw  (v3) edge (v1);
\end{tikzpicture}
\ee
\item Consider the $6d$ SCFT configuration
\be
\begin{tikzpicture}
\node (v1) at (-0.5,0.4) {1};
\node (v7) at (-0.5,0.85) {$\sp(0)$};
\begin{scope}[shift={(1.7,-0.05)}]
\node (v2) at (-0.5,0.45) {$k$};
\node at (-0.45,0.9) {$\su(3)$};
\end{scope}
\draw  (v1) edge (v2);
\node (v3) at (-2.3,0.4) {$3$};
\draw  (v3) edge (v1);
\node (v4) at (-0.5,-0.9) {$\left[\su(3)\right]$};
\draw  (v1) edge (v4);
\node at (-2.3,0.85) {$\su(3)$};
\node (v5) at (-0.5,2.2) {$\left[\su(3)\right]$};
\draw  (v5) edge (v7);
\end{tikzpicture}
\ee
There are multiple ways to twist this configuration, all of which lead to different flavor symmetries. The adjoint of $\fe_8$ decomposes under $\fe_8\to\su(3)^4$ as
\begin{align}
\A\to&(\A,1,1,1)+(1,\A,1,1)+(1,1,\A,1)+(1,1,1,\A)+(1,\F,\bar\F,\bar\F)+(1,\bar\F,\F,\F)+\nn\\
&(\F,\F,1,\F)+(\F,\bar\F,\bar\F,1)+(\F,1,\F,\bar\F)+(\bar\F,\F,\F,1)+(\bar\F,\bar\F,1,\bar\F)+(\bar\F,1,\bar\F,\F)
\end{align}
Performing outer-automorphism on two of the $\su(3)$s forces us to perform outer-automorphisms on the other two $\su(3)$s as well, thus leading to the $5d$ KK theory configuration
\be
\begin{tikzpicture}
\node (v1) at (-0.5,0.4) {1};
\node (v7) at (-0.5,0.85) {$\sp(0)^{(1)}$};
\begin{scope}[shift={(1.7,-0.05)}]
\node (v2) at (-0.5,0.45) {$k$};
\node at (-0.45,0.9) {$\su(3)^{(2)}$};
\end{scope}
\draw  (v1) edge (v2);
\node (v3) at (-2.3,0.4) {$3$};
\draw  (v3) edge (v1);
\node (v4) at (-0.5,-0.9) {$\left[\su(3)^{(2)}\right]$};
\draw  (v1) edge (v4);
\node at (-2.3,0.85) {$\su(3)^{(2)}$};
\node (v5) at (-0.5,2.2) {$\left[\su(3)^{(2)}\right]$};
\draw  (v5) edge (v7);
\end{tikzpicture}
\ee
Performing outer-automorphism on only one of the $\su(3)$s while performing no outer-automorphism on another $\su(3)$ forces us to exchange the remaining two $\su(3)$s leading to the $5d$ KK theory configurations
\be
\begin{tikzpicture}
\node (v1) at (-0.5,0.4) {1};
\node (v7) at (-0.5,0.85) {$\sp(0)^{(1)}$};
\begin{scope}[shift={(1.7,-0.05)}]
\node (v2) at (-0.5,0.45) {$l$};
\node at (-0.45,0.9) {$\su(3)^{(1)}$};
\end{scope}
\draw  (v1) edge (v2);
\node (v3) at (-2.3,0.4) {$k$};
\draw  (v3) edge (v1);
\node at (-2.3,0.85) {$\su(3)^{(2)}$};
\node (v5) at (-0.5,2.6) {$\left[\su(3)^{(1)}\right]$};
\node (v6) at (-0.5,1.7) {\tiny{2}};
\draw  (v7) edge (v6);
\draw [->] (v6) edge (v5);
\end{tikzpicture}
\ee
Similarly, exchanging two $\su(3)$s forces us to perform outer-automorphism on one of the remaining two $\su(3)$s leading to the $5d$ KK theory configuration
\be
\begin{tikzpicture}
\node (v1) at (-0.5,0.4) {1};
\node (v7) at (-0.5,0.85) {$\sp(0)^{(1)}$};
\begin{scope}[shift={(1.8,-0.05)}]
\node (v2) at (-0.5,0.45) {$3$};
\node at (-0.45,0.9) {$\su(3)^{(1)}$};
\end{scope}
\node (v4) at (-2.6,0.4) {$\left[\su(3)^{(1)}\right]$};
\draw  (v1) edge (v4);
\node (v5) at (-0.5,2.2) {$\left[\su(3)^{(2)}\right]$};
\draw  (v5) edge (v7);
\node (v6) at (0.4,0.4) {\tiny{2}};
\draw  (v1) edge (v6);
\draw [->] (v6) edge (v2);
\end{tikzpicture}
\ee
Finally, we want to twist the gauge $\su(3)$s to obtain a $5d$ KK theory configuration of the form
\be
\begin{tikzpicture}
\node (v1) at (-0.5,0.4) {1};
\node (v7) at (-0.5,0.85) {$\sp(0)^{(1)}$};
\begin{scope}[shift={(1.8,-0.05)}]
\node (v2) at (-0.5,0.45) {$3$};
\node at (-0.45,0.9) {$\su(3)^{(2)}$};
\end{scope}
\node (v6) at (0.4,0.4) {\tiny{2}};
\draw  (v1) edge (v6);
\draw [->] (v6) edge (v2);
\end{tikzpicture}
\ee
This can be done by sending the left gauge $\su(3)$ to the right gauge $\su(3)$ along with an outer-automorphism, and the right gauge $\su(3)$ to the left gauge $\su(3)$ without an outer-automorphism, thus exchanging the two $\su(3)$s with a \emph{relative} outer-automorphism involved in the exchange. This operation forces us to also exchange the two flavor $\su(3)$s in the same fashion implying that the full $5d$ KK theory configuration involving the flavor algebras is
\be
\begin{tikzpicture}
\node (v1) at (-0.5,0.4) {1};
\node (v7) at (-0.5,0.85) {$\sp(0)^{(1)}$};
\begin{scope}[shift={(1.8,-0.05)}]
\node (v2) at (-0.5,0.45) {$3$};
\node at (-0.45,0.9) {$\su(3)^{(2)}$};
\end{scope}
\node (v4) at (-3,0.4) {$\left[\su(3)^{(2)}\right]$};
\node (v6) at (0.4,0.4) {\tiny{2}};
\draw  (v1) edge (v6);
\draw [->] (v6) edge (v2);
\node (v3) at (-1.4,0.4) {\tiny{2}};
\draw  (v1) edge (v3);
\draw [->] (v3) edge (v4);
\end{tikzpicture}
\ee
\item Consider the $6d$ SCFT configuration
\be
\begin{tikzpicture}
\node (v1) at (-0.5,0.4) {1};
\node (v7) at (-0.5,0.85) {$\sp(0)$};
\begin{scope}[shift={(1.5,-0.05)}]
\node (v2) at (-0.5,0.45) {$k$};
\node at (-0.45,0.9) {$\so(8)$};
\end{scope}
\draw  (v1) edge (v2);
\node (v3) at (-2.1,0.4) {$l$};
\draw  (v3) edge (v1);
\node (v4) at (-0.5,-0.9) {$\left[\u(1)\right]$};
\draw  (v1) edge (v4);
\node at (-2.1,0.85) {$\su(3)$};
\node (v5) at (-0.5,2.2) {$\left[\u(1)\right]$};
\draw  (v5) edge (v7);
\end{tikzpicture}
\ee
The adjoint of $\fe_8$ decomposes as
\begin{align}
\A\to&(\A,1)_{0,0}+(1,\A)_{0,0}+(\F,\F)_{-2,0}+(\F,1)_{-2,2}+(\F,1)_{-2,-2}+(1,\F)_{0,2}+\nn\\
&(1,\F)_{0,-2}+(\F,\S)_{1,1}+(\F,\C)_{1,-1}+(\bar\F,\S)_{-1,-1}+(\bar\F,\C)_{-1,1}+\nn\\
&(\F,1)_{4,0}+(\bar\F,1)_{-4,0}+(\bar\F,\F)_{2,0}+(\bar\F,1)_{2,2}+(\bar\F,1)_{2,-2}+\nn\\
&(1,\S)_{-3,1}+(1,\C)_{-3,-1}+(1,\S)_{3,-1}+(1,\C)_{3,1}
\end{align}
under $\fe_8\to\su(3)\oplus\so(8)\oplus\u(1)^2$. From this, we see that if we perform an outer-automorphism twist on $\so(8)$ and no outer-automorphism twist, then one of the $\u(1)$s is projected out. Similarly, if we perform outer-automorphism twist on both $\so(8)$ and $\su(3)$, then again one of the $\u(1)$s is projected out. However, if we perform outer-automorphims twist on $\su(3)$ but no outer-automorphism twist on $\so(8)$, then both the $\u(1)$s are projected out, and the resulting $5d$ KK theory configuration has no continuous flavor symmetry, thus this configuration does not appear in the tables.
\item A $6d$ SCFT configuration of the form
\be
\begin{tikzpicture}
\node (v1) at (-0.5,0.4) {1};
\node (v7) at (-0.5,0.85) {$\sp(0)^{(1)}$};
\begin{scope}[shift={(1.8,-0.05)}]
\node (v2) at (-0.5,0.45) {$k$};
\node at (-0.45,0.9) {$\fg^{(1)}$};
\end{scope}
\node (v3) at (-2.7,0.4) {$\left[\fh_2^{(1)}\right]$};
\node (v5) at (-0.5,2.7) {$\left[\fh_1^{(1)}\right]$};
\node (v6) at (-0.5,1.7) {\tiny{$f_1$}};
\draw  (v7) edge (v6);
\draw (v6) edge (v5);
\node (v4) at (-1.4,0.4) {\tiny{$f_2$}};
\draw  (v4) edge (v1);
\draw  (v3) edge (v4);
\begin{scope}[shift={(2.6,0)}]
\node (v1_1) at (0.5,0.4) {1};
\node (v7_1) at (0.5,0.85) {$\sp(0)^{(1)}$};
\node (v3_1) at (2.7,0.4) {$\left[\fh_2^{(1)}\right]$};
\node (v5_1) at (0.5,2.7) {$\left[\fh_1^{(1)}\right]$};
\node (v6_1) at (0.5,1.7) {\tiny{$f_1$}};
\node (v4_1) at (1.4,0.4) {\tiny{$f_2$}};
\end{scope}
\draw  (v1_1) edge (v2);
\draw  (v1) edge (v2);
\draw  (v6_1) edge (v5_1);
\draw  (v6_1) edge (v7_1);
\draw  (v1_1) edge (v4_1);
\draw  (v4_1) edge (v3_1);
\end{tikzpicture}
\ee
can simply be folded to result in a $5d$ KK theory configuration of the form
\be
\begin{tikzpicture}
\node (v1) at (-0.5,0.4) {1};
\node (v7) at (-0.5,0.85) {$\sp(0)^{(1)}$};
\begin{scope}[shift={(1.8,-0.05)}]
\node (v2) at (-0.5,0.45) {$k$};
\node at (-0.45,0.9) {$\fg^{(1)}$};
\end{scope}
\node (v3) at (-2.7,0.4) {$\left[\fh_2^{(1)}\right]$};
\node (v5) at (-0.5,2.7) {$\left[\fh_1^{(1)}\right]$};
\node (v6) at (-0.5,1.7) {\tiny{$f_1$}};
\draw  (v7) edge (v6);
\draw (v6) edge (v5);
\node (v6) at (0.4,0.4) {\tiny{$2$}};
\draw [<-] (v1) edge (v6);
\draw  (v6) edge (v2);
\node (v4) at (-1.4,0.4) {\tiny{$f_2$}};
\draw  (v4) edge (v1);
\draw  (v3) edge (v4);
\end{tikzpicture}
\ee
which is $e=2$ version of the first entry in Table \ref{TR11}. Similarly, the $e=3$ version is produced by folding the $6d$ SCFT configuration of the form
\be
\begin{tikzpicture}
\node (v1) at (-0.5,0.4) {1};
\node (v7) at (-0.5,0.85) {$\sp(0)^{(1)}$};
\begin{scope}[shift={(1.8,-0.05)}]
\node (v2) at (-0.5,0.45) {$k$};
\node at (-0.45,0.9) {$\fg^{(1)}$};
\end{scope}
\node (v3) at (-2.7,0.4) {$\left[\fh_2^{(1)}\right]$};
\node (v5) at (-0.5,2.7) {$\left[\fh_1^{(1)}\right]$};
\node (v6) at (-0.5,1.7) {\tiny{$f_1$}};
\draw  (v7) edge (v6);
\draw (v6) edge (v5);
\node (v4) at (-1.4,0.4) {\tiny{$f_2$}};
\draw  (v4) edge (v1);
\draw  (v3) edge (v4);
\begin{scope}[shift={(2.6,0)}]
\node (v1_1) at (0.5,0.4) {1};
\node (v7_1) at (0.5,0.85) {$\sp(0)^{(1)}$};
\node (v3_1) at (2.7,0.4) {$\left[\fh_2^{(1)}\right]$};
\node (v5_1) at (0.5,2.7) {$\left[\fh_1^{(1)}\right]$};
\node (v6_1) at (0.5,1.7) {\tiny{$f_1$}};
\node (v4_1) at (1.4,0.4) {\tiny{$f_2$}};
\draw  (v1_1) edge (v2);
\draw  (v6_1) edge (v5_1);
\draw  (v6_1) edge (v7_1);
\draw  (v1_1) edge (v4_1);
\draw  (v4_1) edge (v3_1);
\end{scope}
\draw  (v1) edge (v2);
\begin{scope}[shift={(0.9,-0.8)},rotate=-90]
\node (v1_1) at (0.5,0.4) {1};
\node (v7_1) at (1,0.5) {$\sp(0)^{(1)}$};
\node (v3_1) at (0.5,-2) {$\left[\fh_2^{(1)}\right]$};
\node (v5_1) at (0.5,2.7) {$\left[\fh_1^{(1)}\right]$};
\node (v6_1) at (0.5,1.4) {\tiny{$f_1$}};
\node (v4_1) at (0.5,-0.6) {\tiny{$f_2$}};
\draw  (v1_1) edge (v2);
\draw  (v6_1) edge (v5_1);
\draw  (v4_1) edge (v3_1);
\end{scope}
\draw  (v1_1) edge (v6_1);
\draw  (v4_1) edge (v1_1);
\end{tikzpicture}
\ee
Similarly, one can also obtain $e>3$ versions of the entry. In a similar fashion, one can construct the first entry of Table \ref{TR12}.
\item The $5d$ KK theory configuration
\be
\begin{tikzpicture}
\node (v1) at (-0.5,0.4) {1};
\node (v7) at (-0.5,0.85) {$\sp(0)^{(1)}$};
\begin{scope}[shift={(1.8,-0.05)}]
\node (v2) at (-0.5,0.45) {$k$};
\node at (-0.45,0.9) {$\fe_6^{(2)}$};
\end{scope}
\node (v6) at (0.4,0.4) {\tiny{2}};
\draw [<-] (v1) edge (v6);
\draw (v6) edge (v2);
\end{tikzpicture}
\ee
is constructed from the $6d$ SCFT configuration
\be
\begin{tikzpicture}
\node (v1) at (-0.5,0.4) {1};
\node (v7) at (-0.5,0.85) {$\sp(0)$};
\begin{scope}[shift={(1.8,-0.05)}]
\node (v2) at (-0.5,0.45) {$k$};
\node at (-0.45,0.9) {$\fe_6$};
\end{scope}
\node (v3) at (-2.6,0.4) {$\left[\su(3)\right]$};
\node (v5) at (5.2,0.4) {$\left[\su(3)\right]$};
\begin{scope}[shift={(2.6,0)}]
\node (v1_1) at (0.5,0.4) {1};
\node (v7_1) at (0.5,0.85) {$\sp(0)$};
\end{scope}
\draw  (v1) edge (v2);
\draw  (v2) edge (v1_1);
\draw  (v3) edge (v1);
\draw  (v1_1) edge (v5);
\end{tikzpicture}
\ee
where we now have two BPS strings to account for. Let us choose an outer-automorphism frame for the left $\su(3)$ so that the left BPS string transforms as
\be
(\A,1)+(1,\A)+(\F,\F)+(\bar\F,\bar\F)
\ee
under $\su(3)_\text{left}\oplus\fe_{6}$ and an outer-automorphism frame for the right $\su(3)$ so that the right BPS string transforms as
\be
(\A,1)+(1,\A)+(\bar\F,\F)+(\F,\bar\F)
\ee
under $\su(3)_\text{right}\oplus\fe_{6}$. Then, it is clear that exchanging the two strings and the two $\su(3)$ while simultaneously performing outer-automorphism on $\fe_6$ is a symmetry and no outer-automorphism of $\su(3)$ is needed. Thus, the full $5d$ KK theory configuration displaying the flavor symmetry is
\be
\begin{tikzpicture}
\node (v1) at (-0.5,0.4) {1};
\node (v7) at (-0.5,0.85) {$\sp(0)^{(1)}$};
\begin{scope}[shift={(1.8,-0.05)}]
\node (v2) at (-0.5,0.45) {$k$};
\node at (-0.45,0.95) {$\fe_6^{(2)}$};
\end{scope}
\node (v4) at (-2.7,0.4) {$\left[\su(3)^{(1)}\right]$};
\draw  (v1) edge (v4);
\node (v6) at (0.4,0.4) {\tiny{2}};
\draw [<-] (v1) edge (v6);
\draw (v6) edge (v2);
\end{tikzpicture}
\ee
Now, the $5d$ KK theory configuration
\be
\begin{tikzpicture}
\node (v1) at (-0.5,0.4) {1};
\node (v7) at (-0.5,0.85) {$\sp(0)^{(1)}$};
\begin{scope}[shift={(1.8,-0.05)}]
\node (v2) at (-0.5,0.45) {$k$};
\node at (-0.45,0.9) {$\fe_6^{(2)}$};
\end{scope}
\node (v6) at (0.4,0.4) {\tiny{3}};
\draw [<-] (v1) edge (v6);
\draw (v6) edge (v2);
\end{tikzpicture}
\ee
is constructed from the $6d$ SCFT configuration
\be
\begin{tikzpicture}
\node (v1) at (-0.5,0.4) {1};
\node (v7) at (-0.5,0.85) {$\sp(0)$};
\begin{scope}[shift={(1.8,-0.05)}]
\node (v2) at (-0.5,0.45) {$k$};
\node (v4) at (-0.5,0.9) {$\fe_6$};
\end{scope}
\node (v3) at (-2.6,0.4) {$\left[\su(3)\right]$};
\node (v5) at (5.2,0.4) {$\left[\su(3)\right]$};
\begin{scope}[shift={(2.6,0)}]
\node (v1_1) at (0.5,0.4) {1};
\node (v7_1) at (0.5,0.85) {$\sp(0)$};
\end{scope}
\draw  (v1) edge (v2);
\draw  (v2) edge (v1_1);
\draw  (v3) edge (v1);
\draw  (v1_1) edge (v5);
\begin{scope}[rotate=270, shift={(-1.9,0.7)}]
\node (v1) at (-0.2,0.6) {1};
\node (v7) at (-0.65,0.6) {$\sp(0)$};
\node (v3) at (-0.2,-1.4) {$\left[\su(3)\right]$};
\end{scope}
\draw  (v1) edge (v4);
\draw  (v3) edge (v1);
\end{tikzpicture}
\ee
Now there is no way to choose outer-automorphism frame for the three $\su(3)$s so that a cyclic permutation of the three $\su(3)$ combined with an outer-automorphism of $\fe_6$ is a symmetry of the spectrum of BPS strings. to convert the above operation into a symmetry, one is forced to insert a relative outer-automorphism between the three $\su(3)$s while cyclically permuting them. Thus, the corresponding $5d$ KK theory configuration with flavor symmetry included is
\be
\begin{tikzpicture}
\node (v1) at (-0.5,0.4) {1};
\node (v7) at (-0.5,0.85) {$\sp(0)^{(1)}$};
\begin{scope}[shift={(1.8,-0.05)}]
\node (v2) at (-0.5,0.45) {$k$};
\node at (-0.45,0.95) {$\fe_6^{(2)}$};
\end{scope}
\node (v4) at (-2.7,0.4) {$\left[\su(3)^{(2)}\right]$};
\draw  (v1) edge (v4);
\node (v6) at (0.4,0.4) {\tiny{3}};
\draw [<-] (v1) edge (v6);
\draw (v6) edge (v2);
\end{tikzpicture}
\ee
\eit

\subsubsection{$\Omega$ Matrix associated to a $5d$ KK theory}
As we saw above, there is a graph that can be associated to any $5d$ KK theory which also incorporates the flavor nodes. In the case of $6d$ SCFTs, we associated matrix $\Omega$ to such a graph where the diagonal entry of $\Omega$ for flavor nodes was not specified since it is not well-defined. In the same fashion, we can associate a matrix $\Omega$ to the graph of a $5d$ KK theory, which was done for non-flavor nodes in \cite{Bhardwaj:2019fzv}. Their analysis can be extended to incorporate flavor nodes. If we have a $5d$ KK theory configuration of the form
\be
\begin{tikzpicture}
\node (v1) at (-0.5,0.4) {$k_i$};
\node (v7) at (-0.35,0.95) {$\fg_i^{(q_i)}$};
\begin{scope}[shift={(2.1,-0.05)}]
\node (v2) at (-0.5,0.45) {$k_j$};
\node at (-0.35,1) {$\fg_j^{(q_j)}$};
\end{scope}
\node (v3) at (0.55,0.4) {\tiny{$e$}};
\draw  (v1) edge (v3);
\draw[->]  (v3) -- (v2);
\end{tikzpicture}
\ee
then we associate to it $\Omega^{ii}=k_i,\Omega^{jj}=k_j,\Omega^{ij}=-e,\Omega^{ji}=-1$. This rule extends over to flavor nodes, and for the $5d$ KK theory configuration of the form
\be
\begin{tikzpicture}
\node (v1) at (-0.5,0.4) {$k_i$};
\node (v7) at (-0.4,0.95) {$\fg_i^{(q_i)}$};
\begin{scope}[shift={(2.5,-0.05)}]
\node (v2) at (-0.5,0.45) {$\left[\fg_j^{(q_j)}\right]$};
\end{scope}
\node (v3) at (0.55,0.4) {\tiny{$e$}};
\draw  (v1) edge (v3);
\draw[->]  (v3) -- (v2);
\end{tikzpicture}
\ee
we associate $\Omega^{ii}=k_i,\Omega^{ij}=-e,\Omega^{ji}=-1$ where $\Omega^{jj}$ is not well-defined. Other rules for converting the graph to $\Omega$ are the same as in the case of $6d$ SCFTs, with the only exception being that to the node
\be
\begin{tikzpicture}
\node (v1) at (-0.5,0.4) {2};
\node (v7) at (-0.5,0.9) {$\su(n)^{(1)}$};
\draw (v1) .. controls (-1.05,-0.45) and (0.05,-0.45) .. (v1);
\end{tikzpicture}
\ee
we associate $\Omega^{ii}=1$ rather than $\Omega^{ii}=2$.

\subsection{Delocalized flavor symmetries}\label{delocal5d}
Now we turn to a discussion of delocalized flavor symmetries in $5d$ KK theories that descend from delocalized flavor symmetries in $6d$ SCFTs (see Section \ref{delocal6d}). The nodes responsible for the delocalization are of the form
\be\label{dnodeK}
\begin{tikzpicture}
\node (v1) at (-0.5,0.45) {$k$};
\node at (-0.45,0.9) {$\su(n)^{(1)}$};
\end{tikzpicture}
\ee
except for the $n=2,k=1$ case. These nodes remove $\u(1)^{(1)}$ and $\su(2)^{(1)}$ factors from the naive flavor symmetry associated to a $5d$ KK theory.

The counting of $\u(1)^{(1)}$ factors in the naive flavor symmetry algebra deserves some comments since the outer-automorphism twists may project out some of the $\u(1)$ factors in the flavor symmetry of the parent $6d$ SCFT. For example, in the $5d$ KK theory configuration
\be
\begin{tikzpicture}
\node (v1) at (-0.5,0.4) {$k$};
\node (v7) at (-0.5,0.9) {$\su(m)^{(2)}$};
\begin{scope}[shift={(2.1,-0.05)}]
\node (v2) at (-0.5,0.45) {$l$};
\node at (-0.45,0.95) {$\su(n)^{(2)}$};
\end{scope}
\draw  (v1) edge (v2);
\end{tikzpicture}
\ee
there is no $\u(1)^{(1)}$ flavor factor associated to the edge since the $\u(1)$ flavor factor associated to the edge in the parent $6d$ SCFT configuration
\be
\begin{tikzpicture}
\node (v1) at (-0.5,0.4) {$k$};
\node (v7) at (-0.5,0.85) {$\su(m)$};
\begin{scope}[shift={(2.1,-0.05)}]
\node (v2) at (-0.5,0.45) {$l$};
\node at (-0.45,0.9) {$\su(n)$};
\end{scope}
\draw  (v1) edge (v2);
\end{tikzpicture}
\ee
is complex conjugated by the outer-automorphism implementing the twist, and hence the $\u(1)$ factor is projected out upon carrying out the twist. Similarly, there is not $\u(1)^{(1)}$ factor associated to the edge in the following $5d$ KK theory configuration
\be
\begin{tikzpicture}
\node (v1) at (-0.5,0.4) {$k$};
\node (v7) at (-0.5,0.9) {$\su(m)^{(2)}$};
\begin{scope}[shift={(2.1,-0.05)}]
\node (v2) at (-0.5,0.45) {$l$};
\node at (-0.45,0.95) {$\su(n)^{(2)}$};
\end{scope}
\node (v3) at (0.55,0.4) {\tiny{$e$}};
\draw  (v1) edge (v3);
\draw[->]  (v3) -- (v2);
\end{tikzpicture}
\ee
On the other hand, there is a single $\u(1)^{(1)}$ factor associated to the edge in
\be
\begin{tikzpicture}
\node (v1) at (-0.5,0.4) {$k$};
\node (v7) at (-0.5,0.9) {$\su(m)^{(1)}$};
\begin{scope}[shift={(2.1,-0.05)}]
\node (v2) at (-0.5,0.45) {$l$};
\node at (-0.45,0.95) {$\su(n)^{(1)}$};
\end{scope}
\node (v3) at (0.55,0.4) {\tiny{$e$}};
\draw  (v1) edge (v3);
\draw[->]  (v3) -- (v2);
\end{tikzpicture}
\ee
For example for $e=2$, the above configuration is produced by applying a permutation twist on the following $6d$ SCFT configuration
\be
\begin{tikzpicture}
\node (v1) at (-0.5,0.4) {$k$};
\node (v7) at (-0.5,0.85) {$\su(m)$};
\begin{scope}[shift={(2.1,-0.05)}]
\node (v2) at (-0.5,0.45) {$l$};
\node at (-0.45,0.9) {$\su(n)$};
\end{scope}
\draw  (v1) edge (v2);
\begin{scope}[shift={(-2.25,-0.05)}]
\node (v3) at (-0.5,0.45) {$l$};
\node at (-0.45,0.9) {$\su(n)$};
\end{scope}
\draw  (v3) edge (v1);
\end{tikzpicture}
\ee
which has a $\u(1)$ factor associated to each edge, but the two $\u(1)$ factors are combined into a single $\u(1)^{(1)}$ factor by the permutation twist.

Once the naive flavor symmetry algebra for the $5d$ KK theory has been figured out, the rule is that each node of the form (\ref{dnodeK}) for $n\ge3$ removes a $\u(1)^{(1)}$ factor, thus delocalizing other $\u(1)^{(1)}$ factors. The effect of (\ref{dnodeK}) for $n=k=2$ is same as for the case of $6d$ SCFTs. If such a node appears without a gauge theoretic neighbor then it modifies the naive flavor symmetry localized at the node as $\so(8)^{(1)}\to\so(7)^{(1)}$ or $\so(7)^{(1)}\to\fg_2^{(1)}$ or $\so(6)^{(1)}\to\fg_2^{(1)}$ or $\so(5)^{(1)}\to\su(2)^{(1)}$. If such a node appears such that naive flavor symmetry localized at the node is $\so(4)^{(1)}$ or $\so(3)^{(1)}$, then it removes an $\su(2)^{(1)}$. If such a node appears such that naive flavor symmetry localized at the node is $\so(2)^{(1)}$, then it removes a $\u(1)^{(1)}$. If such a node appears such that naive flavor symmetry localized at the node is trivial but the naive flavor symmetry localized at a neighboring edge is $\u(1)^{(1)}$, then it removes a $\u(1)^{(1)}$. If such a node appears such that naive flavor symmetry localized at the node is trivial but the naive flavor symmetry localized at a neighboring edge is $\su(2)^{(1)}$, then it removes an $\su(2)^{(1)}$. If such a node appears such that naive flavor symmetry localized at the node is trivial and the naive flavor symmetry localized at all the neighboring edges is trivial, then it removes nothing. Below we present several examples demonstrating the rules discussed in this paragraph.

So, for example, the configuration
\be
\begin{tikzpicture}
\node (v1) at (-0.5,0.4) {$2$};
\node (v7) at (-0.5,0.9) {$\su(2)^{(1)}$};
\begin{scope}[shift={(2.1,-0.05)}]
\node (v2) at (-0.5,0.45) {$2$};
\node at (-0.45,0.95) {$\su(1)^{(1)}$};
\end{scope}
\draw  (v1) edge (v2);
\end{tikzpicture}
\ee
has a $\fg_2^{(1)}$ flavor symmetry. The configuration
\be
\begin{tikzpicture}
\node (v1) at (-0.5,0.4) {$2$};
\node (v7) at (-0.5,0.9) {$\su(2)^{(1)}$};
\begin{scope}[shift={(2.1,-0.05)}]
\node (v2) at (-0.5,0.45) {$2$};
\node at (-0.45,0.95) {$\su(1)^{(1)}$};
\end{scope}
\draw  (v1) edge (v2);
\begin{scope}[shift={(-2.2,-0.05)}]
\node (v3) at (-0.5,0.45) {$2$};
\node at (-0.45,0.95) {$\su(1)^{(1)}$};
\end{scope}
\draw  (v3) edge (v1);
\end{tikzpicture}
\ee
also has a $\fg_2^{(1)}$ flavor symmetry which is viewed as a non-standard affinization of the $\su(3)$ flavor symmetry of the $6d$ SCFT. This phenomenon is discussed in the detailed analysis for (2.77) in Part 2 of this series of papers. The configuration
\be
\begin{tikzpicture}
\node (v1) at (-0.5,0.4) {$2$};
\node (v7) at (-0.5,0.9) {$\su(2)^{(1)}$};
\begin{scope}[shift={(2.1,-0.05)}]
\node (v2) at (-0.5,0.45) {$2$};
\node at (-0.45,0.95) {$\su(1)^{(1)}$};
\end{scope}
\draw  (v1) edge (v2);
\begin{scope}[shift={(-2.2,-0.05)}]
\node (v3) at (-0.5,0.45) {$2$};
\node at (-0.45,0.95) {$\su(1)^{(1)}$};
\end{scope}
\draw  (v3) edge (v1);
\begin{scope}[shift={(0,2.05)}]
\node (v4) at (-0.5,0.45) {$2$};
\node at (-0.45,0.95) {$\su(1)^{(1)}$};
\end{scope}
\draw  (v4) edge (v7);
\end{tikzpicture}
\ee
has an $\su(2)^{(1)}$ flavor symmetry. The configuration
\be
\begin{tikzpicture}
\node (v1) at (-0.5,0.4) {$2$};
\node (v7) at (-0.5,0.9) {$\su(2)^{(1)}$};
\begin{scope}[shift={(2.1,-0.05)}]
\node (v2) at (-0.5,0.45) {$2$};
\node at (-0.45,0.95) {$\su(2)^{(1)}$};
\end{scope}
\draw  (v1) edge (v2);
\end{tikzpicture}
\ee
has an $(\su(2)^{(1)})^{\oplus 5}$ naive flavor symmetry, but the true flavor symmetry is $(\su(2)^{(1)})^{\oplus 3}$ since two of the $\su(2)^{(1)}$ flavor factors are removed by the two $\su(2)^{(1)}$ gauge nodes. The configuration
\be
\begin{tikzpicture}
\node (v1) at (-0.5,0.4) {$2$};
\node (v7) at (-0.5,0.9) {$\su(2)^{(1)}$};
\begin{scope}[shift={(2.1,-0.05)}]
\node (v2) at (-0.5,0.45) {$2$};
\node at (-0.45,0.95) {$\su(2)^{(1)}$};
\end{scope}
\draw  (v1) edge (v2);
\begin{scope}[shift={(-2.2,-0.05)}]
\node (v3) at (-0.5,0.45) {$2$};
\node at (-0.45,0.95) {$\su(1)^{(1)}$};
\end{scope}
\draw  (v3) edge (v1);
\end{tikzpicture}
\ee
has an $(\su(2)^{(1)})^{\oplus 4}$ naive flavor symmetry, but the true flavor symmetry is $(\su(2)^{(1)})^{\oplus 2}$. The configuration
\be
\begin{tikzpicture}
\node (v1) at (-0.5,0.4) {$2$};
\node (v7) at (-0.5,0.9) {$\su(2)^{(1)}$};
\begin{scope}[shift={(2.1,-0.05)}]
\node (v2) at (-0.5,0.45) {$2$};
\node at (-0.45,0.95) {$\su(2)^{(1)}$};
\end{scope}
\draw  (v1) edge (v2);
\begin{scope}[shift={(-2.2,-0.05)}]
\node (v3) at (-0.5,0.45) {$2$};
\node at (-0.45,0.95) {$\su(1)^{(1)}$};
\end{scope}
\draw  (v3) edge (v1);
\begin{scope}[shift={(0,2.05)}]
\node (v4) at (-0.5,0.45) {$2$};
\node at (-0.45,0.95) {$\su(1)^{(1)}$};
\end{scope}
\draw  (v4) edge (v7);
\end{tikzpicture}
\ee
has an $(\su(2)^{(1)})^{\oplus 3}\oplus\u(1)^{(1)}$ naive flavor symmetry, but the true flavor symmetry is $(\su(2)^{(1)})^{\oplus 2}$. The configuration
\be
\begin{tikzpicture}
\node (v1) at (-0.5,0.4) {$2$};
\node (v7) at (-0.5,0.9) {$\su(2)^{(1)}$};
\begin{scope}[shift={(2.1,-0.05)}]
\node (v2) at (-0.5,0.45) {$2$};
\node at (-0.45,0.95) {$\su(2)^{(1)}$};
\end{scope}
\draw  (v1) edge (v2);
\begin{scope}[shift={(-2.2,-0.05)}]
\node (v3) at (-0.5,0.45) {$2$};
\node at (-0.45,0.95) {$\su(2)^{(1)}$};
\end{scope}
\draw  (v3) edge (v1);
\end{tikzpicture}
\ee
has an $(\su(2)^{(1)})^{\oplus 6}$ naive flavor symmetry, but the true flavor symmetry is $(\su(2)^{(1)})^{\oplus 3}$. The configuration
\be
\begin{tikzpicture}
\node (v1) at (-0.5,0.4) {$2$};
\node (v7) at (-0.5,0.9) {$\su(2)^{(1)}$};
\begin{scope}[shift={(2.1,-0.05)}]
\node (v2) at (-0.5,0.45) {$3$};
\node at (-0.45,0.95) {$\fg_2^{(1)}$};
\end{scope}
\draw  (v1) edge (v2);
\end{tikzpicture}
\ee
has a $\u(1)^{(1)}$ true flavor symmetry arising from the radius of circle compactification. The configuration
\be
\begin{tikzpicture}
\node (v1) at (-0.5,0.4) {$2$};
\node (v7) at (-0.5,0.9) {$\su(2)^{(1)}$};
\begin{scope}[shift={(2.1,-0.05)}]
\node (v2) at (-0.5,0.45) {$2$};
\node at (-0.45,0.95) {$\su(2)^{(1)}$};
\end{scope}
\node (v3) at (0.55,0.4) {\tiny{2}};
\draw  (v1) edge (v3);
\draw [->] (v3) edge (v2);
\end{tikzpicture}
\ee
has an $(\su(2)^{(1)})^{\oplus 3}$ naive flavor symmetry, but the true flavor symmetry is $\su(2)^{(1)}$.

\section{Gluing rules}\label{GR}
To every $5d$ KK theory, one can associated a CY3 such that M-theory compactified on it manufactures the $5d$ KK theory on its Coulomb branch. The CY3 surface geometry can be constructed using the data of the graph (restricted to non-flavor nodes) associated to the $5d$ KK theory. To each non-flavor node one associates a collection of compact surfaces intersecting each other. To an edge between two non-flavor nodes, one associates a set of ``gluing rules'' which describe how the two collections of surfaces associated to the two nodes intersect each other. See \cite{Bhardwaj:2019fzv,Bhardwaj:2020kim,Bhardwaj:2018vuu,Bhardwaj:2018yhy} for more details.

In this section, we expand their procedure to incorporate flavor nodes (carrying non-abelian flavor symmetries) into the geometry. For localized flavor symmetries, the story is the same as above. One associates a collection of non-compact $\P^1$ fibered surfaces (intersecting each other) to each such flavor node. For each edge between a flavor and a non-flavor node, one associates a set of gluing rules which describe how the collection of non-compact $\P^1$ fibered surfaces associated to the flavor node intersect the collection of compact surfaces associated to the non-flavor node. For delocalized flavor symmetries, the gluing rules describe how the collection of non-compact $\P^1$ fibered surfaces associated to the flavor symmetry intersect the collection of compact surfaces associated to the non-flavor \emph{nodes} over which the flavor symmetry is delocalized.

From this point on, we use a lot of notation and geometric background which can be found in Section 5 and Appendix A of \cite{Bhardwaj:2019fzv}. We also explain a lot of this notation example by example as we move through this section.

\subsection{Description of $\P^1$ fibered surfaces}\label{NN}
In this subsection, we describe a notation for $\P^1$ fibered surfaces that we use to describe gluing rules in this paper and Part 2 \cite{Bhardwaj:2020avz}. We will denote such surfaces as $\mathbf{N}_i$ or $\mathbf{M}_i$ etc. where $i$ is simply a label to distinguish different surfaces. We also add a superscript sometimes and write $\mathbf{N}^p_i$ which means that the $\P^1$ fibered surface $\mathbf{N}_i$ carries $p$ blowups. We denote the $\P^1$ fibers as $f_i$ and various blowups as $x_a,y_a$ etc.

We label an arbitrary section of $\mathbf{N}$ which intersects $f$ at $p$ points as $pe+\sum \alpha_a x_a$ (where $x_a$ are some blowups and $\alpha_a$ are arbitrary coefficients) instead of displaying it as $pe+\sum \alpha_a x_a+\beta f$. That is, we do not display the number of $f$ involved in the section. The rationale for this is that
\be
f\cdot (pe+\sum \alpha_a x_a)=f\cdot (pe+\sum \alpha_a x_a+\beta f)
\ee
and
\be
x_b\cdot (pe+\sum \alpha_a x_a)=x_b\cdot (pe+\sum \alpha_a x_a+\beta f)
\ee
Since only the above two types of intersection numbers are relevant while discussing gluing rules, we see that we can omit the data about the number of $f$ involved in the section, which can be easily restored using the general consistency conditions discussed in Section 5 and Appendix A of \cite{Bhardwaj:2019fzv}. 

For a collection $\mathbf{N}_i$ of \emph{non-compact} $\P^1$ fibered surfaces, we also do not display individual gluing curves between two non-compact surfaces $\mathbf{N}_i$ and $\mathbf{N}_j$ if those gluing curves are non-compact. So, if $\mathbf{N}_0$ and $\mathbf{N}_1$ are non-compact, the geometry
\be
\begin{tikzpicture} [scale=1.9]
\node (v3) at (2.7,0.2) {$\mathbf{N}^4_0$};
\node (v4) at (4.7,0.2) {$\mathbf{N}_1$};
\draw  (v3) edge (v4);
\node[rotate=0] at (3.45,0.3) {\scriptsize{$e,e+2f$-$\sum x_i$}};
\node[rotate=0] at (4.3,0.3) {\scriptsize{$e,e$}};
\end{tikzpicture}
\ee
is represented as the geometry
\be
\begin{tikzpicture} [scale=1.9]
\node (v3) at (2.9,0.2) {$\mathbf{N}^4_0$};
\node (v4) at (4.7,0.2) {$\mathbf{N}_1$};
\draw  (v3) edge (v4);
\node[rotate=0] at (3.4,0.3) {\scriptsize{$2e$-$\sum x_i$}};
\node[rotate=0] at (4.3,0.3) {\scriptsize{$2e$}};
\end{tikzpicture}
\ee
where we have replaced $e+2f$ by $e$ and combined the two gluing curves together. This is because the splitting of a non-compact gluing curve between two non-compact surfaces into multiple non-compact gluing curves is not relevant for any physical information about the resulting $5d$ theory.

The above notation also allows us to present the gluing rules between two compact $\P^1$ fibered surfaces, or a compact $\P^1$ fibered and a non-compact $\P^1$ fibered surface, or two non-compact $\P^1$ fibered surfaces on an equal footing in what follows.

\subsection{Condition on the gluing rules}
In this subsection, we recall an important condition that gluing rules are supposed to satisfy \cite{Bhardwaj:2020gyu,Bhardwaj:2020phs}. Let us label the surfaces associated to node $i$ (which could be flavor or non-flavor) as $S_{i,a}$ where different values of $a$ label different surfaces associated to this node $i$. Then to this node we assign a linear combination of surfaces
\be
S_i:=\sum_ad^\vee_{i,a}S_{i,a}
\ee
where $d^\vee_{i,a}$ are dual Coxeter numbers\footnote{These numbers are the smallest positive integers which satisfy $\sum_bM_{i,ab}d^\vee_{i,b}=0$ where $M_{i,ab}$ is the Cartan matrix of $\fg_i^{(q_i)}$.} associated to the affine algebra $\fg_i^{(q_i)}$ associated to this node. If the node does not have a non-trivial algebra associated to it, then there is only a single surface $S_{i,0}$ and we define $S_i=S_{i,0}$.

Moreover, a curve $\tilde e_i$ for each collection of surfaces $S_{i,a}$ was defined in Section 3.3.3 of \cite{Bhardwaj:2020phs}. Then, we must have
\be
-S_{i}\cdot\tilde e_j=\Omega^{ij}
\ee
for all $i$ and $j$ including localized flavor nodes. This curve $\tilde e_i$ is defined as follows. Due to the gluing rules among different surfaces $S_{i,a}$ for fixed $i$ and different $a$, we can in general write a multiple of $e_{i,a}$ for arbitrary $a$ in terms of $e_{i,b}$ for a fixed $b$ as
\be
m_{ab}e_{i,a}=n_{ab}e_{i,b}+\sum_c p_{abc}f_{i,c}+\sum_\alpha q_{ab\alpha} x_\alpha
\ee
as homology classes in the full CY3, where $x_\alpha$ are arbitrary blowups living in surfaces $S_{i,a}$, and $m_{ab},n_{ab},p_{abc},q_{ab\alpha}$ are some coefficients determined by the gluing rules, and we impose the condition that $\text{gcd}(m_{ab},n_{ab},p_{abc},q_{ab\alpha})=1$. Now $\tilde e_i:=e_{i,b}$ for that particular $b$ such that $m_{ab}=1$ for all $a$. There are be multiple $b$ which satisfies this condition, and in such a case we can choose any such $b$ to define $\tilde e_i$ without any change in the consequences. Another fact to note is that if the algebra (flavor or non-flavor) $\fg_i^{(q_i)}$ associated to the node $i$ has $q_i=1$, that is if $\fg_i^{(q_i)}$ is an untwisted affine algebra, then we can always choose $\tilde e_{i}=e_{i,0}$ where $S_{i,0}$ corresponds to the affine node in the Dynkin diagram of $\fg_i^{(q_i)}=\fg_i^{(1)}$.

Another consistency condition for the introduction of a blowup $x$ living in any surface $S_{j,b}$ is that, we must have
\be
x\cdot S_i=0
\ee
for all $i$.

The Cartan matrix $M_{ab}$ of the algebra $\fg_i^{(q_i)}$ associated to a node $i$ can be be recovered by computing
\be
M_{ab}=-f_{i,a}\cdot S_{i,b}
\ee
and we must also have
\be
f_{i,a}\cdot S_{j,b}=0
\ee
for $i\neq j$.

\subsection{Gluing rules for localized flavor symmetries: without $\sp(0)^{(1)}$ nodes}
In this subsection, we will describe gluing rules of the form
\be\label{ge1}
\begin{tikzpicture}
\node (v1) at (-0.5,0.4) {$k_i$};
\node (v7) at (-0.4,0.95) {$\fg_i^{(q_i)}$};
\begin{scope}[shift={(2.5,-0.05)}]
\node (v2) at (-0.5,0.45) {$\left[\fg_j^{(q_j)}\right]$};
\end{scope}
\node (v3) at (0.55,0.4) {\tiny{$e$}};
\draw  (v1) edge (v3);
\draw[->]  (v3) -- (v2);
\end{tikzpicture}
\ee
or of the form
\be\label{ge2}
\begin{tikzpicture}
\node (v1) at (-0.5,0.4) {$k_i$};
\node (v7) at (-0.4,0.95) {$\fg_i^{(q_i)}$};
\begin{scope}[shift={(2.5,-0.05)}]
\node (v2) at (-0.5,0.45) {$\left[\fg_j^{(q_j)}\right]$};
\end{scope}
\node (v3) at (0.55,0.4) {\tiny{$e$}};
\draw  (v1) edge (v3);
\draw[-]  (v3) -- (v2);
\end{tikzpicture}
\ee
Actually, the gluing rules do not depend on the data of diagonal entries of $\Omega^{ii}$ or on whether the node $i$ or $j$ is a flavor or a non-flavor node, we will denote the gluing rules for (\ref{ge1}) as
\be
\begin{tikzpicture}
\node (v1) at (-0.5,0.4) {$\fg_i^{(q_i)}$};
\begin{scope}[shift={(2.5,-0.05)}]
\node (v2) at (-0.5,0.45) {$\fg_j^{(q_j)}$};
\end{scope}
\node (v3) at (0.75,0.4) {\tiny{$e$}};
\draw  (v1) edge (v3);
\draw[->]  (v3) -- (v2);
\end{tikzpicture}
\ee
and the gluing rules for (\ref{ge2}) as
\be
\begin{tikzpicture}
\node (v1) at (-0.5,0.4) {$\fg_i^{(q_i)}$};
\begin{scope}[shift={(2.5,-0.05)}]
\node (v2) at (-0.5,0.45) {$\fg_j^{(q_j)}$};
\end{scope}
\node (v3) at (0.75,0.4) {\tiny{$e$}};
\draw  (v1) edge (v3);
\draw[-]  (v3) -- (v2);
\end{tikzpicture}
\ee

Some of the gluing rules have been previously discussed in \cite{Bhardwaj:2019fzv,Bhardwaj:2020kim,Bhardwaj:2018vuu,Bhardwaj:2018yhy} as the gluing rules for two gauge nodes. We thus don't need to discuss them again, and hence only display the gluing rules which do not appear in those works. Unfortunately, we will not be able to provide the gluing rules for cases where the non-flavor node is $\fe_7^{(1)}$ since these cases are computationally quite intensive. However, we do not expect these cases to be special in any way: it should be possible to figure out the gluing rules for these cases with enough effort.

\noindent\ubf{Gluing rules for \raisebox{-.25\height}{\begin{tikzpicture}
\node (w1) at (-0.5,0.9) {$\su(4)^{(1)}$};
\begin{scope}[shift={(3,0)}]
\node (w2) at (-0.5,0.9) {$\su(2)^{(1)}$};
\end{scope}
\draw [dashed] (w1)--(w2);
\end{tikzpicture}}}:\\
Here, as discussed before, the dashed edge means that $\su(2)$ gauges an antisymmetric of $\su(4)$ rather than gauging two fundamentals of $\su(4)$ (which would be represented by a solid edge). 

The $\P^1$ fibered surfaces corresponding to $\su(4)^{(1)}$ and $\su(2)^{(1)}$ can be represented as
\be
\begin{tikzpicture} [scale=1.9]
\node (v1) at (-0.5,0.2) {$\mathbf{M}_0$};
\node (v2) at (0.6,1) {$\mathbf{M}_{3}$};
\node (v10) at (0.6,-0.6) {$\mathbf{M}_1^{1+1}$};
\node (v6) at (1.7,0.2) {$\mathbf{M}_{2}$};
\draw  (v1) edge (v2);
\draw  (v10) edge (v1);
\node[rotate=0] at (-0.3,0.5) {\scriptsize{$e$}};
\node at (-0.3,-0.1) {\scriptsize{$e$}};
\node[rotate=0] at (0.2,0.9) {\scriptsize{$e$}};
\node[rotate=0] at (1.5,0.5) {\scriptsize{$e$}};
\node at (1.5,-0.1) {\scriptsize{$e$}};
\node at (0.2,-0.5) {\scriptsize{$e$}};
\draw  (v2) edge (v10);
\node at (0.5,0.7) {\scriptsize{$f$}};
\node at (0.4,-0.1) {\scriptsize{$f$-$x$-$y$}};
\draw  (v2) edge (v6);
\draw  (v6) edge (v10);
\node at (1,0.9) {\scriptsize{$e$}};
\node at (1,-0.5) {\scriptsize{$e$}};
\node (v3) at (2.9,0.2) {$\mathbf{N}^3_0$};
\node (v4) at (4.7,0.2) {$\mathbf{N}_1$};
\draw  (v3) edge (v4);
\node[rotate=0] at (3.4,0.3) {\scriptsize{$2e$-$\sum x_i$}};
\node[rotate=0] at (4.3,0.3) {\scriptsize{$2e$}};
\end{tikzpicture}
\ee
where $\mathbf{M}_i$ parametrize $\su(4)^{(1)}$ and $\mathbf{N}_i$ parametrize $\su(2)^{(1)}$. We are using the notation for $\P^1$ fibered (compact or non-compact) surfaces described in Section \ref{NN}. Moreover, we only display blowups which are used in the process of gluing. Other blowups are omitted. We represent a surface $\mathbf{N}_i$ as
\be
\mathbf{N}_i^{b_1+b_2+b_3+b_4+\cdots}
\ee
when we want to label the $\sum_i b_i$ number of blowups of $\mathbf{N}_i$ by different alphabets. We label the first $b_1$ blowups as $x_i$ where $i=1,\cdots,b_1$, the next $b_2$ blowups as $y_i$ where $i=1,\cdots,b_2$, the next $b_3$ blowups as $z_i$ where $i=1,\cdots,b_3$, the next $b_4$ blowups as $w_i$ where $i=1,\cdots,b_4$, and so on. In keeping line with this notation, we label $\mathbf{M}_1$ in the above set of surfaces as $\mathbf{M}_1^{1+1}$ and label the two blowups as $x$ and $y$. We do not put a subscript on the blowups if there is only a single blowup of that type.

Each edge describes a gluing between the surfaces that the edge is subtended between. The labels at the end of each edge describe the gluing curve in the corresponding participating in that gluing. Whenever we write $\sum x_i$ or $\sum y_i$, we mean a sum over all the blowups of that type. So, the gluing curve $2e-\sum x_i$ living in $\mathbf{N}_0$ for its gluing with $\mathbf{N}_1$ should be read as $2e-x_1-x_2-x_3$.

The gluings between $\su(4)^{(1)}$ and $\su(2)^{(1)}$ are referred to as the \emph{gluing rules}, and are listed below:
\bit
\item $f-x,y$ in $\mathbf{M}_{1}$ are glued to $x_2,x_1$ in $\mathbf{N}_0$.
\item $x-y$ in $\mathbf{M}_{1}$ is glued to $f$ in $\mathbf{N}_1$.
\item $f$ in $\mathbf{M}_{0}$ is glued to $x_3-x_2$ in $\mathbf{N}_0$.
\item $f$ in $\mathbf{M}_{3}$ is glued to $x_2-x_1$ in $\mathbf{N}_0$.
\item $f$ in $\mathbf{M}_{2}$ is glued to $f-x_3-x_2$ in $\mathbf{N}_0$.
\eit
Whenever we say that $C_i,D_i,E_i,\cdots$ in $\mathbf{M}_i$ are glued to $C_j,D_j,E_j,\cdots$ in $\mathbf{M}_j$, the gluings are supposed to be read in order. That is, $C_i$ in $\mathbf{M}_i$ is glued to $C_j$ in $\mathbf{M}_j$, $D_i$ in $\mathbf{M}_i$ is glued to $D_j$ in $\mathbf{M}_j$, $E_i$ in $\mathbf{M}_i$ is glued to $E_j$ in $\mathbf{M}_j$, and so on.

\noindent\ubf{Gluing rules for \raisebox{-.25\height}{\begin{tikzpicture}
\node (w1) at (-0.5,0.9) {$\so(9)^{(1)}$};
\begin{scope}[shift={(3,0)}]
\node (w2) at (-0.5,0.9) {$\sp(m)^{(1)}$};
\end{scope}
\node (v1) at (1,0.9) {\tiny{2}};
\draw [dashed] (w1) edge (v1);
\draw [dashed] (v1) edge (w2);
\end{tikzpicture}}}:
\be
\begin{tikzpicture} [scale=1.9]
\node (v2) at (0.3,1) {$\mathbf{M}_{0}$};
\node (v10) at (0.3,-0.6) {$\mathbf{M}_1$};
\node (v6) at (1.4,0.2) {$\mathbf{M}_{2}$};
\node[rotate=0] at (1.2,0.5) {\scriptsize{$e$}};
\node at (1,0.1) {\scriptsize{$e$}};
\draw  (v2) edge (v6);
\draw  (v6) edge (v10);
\node at (0.7,0.9) {\scriptsize{$e$}};
\node at (0.5,-0.3) {\scriptsize{$e$}};
\node (v3) at (0.7,-1.5) {$\mathbf{N}_0$};
\node (v4) at (1.9,-1.5) {$\mathbf{N}_1$};
\draw  (v3) edge (v4);
\node[rotate=0] at (4.4,-1.4) {\scriptsize{$2e$-$\sum x_i$}};
\node[rotate=0] at (1,-1.4) {\scriptsize{$2e$}};
\node (v1) at (2.9,0.2) {$\mathbf{M}^{m+m}_{3}$};
\node (v5) at (5.2,0.2) {$\mathbf{M}_{4}$};
\draw  (v6) edge (v1);
\node[rotate=0] at (1.7,0.3) {\scriptsize{$e$}};
\node[rotate=0] at (2.3,0.3) {\scriptsize{$e$-$\sum x_i$}};
\draw  (v1) edge (v5);
\node[rotate=0] at (3.8,0.3) {\scriptsize{$2e$-$\sum x_i$-$\sum y_i$}};
\node[rotate=0] at (4.8,0.3) {\scriptsize{$e$}};
\node (v7) at (1.6,-0.2) {\scriptsize{$m$}};
\draw  (v10) edge (v7);
\draw  (v7) edge (v1);
\node at (0.7,-0.6) {\scriptsize{$f$}};
\node at (2.4,-0.1) {\scriptsize{$x_i$-$y_i$}};
\node at (1.6,-1.4) {\scriptsize{$e$}};
\node (v8) at (2.6,-1.5) {$\cdots$};
\draw  (v4) edge (v8);
\node at (2.2,-1.4) {\scriptsize{$e$}};
\node (v9) at (3.4,-1.5) {$\mathbf{N}_{m-1}$};
\draw  (v8) edge (v9);
\node at (3,-1.4) {\scriptsize{$e$}};
\node (v11) at (4.9,-1.5) {$\mathbf{N}^8_{m}$};
\draw  (v9) edge (v11);
\node at (3.8,-1.4) {\scriptsize{$e$}};
\end{tikzpicture}
\ee
A subscript in the middle of an edge between two surfaces $\mathbf{N}_i$ and $\mathbf{N}_j$ denotes the number of gluings between the two surfaces. In the above case, there are $m$ gluings between $\mathbf{M}_1$ and $\mathbf{M}_3$. The gluing is represented as $x_i-y_i$ being glued to $f$, which means that $x_i-y_i$ for each $i$ is glued to a copy of $f$, leading to a total of $m$ gluings since $i=1,\cdots,m$ as is visible from the notation $\mathbf{M}_3^{m+m}$. The gluing rules between $\mathbf{M}_i$ and $\mathbf{N}_j$ are:
\bit
\item $f-x_1-y_1$ in $\mathbf{M}_{3}$ is glued to $f$ in $\mathbf{N}_0$.
\item $x_i-x_{i+1},y_i-y_{i+1}$ in $\mathbf{M}_{3}$ are glued to $f,f$ in $\mathbf{N}_i$ for $i=1,\cdots,m-1$.
\item $f,x_m,y_m$ in $\mathbf{M}_{3}$ are glued to $x_2-x_3,x_6,x_8$ in $\mathbf{N}_m$.
\item $f,f,f,f$ in $\mathbf{M}_{4}$ are glued to $x_1-x_2,x_3-x_5,x_4-x_6,x_7-x_8$ in $\mathbf{N}_m$.
\item $f,f$ in $\mathbf{M}_{2}$ are glued to $x_3-x_4,x_5-x_6$ in $\mathbf{N}_m$.
\item $f,f$ in $\mathbf{M}_{1}$ are glued to $x_4-x_7,x_6-x_8$ in $\mathbf{N}_m$.
\item $f,f$ in $\mathbf{M}_{0}$ are glued to $f-x_1-x_5,f-x_2-x_3$ in $\mathbf{N}_m$.
\eit

\noindent\ubf{Gluing rules for \raisebox{-.25\height}{\begin{tikzpicture}
\node (w1) at (-0.5,0.9) {$\so(10)^{(1)}$};
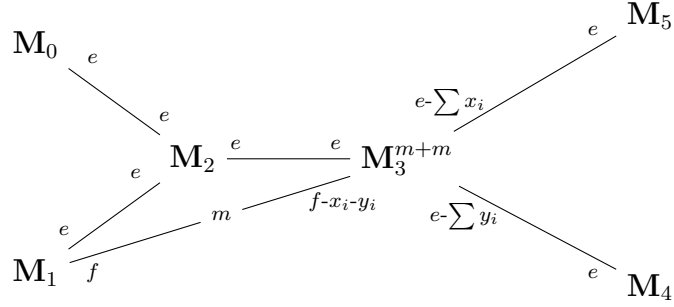
\begin{scope}[shift={(3,0)}]
\node (w2) at (-0.5,0.9) {$\su(m)^{(1)}$};
\end{scope}
\node (v1) at (1,0.9) {\tiny{4}};
\draw [dashed] (w1) edge (v1);
\draw [dashed] (v1) edge (w2);
\end{tikzpicture}}}:
\be
\begin{tikzpicture} [scale=1.9]
\node (v2) at (0.3,1) {$\mathbf{M}_{0}$};
\node (v10) at (0.3,-0.6) {$\mathbf{M}_1$};
\node (v6) at (1.4,0.2) {$\mathbf{M}_{2}$};
\node[rotate=0] at (1.2,0.5) {\scriptsize{$e$}};
\node at (1,0.1) {\scriptsize{$e$}};
\draw  (v2) edge (v6);
\draw  (v6) edge (v10);
\node at (0.7,0.9) {\scriptsize{$e$}};
\node at (0.5,-0.3) {\scriptsize{$e$}};
\node (v3) at (0.9,-1.5) {$\mathbf{N}^{8+8}_0$};
\node (v4) at (2.4,-1.5) {$\mathbf{N}_1$};
\draw  (v3) edge (v4);
\node[rotate=0] at (0.7,-1.8) {\scriptsize{$e$-$\sum x_i$}};
\node[rotate=0] at (1.5,-1.4) {\scriptsize{$e$-$\sum y_i$}};
\node (v1) at (2.9,0.2) {$\mathbf{M}^{m+m}_{3}$};
\node (v5) at (4.6,1.2) {$\mathbf{M}_{5}$};
\draw  (v6) edge (v1);
\node[rotate=0] at (1.7,0.3) {\scriptsize{$e$}};
\node[rotate=0] at (2.4,0.3) {\scriptsize{$e$}};
\draw  (v1) edge (v5);
\node[rotate=0] at (3.2,0.6) {\scriptsize{$e$-$\sum x_i$}};
\node[rotate=0] at (4.2,1.1) {\scriptsize{$e$}};
\node (v7) at (1.6,-0.2) {\scriptsize{$m$}};
\draw  (v10) edge (v7);
\draw  (v7) edge (v1);
\node at (0.7,-0.6) {\scriptsize{$f$}};
\node at (2.45,-0.1) {\scriptsize{$f$-$x_i$-$y_i$}};
\node at (2.1,-1.4) {\scriptsize{$e$}};
\node (v8) at (3.1,-1.5) {$\cdots$};
\draw  (v4) edge (v8);
\node at (2.7,-1.4) {\scriptsize{$e$}};
\node (v9) at (3.9,-1.5) {$\mathbf{N}_{m-1}$};
\draw  (v8) edge (v9);
\node at (3.5,-1.4) {\scriptsize{$e$}};
\node at (3.9,-1.8) {\scriptsize{$e$}};
\node (v12) at (4.6,-0.7) {$\mathbf{M}_{4}$};
\draw  (v1) edge (v12);
\node[rotate=0] at (3.3,-0.2) {\scriptsize{$e$-$\sum y_i$}};
\node[rotate=0] at (4.2,-0.6) {\scriptsize{$e$}};
\draw (v3) .. controls (0.9,-2) and (3.9,-2) .. (v9);
\end{tikzpicture}
\ee
\bit
\item $f,f,f-y_1,f-x_2,x_1,y_2$ in $\mathbf{M}_{3}$ are glued to $x_2-x_3,y_2-y_3,x_6,y_6,x_8,y_8$ in $\mathbf{N}_0$.
\item $x_{i+1}-x_{i+2},y_{i+2}-y_{i+1}$ in $\mathbf{M}_{3}$ are glued to $f,f$ in $\mathbf{N}_i$ for $i=1,\cdots,m-2$.
\item $x_{m}-x_{1},y_{1}-y_{m}$ in $\mathbf{M}_{3}$ are glued to $f,f$ in $\mathbf{N}_{m-1}$.
\item $f,f,f,f$ in $\mathbf{M}_{4}$ are glued to $x_1-x_2,y_3-y_5,y_4-y_6,y_7-y_8$ in $\mathbf{N}_0$.
\item $f,f,f,f$ in $\mathbf{M}_{5}$ are glued to $y_1-y_2,x_3-x_5,x_4-x_6,x_7-x_8$ in $\mathbf{N}_0$.
\item $f,f,f,f$ in $\mathbf{M}_{2}$ are glued to $x_3-x_4,y_3-y_4,x_5-x_6,y_5-y_6$ in $\mathbf{N}_0$.
\item $f,f,f,f$ in $\mathbf{M}_{1}$ are glued to $x_4-x_7,y_4-y_7,x_6-x_8,y_6-y_8$ in $\mathbf{N}_0$.
\item $f,f,f,f$ in $\mathbf{M}_{0}$ are glued to $f-x_1-y_5,f-y_1-x_5,f-x_2-y_3,f-y_2-x_3$ in $\mathbf{N}_0$.
\eit

\noindent\ubf{Gluing rules for \raisebox{-.25\height}{\begin{tikzpicture}
\node (w1) at (-0.5,0.9) {$\so(11)^{(1)}$};
\begin{scope}[shift={(3,0)}]
\node (w2) at (-0.5,0.9) {$\su(2)^{(1)}$};
\end{scope}
\node (v1) at (1,0.9) {\tiny{8}};
\draw [dashed] (w1) edge (v1);
\draw [dashed] (v1) edge (w2);
\end{tikzpicture}}}:
\be
\begin{tikzpicture} [scale=1.9]
\node (v2) at (1.4,1.5) {$\mathbf{M}^{2+2+2+2}_{0}$};
\node (v10) at (1.4,-1.1) {$\mathbf{M}^{1+1+1+1}_1$};
\node (v6) at (1.4,0.2) {$\mathbf{M}_{2}$};
\node[rotate=0] at (1.3,0.5) {\scriptsize{$e$}};
\node at (1.3,-0.1) {\scriptsize{$e$}};
\draw  (v2) edge (v6);
\draw  (v6) edge (v10);
\node at (1.3,1.2) {\scriptsize{$e$}};
\node at (1.3,-0.8) {\scriptsize{$e$}};
\node (v3) at (1.8,-1.9) {$\mathbf{N}^{16+16}_0$};
\node (v4) at (5.3,-1.9) {$\mathbf{N}^{16+16}_1$};
\draw  (v3) edge (v4);
\node[rotate=0] at (4.3,-1.8) {\scriptsize{$2e$-$2\sum x_i$-$2\sum y_i$}};
\node[rotate=0] at (2.8,-1.8) {\scriptsize{$2e$-$2\sum x_i$-$2\sum y_i$}};
\node (v1) at (2.9,0.2) {$\mathbf{M}_{3}$};
\node (v5) at (4.3,0.2) {$\mathbf{M}_{4}$};
\draw  (v6) edge (v1);
\node[rotate=0] at (1.7,0.3) {\scriptsize{$e$}};
\node[rotate=0] at (2.5,0.3) {\scriptsize{$e$}};
\draw  (v1) edge (v5);
\node[rotate=0] at (3.2,0.3) {\scriptsize{$e$}};
\node[rotate=0] at (3.8,0.3) {\scriptsize{$e$}};
\node at (2.5,0) {\scriptsize{$f$}};
\node[rotate=-40] at (1.7826,1.0607) {\scriptsize{$f$-$x_i$-$y_i$}};
\node (v7) at (5.6,0.2) {$\mathbf{M}_{5}$};
\draw  (v5) edge (v7);
\node[rotate=0] at (4.6,0.3) {\scriptsize{$2e$}};
\node[rotate=0] at (5.0947,0.2651) {\scriptsize{$e$}};
\draw  (v10) edge (v1);
\draw  (v10) edge (v5);
\node (v12) at (2.2,0.8) {\scriptsize{2}};
\node (v13) at (2.959,0.7938) {\scriptsize{2}};
\node (v14) at (3.7529,0.7634) {\scriptsize{4}};
\draw  (v2) edge (v12);
\draw  (v12) edge (v1);
\draw  (v2) edge (v13);
\draw  (v13) edge (v5);
\draw  (v2) edge (v14);
\draw  (v14) edge (v7);
\node[rotate=0] at (2.4356,0.4857) {\scriptsize{$f$}};
\node[rotate=-25] at (2.2442,1.0191) {\scriptsize{$y_i$-$z_i$}};
\node[rotate=0] at (3.481,0.4617) {\scriptsize{$f$}};
\node[rotate=-18] at (2.6415,1.221) {\scriptsize{$x_1$-$y_2,x_2$-$y_1,z_i$-$w_i$}};
\node[rotate=0] at (5.3,0.5) {\scriptsize{$f,f,f$}};
\node (v15) at (3.7,-0.4) {\scriptsize{2}};
\draw  (v10) edge (v15);
\draw  (v15) edge (v7);
\node[rotate=40] at (1.7,-0.7) {\scriptsize{$f$-$x$-$y$}};
\node[rotate=20] at (2.1,-0.7) {\scriptsize{$y$-$z$}};
\node[rotate=0] at (3.6,0) {\scriptsize{$f$}};
\node[rotate=15] at (2.2993,-0.9497) {\scriptsize{$x$-$y,z$-$w$}};
\node[rotate=0] at (5.2,-0.1) {\scriptsize{$f,f$}};
\node (w) at (1.1,-1.9) {\scriptsize{16}};
\draw (v3) .. controls (1.7,-1.7) and (1.1,-1.4) .. (w);
\draw (v3) .. controls (1.7,-2.1) and (1.1,-2.4) .. (w);
\node at (1.5,-1.5) {\scriptsize{$x_i$}};
\node at (1.5,-2.3) {\scriptsize{$y_i$}};
\node (w) at (6,-1.9) {\scriptsize{16}};
\draw (v4) .. controls (5.4,-1.7) and (6,-1.4) .. (w);
\draw (v4) .. controls (5.4,-2.1) and (6,-2.4) .. (w);
\node at (5.6,-1.5) {\scriptsize{$x_i$}};
\node at (5.6,-2.3) {\scriptsize{$y_i$}};
\end{tikzpicture}
\ee
An edge starting from a surface $\mathbf{N}_i$ and ending at the \emph{same surface} $\mathbf{N}_i$ represents a self-gluing of the surface $\mathbf{N}_i$. Above, we have 16 self-gluings of $\mathbf{N}_0$ and 16 self-gluings of $\mathbf{N}_1$.
\bit
\item $w,z,y,f-x,x,f-y,f-z,f-w,w,z,y,f-x,x,f-y,f-z,f-w$ in $\mathbf{M}_{0}$ are glued to $f-x_1,f-x_2,f-x_3,f-x_4,f-x_6,f-x_7,f-x_9,f-x_{11},f-y_1,f-y_2,f-y_3,f-y_4,f-y_6,f-y_7,f-y_9,f-y_{11}$ in $\mathbf{N}_i$ for $i=0,1$.
\item $8f$ in $\mathbf{M}_{2}$ are glued to $x_4-x_5,x_7-x_8,x_9-x_{10},x_{11}-x_{12},y_4-y_5,y_7-y_8,y_9-y_{10},y_{11}-y_{12}$ in $\mathbf{N}_i$ for $i=0,1$.
\item $8f$ in $\mathbf{M}_{3}$ are glued to $x_3-x_4,x_6-x_7,x_{10}-x_{13},x_{12}-x_{14},y_3-y_4,y_6-y_7,y_{10}-y_{13},y_{12}-y_{14}$ in $\mathbf{N}_i$ for $i=0,1$.
\item $8f$ in $\mathbf{M}_{4}$ are glued to $x_2-x_3,x_7-x_9,x_{8}-x_{10},x_{14}-x_{15},y_2-y_3,y_7-y_9,y_{8}-y_{10},y_{14}-y_{15}$ in $\mathbf{N}_i$ for $i=0,1$.
\item $16f$ in $\mathbf{M}_{5}$ are glued to $x_1-x_2,x_3-x_6,x_{4}-x_{7},x_{5}-x_{8},x_9-x_{11},x_{10}-x_{12},x_{13}-x_{14},x_{15}-x_{16},y_1-y_2,y_3-y_6,y_{4}-y_{7},y_{5}-y_{8},y_9-y_{11},y_{10}-y_{12},y_{13}-y_{14},y_{15}-y_{16}$ in $\mathbf{N}_i$ for $i=0,1$.
\item $f-w_2,f-z_2,f-y_2,f-x_1,x_2,y_1,z_1,w_1,f-w_2,f-z_2,f-y_2,f-x_1,x_2,y_1,z_1,w_1$ in $\mathbf{M}_{1}$ are glued to $x_5,x_8,x_{10}x_{12},x_{13},x_{14},x_{15},x_{16},y_5,y_8,y_{10}y_{12},y_{13},y_{14},y_{15},y_{16}$ in $\mathbf{N}_1$.
\item $f-w_1,f-z_1,f-y_1,f-x_2,x_1,y_2,z_2,w_2,f-w_1,f-z_1,f-y_1,f-x_2,x_1,y_2,z_2,w_2$ in $\mathbf{M}_{1}$ are glued to $x_5,x_8,x_{10}x_{12},x_{13},x_{14},x_{15},x_{16},y_5,y_8,y_{10}y_{12},y_{13},y_{14},y_{15},y_{16}$ in $\mathbf{N}_0$.
\eit
Here we have started abbreviating $n$ copies of the same curve $C$ in the gluing rules as $nC$. Thus, for example, the second item in the above list should be read as:
\bit
\item $f,f,f,f,f,f,f,f$ in $\mathbf{M}_{2}$ are glued to $x_4-x_5,x_7-x_8,x_9-x_{10},x_{11}-x_{12},y_4-y_5,y_7-y_8,y_9-y_{10},y_{11}-y_{12}$ in $\mathbf{N}_i$ for $i=0,1$.
\eit

\noindent\ubf{Gluing rules for \raisebox{-.25\height}{\begin{tikzpicture}
\node (w1) at (-0.5,0.9) {$\so(12)^{(1)}$};
\begin{scope}[shift={(3,0)}]
\node (w2) at (-0.5,0.9) {$\su(2)^{(1)}$};
\end{scope}
\node (v1) at (1,0.9) {\tiny{8}};
\draw [dashed] (w1) edge (v1);
\draw [dashed] (v1) edge (w2);
\end{tikzpicture}}}:
\be
\begin{tikzpicture} [scale=1.9]
\node (v2) at (1.4,1.5) {$\mathbf{M}^{2+2+2+2}_{0}$};
\node (v10) at (1.4,-1.1) {$\mathbf{M}^{1+1+1+1}_1$};
\node (v6) at (1.4,0.2) {$\mathbf{M}_{2}$};
\node[rotate=0] at (1.3,0.5) {\scriptsize{$e$}};
\node at (1.3,-0.1) {\scriptsize{$e$}};
\draw  (v2) edge (v6);
\draw  (v6) edge (v10);
\node at (1.3,1.2) {\scriptsize{$e$}};
\node at (1.3,-0.8) {\scriptsize{$e$}};
\node (v3) at (1.7,-2) {$\mathbf{N}^{16+16}_0$};
\node (v4) at (5.2,-2) {$\mathbf{N}^{16+16}_1$};
\draw  (v3) edge (v4);
\node[rotate=0] at (4.2,-1.9) {\scriptsize{$2e$-$2\sum x_i$-$2\sum y_i$}};
\node[rotate=0] at (2.7,-1.9) {\scriptsize{$2e$-$2\sum x_i$-$2\sum y_i$}};
\node (v1) at (2.9,0.2) {$\mathbf{M}_{3}$};
\node (v5) at (4.3,0.2) {$\mathbf{M}_{4}$};
\draw  (v6) edge (v1);
\node[rotate=0] at (1.7,0.3) {\scriptsize{$e$}};
\node[rotate=0] at (2.5,0.3) {\scriptsize{$e$}};
\draw  (v1) edge (v5);
\node[rotate=0] at (3.2,0.3) {\scriptsize{$e$}};
\node[rotate=0] at (3.8,0.3) {\scriptsize{$e$}};
\node at (2.5,0) {\scriptsize{$f$}};
\node[rotate=-40] at (1.7826,1.0607) {\scriptsize{$f$-$x_i$-$y_i$}};
\node (v7) at (5.6,1.5) {$\mathbf{M}_{5}$};
\draw  (v5) edge (v7);
\node[rotate=0] at (4.4,0.5) {\scriptsize{$e$}};
\node[rotate=0] at (5.2,1.3) {\scriptsize{$e$}};
\draw  (v10) edge (v1);
\draw  (v10) edge (v5);
\node (v12) at (2.2,0.8) {\scriptsize{2}};
\node (v13) at (2.959,0.7938) {\scriptsize{2}};
\node (v14) at (3.6,1.5) {\scriptsize{2}};
\draw  (v2) edge (v12);
\draw  (v12) edge (v1);
\draw  (v2) edge (v13);
\draw  (v13) edge (v5);
\draw  (v2) edge (v14);
\draw  (v14) edge (v7);
\node[rotate=0] at (2.4356,0.4857) {\scriptsize{$f$}};
\node[rotate=-25] at (2.2442,1.0191) {\scriptsize{$y_i$-$z_i$}};
\node[rotate=0] at (3.481,0.4617) {\scriptsize{$f$}};
\node[rotate=-0] at (2.4,1.6) {\scriptsize{$x_1$-$y_2,x_2$-$y_1$}};
\node[rotate=0] at (5.1,1.6) {\scriptsize{$f,f$}};
\node[rotate=40] at (1.7,-0.7) {\scriptsize{$f$-$x$-$y$}};
\node[rotate=20] at (2.1,-0.7) {\scriptsize{$y$-$z$}};
\node[rotate=0] at (3.6,0) {\scriptsize{$f$}};
\node[rotate=0] at (2.2,-1.2) {\scriptsize{$x$-$y$}};
\node[rotate=0] at (5.2,-1.2) {\scriptsize{$f$}};
\node (w) at (1,-2) {\scriptsize{16}};
\draw (v3) .. controls (1.6,-1.8) and (1,-1.5) .. (w);
\draw (v3) .. controls (1.6,-2.2) and (1,-2.5) .. (w);
\node at (1.4,-1.6) {\scriptsize{$x_i$}};
\node at (1.4,-2.4) {\scriptsize{$y_i$}};
\node (w) at (5.9,-2) {\scriptsize{16}};
\draw (v4) .. controls (5.3,-1.8) and (5.9,-1.5) .. (w);
\draw (v4) .. controls (5.3,-2.2) and (5.9,-2.5) .. (w);
\node at (5.5,-1.6) {\scriptsize{$x_i$}};
\node at (5.5,-2.4) {\scriptsize{$y_i$}};
\node (v8) at (5.6,-1.1) {$\mathbf{M}_{6}$};
\draw  (v5) edge (v8);
\draw  (v10) edge (v8);
\node[rotate=0] at (4.4,-0.1) {\scriptsize{$e$}};
\node[rotate=0] at (5.2,-0.9) {\scriptsize{$e$}};
\draw (v10) .. controls (5,-0.8) and (5.6,-0.1) .. (v7);
\node[rotate=0] at (2.1412,-0.9574) {\scriptsize{$z$-$w$}};
\node[rotate=0] at (5.5,1.1) {\scriptsize{$f$}};
\node (v9) at (3.5,1.1) {\scriptsize{2}};
\draw  (v2) edge (v9);
\draw (v9) .. controls (4.7,0.8) and (5.6,0.3) .. (v8);
\node[rotate=-10] at (2.553,1.3588) {\scriptsize{$z_i$-$w_i$}};
\node[rotate=0] at (5.7,-0.8) {\scriptsize{$f$}};
\end{tikzpicture}
\ee
\bit
\item $w,z,y,f-x,x,f-y,f-z,f-w,w,z,y,f-x,x,f-y,f-z,f-w$ in $\mathbf{M}_{0}$ are glued to $f-x_1,f-x_2,f-x_3,f-x_4,f-x_6,f-x_7,f-x_9,f-x_{11},f-y_1,f-y_2,f-y_3,f-y_4,f-y_6,f-y_7,f-y_9,f-y_{11}$ in $\mathbf{N}_i$ for $i=0,1$.
\item $8f$ in $\mathbf{M}_{2}$ are glued to $x_4-x_5,x_7-x_8,x_9-x_{10},x_{11}-x_{12},y_4-y_5,y_7-y_8,y_9-y_{10},y_{11}-y_{12}$ in $\mathbf{N}_i$ for $i=0,1$.
\item $8f$ in $\mathbf{M}_{3}$ are glued to $x_3-x_4,x_6-x_7,x_{10}-x_{13},x_{12}-x_{14},y_3-y_4,y_6-y_7,y_{10}-y_{13},y_{12}-y_{14}$ in $\mathbf{N}_i$ for $i=0,1$.
\item $8f$ in $\mathbf{M}_{4}$ are glued to $x_2-x_3,x_7-x_9,x_{8}-x_{10},x_{14}-x_{15},y_2-y_3,y_7-y_9,y_{8}-y_{10},y_{14}-y_{15}$ in $\mathbf{N}_i$ for $i=0,1$.
\item $8f$ in $\mathbf{M}_{5}$ are glued to $x_1-x_2,x_9-x_{11},x_{10}-x_{12},x_{13}-x_{14},y_1-y_2,y_9-y_{11},y_{10}-y_{12},y_{13}-y_{14}$ in $\mathbf{N}_i$ for $i=0,1$.
\item $8f$ in $\mathbf{M}_{6}$ are glued to $x_3-x_6,x_{4}-x_{7},x_{5}-x_{8},x_{15}-x_{16},y_3-y_6,y_{4}-y_{7},y_{5}-y_{8},y_{15}-y_{16}$ in $\mathbf{N}_i$ for $i=0,1$.
\item $f-w_2,f-z_2,f-y_2,f-x_1,x_2,y_1,z_1,w_1,f-w_2,f-z_2,f-y_2,f-x_1,x_2,y_1,z_1,w_1$ in $\mathbf{M}_{1}$ are glued to $x_5,x_8,x_{10}x_{12},x_{13},x_{14},x_{15},x_{16},y_5,y_8,y_{10}y_{12},y_{13},y_{14},y_{15},y_{16}$ in $\mathbf{N}_1$.
\item $f-w_1,f-z_1,f-y_1,f-x_2,x_1,y_2,z_2,w_2,f-w_1,f-z_1,f-y_1,f-x_2,x_1,y_2,z_2,w_2$ in $\mathbf{M}_{1}$ are glued to $x_5,x_8,x_{10}x_{12},x_{13},x_{14},x_{15},x_{16},y_5,y_8,y_{10}y_{12},y_{13},y_{14},y_{15},y_{16}$ in $\mathbf{N}_0$.
\eit

\noindent\ubf{Gluing rules for \raisebox{-.25\height}{\begin{tikzpicture}
\node (w1) at (-0.5,0.9) {$\ff_4^{(1)}$};
\begin{scope}[shift={(2.9,0)}]
\node (w2) at (-0.5,0.9) {$\sp(m)^{(1)}$};
\end{scope}
\node (v1) at (1,0.9) {\tiny{3}};
\draw  (w1) edge (v1);
\draw  (v1) edge (w2);
\end{tikzpicture}}}:
\be
\begin{tikzpicture} [scale=1.9]
\node (v2) at (0.2,0.2) {$\mathbf{M}_{0}$};
\node (v6) at (1.4,0.2) {$\mathbf{M}_{4}$};
\node[rotate=0] at (1.1,0.3) {\scriptsize{$e$}};
\draw  (v2) edge (v6);
\node at (0.5,0.3) {\scriptsize{$e$}};
\node (v3) at (1.3,-0.7) {$\mathbf{N}^1_0$};
\node (v4) at (2.6,-0.7) {$\mathbf{N}_1$};
\draw  (v3) edge (v4);
\node[rotate=0] at (5.1,-0.6) {\scriptsize{$2e$-$\sum x_i$}};
\node[rotate=0] at (1.7,-0.6) {\scriptsize{$2e$-$x$}};
\node (v1) at (2.6,0.2) {$\mathbf{M}_{3}$};
\node (v5) at (4.6,0.2) {$\mathbf{M}^{m+m}_{2}$};
\draw  (v6) edge (v1);
\node[rotate=0] at (1.7,0.3) {\scriptsize{$e$}};
\node[rotate=0] at (2.3,0.3) {\scriptsize{$e$}};
\draw  (v1) edge (v5);
\node[rotate=0] at (3,0.3) {\scriptsize{$2e$}};
\node[rotate=0] at (3.8,0.3) {\scriptsize{$e$-$\sum x_i$-$\sum y_i$}};
\node at (2.3,-0.6) {\scriptsize{$e$}};
\node (v8) at (3.3,-0.7) {$\cdots$};
\draw  (v4) edge (v8);
\node at (2.9,-0.6) {\scriptsize{$e$}};
\node (v9) at (4.1,-0.7) {$\mathbf{N}_{m-1}$};
\draw  (v8) edge (v9);
\node at (3.7,-0.6) {\scriptsize{$e$}};
\node (v11) at (5.7,-0.7) {$\mathbf{N}^{12}_{m}$};
\draw  (v9) edge (v11);
\node at (4.5,-0.6) {\scriptsize{$e$}};
\node (v7) at (6.2,0.2) {\scriptsize{$m$+1}};
\draw  (v5) edge (v7);
\node[rotate=0] at (5.4,0.3) {\scriptsize{$e$-$\sum x_i,f$-$y_i$}};
\node (v10) at (7.8,0.2) {$\mathbf{M}^{m+m}_{1}$};
\draw  (v7) edge (v10);
\node[rotate=0] at (7,0.3) {\scriptsize{$e$-$\sum x_i,f$-$y_i$}};
\node (w) at (4.6,0.8) {\scriptsize{$m$}};
\draw (v5) .. controls (4.5,0.4) and (4.2,0.7) .. (w);
\draw (v5) .. controls (4.7,0.4) and (5,0.7) .. (w);
\node[rotate=0] at (4.3,0.5) {\scriptsize{$x_i$}};
\node[rotate=0] at (4.9,0.5) {\scriptsize{$y_i$}};
\node (w) at (7.8,0.8) {\scriptsize{$m$}};
\draw (v10) .. controls (7.7,0.4) and (7.4,0.7) .. (w);
\draw (v10) .. controls (7.9,0.4) and (8.2,0.7) .. (w);
\node[rotate=0] at (7.5,0.5) {\scriptsize{$x_i$}};
\node[rotate=0] at (8.1,0.5) {\scriptsize{$y_i$}};
\end{tikzpicture}
\ee
\bit
\item $f-x_1,f-y_1$ in $\mathbf{M}_{i}$ are glued to $f-x,x$ in $\mathbf{N}_0$ for $i=1,2$.
\item $x_i-x_{i+1},y_i-y_{i+1}$ in $\mathbf{M}_{j}$ are glued to $f,f$ in $\mathbf{N}_i$ for $i=1,\cdots,m-1$ and $j=1,2$.
\item $f,f,f,f,x_m,y_m$ in $\mathbf{M}_{1}$ are glued to $x_1-x_2,x_6-x_8,x_7-x_9,x_{10}-x_{11},x_{12},x_{12}$ in $\mathbf{N}_m$.
\item $f,f,f,f,x_m,y_m$ in $\mathbf{M}_{2}$ are glued to $x_2-x_3,x_4-x_6,x_5-x_7,x_{10}-x_{12},x_{11},x_{11}$ in $\mathbf{N}_m$.
\item $f,f,f$ in $\mathbf{M}_{3}$ are glued to $x_3-x_4,x_7-x_{10},x_9-x_{11}$ in $\mathbf{N}_m$.
\item $f,f,f$ in $\mathbf{M}_{4}$ are glued to $x_4-x_5,x_6-x_{7},x_8-x_{9}$ in $\mathbf{N}_m$.
\item $f,f,f$ in $\mathbf{M}_{0}$ are glued to $f-x_1-x_8,f-x_2-x_{6},f-x_3-x_{4}$ in $\mathbf{N}_m$.
\eit

\noindent\ubf{Gluing rules for \raisebox{-.25\height}{\begin{tikzpicture}
\node (w1) at (-0.5,0.9) {$\fe_6^{(1)}$};
\begin{scope}[shift={(2.9,0)}]
\node (w2) at (-0.5,0.9) {$\su(m)^{(1)}$};
\end{scope}
\node (v1) at (1,0.9) {\tiny{6}};
\draw  (w1) edge (v1);
\draw  (v1) edge (w2);
\end{tikzpicture}}}:
\be
\begin{tikzpicture} [scale=1.9]
\node (v10) at (0.4,0.2) {$\mathbf{M}_0$};
\node (v6) at (1.6,0.2) {$\mathbf{M}_{4}$};
\node[rotate=0] at (3.7,1) {\scriptsize{$e$-$\sum x_i$-$\sum y_i$}};
\node[rotate=0] at (4.4,0.8) {\scriptsize{$x_i,y_i$}};
\node at (1.3,0.3) {\scriptsize{$e$}};
\draw  (v6) edge (v10);
\node at (0.7,0.3) {\scriptsize{$e$}};
\node (v3) at (0.9,-1.5) {$\mathbf{N}^{12+12}_0$};
\node (v4) at (2.5,-1.5) {$\mathbf{N}^3_1$};
\draw  (v3) edge (v4);
\node[rotate=0] at (0.7,-1.8) {\scriptsize{$e$-$\sum x_i$}};
\node[rotate=0] at (1.5,-1.4) {\scriptsize{$e$-$\sum y_i$}};
\node (v1) at (2.9,0.2) {$\mathbf{M}_{3}$};
\node (v5) at (4.6,1.2) {$\mathbf{M}^{2+(m-2)}_{5}$};
\draw  (v6) edge (v1);
\node[rotate=0] at (1.9,0.3) {\scriptsize{$e$}};
\node[rotate=0] at (2.6,0.3) {\scriptsize{$e$}};
\draw  (v1) edge (v5);
\node (v2) at (7.1,1.2) {$\mathbf{M}^{2+(m-2)}_{6}$};
\node[rotate=0] at (3.2,0.5) {\scriptsize{$e$}};
\node (v7) at (7.1,-0.7) {$\mathbf{M}^{2+(m-2)}_{1}$};
\node at (2.1,-1.4) {\scriptsize{$e$-$x_2$}};
\node (v8) at (4.6,-1.5) {$\cdots$};
\node at (3,-1.4) {\scriptsize{$e$-$x_1$-$x_3$}};
\node (v9) at (5.4,-1.5) {$\mathbf{N}_{m-1}$};
\draw  (v8) edge (v9);
\node at (5,-1.4) {\scriptsize{$e$}};
\node at (5.4,-1.8) {\scriptsize{$e$}};
\node (v12) at (4.6,-0.7) {$\mathbf{M}^{2+(m-2)}_{2}$};
\draw  (v1) edge (v12);
\node[rotate=0] at (3.1,-0.1) {\scriptsize{$e$}};
\draw (v3) .. controls (0.9,-2) and (5.4,-2) .. (v9);
\draw  (v5) edge (v2);
\draw  (v12) edge (v7);
\node[rotate=0] at (3.7,-0.5) {\scriptsize{$e$-$\sum x_i$-$\sum y_i$}};
\node (v11) at (4.6,0.2) {\scriptsize{$m$}};
\node (v13) at (7.1,0.2) {\scriptsize{$m$}};
\node[rotate=0] at (4.4,-0.3) {\scriptsize{$x_i,y_i$}};
\node[rotate=0] at (7.4,0.8) {\scriptsize{$x_i,y_i$}};
\node[rotate=0] at (7.4,-0.4) {\scriptsize{$x_i,y_i$}};
\node[rotate=0] at (5.2,1) {\scriptsize{$f$-$x_1$}};
\node[rotate=0] at (6.8,-0.2) {\scriptsize{$f$-$x_1$}};
\node[rotate=0] at (5.5,1.3) {\scriptsize{$e$-$x_2$-$\sum y_i$}};
\node[rotate=0] at (6.4,1.3) {\scriptsize{$e$-$x_1$}};
\node[rotate=0] at (5.3,-0.8) {\scriptsize{$e$-$x_1$}};
\node[rotate=0] at (6.2,-0.8) {\scriptsize{$e$-$x_2$-$\sum y_i$}};
\node[rotate=40] at (5.048,-0.2213) {\scriptsize{$f$-$x_2,f$-$y_i$}};
\node[rotate=40] at (6.4185,0.8196) {\scriptsize{$f$-$x_2,f$-$y_i$}};
\node (v15) at (3.9,-1.5) {$\mathbf{N}_2$};
\draw  (v5) edge (v11);
\draw  (v11) edge (v12);
\draw  (v2) edge (v13);
\draw  (v13) edge (v7);
\draw  (v5) edge (v7);
\node[rotate=40] (v14) at (5.6091,0.083) {\scriptsize{$m$-$1$}};
\draw  (v12) edge (v14);
\draw  (v14) edge (v2);
\draw  (v4) edge (v15);
\draw  (v15) edge (v8);
\node at (4.2,-1.4) {\scriptsize{$e$}};
\node at (3.6,-1.4) {\scriptsize{$e$}};
\end{tikzpicture}
\ee
\bit
\item $6f$ in $\mathbf{M}_{0}$ are glued to $f-y_1-x_8,f-x_1-y_8,f-y_2-x_6,f-x_2-y_6,f-y_3-x_4,f-x_3-y_4$ in $\mathbf{N}_0$.
\item $6f$ in $\mathbf{M}_{4}$ are glued to $x_4-x_5,x_6-x_7,x_8-x_9,y_4-y_5,y_6-y_7,y_8-y_9$ in $\mathbf{N}_0$.
\item $6f$ in $\mathbf{M}_{3}$ are glued to $x_3-x_4,x_7-x_{10},x_9-x_{11},y_3-y_4,y_7-y_{10},y_9-y_{11}$ in $\mathbf{N}_0$.
\item $4f,x_1,x_2$ in $\mathbf{M}_{2}$ are glued to $y_2-y_3,x_5-x_7,x_4-x_6,y_{10}-y_{12},x_{11},y_{11}$ in $\mathbf{N}_0$.
\item $4f,x_1,x_2$ in $\mathbf{M}_{5}$ are glued to $x_2-x_3,y_5-y_7,y_4-y_6,x_{10}-x_{12},x_{11},y_{11}$ in $\mathbf{N}_0$.
\item $4f,x_1,x_2$ in $\mathbf{M}_{1}$ are glued to $y_1-y_2,x_7-x_9,x_6-x_8,x_{10}-x_{11},x_{12},y_{12}$ in $\mathbf{N}_0$.
\item $4f,x_1,x_2$ in $\mathbf{M}_{6}$ are glued to $x_1-x_2,y_7-y_9,y_6-y_8,y_{10}-y_{11},x_{12},y_{12}$ in $\mathbf{N}_0$.
\item $f-x_1,f-y_1$ in $\mathbf{M}_{1}$ are glued to $x_2,x_1$ in $\mathbf{N}_1$.
\item $f-x_1,f-y_1$ in $\mathbf{M}_{2}$ are glued to $f-x_3,f-x_2$ in $\mathbf{N}_1$.
\item $f-x_1,f-y_1$ in $\mathbf{M}_{5}$ are glued to $x_2,x_3$ in $\mathbf{N}_1$.
\item $f-x_1,f-y_1$ in $\mathbf{M}_{6}$ are glued to $f-x_1,f-x_2$ in $\mathbf{N}_1$.
\item $y_i-y_{i+1}$ in $\mathbf{M}_{j}$ is glued to $f$ in $\mathbf{N}_{i+1}$ for $i=1,\cdots,m-3$ and $j=1,2,5,6$.
\item $y_{m-2}-x_2$ in $\mathbf{M}_{j}$ is glued to $f$ in $\mathbf{N}_{m-1}$ for $j=1,2,5,6$.
\eit

\noindent\ubf{Gluing rules for \raisebox{-.25\height}{\begin{tikzpicture}
\node (w1) at (-0.5,0.9) {$\su(4)^{(2)}$};
\begin{scope}[shift={(3,0)}]
\node (w2) at (-0.5,0.9) {$\su(2)^{(1)}$};
\end{scope}
\draw [dashed] (w1)--(w2);
\end{tikzpicture}}}:
\be
\begin{tikzpicture} [scale=1.9]
\node (v1) at (0.1,-0.7) {$\mathbf{M}^{6+6+1+1}_0$};
\node (v2) at (0.1,0.2) {$\mathbf{M}_{1}$};
\node (v6) at (1.7,0.2) {$\mathbf{M}_{2}$};
\draw  (v1) edge (v2);
\node[rotate=0] at (-0.15,-0.4) {\scriptsize{$e$-$z$-$w$}};
\node[rotate=0] at (-0.1,-0.1) {\scriptsize{$2e$}};
\node[rotate=0] at (1.3,0.3) {\scriptsize{$e$}};
\draw  (v2) edge (v6);
\node at (0.5,0.3) {\scriptsize{$2e$}};
\node (v3) at (2.8,0.2) {$\mathbf{N}^3_0$};
\node (v4) at (4.6,0.2) {$\mathbf{N}^1_1$};
\draw  (v3) edge (v4);
\node[rotate=0] at (3.3,0.3) {\scriptsize{$2e$-$\sum x_i$}};
\node[rotate=0] at (4.2,0.3) {\scriptsize{$2e$-$x$}};
\node (v5) at (1,-0.2) {\scriptsize{6}};
\draw  (v1) edge (v5);
\draw  (v5) edge (v6);
\node[rotate=0] at (0.9,-0.5) {\scriptsize{$f$-$x_i$-$y_i$}};
\node[rotate=0] at (1.5,-0.1) {\scriptsize{$f$}};
\draw (v1) .. controls (-0.3,-1.3) and (0.5,-1.3) .. (v1);
\node[rotate=0] at (-0.2,-1.1) {\scriptsize{$z$}};
\node[rotate=0] at (0.4,-1.1) {\scriptsize{$w$}};
\end{tikzpicture}
\ee
\bit
\item $z,w$ in $\mathbf{M}_{0}$ are glued to $x_3,x_3$ in $\mathbf{N}_0$.
\item $f-z,f-w$ in $\mathbf{M}_{0}$ are glued to $f-x,x$ in $\mathbf{N}_1$.
\item $f$ in $\mathbf{M}_{1}$ is glued to $x_2-x_3$ in $\mathbf{N}_0$.
\item $f,f$ in $\mathbf{M}_{2}$ are glued to $f-x_1-x_2,x_1-x_2$ in $\mathbf{N}_0$.
\eit

\noindent\ubf{Gluing rules for \raisebox{-.25\height}{\begin{tikzpicture}
\node (w1) at (-0.5,0.9) {$\so(8)^{(2)}$};
\begin{scope}[shift={(3,0)}]
\node (w2) at (-0.5,0.9) {$\sp(m)^{(1)}$};
\end{scope}
\node (v1) at (1,0.9) {\scriptsize{2}};
\draw [densely dashed] (w1) edge (v1);
\draw [->] (v1) edge (w2);
\end{tikzpicture}}}:
\be
\begin{tikzpicture} [scale=1.9]
\node (v1) at (0.1,-0.7) {$\mathbf{M}^{m+m}_1$};
\node (v2) at (0.1,0.2) {$\mathbf{M}_{2}$};
\node (v6) at (1.7,0.2) {$\mathbf{M}_{3}$};
\draw  (v1) edge (v2);
\node[rotate=0] at (0,-0.4) {\scriptsize{$e$}};
\node[rotate=0] at (0,-0.1) {\scriptsize{$e$}};
\node[rotate=0] at (1.3,0.3) {\scriptsize{$e$}};
\draw  (v2) edge (v6);
\node at (0.5,0.3) {\scriptsize{$2e$}};
\node (v3) at (-1.8,-1.5) {$\mathbf{N}^4_0$};
\node (v4) at (-0.3,-1.5) {$\mathbf{N}_1$};
\draw  (v3) edge (v4);
\node[rotate=0] at (-1.3,-1.4) {\scriptsize{$2e$-$\sum x_i$}};
\node[rotate=0] at (-0.6,-1.4) {\scriptsize{$e$}};
\node (v5) at (1,-0.2) {\scriptsize{$m$}};
\draw  (v1) edge (v5);
\draw  (v5) edge (v6);
\node[rotate=0] at (0.9,-0.5) {\scriptsize{$f$-$x_i$-$y_i$}};
\node[rotate=0] at (1.5,-0.1) {\scriptsize{$f$}};
\node (v7) at (-1.8,-0.7) {$\mathbf{M}_{0}$};
\node[rotate=0] at (-1.5,-0.6) {\scriptsize{$e$}};
\node[rotate=0] at (-0.7,-0.6) {\scriptsize{$2e$-$\sum x_i$-$\sum y_i$}};
\node (v8) at (0.4,-1.5) {$\cdots$};
\draw  (v7) edge (v1);
\node[rotate=0] at (0,-1.4) {\scriptsize{$e$}};
\node[rotate=0] at (2,-1.4) {\scriptsize{$2e$}};
\node (v9) at (1.2,-1.5) {$\mathbf{N}_{m-1}$};
\node (v10) at (2.3,-1.5) {$\mathbf{N}_{m}$};
\draw  (v4) edge (v8);
\draw  (v8) edge (v9);
\draw  (v9) edge (v10);
\node[rotate=0] at (0.8,-1.4) {\scriptsize{$e$}};
\node[rotate=0] at (1.6,-1.4) {\scriptsize{$e$}};
\end{tikzpicture}
\ee
\bit
\item $f-x_1,y_1$ in $\mathbf{M}_{1}$ are glued to $x_3,x_4$ in $\mathbf{N}_0$.
\item $x_i-x_{i+1},y_{i+1}-y_i$ in $\mathbf{M}_{1}$ are glued to $f,f$ in $\mathbf{N}_i$ for $i=1,\cdots,m-1$.
\item $x_m-y_m$ in $\mathbf{M}_{1}$ is glued to $f$ in $\mathbf{N}_m$.
\item $f,f$ in $\mathbf{M}_{0}$ are glued to $f-x_2-x_3,x_1-x_4$ in $\mathbf{N}_0$.
\item $f$ in $\mathbf{M}_{2}$ is glued to $x_2-x_3$ in $\mathbf{N}_0$.
\item $f,f$ in $\mathbf{M}_{3}$ are glued to $x_3-x_4,x_1-x_2$ in $\mathbf{N}_0$.
\eit

\noindent\ubf{Gluing rules for \raisebox{-.25\height}{\begin{tikzpicture}
\node (v1) at (-1.1,0.4) {$\so(8)^{(3)}$};
\begin{scope}[shift={(3.1,-0.05)}]
\node (v2) at (-0.5,0.45) {$\sp(m)^{(1)}$};
\end{scope}
\node (v4) at (0.7,0.4) {\tiny{3}};
\draw [densely dashed]  (v1) edge (0.2,0.4);
\draw [densely dotted]  (0.2,0.4) edge (v4);
\draw [densely dotted]  (v4) edge (1.2,0.4);
\draw [->] (1.2,0.4) -- (v2);
\end{tikzpicture}}}:
\be
\begin{tikzpicture} [scale=1.9]
\node (v1) at (0.1,-0.7) {$\mathbf{M}^{m+m}_1$};
\node (v6) at (3.3,-0.7) {$\mathbf{M}^{m+m}_{0}$};
\node (v3) at (-1.8,-1.5) {$\mathbf{N}^3_0$};
\node (v4) at (-0.3,-1.5) {$\mathbf{N}_1$};
\draw  (v3) edge (v4);
\node[rotate=0] at (-1.3,-1.4) {\scriptsize{$2e$-$\sum x_i$}};
\node[rotate=0] at (-0.6,-1.4) {\scriptsize{$e$}};
\node (v5) at (1.7,-0.7) {\scriptsize{$m$+1}};
\draw  (v1) edge (v5);
\draw  (v5) edge (v6);
\node[rotate=0] at (0.9,-0.6) {\scriptsize{$e$-$\sum x_i,f$-$y_i$}};
\node[rotate=0] at (2.5,-0.6) {\scriptsize{$e$-$\sum x_i,f$-$y_i$}};
\node (v7) at (-1.8,-0.7) {$\mathbf{M}_{2}$};
\node[rotate=0] at (-1.5,-0.6) {\scriptsize{$3e$}};
\node[rotate=0] at (-0.7,-0.6) {\scriptsize{$e$-$\sum x_i$-$\sum y_i$}};
\node (v8) at (0.4,-1.5) {$\cdots$};
\draw  (v7) edge (v1);
\node[rotate=0] at (0,-1.4) {\scriptsize{$e$}};
\node[rotate=0] at (2.1,-1.4) {\scriptsize{$2e$-$x$}};
\node (v9) at (1.2,-1.5) {$\mathbf{N}_{m-1}$};
\node (v10) at (2.5,-1.5) {$\mathbf{N}^1_{m}$};
\draw  (v4) edge (v8);
\draw  (v8) edge (v9);
\draw  (v9) edge (v10);
\node[rotate=0] at (0.8,-1.4) {\scriptsize{$e$}};
\node[rotate=0] at (1.6,-1.4) {\scriptsize{$e$}};
\node (w) at (3.3,-0.1) {\scriptsize{$m$}};
\draw (v6) .. controls (3.2,-0.5) and (2.9,-0.2) .. (w);
\draw (v6) .. controls (3.4,-0.5) and (3.7,-0.2) .. (w);
\node[rotate=0] at (3,-0.4) {\scriptsize{$x_i$}};
\node[rotate=0] at (3.6,-0.4) {\scriptsize{$y_i$}};
\node (w) at (0.1,-0.1) {\scriptsize{$m$}};
\draw (v1) .. controls (0,-0.5) and (-0.3,-0.2) .. (w);
\draw (v1) .. controls (0.2,-0.5) and (0.5,-0.2) .. (w);
\node[rotate=0] at (-0.2,-0.4) {\scriptsize{$x_i$}};
\node[rotate=0] at (0.4,-0.4) {\scriptsize{$y_i$}};
\end{tikzpicture}
\ee
\bit
\item $f,x_1,y_1$ in $\mathbf{M}_{1}$ are glued to $x_1-x_2,x_3,x_3$ in $\mathbf{N}_0$.
\item $f,x_1,y_1$ in $\mathbf{M}_{0}$ are glued to $f-x_3-x_2,f-x_1,f-x_1$ in $\mathbf{N}_0$.
\item $x_{i+1}-x_{i},y_{i+1}-y_i$ in $\mathbf{M}_{j}$ are glued to $f,f$ in $\mathbf{N}_i$ for $i=1,\cdots,m-1$ and $j=0,1$.
\item $f-x_m,f-y_m$ in $\mathbf{M}_{j}$ are glued to $f-x,x$ in $\mathbf{N}_m$ for $j=0,1$.
\item $f$ in $\mathbf{M}_{2}$ is glued to $x_2-x_3$ in $\mathbf{N}_0$.
\eit

\noindent\ubf{Gluing rules for \raisebox{-.25\height}{\begin{tikzpicture}
\node (w1) at (-0.5,0.9) {$\so(10)^{(2)}$};
\begin{scope}[shift={(3,0)}]
\node (w2) at (-0.5,0.9) {$\su(2)^{(1)}$};
\end{scope}
\node (v1) at (1,0.9) {\tiny{4}};
\draw [dashed] (w1) edge (v1);
\draw [dashed] (v1) edge (w2);
\end{tikzpicture}}}:
\be
\begin{tikzpicture} [scale=1.9]
\node (v1) at (0.1,-0.7) {$\mathbf{M}_1$};
\node (v2) at (0.1,0.2) {$\mathbf{M}_{2}$};
\node (v6) at (1.7,0.2) {$\mathbf{M}^{1+1}_{3}$};
\draw  (v1) edge (v2);
\node[rotate=0] at (0,-0.4) {\scriptsize{$e$}};
\node[rotate=0] at (0,-0.1) {\scriptsize{$e$}};
\node[rotate=0] at (1.2,0.3) {\scriptsize{$e$-$x$}};
\draw  (v2) edge (v6);
\node at (0.4,0.3) {\scriptsize{$e$}};
\node (v3) at (0.4,-2.3) {$\mathbf{N}^8_0$};
\node (v4) at (2.2,-2.3) {$\mathbf{N}^8_1$};
\draw  (v3) edge (v4);
\node[rotate=0] at (0.9,-2.2) {\scriptsize{$2e$-$\sum x_i$}};
\node[rotate=0] at (1.7,-2.2) {\scriptsize{$2e$-$\sum x_i$}};
\node[rotate=50] at (0.4177,-1.098) {\scriptsize{$f$-$x_1,f$-$x_2$}};
\node[rotate=0] at (0.3476,-0.4382) {\scriptsize{$f$}};
\node (v7) at (0.1,-1.6) {$\mathbf{M}^{2+2}_{0}$};
\node[rotate=0] at (-0.05,-1) {\scriptsize{$2e$}};
\node[rotate=0] at (0,-1.35) {\scriptsize{$e$}};
\node[rotate=0] at (1.1,0) {\scriptsize{$x$-$y$}};
\node[rotate=50] at (1.1549,-0.2671) {\scriptsize{$f$-$x,y$}};
\node[rotate=0] at (2.9,-0.1) {\scriptsize{$f$}};
\draw  (v7) edge (v1);
\node[rotate=30] at (0.686,-1.3795) {\scriptsize{$x_i$-$y_i$}};
\node (v11) at (3.2,0.2) {$\mathbf{M}_{4}$};
\draw  (v6) edge (v11);
\node[rotate=0] at (2.3,0.3) {\scriptsize{$2e$-$x$-$y$}};
\node[rotate=0] at (2.9,0.3) {\scriptsize{$e$}};
\draw  (v1) edge (v6);
\node (v5) at (0.9,-0.7) {\scriptsize{2}};
\node (v8) at (2,-0.5) {\scriptsize{2}};
\draw  (v7) edge (v5);
\draw  (v5) edge (v6);
\draw  (v7) edge (v8);
\draw  (v8) edge (v11);
\end{tikzpicture}
\ee
\bit
\item $f-x_2,x_1,f-y_2,y_1,f,f$ in $\mathbf{M}_{0}$ are glued to $x_8,f-x_2,x_7,f-x_1,f-x_4-x_5,f-x_3-x_6$ in $\mathbf{N}_0$.
\item $f-x_1,x_2,f-y_1,y_2,f,f$ in $\mathbf{M}_{0}$ are glued to $x_8,f-x_2,x_7,f-x_1,f-x_4-x_5,f-x_3-x_6$ in $\mathbf{N}_1$.
\item $f,f$ in $\mathbf{M}_{1}$ are glued to $x_4-x_7,x_6-x_8$ in $\mathbf{N}_i$ for $i=0,1$.
\item $f,f$ in $\mathbf{M}_{2}$ are glued to $x_3-x_4,x_5-x_6$ in $\mathbf{N}_i$ for $i=0,1$.
\item $f,f,f,f$ in $\mathbf{M}_{4}$ are glued to $x_1-x_2,x_3-x_5,x_4-x_6,x_7-x_8$ in $\mathbf{N}_i$ for $i=0,1$.
\item $f,x,y$ in $\mathbf{M}_{3}$ are glued to $x_2-x_3,x_6,x_8$ in $\mathbf{N}_0$.
\item $f,f-x,f-y$ in $\mathbf{M}_{3}$ are glued to $x_2-x_3,x_8,x_6$ in $\mathbf{N}_1$.
\eit

\noindent\ubf{Gluing rules for \raisebox{-.25\height}{\begin{tikzpicture}
\node (w1) at (-0.5,0.9) {$\so(10)^{(2)}$};
\begin{scope}[shift={(3,0)}]
\node (w2) at (-0.5,0.9) {$\su(3)^{(2)}$};
\end{scope}
\node (v1) at (1,0.9) {\tiny{4}};
\draw [dashed] (w1) edge (v1);
\draw [dashed] (v1) edge (w2);
\end{tikzpicture}}}:
\be
\begin{tikzpicture} [scale=1.9]
\node (v1) at (0.1,-0.7) {$\mathbf{M}_1$};
\node (v2) at (0.1,0.2) {$\mathbf{M}_{2}$};
\node (v6) at (1.7,0.2) {$\mathbf{M}^{1+1+1}_{3}$};
\draw  (v1) edge (v2);
\node[rotate=0] at (0,-0.4) {\scriptsize{$e$}};
\node[rotate=0] at (0,-0.1) {\scriptsize{$e$}};
\node[rotate=0] at (1.1,0.3) {\scriptsize{$e$-$x$}};
\draw  (v2) edge (v6);
\node at (0.4,0.3) {\scriptsize{$e$}};
\node (v3) at (0.7,-3.1) {$\mathbf{N}^{8+8}_0$};
\node (v4) at (2.8,-3.1) {$\mathbf{N}_1$};
\draw  (v3) edge (v4);
\node[rotate=0] at (1.5,-3) {\scriptsize{$e$-$\sum x_i$-$\sum y_i$}};
\node[rotate=0] at (2.4,-3) {\scriptsize{$4e$}};
\node[rotate=55] at (0.4574,-1.3423) {\scriptsize{$z_2,f$-$x_1,f$-$x_2$}};
\node[rotate=0] at (0.3476,-0.4382) {\scriptsize{$f$}};
\node (v7) at (0.1,-2) {$\mathbf{M}^{2+2+2}_{0}$};
\node[rotate=0] at (-0.05,-1) {\scriptsize{$2e$}};
\node[rotate=0] at (0,-1.75) {\scriptsize{$e$}};
\node[rotate=0] at (1.1,0) {\scriptsize{$x$-$y$}};
\node[rotate=53] at (1.1148,-0.3961) {\scriptsize{$f$-$z,f$-$x,y$}};
\node[rotate=0] at (2.9,-0.2) {\scriptsize{$f,f$}};
\draw  (v7) edge (v1);
\node[rotate=35] at (0.9,-1.6) {\scriptsize{$z_1$-$z_2,x_i$-$y_i$}};
\node (v11) at (3.2,0.2) {$\mathbf{M}_{4}$};
\draw  (v6) edge (v11);
\node[rotate=0] at (2.4,0.3) {\scriptsize{$2e$-$x$-$y$-$z$}};
\node[rotate=0] at (2.9,0.3) {\scriptsize{$e$}};
\draw  (v1) edge (v6);
\node (v5) at (0.9,-0.9) {\scriptsize{3}};
\node (v8) at (2,-0.7) {\scriptsize{3}};
\draw  (v7) edge (v5);
\draw  (v5) edge (v6);
\draw  (v7) edge (v8);
\draw  (v8) edge (v11);
\node (w) at (0.7,-2.5) {\scriptsize{$8$}};
\draw (v3) .. controls (0.6,-2.9) and (0.3,-2.6) .. (w);
\draw (v3) .. controls (0.8,-2.9) and (1.1,-2.6) .. (w);
\node[rotate=0] at (0.4,-2.8) {\scriptsize{$x_i$}};
\node[rotate=0] at (1,-2.8) {\scriptsize{$y_i$}};
\draw (v7) .. controls (-0.7,-2) and (-0.7,0.2) .. (v2);
\node[rotate=0] at (-0.8,-1.8) {\scriptsize{$f$-$z_1$-$z_2$}};
\node[rotate=0] at (-0.4,0.2) {\scriptsize{$f$}};
\end{tikzpicture}
\ee
\bit
\item $f-x_2,x_1,f-y_2,y_1,z_2,f-z_2,z_1,f-z_1,f-x_2,x_1,f-y_2,y_1,f-z_2,z_2,f-z_1,z_1$ in $\mathbf{M}_{0}$ are glued to $x_8,x_2,x_7,x_1,f-x_3,f-x_6,f-x_5,f-x_4,y_8,y_2,y_7,y_1,f-y_3,f-y_6,f-y_5,f-y_4$ in $\mathbf{N}_0$.
\item $x_2-x_1,y_2-y_1$ in $\mathbf{M}_{0}$ are glued to $f,f$ in $\mathbf{N}_1$.
\item $f,f,f,f$ in $\mathbf{M}_{1}$ are glued to $x_4-x_7,x_6-x_8,y_4-y_7,y_6-y_8$ in $\mathbf{N}_0$.
\item $f,f,f,f$ in $\mathbf{M}_{2}$ are glued to $x_3-x_4,x_5-x_6,y_3-y_4,y_5-y_6$ in $\mathbf{N}_0$.
\item $8f$ in $\mathbf{M}_{4}$ are glued to $x_2-x_1,x_3-x_5,x_4-x_6,x_7-x_8,y_2-y_1,y_3-y_5,y_4-y_6,y_7-y_8$ in $\mathbf{N}_0$.
\item $f-z,x,y,z,z,x,y,f-z$ in $\mathbf{M}_{3}$ are glued to $f-x_3,x_6,x_8,f-x_2,f-y_3,y_6,y_8,f-y_2$ in $\mathbf{N}_0$.
\item $f-x-y$ in $\mathbf{M}_{3}$ is glued to $f$ in $\mathbf{N}_1$.
\eit

\noindent\ubf{Gluing rules for \raisebox{-.25\height}{\begin{tikzpicture}
\node (w1) at (-0.5,0.9) {$\fe_6^{(2)}$};
\begin{scope}[shift={(2.9,0)}]
\node (w2) at (-0.5,0.9) {$\su(2)^{(1)}$};
\end{scope}
\node (v1) at (1,0.9) {\tiny{6}};
\draw  (w1) edge (v1);
\draw  (v1) edge (w2);
\end{tikzpicture}}}:
\be
\begin{tikzpicture} [scale=1.9]
\node (v2) at (0.1,0.2) {$\mathbf{M}_{4}$};
\node (v6) at (1.3,0.2) {$\mathbf{M}_{3}$};
\node[rotate=0] at (0.9,0.3) {\scriptsize{$e$}};
\draw  (v2) edge (v6);
\node at (0.5,0.3) {\scriptsize{$e$}};
\node (v3) at (2.8,-0.7) {$\mathbf{N}^{12}_0$};
\node (v4) at (4.9,-0.7) {$\mathbf{N}^2_1$};
\draw  (v3) edge (v4);
\node[rotate=0] at (3.4,-0.6) {\scriptsize{$2e$-$\sum x_i$}};
\node[rotate=0] at (4.3,-0.6) {\scriptsize{$2e$-$\sum x_i$}};
\node (v11) at (2.8,0.2) {$\mathbf{M}^{1+1}_{2}$};
\draw  (v6) edge (v11);
\node[rotate=0] at (1.7,0.3) {\scriptsize{$2e$}};
\node[rotate=0] at (2.2,0.3) {\scriptsize{$e$-$x$-$y$}};
\node (v1) at (3.9,0.2) {\scriptsize{2}};
\node (v5) at (5,0.2) {$\mathbf{M}^{1+1}_{1}$};
\node (v7) at (6.1,0.2) {\scriptsize{2}};
\node (v8) at (7.2,0.2) {$\mathbf{M}^{1+1}_{0}$};
\draw  (v11) edge (v1);
\draw  (v1) edge (v5);
\draw  (v5) edge (v7);
\draw  (v7) edge (v8);
\node[rotate=0] at (3.4,0.3) {\scriptsize{$e$-$x,f$-$y$}};
\node[rotate=0] at (4.4,0.3) {\scriptsize{$e$-$x,f$-$y$}};
\node[rotate=0] at (5.6,0.3) {\scriptsize{$e$-$y,f$-$x$}};
\node[rotate=0] at (6.6,0.3) {\scriptsize{$e$-$y,f$-$x$}};
\draw (v11) .. controls (2.2,0.9) and (3.4,0.9) .. (v11);
\node at (2.4,0.6) {\scriptsize{$x$}};
\node at (3.2,0.6) {\scriptsize{$y$}};
\draw (v5) .. controls (4.4,0.9) and (5.6,0.9) .. (v5);
\node at (4.6,0.6) {\scriptsize{$x$}};
\node at (5.4,0.6) {\scriptsize{$y$}};
\draw (v8) .. controls (6.6,0.9) and (7.8,0.9) .. (v8);
\node at (6.8,0.6) {\scriptsize{$x$}};
\node at (7.6,0.6) {\scriptsize{$y$}};
\end{tikzpicture}
\ee
\bit
\item $f,f,f$ in $\mathbf{M}_{4}$ are glued to $x_4-x_5,x_6-x_7,x_8-x_9$ in $\mathbf{N}_0$.
\item $f,f,f$ in $\mathbf{M}_{3}$ are glued to $x_3-x_4,x_7-x_{10},x_9-x_{11}$ in $\mathbf{N}_0$.
\item $f,f,f,f,x,y$ in $\mathbf{M}_{2}$ are glued to $x_2-x_3,x_4-x_6,x_5-x_7,x_{10}-x_{12},x_{11},x_{11}$ in $\mathbf{N}_0$.
\item $f,f,f,f,x,y$ in $\mathbf{M}_{1}$ are glued to $x_1-x_2,x_6-x_8,x_7-x_9,x_{10}-x_{11},x_{12},x_{12}$ in $\mathbf{N}_0$.
\item $f,f,f,f,x,y$ in $\mathbf{M}_{0}$ are glued to $f-x_6-x_5,f-x_7-x_4,f-x_{10}-x_3,f-x_{2}-x_{12},f-x_{1},f-x_{1}$ in $\mathbf{N}_0$.
\item $f-x,f-y$ in $\mathbf{M}_{2}$ are glued to $x_1,f-x_1$ in $\mathbf{N}_1$.
\item $f-x,f-y$ in $\mathbf{M}_{1}$ are glued to $x_1,f-x_1$ in $\mathbf{N}_1$.
\item $f-x,f-y$ in $\mathbf{M}_{0}$ are glued to $x_2,f-x_2$ in $\mathbf{N}_1$.
\eit

\noindent\ubf{Gluing rules for \raisebox{-.25\height}{\begin{tikzpicture}
\node (w1) at (-0.5,0.9) {$\fe_6^{(2)}$};
\begin{scope}[shift={(2.9,0)}]
\node (w2) at (-0.5,0.9) {$\su(3)^{(2)}$};
\end{scope}
\node (v1) at (1,0.9) {\tiny{6}};
\draw  (w1) edge (v1);
\draw  (v1) edge (w2);
\end{tikzpicture}}}:
\be
\begin{tikzpicture} [scale=1.9]
\node (v2) at (0.1,0.2) {$\mathbf{M}_{4}$};
\node (v6) at (1.3,0.2) {$\mathbf{M}_{3}$};
\node[rotate=0] at (0.9,0.3) {\scriptsize{$e$}};
\draw  (v2) edge (v6);
\node at (0.5,0.3) {\scriptsize{$e$}};
\node (v3) at (1.7,-2.2) {$\mathbf{N}^{12+12}_0$};
\node (v4) at (4.3,-2.2) {$\mathbf{N}^2_1$};
\draw  (v3) edge (v4);
\node[rotate=0] at (2.6,-2.1) {\scriptsize{$e$-$\sum x_i$-$\sum y_i$}};
\node[rotate=0] at (3.7,-2.1) {\scriptsize{$4e$-$2\sum x_i$}};
\node (v11) at (2.8,0.2) {$\mathbf{M}^{1+1}_{2}$};
\draw  (v6) edge (v11);
\node[rotate=0] at (1.7,0.3) {\scriptsize{$2e$}};
\node[rotate=0] at (2.2,0.3) {\scriptsize{$e$-$x$-$y$}};
\node (v1) at (3.85,0.2) {\scriptsize{2}};
\node (v5) at (5,0.2) {$\mathbf{M}^{1+1+1}_{1}$};
\node (v7) at (5,-0.5) {\scriptsize{3}};
\node (v8) at (5,-1.3) {$\mathbf{M}^{1+1+4}_{0}$};
\draw  (v11) edge (v1);
\draw  (v1) edge (v5);
\draw  (v5) edge (v7);
\draw  (v7) edge (v8);
\node[rotate=0] at (3.4,0.3) {\scriptsize{$e$-$x,f$-$y$}};
\node[rotate=0] at (4.3,0.3) {\scriptsize{$e$-$x,f$-$y$}};
\node[rotate=0] at (5.5,-0.1) {\scriptsize{$e$-$y$-$z,f$-$x,f$-$z$}};
\node[rotate=0] at (5.4,-0.8) {\scriptsize{$e$-$y,f$-$x,z_4$}};
\draw (v11) .. controls (2.2,0.9) and (3.4,0.9) .. (v11);
\node at (2.4,0.6) {\scriptsize{$x$}};
\node at (3.2,0.6) {\scriptsize{$y$}};
\draw (v5) .. controls (4.4,0.9) and (5.6,0.9) .. (v5);
\node at (4.6,0.6) {\scriptsize{$x$}};
\node at (5.4,0.6) {\scriptsize{$y$}};
\begin{scope}[rotate=-90, shift={(-5.9,5.2)}]
\draw (v8) .. controls (6.6,0.9) and (7.8,0.9) .. (v8);
\node at (6.9,0.7) {\scriptsize{$x$}};
\node at (7.5,0.7) {\scriptsize{$y$}};
\end{scope}
\draw  (v6) edge (v8);
\draw  (v2) edge (v8);
\node (v9) at (3.8,-0.5) {\scriptsize{2}};
\draw  (v11) edge (v9);
\draw  (v9) edge (v8);
\node[rotate=0] at (0.4,0) {\scriptsize{$f$}};
\node[rotate=0] at (2.0118,0.0299) {\scriptsize{$f$}};
\node[rotate=0] at (3.4,0) {\scriptsize{$f,f$}};
\node[rotate=-32] at (4.4051,-0.7667) {\scriptsize{$f$-$z_1$-$z_2,z_3$-$z_4$}};
\node[rotate=-22] at (3.9933,-0.8027) {\scriptsize{$z_2$-$z_3$}};
\node[rotate=-0] at (4.2,-1.2) {\scriptsize{$z_1$-$z_2$}};
\node (w) at (1.7,-1.6) {\scriptsize{$12$}};
\draw (v3) .. controls (1.6,-2) and (1.3,-1.7) .. (w);
\draw (v3) .. controls (1.8,-2) and (2.1,-1.7) .. (w);
\node[rotate=0] at (1.4,-1.9) {\scriptsize{$x_i$}};
\node[rotate=0] at (2,-1.9) {\scriptsize{$y_i$}};
\end{tikzpicture}
\ee
\bit
\item $6f$ in $\mathbf{M}_{4}$ are glued to $x_4-x_5,x_6-x_7,x_8-x_9,y_4-y_5,y_6-y_7,y_8-y_9$ in $\mathbf{N}_0$.
\item $6f$ in $\mathbf{M}_{3}$ are glued to $x_3-x_4,x_7-x_{10},x_9-x_{11},y_3-y_4,y_7-y_{10},y_9-y_{11}$ in $\mathbf{N}_0$.
\item $4f,x,y,4f,x,y$ in $\mathbf{M}_{2}$ are glued to $x_2-x_3,x_4-x_6,x_5-x_7,x_{10}-x_{12},x_{11},x_{11},y_2-y_3,y_4-y_6,y_5-y_7,y_{10}-y_{12},y_{11},y_{11}$ in $\mathbf{N}_0$.
\item $z,f-z,3f,x,y,z,f-z,3f,x,y$ in $\mathbf{M}_{1}$ are glued to $f-x_1,f-x_2,x_6-x_8,x_7-x_9,x_{10}-x_{11},x_{12},x_{12},f-y_1,f-y_2,y_6-y_8,y_7-y_9,y_{10}-y_{11},y_{12},y_{12}$ in $\mathbf{N}_0$.
\item $f-z_1,z_1,f-z_2,z_2,f-z_3,z_3,f-z_4,z_4,x,y,f-z_1,z_1,f-z_2,z_2,f-z_3,z_3,f-z_4,z_4,x,y$ in $\mathbf{M}_{0}$ are glued to $f-x_6,f-x_5,f-x_7,f-x_4,f-x_{10},f-x_3,f-x_{12},f-x_{2},x_{1},x_{1},f-y_6,f-y_5,f-y_7,f-y_4,f-y_{10},f-y_3,f-y_{12},f-y_{2}y_{1},y_{1}$ in $\mathbf{N}_0$.
\item $f-x,f-y$ in $\mathbf{M}_{2}$ are glued to $f-x_1,x_1$ in $\mathbf{N}_1$.
\item $f-x,f-y$ in $\mathbf{M}_{1}$ are glued to $f-x_1,x_1$ in $\mathbf{N}_1$.
\item $f-x,f-y$ in $\mathbf{M}_{0}$ are glued to $f-x_2,x_2$ in $\mathbf{N}_1$.
\eit

\noindent\ubf{Gluing rules for \raisebox{-.25\height}{\begin{tikzpicture}
\node (w1) at (-0.5,0.9) {$\fe_6^{(2)}$};
\begin{scope}[shift={(2.9,0)}]
\node (w2) at (-0.5,0.9) {$\su(4)^{(2)}$};
\end{scope}
\node (v1) at (1,0.9) {\tiny{6}};
\draw  (w1) edge (v1);
\draw  (v1) edge (w2);
\end{tikzpicture}}}:
\be
\begin{tikzpicture} [scale=1.9]
\node (v2) at (0.1,0.2) {$\mathbf{M}_{4}$};
\node (v6) at (1.3,0.2) {$\mathbf{M}_{3}$};
\node[rotate=0] at (0.9,0.3) {\scriptsize{$e$}};
\draw  (v2) edge (v6);
\node at (0.5,0.3) {\scriptsize{$e$}};
\node (v3) at (0.8,-0.8) {$\mathbf{N}^{2+2}_0$};
\node (v4) at (2.7,-0.8) {$\mathbf{N}^{12}_2$};
\draw  (v3) edge (v4);
\node[rotate=0] at (1.2,-0.7) {\scriptsize{$e$}};
\node[rotate=0] at (2.2,-0.7) {\scriptsize{$2e$-$\sum x_i$}};
\node (v11) at (3.2,0.2) {$\mathbf{M}^{2+2}_{2}$};
\draw  (v6) edge (v11);
\node[rotate=0] at (1.7,0.3) {\scriptsize{$2e$}};
\node[rotate=0] at (2.4,0.3) {\scriptsize{$e$-$\sum x_i$-$\sum y_i$}};
\node (v1) at (4.6,0.2) {\scriptsize{3}};
\node (v5) at (6.1,0.2) {$\mathbf{M}^{2+2}_{1}$};
\node (v7) at (6.1,-0.5) {\scriptsize{3}};
\node (v8) at (6.1,-1.2) {$\mathbf{M}^{2+2}_{0}$};
\draw  (v11) edge (v1);
\draw  (v1) edge (v5);
\draw  (v5) edge (v7);
\draw  (v7) edge (v8);
\node[rotate=0] at (4,0.3) {\scriptsize{$e$-$\sum x_i,f$-$y_i$}};
\node[rotate=0] at (5.3,0.3) {\scriptsize{$e$-$\sum x_i,f$-$y_i$}};
\node[rotate=0] at (6.6,-0.1) {\scriptsize{$e$-$\sum y_i,f$-$x_i$}};
\node[rotate=0] at (6.6,-0.9) {\scriptsize{$e$-$\sum y_i,f$-$x_i$}};
\node (w) at (3.2,0.8) {\scriptsize{$2$}};
\draw (v11) .. controls (3.1,0.4) and (2.8,0.7) .. (w);
\draw (v11) .. controls (3.3,0.4) and (3.6,0.7) .. (w);
\node[rotate=0] at (2.9,0.5) {\scriptsize{$x_i$}};
\node[rotate=0] at (3.5,0.5) {\scriptsize{$y_i$}};
\node (w) at (6.1,0.8) {\scriptsize{$2$}};
\draw (v5) .. controls (6,0.4) and (5.7,0.7) .. (w);
\draw (v5) .. controls (6.2,0.4) and (6.5,0.7) .. (w);
\node[rotate=0] at (5.8,0.5) {\scriptsize{$x_i$}};
\node[rotate=0] at (6.4,0.5) {\scriptsize{$y_i$}};
\begin{scope}[rotate=-180, shift={(-11.1,1)}]
\node (w) at (5,0.8) {\scriptsize{$2$}};
\draw (v8) .. controls (4.9,0.4) and (4.6,0.7) .. (w);
\draw (v8) .. controls (5.1,0.4) and (5.4,0.7) .. (w);
\node[rotate=0] at (4.7,0.5) {\scriptsize{$x_i$}};
\node[rotate=0] at (5.3,0.5) {\scriptsize{$y_i$}};
\end{scope}
\node (v9) at (2.7,-1.9) {$\mathbf{N}_1$};
\draw  (v4) edge (v9);
\node[rotate=0] at (3.1,-1.1) {\scriptsize{$2e$-$\sum x_i$}};
\node[rotate=0] at (2.9,-1.6) {\scriptsize{$e$}};
\node (v10) at (1.8,-1.4) {\scriptsize{2}};
\draw  (v3) edge (v10);
\draw  (v10) edge (v9);
\node[rotate=0] at (1.1,-1.2) {\scriptsize{$f$-$x_i$-$y_i$}};
\node[rotate=0] at (2.2,-1.8) {\scriptsize{$f$}};
\end{tikzpicture}
\ee
\bit
\item $f,f,f$ in $\mathbf{M}_{4}$ are glued to $x_4-x_5,x_6-x_7,x_8-x_9$ in $\mathbf{N}_0$.
\item $f,f,f$ in $\mathbf{M}_{3}$ are glued to $x_3-x_4,x_7-x_{10},x_9-x_{11}$ in $\mathbf{N}_0$.
\item $f,f,f,f,x_2,y_2$ in $\mathbf{M}_{2}$ are glued to $x_2-x_3,x_4-x_6,x_5-x_7,x_{10}-x_{12},x_{11},x_{11}$ in $\mathbf{N}_0$.
\item $f,f,f,f,x_2,y_2$ in $\mathbf{M}_{1}$ are glued to $x_1-x_2,x_6-x_8,x_7-x_9,x_{10}-x_{11},x_{12},x_{12}$ in $\mathbf{N}_0$.
\item $f,f,f,f,x_2,y_2$ in $\mathbf{M}_{0}$ are glued to $f-x_6-x_5,f-x_7-x_4,f-x_{10}-x_3,f-x_{2}-x_{12},f-x_{1},f-x_{1}$ in $\mathbf{N}_0$.
\item $x_1-x_2,y_1-y_2$ in $\mathbf{M}_{i}$ are glued to $f,f$ in $\mathbf{N}_1$ for $i=0,1,2$.
\item $f-x_1,f-x_2,f-y_1,f-y_2$ in $\mathbf{M}_{i}$ are glued to $y_1,f-x_1,x_1,f-y_1$ in $\mathbf{N}_0$ for $i=1,2$.
\item $f-x_1,f-x_2,f-y_1,f-y_2$ in $\mathbf{M}_{0}$ are glued to $y_2,f-x_2,x_2,f-y_2$ in $\mathbf{N}_0$.
\eit

\noindent\ubf{Gluing rules for \raisebox{-.25\height}{\begin{tikzpicture}
\node (w1) at (-0.5,0.9) {$\fe_6^{(2)}$};
\begin{scope}[shift={(2.9,0)}]
\node (w2) at (-0.5,0.9) {$\su(5)^{(2)}$};
\end{scope}
\node (v1) at (1,0.9) {\tiny{6}};
\draw  (w1) edge (v1);
\draw  (v1) edge (w2);
\end{tikzpicture}}}:
\be
\begin{tikzpicture} [scale=1.9]
\node (v2) at (0.1,0.2) {$\mathbf{M}_{4}$};
\node (v6) at (1.3,0.2) {$\mathbf{M}_{3}$};
\node[rotate=0] at (0.9,0.3) {\scriptsize{$e$}};
\draw  (v2) edge (v6);
\node at (0.5,0.3) {\scriptsize{$e$}};
\node (v3) at (0.8,-1.5) {$\mathbf{N}^{12+12}_0$};
\node (v4) at (2.7,-1.5) {$\mathbf{N}_1$};
\draw  (v3) edge (v4);
\node[rotate=0] at (1.6,-1.4) {\scriptsize{$e$-$\sum x_i$-$\sum y_i$}};
\node[rotate=0] at (2.35,-1.4) {\scriptsize{$2e$}};
\node (v11) at (3.2,0.2) {$\mathbf{M}^{2+2}_{2}$};
\draw  (v6) edge (v11);
\node[rotate=0] at (1.7,0.3) {\scriptsize{$2e$}};
\node[rotate=0] at (2.4,0.3) {\scriptsize{$e$-$\sum x_i$-$\sum y_i$}};
\node (v1) at (4.6,0.2) {\scriptsize{3}};
\node (v5) at (6.1,0.2) {$\mathbf{M}^{2+2+1}_{1}$};
\node (v7) at (6.1,-0.5) {\scriptsize{4}};
\node (v8) at (6.1,-1.2) {$\mathbf{M}^{2+2+4}_{0}$};
\draw  (v11) edge (v1);
\draw  (v1) edge (v5);
\draw  (v5) edge (v7);
\draw  (v7) edge (v8);
\node[rotate=0] at (4,0.3) {\scriptsize{$e$-$\sum x_i,f$-$y_i$}};
\node[rotate=0] at (5.3,0.3) {\scriptsize{$e$-$\sum x_i,f$-$y_i$}};
\node[rotate=0] at (6.8,-0.1) {\scriptsize{$e$-$\sum y_i$-$z,f$-$x_i,f$-$z$}};
\node[rotate=0] at (6.7,-0.9) {\scriptsize{$e$-$\sum y_i,f$-$x_i,z_4$}};
\node (w) at (3.2,0.8) {\scriptsize{$2$}};
\draw (v11) .. controls (3.1,0.4) and (2.8,0.7) .. (w);
\draw (v11) .. controls (3.3,0.4) and (3.6,0.7) .. (w);
\node[rotate=0] at (2.9,0.5) {\scriptsize{$x_i$}};
\node[rotate=0] at (3.5,0.5) {\scriptsize{$y_i$}};
\node (w) at (6.1,0.8) {\scriptsize{$2$}};
\draw (v5) .. controls (6,0.4) and (5.7,0.7) .. (w);
\draw (v5) .. controls (6.2,0.4) and (6.5,0.7) .. (w);
\node[rotate=0] at (5.8,0.5) {\scriptsize{$x_i$}};
\node[rotate=0] at (6.4,0.5) {\scriptsize{$y_i$}};
\begin{scope}[rotate=-180, shift={(-11.1,1)}]
\node (w) at (5,0.8) {\scriptsize{$2$}};
\draw (v8) .. controls (4.9,0.4) and (4.6,0.7) .. (w);
\draw (v8) .. controls (5.1,0.4) and (5.4,0.7) .. (w);
\node[rotate=0] at (4.7,0.5) {\scriptsize{$x_i$}};
\node[rotate=0] at (5.3,0.5) {\scriptsize{$y_i$}};
\end{scope}
\node (v9) at (4.1,-1.5) {$\mathbf{N}^2_2$};
\draw  (v4) edge (v9);
\node[rotate=0] at (3.6,-1.4) {\scriptsize{$2e$-$\sum x_i$}};
\node[rotate=0] at (3,-1.4) {\scriptsize{$e$}};
\node (w) at (0.8,-0.9) {\scriptsize{$12$}};
\draw (v3) .. controls (0.7,-1.3) and (0.4,-1) .. (w);
\draw (v3) .. controls (0.9,-1.3) and (1.2,-1) .. (w);
\node[rotate=0] at (0.5,-1.2) {\scriptsize{$x_i$}};
\node[rotate=0] at (1.1,-1.2) {\scriptsize{$y_i$}};
\draw  (v8) edge (v6);
\draw  (v2) edge (v8);
\node (v10) at (4.2,-0.3) {\scriptsize{2}};
\draw  (v11) edge (v10);
\draw  (v10) edge (v8);
\node at (0.5,0) {\scriptsize{$f$}};
\node at (2.2668,0.0238) {\scriptsize{$f$}};
\node at (3.9,0) {\scriptsize{$f,f$}};
\node[rotate=-25] at (5.3,-0.7) {\scriptsize{$f$-$z_1$-$z_2,z_3$-$z_4$}};
\node[rotate=-17] at (4.7,-0.7) {\scriptsize{$z_2$-$z_3$}};
\node[rotate=-0] at (5.1,-1.1) {\scriptsize{$z_1$-$z_2$}};
\end{tikzpicture}
\ee
\bit
\item $6f$ in $\mathbf{M}_{4}$ are glued to $x_4-x_5,x_6-x_7,x_8-x_9,y_4-y_5,y_6-y_7,y_8-y_9$ in $\mathbf{N}_0$.
\item $6f$ in $\mathbf{M}_{3}$ are glued to $x_3-x_4,x_7-x_{10},x_9-x_{11},y_3-y_4,y_7-y_{10},y_9-y_{11}$ in $\mathbf{N}_0$.
\item $4f,x_2,y_2,4f,x_2,y_2$ in $\mathbf{M}_{2}$ are glued to $x_2-x_3,x_4-x_6,x_5-x_7,x_{10}-x_{12},x_{11},x_{11},y_2-y_3,y_4-y_6,y_5-y_7,y_{10}-y_{12},y_{11},y_{11}$ in $\mathbf{N}_0$.
\item $z,f-z,3f,x_2,y_2,z,f-z,3f,x_2,y_2$ in $\mathbf{M}_{1}$ are glued to $f-x_1,f-x_2,x_6-x_8,x_7-x_9,x_{10}-x_{11},x_{12},x_{12},f-y_1,f-y_2,y_6-y_8,y_7-y_9,y_{10}-y_{11},y_{12},y_{12}$ in $\mathbf{N}_0$.
\item $f-z_1,z_1,f-z_2,z_2,f-z_3,z_3,f-z_4,z_4,x_2,y_2,f-z_1,z_1,f-z_2,z_2,f-z_3,z_3,f-z_4,z_4,x_2,y_2$ in $\mathbf{M}_{0}$ are glued to $f-x_6,f-x_5,f-x_7,f-x_4,f-x_{10},f-x_3,f-x_{12},f-x_{2},x_{1},x_{1},f-y_6,f-y_5,f-y_7,f-y_4,f-y_{10},f-y_3,f-y_{12},f-y_{2}y_{1},y_{1}$ in $\mathbf{N}_0$.
\item $f-x_1,f-y_1$ in $\mathbf{M}_{2}$ are glued to $f-x_1,x_1$ in $\mathbf{N}_2$.
\item $f-x_1,f-y_1$ in $\mathbf{M}_{1}$ are glued to $f-x_1,x_1$ in $\mathbf{N}_2$.
\item $f-x_1,f-y_1$ in $\mathbf{M}_{0}$ are glued to $f-x_2,x_2$ in $\mathbf{N}_2$.
\item $x_1-x_2,y_1-y_2$ in $\mathbf{M}_{i}$ are glued to $f,f$ in $\mathbf{N}_1$ for $i=0,1,2$.
\eit

\subsection{Gluing rules for localized flavor symmetries: with $\sp(0)^{(1)}$ nodes}
In this subsection, we discuss gluing rules when $\sp(0)^{(1)}$ non-flavor nodes are involved except for the case of
\be
\begin{tikzpicture}
\node (v1) at (-0.5,0.4) {1};
\node (v7) at (-0.5,0.85) {$\sp(0)^{(1)}$};
\begin{scope}[shift={(1.8,-0.05)}]
\node (v2) at (-0.5,0.45) {$3$};
\node at (-0.45,0.9) {$\su(3)^{(2)}$};
\end{scope}
\node (v4) at (-3,0.4) {$\left[\su(3)^{(2)}\right]$};
\node (v6) at (0.4,0.4) {\tiny{2}};
\draw  (v1) edge (v6);
\draw [->] (v6) edge (v2);
\node (v3) at (-1.4,0.4) {\tiny{2}};
\draw  (v1) edge (v3);
\draw [->] (v3) edge (v4);
\end{tikzpicture}
\ee
for which we have been unable to find a consistent geometry. It would be interesting to investigate if this simply reflects the lack of ingenuity of the author or if something is pathological about this $5d$ KK theory (or its M-theory construction).

The gluing rules are again independent of whether the nodes are flavor or non-flavor, and hence we do not make this distinction in what follows.

\be
\begin{tikzpicture}
\node (v1) at (0,0) {\scriptsize{$x_8-x_9$}};
\node (v2) at (2,0) {\scriptsize{$x_7-x_8$}};
\node (v3) at (4,0) {\scriptsize{$x_6-x_7$}};
\node (v4) at (6,0) {\scriptsize{$x_5-x_6$}};
\node (v5) at (8,0) {\scriptsize{$x_4-x_5$}};
\node (v1_1) at (10,0) {\scriptsize{$x_1-x_4$}};
\node (v1_2) at (12.4,0) {\scriptsize{$l-x_1-x_2-x_3$}};
\node (v1_4) at (6,2) {\scriptsize{$2l-x_1-x_4-x_5-x_6-x_7-x_8$}};
\draw  (v1) edge (v2);
\draw  (v2) edge (v3);
\draw  (v3) edge (v4);
\draw  (v4) edge (v5);
\draw  (v5) edge (v1_1);
\draw  (v1_1) edge (v1_2);
\draw  (v1_4) edge (v1);
\draw  (v1_4) edge (v1_2);
\begin{scope}[shift={(4.4,2.6)}]
\begin{scope}[shift={(2,0)}]
\node (w2) at (-0.5,0.9) {$\sp(0)_0^{(1)}$};
\end{scope}
\begin{scope}[shift={(4.4,0)}]
\node (w3) at (-0.5,0.9) {$\su(8)^{(1)}$};
\end{scope}
\draw (w2)--(w3);
\draw  (1.4,0.9) ellipse (3.6 and 0.8);
\begin{scope}[shift={(-0.5,0)}]
\node (w1) at (-0.5,0.9) {$\su(2)^{(1)}$};
\end{scope}
\end{scope}
\node (v1_2) at (3.4,-1.4) {\scriptsize{$3l-x_1-2x_2-x_4-x_5-x_6-x_7-x_8-x_9$}};
\node (v1_1) at (9.2,-1.4) {\scriptsize{$x_2-x_3$}};
\draw [double] (v1_2) -- (v1_1);
\draw  (w1) edge (w2);
\end{tikzpicture}
\ee
where each node represents a curve in the compact surface $\dP^9$ associated to $\sp(0)^{(1)}$. The $\P^1$ hyperplane class is denoted as $l$ and blowups are denoted as $x_i$ with $i=1,\cdots 9$. Number of edges between two nodes displays the intersection number of the two curves inside $\dP^9$. Each displayed curve is glued to a fiber of a $\P^1$ fibered surface among the collection of $\P^1$ fibered surfaces parametrizing $\su(2)^{(1)}\oplus\su(8)^{(1)}$. Since these fibers intersect in the form of Dynkin diagram of $\su(2)^{(1)}\oplus\su(8)^{(1)}$, the above displayed curves also intersect in the form of this Dynkin diagram. For example, we can see from the above that the curves $3l-x_1-2x_2-x_4-x_5-x_6-x_7-x_8-x_9$ and $x_2-x_3$ form a Dynkin diagram of $\su(2)^{(1)}$.

The above gluing rule can be obtained from the gluing rules of $\sp(0)^{(1)}$ with $\fe_8^{(1)}$ since the Dynkin diagram for $\su(8)\oplus\su(2)$ embeds into the Dynkin diagram for $\fe_8^{(1)}$. See \cite{Bhardwaj:2018vuu,Bhardwaj:2019fzv} for more details. Now we can add two of the curves $2l-x_1-x_4-x_5-x_6-x_7-x_8$ and $l-x_1-x_2-x_3$ to obtain
\be
\begin{tikzpicture}
\node (v1) at (0,0) {\scriptsize{$x_8-x_9$}};
\node (v2) at (2,0) {\scriptsize{$x_7-x_8$}};
\node (v3) at (4,0) {\scriptsize{$x_6-x_7$}};
\node (v4) at (6,0) {\scriptsize{$x_5-x_6$}};
\node (v5) at (8,0) {\scriptsize{$x_4-x_5$}};
\node (v1_1) at (10,0) {\scriptsize{$x_1-x_4$}};
\node (v1_4) at (5,2) {\scriptsize{$3l-2x_1-x_2-x_3-x_4-x_5-x_6-x_7-x_8$}};
\draw  (v1) edge (v2);
\draw  (v2) edge (v3);
\draw  (v3) edge (v4);
\draw  (v4) edge (v5);
\draw  (v5) edge (v1_1);
\draw  (v1_4) edge (v1);
\begin{scope}[shift={(3.4,2.6)}]
\begin{scope}[shift={(2,0)}]
\node (w2) at (-0.5,0.9) {$\sp(0)^{(1)}$};
\end{scope}
\begin{scope}[shift={(4.4,0)}]
\node (w3) at (-0.5,0.9) {$\su(7)^{(1)}$};
\end{scope}
\draw (w2)--(w3);
\draw  (1.4,0.9) ellipse (3.6 and 0.8);
\begin{scope}[shift={(-0.5,0)}]
\node (w1) at (-0.5,0.9) {$\su(2)^{(1)}$};
\end{scope}
\end{scope}
\node (v1_2_1) at (2.7,-1.3) {\scriptsize{$3l-x_1-2x_2-x_4-x_5-x_6-x_7-x_8-x_9$}};
\node (v1_1_1) at (8.5,-1.3) {\scriptsize{$x_2-x_3$}};
\draw [double] (v1_2_1) -- (v1_1_1);
\draw  (w1) edge (w2);
\draw  (v1_4) edge (v1_1);
\end{tikzpicture}
\ee
Moving on, we have the gluing rule
\be
\begin{tikzpicture}
\node (v1) at (0,0) {\scriptsize{$x_5-x_6$}};
\node (v2) at (2,0) {\scriptsize{$x_4-x_5$}};
\node (v3) at (4,0) {\scriptsize{$x_1-x_4$}};
\node (v4) at (6,0) {\scriptsize{$x_2-x_1$}};
\node (v5) at (8,0) {\scriptsize{$x_3-x_2$}};
\node (v1_4) at (4,2) {\scriptsize{$3l-x_1-x_2-2x_3-x_4-x_5-x_7-x_8-x_9$}};
\draw  (v1) edge (v2);
\draw  (v2) edge (v3);
\draw  (v3) edge (v4);
\draw  (v4) edge (v5);
\draw  (v1_4) edge (v1);
\draw  (v1_4) edge (v5);
\node (v1) at (2,-2.3) {\scriptsize{$x_8-x_9$}};
\node (v1_1) at (6,-2.3) {\scriptsize{$x_7-x_8$}};
\node (v1_1_1) at (4,-1.3) {\scriptsize{$3l-x_1-x_2-x_3-x_4-x_5-x_6-2x_7-x_8$}};
\draw  (v1) edge (v1_1_1);
\draw  (v1) edge (v1_1);
\draw  (v1_1_1) -- (v1_1);
\begin{scope}[shift={(2.4,2.8)}]
\begin{scope}[shift={(4.4,0)}]
\node (w1) at (-0.5,0.9) {$\su(3)^{(1)}$};
\end{scope}
\begin{scope}[shift={(2,0)}]
\node (w2) at (-0.5,0.9) {$\sp(0)^{(1)}$};
\end{scope}
\begin{scope}[shift={(-0.4,0)}]
\node (w3) at (-0.5,0.9) {$\su(6)^{(1)}$};
\end{scope}
\draw (w2)--(w3);
\draw  (w1) edge (w2);
\draw  (1.5,1.4) node (v6) {} ellipse (3.8 and 1.4);
\node (w4) at (1.5,2.4) {$\su(2)^{(1)}$};
\end{scope}
\node (v1_2_1) at (1.5,-3.2) {\scriptsize{$2l-x_1-x_2-x_3-x_4-x_5-x_6$}};
\node (v1_1_1) at (7.3,-3.2) {\scriptsize{$l-x_7-x_8-x_9$}};
\draw [double] (v1_2_1) -- (v1_1_1);
\draw  (w4) edge (w2);
\end{tikzpicture}
\ee
which can be obtained from the gluing rules of $\sp(0)^{(1)}$ with $\fe_6^{(1)}\oplus\su(3)^{(1)}$ since the Dynkin diagram of $\su(2)\oplus\su(6)$ embeds into the Dynkin diagram of $\fe_6^{(1)}$. From the above, we can combine the two curves $3l-x_1-x_2-2x_3-x_4-x_5-x_7-x_8-x_9$ and $x_3-x_2$ to obtain
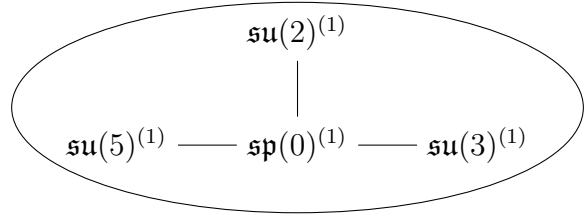
\be
\begin{tikzpicture}
\node (v1) at (0,0) {\scriptsize{$x_5-x_6$}};
\node (v2) at (2,0) {\scriptsize{$x_4-x_5$}};
\node (v3) at (4,0) {\scriptsize{$x_1-x_4$}};
\node (v4) at (6,0) {\scriptsize{$x_2-x_1$}};
\node (v1_4) at (3,1.3) {\scriptsize{$3l-x_1-2x_2-x_3-x_4-x_5-x_7-x_8-x_9$}};
\draw  (v1) edge (v2);
\draw  (v2) edge (v3);
\draw  (v3) edge (v4);
\draw  (v1_4) edge (v1);
\node (v1) at (1,-2.1) {\scriptsize{$x_8-x_9$}};
\node (v1_1) at (5,-2.1) {\scriptsize{$x_7-x_8$}};
\node (v1_1_1) at (3,-1.1) {\scriptsize{$3l-x_1-x_2-x_3-x_4-x_5-x_6-2x_7-x_8$}};
\draw  (v1) edge (v1_1_1);
\draw  (v1) edge (v1_1);
\draw  (v1_1_1) -- (v1_1);
\begin{scope}[shift={(1.4,2)}]
\begin{scope}[shift={(4.4,0)}]
\node (w1) at (-0.5,0.9) {$\su(3)^{(1)}$};
\end{scope}
\begin{scope}[shift={(2,0)}]
\node (w2) at (-0.5,0.9) {$\sp(0)^{(1)}$};
\end{scope}
\begin{scope}[shift={(-0.4,0)}]
\node (w3) at (-0.5,0.9) {$\su(5)^{(1)}$};
\end{scope}
\draw (w2)--(w3);
\draw  (w1) edge (w2);
\draw  (1.5,1.4) node (v6) {} ellipse (3.8 and 1.4);
\node (w4) at (1.5,2.4) {$\su(2)^{(1)}$};
\end{scope}
\node (v1_2_1) at (1.4,-3.2) {\scriptsize{$2l-x_1-x_2-x_3-x_4-x_5-x_6$}};
\node (v1_1_1) at (5.6,-3.2) {\scriptsize{$l-x_7-x_8-x_9$}};
\draw [double] (v1_2_1) -- (v1_1_1);
\draw  (w4) edge (w2);
\draw  (v1_4) edge (v4);
\end{tikzpicture}
\ee
From the $\sp(0)^{(1)}$ gluing rules for $\fe_8^{(1)}$, we can obtain the gluing rule
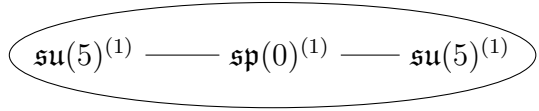
\be
\begin{tikzpicture}
\node (v1) at (0,0) {\scriptsize{$x_8-x_9$}};
\node (v2) at (2,0) {\scriptsize{$x_7-x_8$}};
\node (v3) at (4,0) {\scriptsize{$x_6-x_7$}};
\node (v4) at (6,0) {\scriptsize{$x_5-x_6$}};
\node (v1_4) at (3,1.5) {\scriptsize{$3l-x_1-x_2-x_3-x_4-2x_5-x_6-x_7-x_8$}};
\draw  (v1) edge (v2);
\draw  (v2) edge (v3);
\draw  (v3) edge (v4);
\draw  (v1_4) edge (v1);
\draw  (v1_4) edge (v4);
\begin{scope}[shift={(1.6,2.1)}]
\begin{scope}[shift={(2,0)}]
\node (w2) at (-0.5,0.9) {$\sp(0)^{(1)}$};
\end{scope}
\begin{scope}[shift={(4.4,0)}]
\node (w3) at (-0.5,0.9) {$\su(5)^{(1)}$};
\end{scope}
\draw (w2)--(w3);
\draw  (1.4,0.9) ellipse (3.5 and 0.7);
\begin{scope}[shift={(-0.6,0)}]
\node (w3) at (-0.5,0.9) {$\su(5)^{(1)}$};
\end{scope}
\end{scope}
\begin{scope}[shift={(0.2,-2.3)}]
\node (v1_1) at (0,0) {\scriptsize{$l-x_1x_2-x_3$}};
\node (v2_1) at (2,0) {\scriptsize{$x_1-x_4$}};
\node (v3_1) at (4,0) {\scriptsize{$x_2-x_1$}};
\node (v4_1) at (6,0) {\scriptsize{$x_3-x_2$}};
\node (v1_4_1) at (3,1.5) {\scriptsize{$2l-x_3-x_5-x_6-x_7-x_8-x_9$}};
\end{scope}
\draw  (v1_4_1) edge (v1_1);
\draw  (v1_1) edge (v2_1);
\draw  (v2_1) edge (v3_1);
\draw  (v3_1) edge (v4_1);
\draw  (v4_1) edge (v1_4_1);
\draw  (w3) edge (w2);
\end{tikzpicture}
\ee
Another gluing rule is
\be\label{doub}
\begin{tikzpicture}
\node (v1) at (0,0) {\scriptsize{$x_8-x_9$}};
\node (v2) at (2,0) {\scriptsize{$x_7-x_8$}};
\node (v3) at (4,0) {\scriptsize{$x_6-x_7$}};
\node (v4) at (6,0) {\scriptsize{$x_5-x_6$}};
\node (v5) at (8,0) {\scriptsize{$x_4-x_5$}};
\node (v1_1) at (11,0) {\scriptsize{$x_1-x_4$,\:\:$l-x_1-x_3-x_4$}};
\node (v1_3) at (2,1) {\scriptsize{$2l-x_1-x_2-x_4-x_5-x_6-x_7$}};
\draw  (v1) edge (v2);
\draw  (v2) edge (v3);
\draw  (v3) edge (v4);
\draw  (v4) edge (v5);
\draw[double]  (v5) -- (v1_1);
\draw  (v1_3) edge (v2);
\begin{scope}[shift={(3.7,1.5)}]
\begin{scope}[shift={(2,0)}]
\node (w2) at (-0.5,0.9) {$\sp(0)^{(1)}$};
\end{scope}
\begin{scope}[shift={(4.4,0)}]
\node (w3) at (-0.5,0.9) {$\so(13)^{(1)}$};
\end{scope}
\draw (w2)--(w3);
\draw  (1.3,0.9) ellipse (3.6 and 0.7);
\begin{scope}[shift={(-0.8,0)}]
\node (w1) at (-0.5,0.9) {$\su(2)^{(1)}$};
\end{scope}
\end{scope}
\node (v6) at (5.2,-1.2) {\scriptsize{$2l-x_4-x_5-x_6-x_7-x_8-x_9,x_2-x_1$}};
\node (v7) at (5.2,-2.2) {\scriptsize{$3l-2x_2-x_3-x_4-x_5-x_6-x_7-x_8-x_9,l-x_1-x_2-x_3$}};
\draw (5.175,-1.35) -- (5.175,-2.075) (5.225,-2.075) -- (5.225,-1.35) (5.275,-1.35) -- (5.275,-2.075) (5.125,-2.075) -- (5.125,-1.35);
\node (v8) at (3.8,2.4) {\tiny{2}};
\draw  (w1) edge (v8);
\draw  (v8) edge (w2);
\end{tikzpicture}
\ee
A node with multiple curves $C,D,\cdots$ indicates that the curves $C,D,\cdots$ inside $\dP^9$ glue to different copies of $\P^1$ fiber of the $\P^1$ fibered surface corresponding to that node. Thus, in the above gluing rule, $x_1-x_4$ and $l-x_1-x_3-x_4$ glue to two copies of $\P^1$ fiber of the surface associated to that node of Dynkin diagram of $\so(13)^{(1)}$ which corresponds to spinor representation of $\so(13)$. In such a case, the number of edges between two nodes denotes the total intersection number. Thus, we have spanned two edges between node corresponding to $x_4-x_5$ and the node corresponding to $x_1-x_4,l-x_1-x_3-x_4$ since $(x_4-x_5)\cdot((x_1-x_4)+(l-x_1-x_3-x_4))=2$.

From the gluing rules for $\fe_7^{(1)}\oplus\su(2)^{(1)}$, one can obtain
\be\label{dkwtp}

\ee
Using the gluing rules (\ref{doub}) we can obtain
\be\label{so7f}
%
\ee
since the Dynkin digram of $\so(7)\oplus\su(4)$ embeds into the Dynkin diagram of $\so(13)^{(1)}$. Combining two of the curves, we also obtain
\be
%
\ee
Similarly, from (\ref{dkwtp}) we can deduce the following gluing rule
\be
%
\ee
From the gluing rules for $\fe_6^{(1)}\oplus\su(3)^{(1)}$, we obtain
\be
%
\ee
Combining two curves in $\so(12)^{(1)}$ part of (\ref{dkwtp}), we obtain
\be
%
\ee
Folding the $\su(8)^{(2)}$ part in the above gluing rule, we obtain
\be
%
\ee

The gluing rules for configurations of the form
\be
%
\ee
can be obtained from the gluing rules for the corresponding configurations of the form
\be
%
\ee
by adding $3l-\sum x_i$ to $e-1$ number of gluing curves for $\fg^{(1)}$. Similarly, the gluing rules for configurations of the form
\be
%
\ee
can be obtained from the gluing rules for the corresponding configurations of the form
\be
%
\ee
by adding $3l-\sum x_i$ to $e-1$ number of gluing curves for $\fg_1^{(1)}$.

Using the same procedure, we obtain the gluing rules
\be
%
\ee
from the gluing rules (\ref{e6gr}) by adding $3l-\sum x_i$ to two of the gluing curves for $\fe_6^{(2)}$. Similarly, the gluing rules for
\be
%
\ee
can be obtained from the gluing rules for
\be
%
\ee
by adding $3l-\sum x_i$ to two of the gluing curves for $\fe_6^{(2)}$. And, the gluing rules for
\be
%
\ee
can be obtained from the gluing rules for
\be
%
\ee
by adding $3l-\sum x_i$ to two of the gluing curves for $\fe_6^{(2)}$.

\subsection{Gluing rules for delocalized flavor symmetries}
In this section we will be using $\mathbf{N}_i,\mathbf{M}_i$ etc. to denote only \emph{non-compact} $\P^1$ fibered surfaces. The compact surfaces have a $\P^1$ base and thus are Hirzebruch surfaces. They are denoted as
\be
\mathbf{i}_n^b
\ee
where $i$ is simply a label, $n$ is the degree of the $\P^1$ fibration and $b$ is the number of blowups. Such a surface is also denoted as $\bF_n^b$, referred to as a Hirzebruch surface of degree $n$ and carrying $b$ blowups.

The gluing rules for the $\su(2)^{(1)}$ delocalized flavor symmetry of
\be

\ee
are described by
\be
%
\ee
where $\lceil r\rceil$ denotes the smallest integer greater than or equal to $r$.
\bit
\item $f$ in $\mathbf{N}_{1}$ is glued to $f-x-y$ in $\mathbf{S}_i$ for $i=1,\cdots,m$.
\item $f-x_i,x_i$ in $\mathbf{N}_{0}$ are glued to $x,y$ in $\mathbf{S}_{2i-1}$ for $i=1,\cdots,\lceil m/2\rceil$.
\item $f-x_i,x_i$ in $\mathbf{N}_{0}$ are glued to $y,x$ in $\mathbf{S}_{2i}$ for $i=1,\cdots,\lceil m/2\rceil$.
\eit
where we refer to the Hirzebruch surface $\mathbf{i}_n^b$ as $\mathbf{S}_i$.

The gluing rules for the $\su(2)^{(1)}$ delocalized flavor symmetry of
\be
%
\ee
are described by
\be
%
\ee
\bit
\item $f$ in $\mathbf{N}_{1}$ is glued to $f-x-y$ in $\mathbf{S}_i$ for $i=1,\cdots,m+1$.
\item $f-x_i,x_i$ in $\mathbf{N}_{0}$ are glued to $x,y$ in $\mathbf{S}_{2i-1}$ for $i=1,\cdots,\lceil m/2\rceil$.
\item $f-x_i,x_i$ in $\mathbf{N}_{0}$ are glued to $y,x$ in $\mathbf{S}_{2i}$ for $i=1,\cdots,\lceil m/2\rceil$.
\item $f-x_{\lceil m/2\rceil+1},x_{\lceil m/2\rceil+1}$ in $\mathbf{N}_{0}$ are glued to $x,y$ in $\mathbf{S}_{m+1}$.
\eit

The gluing rules for the $\su(2)^{(1)}$ delocalized flavor symmetry of
\be
%
\ee
are described by
\be
%
\ee
\bit
\item $f$ in $\mathbf{N}_{1}$ is glued to $f-x-y$ in $\mathbf{S}_i$ for $i=1,\cdots,6$.
\item $f-x_i,x_i$ in $\mathbf{N}_{0}$ are glued to $x,y$ in $\mathbf{S}_{2i-1}$ for $i=1,\cdots,3$.
\item $f-x_i,x_i$ in $\mathbf{N}_{0}$ are glued to $y,x$ in $\mathbf{S}_{2i}$ for $i=1,\cdots,2$.
\item $f-x_{4},x_{4}$ in $\mathbf{N}_{0}$ are glued to $x,y$ in $\mathbf{S}_{6}$.
\eit

The gluing rules for the $\su(2)^{(1)}$ delocalized flavor symmetry of
\be
%
\ee
\bit
\item $f$ in $\mathbf{N}_{1}$ is glued to $f-x-y$ in $\mathbf{S}_i$ for $i=1,\cdots,7$.
\item $f-x_i,x_i$ in $\mathbf{N}_{0}$ are glued to $x,y$ in $\mathbf{S}_{2i-1}$ for $i=1,\cdots,3$.
\item $f-x_i,x_i$ in $\mathbf{N}_{0}$ are glued to $y,x$ in $\mathbf{S}_{2i}$ for $i=1,\cdots,3$.
\item $f-x_{4},x_{4}$ in $\mathbf{N}_{0}$ are glued to $x,y$ in $\mathbf{S}_{7}$.
\eit

The gluing rules for the $\su(2)^{(1)}$ delocalized flavor symmetry of
\be
%
\ee
\bit
\item $f$ in $\mathbf{N}_{1}$ is glued to $f-x-y$ in $\mathbf{S}_i$ for $i=1,\cdots,8$.
\item $f-x_i,x_i$ in $\mathbf{N}_{0}$ are glued to $x,y$ in $\mathbf{S}_{2i-1}$ for $i=1,\cdots,4$.
\item $f-x_i,x_i$ in $\mathbf{N}_{0}$ are glued to $y,x$ in $\mathbf{S}_{2i}$ for $i=1,\cdots,3$.
\item $f-x_{5},x_{5}$ in $\mathbf{N}_{0}$ are glued to $x,y$ in $\mathbf{S}_{8}$.
\eit

The gluing rules for the $(\su(2)^{(1)})^{\oplus 3}$ delocalized flavor symmetry of
\be
%
\ee
for $m$ even are described by
\be
%
\ee
\bit
\item $y_1,y_2$ in $\mathbf{S}_{m+2}$ are glued to $x_1,x_2$ in $\mathbf{P}_0$.
\item $f-y_1-y_2$ in $\mathbf{S}_{m+2}$ is glued to $f$ in $\mathbf{P}_1$.
\item $f$ in $\mathbf{S}_{m}$ is glued to $f-x_1-x_2$ in $\mathbf{P}_0$.
\item $x_1,x_2$ in $\mathbf{S}_{1}$ are glued to $x_1,x_2$ in $\mathbf{Q}_0$.
\item $f-x_1-x_2$ in $\mathbf{S}_{1}$ is glued to $f$ in $\mathbf{Q}_1$.
\item $f$ in $\mathbf{S}_{2m+1}$ is glued to $f-x_1-x_2$ in $\mathbf{Q}_0$.
\item $x_1-x_2,y_2-y_1$ in $\mathbf{S}_{m+2i}$ are glued to $f,f$ in $\mathbf{M}_1$ for $i=1,\cdots,\frac m2$.
\item $x_2-x_1,y_1-y_2$ in $\mathbf{S}_{m+1-2i}$ are glued to $f,f$ in $\mathbf{M}_1$ for $i=1,\cdots,\frac m2$.
\item $f-x_1,x_2,f-y_2,y_1$ in $\mathbf{S}_{m+2i}$ are glued to $f-x_{2i},y_{2i},x_{2i-1},y_{2i-1}$ in $\mathbf{M}_0$ for $i=1,\cdots,\frac m2$.
\item $f-x_2,x_1,f-y_1,y_2$ in $\mathbf{S}_{m+1-2i}$ are glued to $f-x_{2i+1},y_{2i+1},x_{2i},y_{2i}$ in $\mathbf{M}_0$ for $i=1,\cdots,\frac m2$.
\item $f,f$ in $\mathbf{S}_{m+2-2i}$ are glued to $x_{2i}-y_{2i},f-x_{2i-1}-y_{2i-1}$ in $\mathbf{M}_0$ for $i=1,\cdots,\frac m2$.
\item $f,f$ in $\mathbf{S}_{m+1+2i}$ are glued to $x_{2i+1}-y_{2i+1},f-x_{2i}-y_{2i}$ in $\mathbf{M}_0$ for $i=1,\cdots,\frac m2$.
\item $x_2-x_1$ in $\mathbf{P}_{0}$ is glued to $f$ in $\mathbf{M}_1$.
\item $f-x_2,x_1$ in $\mathbf{P}_{0}$ is glued to $f-x_1,y_1$ in $\mathbf{M}_0$.
\item $f$ in $\mathbf{P}_{1}$ is glued to $x_1-y_1$ in $\mathbf{M}_0$.
\item $x_2-x_1$ in $\mathbf{Q}_{0}$ is glued to $f$ in $\mathbf{M}_1$.
\item $f-x_2,x_1$ in $\mathbf{Q}_{0}$ is glued to $x_{m+1},y_{m+1}$ in $\mathbf{M}_0$.
\item $f$ in $\mathbf{Q}_{1}$ is glued to $f-x_{m+1}-y_{m+1}$ in $\mathbf{M}_0$.
\eit
Let us note, in order to avoid confusion, that there is no surface $\mathbf{S}_{m+1}$ in the above geometry. For $m$ odd, the gluing rules are described by
\be
\begin{tikzpicture} [scale=1.9]
\node (v2) at (0.9,1.3) {$\mathbf{m}_{0}$};
\node (v3) at (3.1,1.3) {$\mathbf{(m-1)}_{0}^{2+2}$};
\node (v10) at (0.9,-0.8) {$\mathbf{(m+2)}_0^{2+2}$};
\node (v9) at (3.1,-0.8) {$\mathbf{(m+3)}_0$};
\node (v4) at (4.5,1.3) {$\cdots$};
\node (v8) at (4,-0.8) {$\cdots$};
\node (v5) at (5.3,1.3) {$\mathbf{1}_0$};
\node (v7) at (5.3,-0.8) {$\mathbf{2m+1}_{0}^{2+2}$};
\draw  (v2) edge (v3);
\draw  (v3) edge (v4);
\draw  (v4) edge (v5);
\draw  (v7) edge (v8);
\draw  (v8) edge (v9);
\draw  (v9) edge (v10);
\node at (1.3,1.4) {\scriptsize{$f$}};
\node at (2.3,1.4) {\scriptsize{$f$-$\sum y_i$}};
\node at (3.9,1.4) {\scriptsize{$f$-$\sum x_i$}};
\node at (5,1.4) {\scriptsize{$f$}};
\node at (4.5,-0.9) {\scriptsize{$f$-$\sum y_i$}};
\node at (1.7,-0.9) {\scriptsize{$f$-$\sum x_i$}};
\node at (1.1646,-0.3943) {\scriptsize{$x_i$}};
\node at (2.6,1) {\scriptsize{$y_i$}};
\node at (3.4,0.9) {\scriptsize{$x_i$}};
\node at (2.5,-0.9) {\scriptsize{$f$}};
\node at (0.75,0.8) {\scriptsize{$e$}};
\node at (3.7,-0.9) {\scriptsize{$f$}};
\node at (4.8,-0.5) {\scriptsize{$y_i$}};
\node (v14) at (0.9,0.25) {\scriptsize{2}};
\draw  (v2) edge (v14);
\draw  (v10) edge (v14);
\node at (0.65,-0.15) {\scriptsize{$e$-$x_i$-$y_i$}};
\begin{scope}[shift={(2.2,-0.1)}]
\node at (0.75,-0.3) {\scriptsize{$e$}};
\end{scope}
\begin{scope}[shift={(2.2,0)}]
\node at (0.65,0.65) {\scriptsize{$e$-$x_i$-$y_i$}};
\end{scope}
\begin{scope}[shift={(2.2,0)}]
\node (v14) at (0.9,0.25) {\scriptsize{2}};
\end{scope}
\draw  (v3) edge (v14);
\draw  (v14) edge (v9);
\draw (v3) -- (3.9,0.6);
\draw (v7) -- (4.5,0);
\begin{scope}[shift={(4.4,0.1)}]
\node at (0.75,0.8) {\scriptsize{$e$}};
\end{scope}
\begin{scope}[shift={(4.4,0)}]
\node at (0.65,-0.15) {\scriptsize{$e$-$x_i$-$y_i$}};
\end{scope}
\begin{scope}[shift={(4.4,0)}]
\node (v14) at (0.9,0.25) {\scriptsize{2}};
\end{scope}
\draw  (v5) edge (v14);
\draw  (v14) edge (v7);
\node at (4.3,0.3) {$\cdots$};
\begin{scope}[shift={(0.2,-0.1)}]
\node (v24) at (1,-1.6) {$\mathbf{M}^{(m+1)+(m+1)}_{0}$};
\node (v25) at (3.6,-1.6) {$\mathbf{M}_{1}$};
\node[rotate=0] at (2.1,-1.5) {\scriptsize{$2e$-$\sum x_i$}};
\node[rotate=0] at (3.2,-1.5) {\scriptsize{$2e$}};
\draw  (v24) edge (v25);
\end{scope}
\begin{scope}[shift={(-1.6,-1)}]
\node (v24) at (1.5,-1.6) {$\mathbf{P}^{2}_{0}$};
\node (v25) at (3.6,-1.6) {$\mathbf{P}_{1}$};
\node[rotate=0] at (2.1,-1.5) {\scriptsize{$2e$-$\sum x_i$}};
\node[rotate=0] at (3.2,-1.5) {\scriptsize{$2e$}};
\draw  (v24) edge (v25);
\end{scope}
\begin{scope}[shift={(1.6,-1)}]
\node (v24) at (1.5,-1.6) {$\mathbf{Q}^{2}_{0}$};
\node (v25) at (3.6,-1.6) {$\mathbf{Q}_{1}$};
\node[rotate=0] at (2.1,-1.5) {\scriptsize{$2e$-$\sum x_i$}};
\node[rotate=0] at (3.2,-1.5) {\scriptsize{$2e$}};
\draw  (v24) edge (v25);
\end{scope}
\node (v26) at (1.9941,0.2425) {\scriptsize{2}};
\draw  (v10) edge (v26);
\draw  (v26) edge (v3);
\end{tikzpicture}
\ee
\bit
\item $y_1,y_2$ in $\mathbf{S}_{m+2}$ are glued to $x_1,x_2$ in $\mathbf{P}_0$.
\item $f-y_1-y_2$ in $\mathbf{S}_{m+2}$ is glued to $f$ in $\mathbf{P}_1$.
\item $f$ in $\mathbf{S}_{m}$ is glued to $f-x_1-x_2$ in $\mathbf{P}_0$.
\item $x_1,x_2$ in $\mathbf{S}_{2m+1}$ are glued to $x_1,x_2$ in $\mathbf{Q}_0$.
\item $f-x_1-x_2$ in $\mathbf{S}_{2m+1}$ is glued to $f$ in $\mathbf{Q}_1$.
\item $f$ in $\mathbf{S}_{1}$ is glued to $f-x_1-x_2$ in $\mathbf{Q}_0$.
\item $x_1-x_2,y_2-y_1$ in $\mathbf{S}_{m+2i}$ are glued to $f,f$ in $\mathbf{M}_1$ for $i=1,\cdots,\frac{m+1}2$.
\item $x_2-x_1,y_1-y_2$ in $\mathbf{S}_{m+1-2i}$ are glued to $f,f$ in $\mathbf{M}_1$ for $i=1,\cdots,\frac{m-1}2$.
\item $f-x_1,x_2,f-y_2,y_1$ in $\mathbf{S}_{m+2i}$ are glued to $f-x_{2i},y_{2i},x_{2i-1},y_{2i-1}$ in $\mathbf{M}_0$ for $i=1,\cdots,\frac{m+1}2$.
\item $f-x_2,x_1,f-y_1,y_2$ in $\mathbf{S}_{m+1-2i}$ are glued to $f-x_{2i+1},y_{2i+1},x_{2i},y_{2i}$ in $\mathbf{M}_0$ for $i=1,\cdots,\frac{m-1}2$.
\item $f,f$ in $\mathbf{S}_{m+2-2i}$ are glued to $x_{2i}-y_{2i},f-x_{2i-1}-y_{2i-1}$ in $\mathbf{M}_0$ for $i=1,\cdots,\frac{m+1}2$.
\item $f,f$ in $\mathbf{S}_{m+1+2i}$ are glued to $x_{2i+1}-y_{2i+1},f-x_{2i}-y_{2i}$ in $\mathbf{M}_0$ for $i=1,\cdots,\frac{m-1}2$.
\item $x_2-x_1$ in $\mathbf{P}_{0}$ is glued to $f$ in $\mathbf{M}_1$.
\item $f-x_2,x_1$ in $\mathbf{P}_{0}$ is glued to $f-x_1,y_1$ in $\mathbf{M}_0$.
\item $f$ in $\mathbf{P}_{1}$ is glued to $x_1-y_1$ in $\mathbf{M}_0$.
\item $x_1-x_2$ in $\mathbf{Q}_{0}$ is glued to $f$ in $\mathbf{M}_1$.
\item $f-x_1,x_2$ in $\mathbf{Q}_{0}$ is glued to $x_{m+1},y_{m+1}$ in $\mathbf{M}_0$.
\item $f$ in $\mathbf{Q}_{1}$ is glued to $f-x_{m+1}-y_{m+1}$ in $\mathbf{M}_0$.
\eit
Let us note, in order to avoid confusion, that there is no surface $\mathbf{S}_{m+1}$ in the above geometry.

The gluing rules for the $(\su(2)^{(1)})^{\oplus 2}$ delocalized flavor symmetry of
\be
\begin{tikzpicture} [scale=1.9]
\node (v1) at (-1.5,0.8) {2};
\node at (-1.5,1.1) {$\su(2)^{(1)}$};
\node (v3) at (-2.9,0.8) {2};
\node at (-2.9,1.1) {$\su(2)^{(1)}$};
\node (v2) at (-2.2,0.8) {$\cdots$};
\draw  (v2) edge (v3);
\draw  (v2) edge (v1);
\node (v4) at (-3.8,0.8) {1};
\node at (-3.8,1.1) {$\sp(1)^{(1)}$};
\draw  (v4) edge (v3);
\begin{scope}[shift={(0,0.45)}]
\node at (-2.2,-0.15) {$m$};
\draw (-3.1,0.15) .. controls (-3.1,0.1) and (-3.1,0.05) .. (-3,0.05);
\draw (-3,0.05) -- (-2.3,0.05);
\draw (-2.2,0) .. controls (-2.2,0.05) and (-2.25,0.05) .. (-2.3,0.05);
\draw (-2.2,0) .. controls (-2.2,0.05) and (-2.15,0.05) .. (-2.1,0.05);
\draw (-2.1,0.05) -- (-1.4,0.05);
\draw (-1.3,0.15) .. controls (-1.3,0.1) and (-1.3,0.05) .. (-1.4,0.05);
\end{scope}
\end{tikzpicture}
\ee
for $m$ odd are described by
\be
\begin{tikzpicture} [scale=1.9]
\node (v2) at (0.9,1.3) {$\mathbf{(m+1)^{8}_{0}}$};
\node (v3) at (3.1,1.3) {$\mathbf{m_{0}^{2+2}}$};
\node (v10) at (0.9,-0.8) {$\mathbf{(m+2)_0^{2}}$};
\node (v9) at (3.1,-0.8) {$\mathbf{(m+3)_0}$};
\node (v4) at (4.4,1.3) {$\cdots$};
\node (v8) at (4.1,-0.8) {$\cdots$};
\node (v5) at (5.5,1.3) {$\mathbf{2_0}$};
\node (v7) at (5.5,-0.8) {$\mathbf{(2m+1)_{0}^{2+2}}$};
\node (v6) at (7.8,1.3) {$\mathbf{1_0^{2+2}}$};
\draw  (v2) edge (v3);
\draw  (v3) edge (v4);
\draw  (v4) edge (v5);
\draw  (v5) edge (v6);
\draw  (v7) edge (v8);
\draw  (v8) edge (v9);
\draw  (v9) edge (v10);
\node[rotate=0] at (0.5,0.9) {\scriptsize{$2e$+$f$-$\sum x_i$}};
\node at (0.5,-0.4) {\scriptsize{$2e$+$f$-$\sum x_i$}};
\node at (1.5,1.4) {\scriptsize{$f$}};
\node at (2.5,1.4) {\scriptsize{$f$-$\sum y_i$}};
\node at (3.7,1.4) {\scriptsize{$f$-$\sum x_i$}};
\node at (5.2,1.4) {\scriptsize{$f$}};
\node at (5.8,1.4) {\scriptsize{$f$}};
\node at (6.4,-0.9) {\scriptsize{$f$-$\sum x_i$}};
\node at (4.7,-0.9) {\scriptsize{$f$-$\sum y_i$}};
\node at (1.6,-0.9) {\scriptsize{$f$-$\sum x_i$}};
\node at (1.1683,-0.4022) {\scriptsize{$x_i$}};
\node at (2.6,1) {\scriptsize{$y_i$}};
\node at (3.4,0.9) {\scriptsize{$x_i$}};
\node at (2.5,-0.9) {\scriptsize{$f$}};
\node at (3.7,-0.9) {\scriptsize{$f$}};
\node at (7.1,-0.9) {\scriptsize{$f$}};
\node at (5,-0.5) {\scriptsize{$y_i$}};
\node (v13) at (7.8,-0.8) {$\mathbf{(2m+2)_{0}}$};
\draw  (v7) edge (v13);
\node at (7.2,1.4) {\scriptsize{$f$-$\sum y_i$}};
\node at (5.8,-0.3) {\scriptsize{$x_i$}};
\node at (7.3,1) {\scriptsize{$y_i$}};
\begin{scope}[shift={(2.2,-0.1)}]
\node at (0.75,-0.3) {\scriptsize{$e$}};
\end{scope}
\begin{scope}[shift={(2.2,0)}]
\node at (0.65,0.65) {\scriptsize{$e$-$x_i$-$y_i$}};
\end{scope}
\begin{scope}[shift={(2.2,0)}]
\node (v14) at (0.9,0.25) {\scriptsize{2}};
\end{scope}
\draw  (v3) edge (v14);
\draw  (v14) edge (v9);
\draw (v3) -- (3.9,0.6);
\draw (v7) -- (4.6,0);
\begin{scope}[shift={(4.6,0.1)}]
\node at (0.75,0.8) {\scriptsize{$e$}};
\end{scope}
\begin{scope}[shift={(4.6,0)}]
\node at (0.65,-0.15) {\scriptsize{$e$-$x_i$-$y_i$}};
\end{scope}
\begin{scope}[shift={(4.6,0)}]
\node (v14) at (0.9,0.25) {\scriptsize{2}};
\end{scope}
\draw  (v5) edge (v14);
\draw  (v14) edge (v7);
\begin{scope}[shift={(6.9,0)}]
\node (v14) at (0.9,0.25) {\scriptsize{2}};
\end{scope}
\draw  (v6) edge (v14);
\draw  (v14) edge (v13);
\begin{scope}[shift={(6.9,0)}]
\node at (0.65,0.65) {\scriptsize{$e$-$x_i$-$y_i$}};
\end{scope}
\begin{scope}[shift={(6.9,-0.1)}]
\node at (0.75,-0.3) {\scriptsize{$e$}};
\end{scope}
\node at (4.3,0.3) {$\cdots$};
\draw  (v2) edge (v10);
\node (v1) at (2.0404,0.2768) {\scriptsize{2}};
\node (v11) at (6.5356,0.1577) {\scriptsize{2}};
\draw  (v10) edge (v1);
\draw  (v1) edge (v3);
\draw  (v7) edge (v11);
\draw  (v11) edge (v6);
\begin{scope}[shift={(-0.2,0)}]
\node (v24) at (1,-1.6) {$\mathbf{M}^{(m+1)+(m+1)}_{0}$};
\node (v25) at (3.6,-1.6) {$\mathbf{M}_{1}$};
\node[rotate=0] at (2.1,-1.5) {\scriptsize{$2e$-$\sum x_i$}};
\node[rotate=0] at (3.2,-1.5) {\scriptsize{$2e$}};
\draw  (v24) edge (v25);
\end{scope}
\begin{scope}[shift={(4.3,0)}]
\node (v24) at (1.5,-1.6) {$\mathbf{P}^{2}_{0}$};
\node (v25) at (3.6,-1.6) {$\mathbf{P}_{1}$};
\node[rotate=0] at (2.1,-1.5) {\scriptsize{$2e$-$\sum x_i$}};
\node[rotate=0] at (3.2,-1.5) {\scriptsize{$2e$}};
\draw  (v24) edge (v25);
\end{scope}
\end{tikzpicture}
\ee
\bit
\item $x_1,x_2$ in $\mathbf{S}_{1}$ are glued to $x_1,x_2$ in $\mathbf{P}_0$.
\item $f-x_1-x_2$ in $\mathbf{S}_{1}$ is glued to $f$ in $\mathbf{P}_1$.
\item $f$ in $\mathbf{S}_{2m+2}$ is glued to $f-x_1-x_2$ in $\mathbf{P}_0$.
\item $x_1-x_2$ in $\mathbf{S}_{m+2}$ is glued to $f$ in $\mathbf{M}_1$.
\item $f-x_1,x_2$ in $\mathbf{S}_{m+2}$ are glued to $f-x_{1},y_{1}$ in $\mathbf{M}_0$.
\item $f$ in $\mathbf{S}_{m+1}$ is glued to $x_{1}-y_{1}$ in $\mathbf{M}_0$.
\item $x_2-x_1,y_1-y_2$ in $\mathbf{S}_{m+2-2i}$ are glued to $f,f$ in $\mathbf{M}_1$ for $i=1,\cdots,\frac{m+1}2$.
\item $x_1-x_2,y_2-y_1$ in $\mathbf{S}_{m+2+2i}$ are glued to $f,f$ in $\mathbf{M}_1$ for $i=1,\cdots,\frac{m-1}2$.
\item $f-x_2,x_1,f-y_1,y_2$ in $\mathbf{S}_{m+2-2i}$ are glued to $f-x_{2i},y_{2i},x_{2i-1},y_{2i-1}$ in $\mathbf{M}_0$ for $i=1,\cdots,\frac{m+1}2$.
\item $f-x_1,x_2,f-y_2,y_1$ in $\mathbf{S}_{m+2+2i}$ are glued to $f-x_{2i+1},y_{2i+1},x_{2i},y_{2i}$ in $\mathbf{M}_0$ for $i=1,\cdots,\frac{m-1}2$.
\item $f,f$ in $\mathbf{S}_{m+1+2i}$ are glued to $x_{2i}-y_{2i},f-x_{2i-1}-y_{2i-1}$ in $\mathbf{M}_0$ for $i=1,\cdots,\frac{m+1}2$.
\item $f,f$ in $\mathbf{S}_{m+1-2i}$ are glued to $x_{2i+1}-y_{2i+1},f-x_{2i}-y_{2i}$ in $\mathbf{M}_0$ for $i=1,\cdots,\frac{m-1}2$.
\item $x_2-x_1$ in $\mathbf{P}_{0}$ is glued to $f$ in $\mathbf{M}_1$.
\item $f-x_2,x_1$ in $\mathbf{P}_{0}$ is glued to $x_{m+1},y_{m+1}$ in $\mathbf{M}_0$.
\item $f$ in $\mathbf{P}_{1}$ is glued to $f-x_{m+1}-y_{m+1}$ in $\mathbf{M}_0$.
\eit
For $m$ even, the gluing rules are described by
\be
\begin{tikzpicture} [scale=1.9]
\node (v2) at (0.9,1.3) {$\mathbf{(m+1)^{8}_{0}}$};
\node (v3) at (3.1,1.3) {$\mathbf{m_{0}^{2+2}}$};
\node (v10) at (0.9,-0.8) {$\mathbf{(m+2)_0^{2}}$};
\node (v9) at (3.1,-0.8) {$\mathbf{(m+3)_0}$};
\node (v4) at (4.4,1.3) {$\cdots$};
\node (v8) at (4.1,-0.8) {$\cdots$};
\node (v5) at (5.5,1.3) {$\mathbf{1_0}$};
\node (v7) at (5.5,-0.8) {$\mathbf{(2m+2)_{0}^{2+2}}$};
\draw  (v2) edge (v3);
\draw  (v3) edge (v4);
\draw  (v4) edge (v5);
\draw  (v7) edge (v8);
\draw  (v8) edge (v9);
\draw  (v9) edge (v10);
\node[rotate=0] at (0.5,0.9) {\scriptsize{$2e$+$f$-$\sum x_i$}};
\node at (0.5,-0.4) {\scriptsize{$2e$+$f$-$\sum x_i$}};
\node at (1.5,1.4) {\scriptsize{$f$}};
\node at (2.5,1.4) {\scriptsize{$f$-$\sum y_i$}};
\node at (3.7,1.4) {\scriptsize{$f$-$\sum x_i$}};
\node at (5.2,1.4) {\scriptsize{$f$}};
\node at (4.7,-0.9) {\scriptsize{$f$-$\sum y_i$}};
\node at (1.6,-0.9) {\scriptsize{$f$-$\sum x_i$}};
\node at (1.1683,-0.4022) {\scriptsize{$x_i$}};
\node at (2.6,1) {\scriptsize{$y_i$}};
\node at (3.4,0.9) {\scriptsize{$x_i$}};
\node at (2.5,-0.9) {\scriptsize{$f$}};
\node at (3.7,-0.9) {\scriptsize{$f$}};
\node at (5,-0.5) {\scriptsize{$y_i$}};
\begin{scope}[shift={(2.2,-0.1)}]
\node at (0.75,-0.3) {\scriptsize{$e$}};
\end{scope}
\begin{scope}[shift={(2.2,0)}]
\node at (0.65,0.65) {\scriptsize{$e$-$x_i$-$y_i$}};
\end{scope}
\begin{scope}[shift={(2.2,0)}]
\node (v14) at (0.9,0.25) {\scriptsize{2}};
\end{scope}
\draw  (v3) edge (v14);
\draw  (v14) edge (v9);
\draw (v3) -- (3.9,0.6);
\draw (v7) -- (4.6,0);
\begin{scope}[shift={(4.6,0.1)}]
\node at (0.75,0.8) {\scriptsize{$e$}};
\end{scope}
\begin{scope}[shift={(4.6,0)}]
\node at (0.65,-0.15) {\scriptsize{$e$-$x_i$-$y_i$}};
\end{scope}
\begin{scope}[shift={(4.6,0)}]
\node (v14) at (0.9,0.25) {\scriptsize{2}};
\end{scope}
\draw  (v5) edge (v14);
\draw  (v14) edge (v7);
\node at (4.3,0.3) {$\cdots$};
\draw  (v2) edge (v10);
\node (v1) at (2.0404,0.2768) {\scriptsize{2}};
\draw  (v10) edge (v1);
\draw  (v1) edge (v3);
\begin{scope}[shift={(-0.2,0)}]
\node (v24) at (1,-1.6) {$\mathbf{M}^{(m+1)+(m+1)}_{0}$};
\node (v25) at (3.6,-1.6) {$\mathbf{M}_{1}$};
\node[rotate=0] at (2.1,-1.5) {\scriptsize{$2e$-$\sum x_i$}};
\node[rotate=0] at (3.2,-1.5) {\scriptsize{$2e$}};
\draw  (v24) edge (v25);
\end{scope}
\begin{scope}[shift={(4.3,0)}]
\node (v24) at (1.5,-1.6) {$\mathbf{P}^{2}_{0}$};
\node (v25) at (3.6,-1.6) {$\mathbf{P}_{1}$};
\node[rotate=0] at (2.1,-1.5) {\scriptsize{$2e$-$\sum x_i$}};
\node[rotate=0] at (3.2,-1.5) {\scriptsize{$2e$}};
\draw  (v24) edge (v25);
\end{scope}
\end{tikzpicture}
\ee
\bit
\item $x_1,x_2$ in $\mathbf{S}_{2m+2}$ are glued to $x_1,x_2$ in $\mathbf{P}_0$.
\item $f-x_1-x_2$ in $\mathbf{S}_{2m+2}$ is glued to $f$ in $\mathbf{P}_1$.
\item $f$ in $\mathbf{S}_{1}$ is glued to $f-x_1-x_2$ in $\mathbf{P}_0$.
\item $x_1-x_2$ in $\mathbf{S}_{m+2}$ is glued to $f$ in $\mathbf{M}_1$.
\item $f-x_1,x_2$ in $\mathbf{S}_{m+2}$ are glued to $f-x_{1},y_{1}$ in $\mathbf{M}_0$.
\item $f$ in $\mathbf{S}_{m+1}$ is glued to $x_{1}-y_{1}$ in $\mathbf{M}_0$.
\item $x_2-x_1,y_1-y_2$ in $\mathbf{S}_{m+2-2i}$ are glued to $f,f$ in $\mathbf{M}_1$ for $i=1,\cdots,\frac{m}2$.
\item $x_1-x_2,y_2-y_1$ in $\mathbf{S}_{m+2+2i}$ are glued to $f,f$ in $\mathbf{M}_1$ for $i=1,\cdots,\frac{m}2$.
\item $f-x_2,x_1,f-y_1,y_2$ in $\mathbf{S}_{m+2-2i}$ are glued to $f-x_{2i},y_{2i},x_{2i-1},y_{2i-1}$ in $\mathbf{M}_0$ for $i=1,\cdots,\frac{m}2$.
\item $f-x_1,x_2,f-y_2,y_1$ in $\mathbf{S}_{m+2+2i}$ are glued to $f-x_{2i+1},y_{2i+1},x_{2i},y_{2i}$ in $\mathbf{M}_0$ for $i=1,\cdots,\frac{m}2$.
\item $f,f$ in $\mathbf{S}_{m+1+2i}$ are glued to $x_{2i}-y_{2i},f-x_{2i-1}-y_{2i-1}$ in $\mathbf{M}_0$ for $i=1,\cdots,\frac{m}2$.
\item $f,f$ in $\mathbf{S}_{m+1-2i}$ are glued to $x_{2i+1}-y_{2i+1},f-x_{2i}-y_{2i}$ in $\mathbf{M}_0$ for $i=1,\cdots,\frac{m}2$.
\item $x_1-x_2$ in $\mathbf{P}_{0}$ is glued to $f$ in $\mathbf{M}_1$.
\item $f-x_1,x_2$ in $\mathbf{P}_{0}$ is glued to $x_{m+1},y_{m+1}$ in $\mathbf{M}_0$.
\item $f$ in $\mathbf{P}_{1}$ is glued to $f-x_{m+1}-y_{m+1}$ in $\mathbf{M}_0$.
\eit

The gluing rules for the $(\su(2)^{(1)})^{\oplus 2}$ delocalized flavor symmetry of
\be
\begin{tikzpicture} [scale=1.9]
\node (v1) at (-1.5,0.8) {2};
\node at (-1.5,1.1) {$\su(2)^{(1)}$};
\node (v3) at (-2.9,0.8) {2};
\node at (-2.9,1.1) {$\su(2)^{(1)}$};
\node (v2) at (-2.2,0.8) {$\cdots$};
\draw  (v2) edge (v3);
\draw  (v2) edge (v1);
\begin{scope}[shift={(0,0.45)}]
\node at (-2.2,-0.15) {$m$};
\draw (-3.1,0.15) .. controls (-3.1,0.1) and (-3.1,0.05) .. (-3,0.05);
\draw (-3,0.05) -- (-2.3,0.05);
\draw (-2.2,0) .. controls (-2.2,0.05) and (-2.25,0.05) .. (-2.3,0.05);
\draw (-2.2,0) .. controls (-2.2,0.05) and (-2.15,0.05) .. (-2.1,0.05);
\draw (-2.1,0.05) -- (-1.4,0.05);
\draw (-1.3,0.15) .. controls (-1.3,0.1) and (-1.3,0.05) .. (-1.4,0.05);
\end{scope}
\node (v0) at (-0.6,0.8) {2};
\node at (-0.6,1.1) {$\su(1)^{(1)}$};
\draw  (v1) edge (v0);
\end{tikzpicture}
\ee
for $m$ odd are described by
\be
\begin{tikzpicture} [scale=1.9]
\node (v2) at (0.9,1.3) {$\mathbf{m}_{0}$};
\node (v3) at (3.1,1.3) {$\mathbf{(m-1)}_{0}^{2+2}$};
\node (v10) at (0.9,-0.8) {$\mathbf{(m+2)}_0^{2+2}$};
\node (v9) at (3.1,-0.8) {$\mathbf{(m+3)}_0$};
\node (v4) at (4.5,1.3) {$\cdots$};
\node (v8) at (4,-0.8) {$\cdots$};
\node (v5) at (5.3,1.3) {$\mathbf{1}^1_0$};
\node (v7) at (5.3,-0.8) {$\mathbf{2m+1}_{0}^{1+2}$};
\draw  (v2) edge (v3);
\draw  (v3) edge (v4);
\draw  (v4) edge (v5);
\draw  (v7) edge (v8);
\draw  (v8) edge (v9);
\draw  (v9) edge (v10);
\node at (1.3,1.4) {\scriptsize{$f$}};
\node at (2.3,1.4) {\scriptsize{$f$-$\sum y_i$}};
\node at (3.9,1.4) {\scriptsize{$f$-$\sum x_i$}};
\node at (5,1.4) {\scriptsize{$f$}};
\node at (4.5,-0.9) {\scriptsize{$f$-$\sum y_i$}};
\node at (1.7,-0.9) {\scriptsize{$f$-$\sum x_i$}};
\node at (1.1646,-0.3943) {\scriptsize{$x_i$}};
\node at (2.6,1) {\scriptsize{$y_i$}};
\node at (3.4,0.9) {\scriptsize{$x_i$}};
\node at (2.5,-0.9) {\scriptsize{$f$}};
\node at (0.75,0.8) {\scriptsize{$e$}};
\node at (3.7,-0.9) {\scriptsize{$f$}};
\node at (4.8,-0.5) {\scriptsize{$y_i$}};
\node (v14) at (0.9,0.25) {\scriptsize{2}};
\draw  (v2) edge (v14);
\draw  (v10) edge (v14);
\node at (0.65,-0.15) {\scriptsize{$e$-$x_i$-$y_i$}};
\begin{scope}[shift={(2.2,-0.1)}]
\node at (0.75,-0.3) {\scriptsize{$e$}};
\end{scope}
\begin{scope}[shift={(2.2,0)}]
\node at (0.65,0.65) {\scriptsize{$e$-$x_i$-$y_i$}};
\end{scope}
\begin{scope}[shift={(2.2,0)}]
\node (v14) at (0.9,0.25) {\scriptsize{2}};
\end{scope}
\draw  (v3) edge (v14);
\draw  (v14) edge (v9);
\draw (v3) -- (3.9,0.6);
\draw (v7) -- (4.5,0);
\begin{scope}[shift={(4.3,0.1)}]
\node at (0.75,0.8) {\scriptsize{$e$-$x,e$}};
\end{scope}
\begin{scope}[shift={(4.45,0.15)}]
\node at (0.65,-0.15) {\scriptsize{$e$-$y_1,$}};
\node at (0.6,-0.3) {\scriptsize{$e$-$x$-$y_2$}};
\end{scope}
\begin{scope}[shift={(4.4,0)}]
\node (v14) at (0.9,0.25) {\scriptsize{2}};
\end{scope}
\draw  (v5) edge (v14);
\draw  (v14) edge (v7);
\node at (4.3,0.3) {$\cdots$};
\begin{scope}[shift={(3.5,-0.1)}]
\node (v24) at (1,-1.6) {$\mathbf{M}^{m+m+1+1}_{0}$};
\node (v25) at (3.75,-1.6) {$\mathbf{M}^{1+1}_{1}$};
\node[rotate=0] at (2.1,-1.5) {\scriptsize{$2e$-$\sum x_i$-$2z$-$2w$}};
\node[rotate=0] at (3.1,-1.5) {\scriptsize{$2e$-$2x$-$2y$}};
\draw  (v24) edge (v25);
\end{scope}
\draw (v24) .. controls (4.15,-2.2) and (4.9,-2.2) .. (v24);
\draw (v25) .. controls (6.9,-2.2) and (7.65,-2.2) .. (v25);
\begin{scope}[shift={(-0.75,-0.1)}]
\node (v24) at (1.5,-1.6) {$\mathbf{P}^{2}_{0}$};
\node (v25) at (3.6,-1.6) {$\mathbf{P}_{1}$};
\node[rotate=0] at (2.1,-1.5) {\scriptsize{$2e$-$\sum x_i$}};
\node[rotate=0] at (3.2,-1.5) {\scriptsize{$2e$}};
\draw  (v24) edge (v25);
\end{scope}
\node (v26) at (1.9941,0.2425) {\scriptsize{2}};
\draw  (v10) edge (v26);
\draw  (v26) edge (v3);
\node (v6) at (7,0.3) {$\mathbf{0}_{0}^{1+1}$};
\node at (5.85,-0.55) {\scriptsize{$f$}};
\node at (6.75,-0.05) {\scriptsize{$f$-$x$-$y$}};
\node at (6.7,0.65) {\scriptsize{$x,y$}};
\node at (5.9,1.2) {\scriptsize{$f$-$x,x$}};
\draw (v6) .. controls (7.6,0.8) and (7.6,-0.2) .. (v6);
\node at (7.35,0.6) {\scriptsize{$e$-$x$}};
\node at (7.35,-0.05) {\scriptsize{$e$-$y$}};
\draw  (v7) edge (v6);
\node (v1) at (6.15,0.8) {\scriptsize{2}};
\draw  (v5) edge (v1);
\draw  (v1) edge (v6);
\begin{scope}[shift={(2.25,0.1)}]
\node at (2,-2.1) {\scriptsize{$z$}};
\node at (2.55,-2.1) {\scriptsize{$w$}};
\node at (4.75,-2.05) {\scriptsize{$x$}};
\node at (5.3,-2.05) {\scriptsize{$y$}};
\end{scope}
\end{tikzpicture}
\ee
\bit
\item $y_1,y_2$ in $\mathbf{S}_{m+2}$ are glued to $x_1,x_2$ in $\mathbf{P}_0$.
\item $f-y_1-y_2$ in $\mathbf{S}_{m+2}$ is glued to $f$ in $\mathbf{P}_1$.
\item $f$ in $\mathbf{S}_{m}$ is glued to $f-x_1-x_2$ in $\mathbf{P}_0$.
\item $x_1-x_2,y_2-y_1$ in $\mathbf{S}_{m+2i}$ are glued to $f,f$ in $\mathbf{M}_1$ for $i=1,\cdots,\frac{m-1}2$.
\item $x_2-x_1,y_1-y_2$ in $\mathbf{S}_{m+1-2i}$ are glued to $f,f$ in $\mathbf{M}_1$ for $i=1,\cdots,\frac{m-1}2$.
\item $f-x_1,x_2,f-y_2,y_1$ in $\mathbf{S}_{m+2i}$ are glued to $f-x_{2i},y_{2i},x_{2i-1},y_{2i-1}$ in $\mathbf{M}_0$ for $i=1,\cdots,\frac{m-1}2$.
\item $f-x_2,x_1,f-y_1,y_2$ in $\mathbf{S}_{m+1-2i}$ are glued to $f-x_{2i+1},y_{2i+1},x_{2i},y_{2i}$ in $\mathbf{M}_0$ for $i=1,\cdots,\frac{m-1}2$.
\item $f,f$ in $\mathbf{S}_{m+2-2i}$ are glued to $x_{2i}-y_{2i},f-x_{2i-1}-y_{2i-1}$ in $\mathbf{M}_0$ for $i=1,\cdots,\frac{m-1}2$.
\item $f,f$ in $\mathbf{S}_{m+1+2i}$ are glued to $x_{2i+1}-y_{2i+1},f-x_{2i}-y_{2i}$ in $\mathbf{M}_0$ for $i=1,\cdots,\frac{m-1}2$.
\item $f-x,x$ in $\mathbf{S}_{1}$ are glued to $f-x,f-y$ in $\mathbf{M}_1$.
\item $y_2-y_1,f-x,f-x$ in $\mathbf{S}_{2m+1}$ are glued to $f,x,y$ in $\mathbf{M}_1$.
\item $f,f-x,x$ in $\mathbf{S}_{1}$ are glued to $f-x_m-y_m,f-z,f-w$ in $\mathbf{M}_0$.
\item $f-y_2,y_1,x,x$ in $\mathbf{S}_{2m+1}$ are glued to $x_m,y_m,z,w$ in $\mathbf{M}_0$.
\item $x_2-x_1$ in $\mathbf{P}_{0}$ is glued to $f$ in $\mathbf{M}_1$.
\item $f-x_2,x_1$ in $\mathbf{P}_{0}$ is glued to $f-x_1,y_1$ in $\mathbf{M}_0$.
\item $f$ in $\mathbf{P}_{1}$ is glued to $x_1-y_1$ in $\mathbf{M}_0$.
\eit
Let us note, in order to avoid confusion, that there is no surface $\mathbf{S}_{m+1}$ in the above geometry. For $m$ even, the gluing rules are
\be
\begin{tikzpicture} [scale=1.9]
\node (v2) at (0.6,1.3) {$\mathbf{m}^{2+2}_{0}$};
\node (v3) at (2.8,1.3) {$\mathbf{(m-1)}_{0}$};
\node (v10) at (0.6,-0.8) {$\mathbf{(m+2)}_0$};
\node (v9) at (2.8,-0.8) {$\mathbf{(m+3)}^{2+2}_0$};
\node (v4) at (4.2,1.3) {$\cdots$};
\node (v8) at (4.05,-0.8) {$\cdots$};
\node (v5) at (5.3,1.3) {$\mathbf{1}^1_0$};
\node (v7) at (5.3,-0.8) {$\mathbf{2m+1}_{0}^{1+2}$};
\draw  (v2) edge (v3);
\draw  (v3) edge (v4);
\draw  (v4) edge (v5);
\draw  (v7) edge (v8);
\draw  (v8) edge (v9);
\draw  (v9) edge (v10);
\node at (1.3,-0.9) {\scriptsize{$f$}};
\node at (2.05,-0.9) {\scriptsize{$f$-$\sum y_i$}};
\node at (3.45,1.4) {\scriptsize{$f$}};
\node at (5,1.4) {\scriptsize{$f$}};
\node at (4.5,-0.9) {\scriptsize{$f$-$\sum y_i$}};
\node at (1.15,1.4) {\scriptsize{$f$-$\sum x_i$}};
\node at (0.9,0.85) {\scriptsize{$x_i$}};
\node at (2.25,-0.5) {\scriptsize{$y_i$}};
\node at (3.4,-0.5) {\scriptsize{$x_i$}};
\node at (2.15,1.4) {\scriptsize{$f$}};
\node at (0.4,-0.3) {\scriptsize{$e$}};
\node at (3.6,-0.9) {\scriptsize{$f$-$\sum x_i$}};
\node at (4.8,-0.5) {\scriptsize{$y_i$}};
\node (v14) at (0.6,0.25) {\scriptsize{2}};
\draw  (v2) edge (v14);
\draw  (v10) edge (v14);
\node at (0.3,0.95) {\scriptsize{$e$-$x_i$-$y_i$}};
\begin{scope}[shift={(1.9,1.05)}]
\node at (0.75,-0.3) {\scriptsize{$e$}};
\end{scope}
\begin{scope}[shift={(1.9,-0.85)}]
\node at (0.65,0.65) {\scriptsize{$e$-$x_i$-$y_i$}};
\end{scope}
\begin{scope}[shift={(1.9,0)}]
\node (v14) at (0.9,0.25) {\scriptsize{2}};
\end{scope}
\draw  (v3) edge (v14);
\draw  (v14) edge (v9);
\draw (v9) -- (3.75,0.05);
\draw (v7) -- (4.5,0);
\begin{scope}[shift={(4.3,0.1)}]
\node at (0.75,0.8) {\scriptsize{$e$-$x,e$}};
\end{scope}
\begin{scope}[shift={(4.45,0.15)}]
\node at (0.65,-0.15) {\scriptsize{$e$-$y_1,$}};
\node at (0.6,-0.3) {\scriptsize{$e$-$x$-$y_2$}};
\end{scope}
\begin{scope}[shift={(4.4,0)}]
\node (v14) at (0.9,0.25) {\scriptsize{2}};
\end{scope}
\draw  (v5) edge (v14);
\draw  (v14) edge (v7);
\node at (4.15,0.3) {$\cdots$};
\begin{scope}[shift={(3.5,-0.1)}]
\node (v24) at (1,-1.6) {$\mathbf{M}^{m+m+1+1}_{0}$};
\node (v25) at (3.75,-1.6) {$\mathbf{M}^{1+1}_{1}$};
\node[rotate=0] at (2.1,-1.5) {\scriptsize{$2e$-$\sum x_i$-$2z$-$2w$}};
\node[rotate=0] at (3.1,-1.5) {\scriptsize{$2e$-$2x$-$2y$}};
\draw  (v24) edge (v25);
\end{scope}
\draw (v24) .. controls (4.15,-2.2) and (4.9,-2.2) .. (v24);
\draw (v25) .. controls (6.9,-2.2) and (7.65,-2.2) .. (v25);
\begin{scope}[shift={(-0.75,-0.1)}]
\node (v24) at (1.5,-1.6) {$\mathbf{P}^{2}_{0}$};
\node (v25) at (3.6,-1.6) {$\mathbf{P}_{1}$};
\node[rotate=0] at (2.1,-1.5) {\scriptsize{$2e$-$\sum x_i$}};
\node[rotate=0] at (3.2,-1.5) {\scriptsize{$2e$}};
\draw  (v24) edge (v25);
\end{scope}
\node (v26) at (1.6941,0.2425) {\scriptsize{2}};
\draw  (v9) edge (v26);
\draw  (v26) edge (v2);
\node (v6) at (7,0.3) {$\mathbf{0}_{0}^{1+1}$};
\node at (5.85,-0.55) {\scriptsize{$f$}};
\node at (6.75,-0.05) {\scriptsize{$f$-$x$-$y$}};
\node at (6.7,0.65) {\scriptsize{$x,y$}};
\node at (5.9,1.2) {\scriptsize{$f$-$x,x$}};
\draw (v6) .. controls (7.6,0.8) and (7.6,-0.2) .. (v6);
\node at (7.35,0.6) {\scriptsize{$e$-$x$}};
\node at (7.35,-0.05) {\scriptsize{$e$-$y$}};
\draw  (v7) edge (v6);
\node (v1) at (6.15,0.8) {\scriptsize{2}};
\draw  (v5) edge (v1);
\draw  (v1) edge (v6);
\begin{scope}[shift={(2.25,0.1)}]
\node at (2,-2.1) {\scriptsize{$z$}};
\node at (2.55,-2.1) {\scriptsize{$w$}};
\node at (4.75,-2.05) {\scriptsize{$x$}};
\node at (5.3,-2.05) {\scriptsize{$y$}};
\end{scope}
\end{tikzpicture}
\ee
\bit
\item $y_1,y_2$ in $\mathbf{S}_{m}$ are glued to $x_1,x_2$ in $\mathbf{P}_0$.
\item $f-y_1-y_2$ in $\mathbf{S}_{m}$ is glued to $f$ in $\mathbf{P}_1$.
\item $f$ in $\mathbf{S}_{m+2}$ is glued to $f-x_1-x_2$ in $\mathbf{P}_0$.
\item $x_2-x_1,y_1-y_2$ in $\mathbf{S}_{m+2-2i}$ are glued to $f,f$ in $\mathbf{M}_1$ for $i=1,\cdots,\frac{m}2$.
\item $x_1-x_2,y_2-y_1$ in $\mathbf{S}_{m+1+2i}$ are glued to $f,f$ in $\mathbf{M}_1$ for $i=1,\cdots,\frac{m-2}2$.
\item $f-x_1,x_2,f-y_2,y_1$ in $\mathbf{S}_{m+1+2i}$ are glued to $f-x_{2i+1},y_{2i+1},x_{2i},y_{2i}$ in $\mathbf{M}_0$ for $i=1,\cdots,\frac{m-2}2$.
\item $f-x_2,x_1,f-y_1,y_2$ in $\mathbf{S}_{m+2-2i}$ are glued to $f-x_{2i},y_{2i},x_{2i-1},y_{2i-1}$ in $\mathbf{M}_0$ for $i=1,\cdots,\frac{m}2$.
\item $f,f$ in $\mathbf{S}_{m+1-2i}$ are glued to $x_{2i+1}-y_{2i+1},f-x_{2i}-y_{2i}$ in $\mathbf{M}_0$ for $i=1,\cdots,\frac{m-2}2$.
\item $f,f$ in $\mathbf{S}_{m+2i}$ are glued to $x_{2i}-y_{2i},f-x_{2i-1}-y_{2i-1}$ in $\mathbf{M}_0$ for $i=1,\cdots,\frac{m}2$.
\item $f-x,x$ in $\mathbf{S}_{1}$ are glued to $f-x,f-y$ in $\mathbf{M}_1$.
\item $y_2-y_1,f-x,f-x$ in $\mathbf{S}_{2m+1}$ are glued to $f,x,y$ in $\mathbf{M}_1$.
\item $f,f-x,x$ in $\mathbf{S}_{1}$ are glued to $f-x_m-y_m,f-z,f-w$ in $\mathbf{M}_0$.
\item $f-y_2,y_1,x,x$ in $\mathbf{S}_{2m+1}$ are glued to $x_m,y_m,z,w$ in $\mathbf{M}_0$.
\item $x_1-x_2$ in $\mathbf{P}_{0}$ is glued to $f$ in $\mathbf{M}_1$.
\item $f-x_1,x_2$ in $\mathbf{P}_{0}$ is glued to $f-x_1,y_1$ in $\mathbf{M}_0$.
\item $f$ in $\mathbf{P}_{1}$ is glued to $x_1-y_1$ in $\mathbf{M}_0$.
\eit
Let us note, in order to avoid confusion, that there is no surface $\mathbf{S}_{m+1}$ in the above geometry.

The (partial) gluing rules for the $\su(2)^{(1)}$ delocalized flavor symmetry of
\be
\begin{tikzpicture} [scale=1.9]
\node (v1) at (-1.5,0.8) {2};
\node at (-1.5,1.1) {$\su(2)^{(1)}$};
\node (v3) at (-2.9,0.8) {2};
\node at (-2.9,1.1) {$\su(2)^{(1)}$};
\node (v2) at (-2.2,0.8) {$\cdots$};
\draw  (v2) edge (v3);
\draw  (v2) edge (v1);
\node (v4) at (-3.8,0.8) {1};
\node at (-3.8,1.1) {$\sp(1)^{(1)}$};
\draw  (v4) edge (v3);
\begin{scope}[shift={(0,0.45)}]
\node at (-2.2,-0.15) {$m$};
\draw (-3.1,0.15) .. controls (-3.1,0.1) and (-3.1,0.05) .. (-3,0.05);
\draw (-3,0.05) -- (-2.3,0.05);
\draw (-2.2,0) .. controls (-2.2,0.05) and (-2.25,0.05) .. (-2.3,0.05);
\draw (-2.2,0) .. controls (-2.2,0.05) and (-2.15,0.05) .. (-2.1,0.05);
\draw (-2.1,0.05) -- (-1.4,0.05);
\draw (-1.3,0.15) .. controls (-1.3,0.1) and (-1.3,0.05) .. (-1.4,0.05);
\end{scope}
\node (v0) at (-0.6,0.8) {2};
\node at (-0.6,1.1) {$\su(1)^{(1)}$};
\draw  (v1) edge (v0);
\end{tikzpicture}
\ee
for $m$ even are described by
\be
\begin{tikzpicture} [scale=1.9]
\node (v2) at (0.9,1.3) {$\mathbf{(m+1)}_{0}$};
\node (v3) at (3.1,1.3) {$\mathbf{m}_{0}^{2+2}$};
\node (v10) at (0.9,-0.8) {$\mathbf{(m+2)}_0^{2+8}$};
\node (v9) at (3.1,-0.8) {$\mathbf{(m+3)}_0$};
\node (v4) at (4.5,1.3) {$\cdots$};
\node (v8) at (4,-0.8) {$\cdots$};
\node (v5) at (5.3,1.3) {$\mathbf{1}^1_0$};
\node (v7) at (5.3,-0.8) {$\mathbf{2m+1}_{0}^{1+2}$};
\draw  (v2) edge (v3);
\draw  (v3) edge (v4);
\draw  (v4) edge (v5);
\draw  (v7) edge (v8);
\draw  (v8) edge (v9);
\draw  (v9) edge (v10);
\node at (1.5,1.4) {\scriptsize{$f$}};
\node at (2.5,1.4) {\scriptsize{$f$-$\sum y_i$}};
\node at (3.9,1.4) {\scriptsize{$f$-$\sum x_i$}};
\node at (5,1.4) {\scriptsize{$f$}};
\node at (4.5,-0.9) {\scriptsize{$f$-$\sum y_i$}};
\node at (1.7,-0.9) {\scriptsize{$f$-$\sum x_i$}};
\node at (1.1646,-0.3943) {\scriptsize{$x_i$}};
\node at (2.6,1) {\scriptsize{$y_i$}};
\node at (3.4,0.9) {\scriptsize{$x_i$}};
\node at (2.5,-0.9) {\scriptsize{$f$}};
\node at (0.6,0.95) {\scriptsize{$2e+f$}};
\node at (3.7,-0.9) {\scriptsize{$f$}};
\node at (4.8,-0.5) {\scriptsize{$y_i$}};
\node at (0.45,-0.35) {\scriptsize{$2e$+$f$-$x_i$-$y_i$}};
\begin{scope}[shift={(2.2,-0.1)}]
\node at (0.75,-0.3) {\scriptsize{$e$}};
\end{scope}
\begin{scope}[shift={(2.2,0)}]
\node at (0.65,0.65) {\scriptsize{$e$-$x_i$-$y_i$}};
\end{scope}
\begin{scope}[shift={(2.2,0)}]
\node (v14) at (0.9,0.25) {\scriptsize{2}};
\end{scope}
\draw  (v3) edge (v14);
\draw  (v14) edge (v9);
\draw (v3) -- (3.9,0.6);
\draw (v7) -- (4.5,0);
\begin{scope}[shift={(4.3,0.1)}]
\node at (0.75,0.8) {\scriptsize{$e$-$x,e$}};
\end{scope}
\begin{scope}[shift={(4.45,0.15)}]
\node at (0.65,-0.15) {\scriptsize{$e$-$y_1,$}};
\node at (0.6,-0.3) {\scriptsize{$e$-$x$-$y_2$}};
\end{scope}
\begin{scope}[shift={(4.4,0)}]
\node (v14) at (0.9,0.25) {\scriptsize{2}};
\end{scope}
\draw  (v5) edge (v14);
\draw  (v14) edge (v7);
\node at (4.3,0.3) {$\cdots$};
\begin{scope}[shift={(1.05,-0.1)}]
\node (v24) at (1,-1.6) {$\mathbf{M}^{m+m+1+1}_{0}$};
\node (v25) at (3.75,-1.6) {$\mathbf{M}^{1+1}_{1}$};
\node[rotate=0] at (2.1,-1.5) {\scriptsize{$2e$-$\sum x_i$-$2z$-$2w$}};
\node[rotate=0] at (3.1,-1.5) {\scriptsize{$2e$-$2x$-$2y$}};
\draw  (v24) edge (v25);
\end{scope}
\draw (v24) .. controls (1.7,-2.2) and (2.45,-2.2) .. (v24);
\draw (v25) .. controls (4.45,-2.2) and (5.2,-2.2) .. (v25);
\node (v26) at (1.9941,0.2425) {\scriptsize{2}};
\draw  (v10) edge (v26);
\draw  (v26) edge (v3);
\node (v6) at (7,0.3) {$\mathbf{0}_{0}^{1+1}$};
\node at (5.85,-0.55) {\scriptsize{$f$}};
\node at (6.75,-0.05) {\scriptsize{$f$-$x$-$y$}};
\node at (6.7,0.65) {\scriptsize{$x,y$}};
\node at (5.9,1.2) {\scriptsize{$f$-$x,x$}};
\draw (v6) .. controls (7.6,0.8) and (7.6,-0.2) .. (v6);
\node at (7.35,0.6) {\scriptsize{$e$-$x$}};
\node at (7.35,-0.05) {\scriptsize{$e$-$y$}};
\draw  (v7) edge (v6);
\node (v1) at (6.15,0.8) {\scriptsize{2}};
\draw  (v5) edge (v1);
\draw  (v1) edge (v6);
\draw  (v2) edge (v10);
\begin{scope}[shift={(-0.2,0.1)}]
\node at (2,-2.1) {\scriptsize{$z$}};
\node at (2.55,-2.1) {\scriptsize{$w$}};
\node at (4.75,-2.05) {\scriptsize{$x$}};
\node at (5.3,-2.05) {\scriptsize{$y$}};
\end{scope}
\end{tikzpicture}
\ee
\bit
\item $x_1-x_2$ in $\mathbf{S}_{m+2}$ is glued to $f$ in $\mathbf{M}_1$.
\item $f-x_1,x_2$ in $\mathbf{S}_{m+2}$ are glued to $f-x_1,y_1$ in $\mathbf{M}_0$.
\item $f$ in $\mathbf{S}_{m+1}$ is glued to $x_1-y_1$ in $\mathbf{M}_0$.
\item $x_2-x_1,y_1-y_2$ in $\mathbf{S}_{m+2-2i}$ are glued to $f,f$ in $\mathbf{M}_1$ for $i=1,\cdots,\frac{m}2$.
\item $x_1-x_2,y_2-y_1$ in $\mathbf{S}_{m+2+2i}$ are glued to $f,f$ in $\mathbf{M}_1$ for $i=1,\cdots,\frac{m-2}2$.
\item $f-x_1,x_2,f-y_2,y_1$ in $\mathbf{S}_{m+2+2i}$ are glued to $f-x_{2i+1},y_{2i+1},x_{2i},y_{2i}$ in $\mathbf{M}_0$ for $i=1,\cdots,\frac{m-2}2$.
\item $f-x_2,x_1,f-y_1,y_2$ in $\mathbf{S}_{m+2-2i}$ are glued to $f-x_{2i},y_{2i},x_{2i-1},y_{2i-1}$ in $\mathbf{M}_0$ for $i=1,\cdots,\frac{m}2$.
\item $f,f$ in $\mathbf{S}_{m+1-2i}$ are glued to $x_{2i+1}-y_{2i+1},f-x_{2i}-y_{2i}$ in $\mathbf{M}_0$ for $i=1,\cdots,\frac{m-2}2$.
\item $f,f$ in $\mathbf{S}_{m+1+2i}$ are glued to $x_{2i}-y_{2i},f-x_{2i-1}-y_{2i-1}$ in $\mathbf{M}_0$ for $i=1,\cdots,\frac{m}2$.
\item $f-x,x$ in $\mathbf{S}_{1}$ are glued to $f-x,f-y$ in $\mathbf{M}_1$.
\item $y_2-y_1,f-x,f-x$ in $\mathbf{S}_{2m+1}$ are glued to $f,x,y$ in $\mathbf{M}_1$.
\item $f,f-x,x$ in $\mathbf{S}_{1}$ are glued to $f-x_m-y_m,f-z,f-w$ in $\mathbf{M}_0$.
\item $f-y_2,y_1,x,x$ in $\mathbf{S}_{2m+1}$ are glued to $x_m,y_m,z,w$ in $\mathbf{M}_0$.
\eit
For $m$ odd, the gluing rules are
\be
\begin{tikzpicture} [scale=1.9]
\node (v2) at (0.6,1.3) {$\mathbf{(m+1)}^{2}_{0}$};
\node (v3) at (2.8,1.3) {$\mathbf{m}_{0}$};
\node (v10) at (0.6,-0.8) {$\mathbf{(m+2)}^8_0$};
\node (v9) at (2.8,-0.8) {$\mathbf{(m+3)}^{2+2}_0$};
\node (v4) at (4.2,1.3) {$\cdots$};
\node (v8) at (4.05,-0.8) {$\cdots$};
\node (v5) at (5.3,1.3) {$\mathbf{1}^1_0$};
\node (v7) at (5.3,-0.8) {$\mathbf{2m+1}_{0}^{1+2}$};
\draw  (v2) edge (v3);
\draw  (v3) edge (v4);
\draw  (v4) edge (v5);
\draw  (v7) edge (v8);
\draw  (v8) edge (v9);
\draw  (v9) edge (v10);
\node at (1.3,-0.9) {\scriptsize{$f$}};
\node at (2.05,-0.9) {\scriptsize{$f$-$\sum y_i$}};
\node at (3.2,1.4) {\scriptsize{$f$}};
\node at (5,1.4) {\scriptsize{$f$}};
\node at (4.5,-0.9) {\scriptsize{$f$-$\sum y_i$}};
\node at (1.35,1.4) {\scriptsize{$f$-$\sum x_i$}};
\node at (0.9,0.85) {\scriptsize{$x_i$}};
\node at (2.25,-0.5) {\scriptsize{$y_i$}};
\node at (3.4,-0.5) {\scriptsize{$x_i$}};
\node at (2.4,1.4) {\scriptsize{$f$}};
\node at (3.6,-0.9) {\scriptsize{$f$-$\sum x_i$}};
\node at (4.8,-0.5) {\scriptsize{$y_i$}};
\node (v14) at (0.6,0.25) {\scriptsize{2}};
\draw  (v2) edge (v14);
\draw  (v10) edge (v14);
\node at (0.15,0.95) {\scriptsize{$2e$+$f$-$\sum x_i$}};
\node at (0.15,-0.4) {\scriptsize{$2e$+$f$-$\sum x_i$}};
\begin{scope}[shift={(1.9,1.05)}]
\node at (0.75,-0.3) {\scriptsize{$e$}};
\end{scope}
\begin{scope}[shift={(1.9,-0.85)}]
\node at (0.65,0.65) {\scriptsize{$e$-$x_i$-$y_i$}};
\end{scope}
\begin{scope}[shift={(1.9,0)}]
\node (v14) at (0.9,0.25) {\scriptsize{2}};
\end{scope}
\draw  (v3) edge (v14);
\draw  (v14) edge (v9);
\draw (v9) -- (3.75,0.05);
\draw (v7) -- (4.5,0);
\begin{scope}[shift={(4.3,0.1)}]
\node at (0.75,0.8) {\scriptsize{$e$-$x,e$}};
\end{scope}
\begin{scope}[shift={(4.45,0.15)}]
\node at (0.65,-0.15) {\scriptsize{$e$-$y_1,$}};
\node at (0.6,-0.3) {\scriptsize{$e$-$x$-$y_2$}};
\end{scope}
\begin{scope}[shift={(4.4,0)}]
\node (v14) at (0.9,0.25) {\scriptsize{2}};
\end{scope}
\draw  (v5) edge (v14);
\draw  (v14) edge (v7);
\node at (4.15,0.3) {$\cdots$};
\begin{scope}[shift={(1.25,-0.2)}]
\node (v24) at (1,-1.6) {$\mathbf{M}^{m+m+1+1}_{0}$};
\node (v25) at (3.75,-1.6) {$\mathbf{M}^{1+1}_{1}$};
\node[rotate=0] at (2.1,-1.5) {\scriptsize{$2e$-$\sum x_i$-$2z$-$2w$}};
\node[rotate=0] at (3.1,-1.5) {\scriptsize{$2e$-$2x$-$2y$}};
\draw  (v24) edge (v25);
\end{scope}
\draw (v24) .. controls (1.9,-2.3) and (2.65,-2.3) .. (v24);
\draw (v25) .. controls (4.65,-2.3) and (5.4,-2.3) .. (v25);
\node (v26) at (1.6941,0.2425) {\scriptsize{2}};
\draw  (v9) edge (v26);
\draw  (v26) edge (v2);
\node (v6) at (7,0.3) {$\mathbf{0}_{0}^{1+1}$};
\node at (5.85,-0.55) {\scriptsize{$f$}};
\node at (6.75,-0.05) {\scriptsize{$f$-$x$-$y$}};
\node at (6.7,0.65) {\scriptsize{$x,y$}};
\node at (5.9,1.2) {\scriptsize{$f$-$x,x$}};
\draw (v6) .. controls (7.6,0.8) and (7.6,-0.2) .. (v6);
\node at (7.35,0.6) {\scriptsize{$e$-$x$}};
\node at (7.35,-0.05) {\scriptsize{$e$-$y$}};
\draw  (v7) edge (v6);
\node (v1) at (6.15,0.8) {\scriptsize{2}};
\draw  (v5) edge (v1);
\draw  (v1) edge (v6);
\node at (2,-2.1) {\scriptsize{$z$}};
\node at (2.55,-2.1) {\scriptsize{$w$}};
\node at (4.75,-2.05) {\scriptsize{$x$}};
\node at (5.3,-2.05) {\scriptsize{$y$}};
\end{tikzpicture}
\ee
\bit
\item $x_2-x_1$ in $\mathbf{S}_{m+1}$ is glued to $f$ in $\mathbf{M}_1$.
\item $f-x_2,x_1$ in $\mathbf{S}_{m+1}$ are glued to $f-x_1,y_1$ in $\mathbf{M}_0$.
\item $f$ in $\mathbf{S}_{m+2}$ is glued to $x_1-y_1$ in $\mathbf{M}_0$.
\item $x_2-x_1,y_1-y_2$ in $\mathbf{S}_{m+1-2i}$ are glued to $f,f$ in $\mathbf{M}_1$ for $i=1,\cdots,\frac{m-1}2$.
\item $x_1-x_2,y_2-y_1$ in $\mathbf{S}_{m+1+2i}$ are glued to $f,f$ in $\mathbf{M}_1$ for $i=1,\cdots,\frac{m-1}2$.
\item $f-x_1,x_2,f-y_2,y_1$ in $\mathbf{S}_{m+1+2i}$ are glued to $f-x_{2i},y_{2i},x_{2i-1},y_{2i-1}$ in $\mathbf{M}_0$ for $i=1,\cdots,\frac{m-1}2$.
\item $f-x_2,x_1,f-y_1,y_2$ in $\mathbf{S}_{m+1-2i}$ are glued to $f-x_{2i+1},y_{2i+1},x_{2i},y_{2i}$ in $\mathbf{M}_0$ for $i=1,\cdots,\frac{m-1}2$.
\item $f,f$ in $\mathbf{S}_{m+2-2i}$ are glued to $x_{2i}-y_{2i},f-x_{2i-1}-y_{2i-1}$ in $\mathbf{M}_0$ for $i=1,\cdots,\frac{m-1}2$.
\item $f,f$ in $\mathbf{S}_{m+2+2i}$ are glued to $x_{2i+1}-y_{2i+1},f-x_{2i}-y_{2i}$ in $\mathbf{M}_0$ for $i=1,\cdots,\frac{m-1}2$.
\item $f-x,x$ in $\mathbf{S}_{1}$ are glued to $f-x,f-y$ in $\mathbf{M}_1$.
\item $y_2-y_1,f-x,f-x$ in $\mathbf{S}_{2m+1}$ are glued to $f,x,y$ in $\mathbf{M}_1$.
\item $f,f-x,x$ in $\mathbf{S}_{1}$ are glued to $f-x_m-y_m,f-z,f-w$ in $\mathbf{M}_0$.
\item $f-y_2,y_1,x,x$ in $\mathbf{S}_{2m+1}$ are glued to $x_m,y_m,z,w$ in $\mathbf{M}_0$.
\eit

The gluing rules for the $\su(2)^{(1)}$ delocalized flavor symmetry of
\be
\begin{tikzpicture} [scale=1.9]
\node (v1) at (-1.5,0.8) {2};
\node at (-1.5,1.1) {$\su(1)^{(1)}$};
\node (v3) at (-2.9,0.8) {2};
\node at (-2.9,1.1) {$\su(1)^{(1)}$};
\node (v2) at (-2.2,0.8) {$\cdots$};
\draw  (v2) edge (v3);
\draw  (v2) edge (v1);
\begin{scope}[shift={(0,0.45)}]
\node at (-2.65,-0.15) {$m$};
\draw (-4,0.15) .. controls (-4,0.1) and (-4,0.05) .. (-3.9,0.05);
\draw (-3.9,0.05) -- (-2.75,0.05);
\draw (-2.65,0) .. controls (-2.65,0.05) and (-2.7,0.05) .. (-2.75,0.05);
\draw (-2.65,0) .. controls (-2.65,0.05) and (-2.6,0.05) .. (-2.55,0.05);
\draw (-2.55,0.05) -- (-1.4,0.05);
\draw (-1.3,0.15) .. controls (-1.3,0.1) and (-1.3,0.05) .. (-1.4,0.05);
\end{scope}
\node (v4) at (-3.9,0.8) {2};
\node at (-3.8,1.1) {$\su(1)^{(1)}$};
\draw  (v4) edge (v3);
\draw (v4) .. controls (-4.55,1.2) and (-4.55,0.4) .. (v4);
\end{tikzpicture}
\ee
are described by
\be
\begin{tikzpicture} [scale=1.9]
\begin{scope}[rotate=90]
\begin{scope}[rotate=90, shift={(7.4,-5.6)}]
\node (v10) at (1.05,-2) {$\mathbf{3}^{1+1}_{0}$};
\node[rotate=0] (v9) at (0.1,-2) {$\cdots$};
\node (v8) at (2.7,-2) {$\mathbf{2}^{1+1}_{0}$};
\node (v7_1) at (4.85,-2) {$\mathbf{1}^{1+1}_{1}$};
\node at (1.5,-1.9) {\scriptsize{$f$-$x,x$}};
\node at (4.15,-1.9) {\scriptsize{$2h$-$x$-$2y,f$-$x$}};
\node at (2.25,-1.9) {\scriptsize{$f$-$y,y$}};
\node at (3.15,-1.9) {\scriptsize{$f$-$x,x$}};
\node at (0.6,-1.9) {\scriptsize{$f$-$y,y$}};
\draw (v10) .. controls (0.45,-2.7) and (1.65,-2.7) .. (v10);
\node at (0.65,-2.3) {\scriptsize{$e$-$x$}};
\node at (1.45,-2.3) {\scriptsize{$e$-$y$}};
\node (v3_1) at (3.55,-2) {\scriptsize{2}};
\draw  (v8) edge (v3_1);
\draw  (v3_1) edge (v7_1);
\draw (v8) .. controls (2.1,-2.7) and (3.3,-2.7) .. (v8);
\node at (2.3,-2.3) {\scriptsize{$e$-$x$}};
\node at (3.1,-2.3) {\scriptsize{$e$-$y$}};
\draw (v7_1) .. controls (4.25,-2.7) and (5.45,-2.7) .. (v7_1);
\node at (4.45,-2.3) {\scriptsize{$x$}};
\node at (5.25,-2.3) {\scriptsize{$y$}};
\node (v15) at (-0.8,-2) {$\mathbf{m}^{1+1}_{0}$};
\draw (v15) .. controls (-1.4,-2.7) and (-0.2,-2.7) .. (v15);
\node at (-1.2,-2.3) {\scriptsize{$e$-$x$}};
\node at (-0.4,-2.3) {\scriptsize{$e$-$y$}};
\node at (-0.3,-1.9) {\scriptsize{$f$-$x,x$}};
\end{scope}
\node (v12) at (6.6,8.55) {$\mathbf{N}^{\lceil m/2\rceil}_0$};
\node (v20) at (6.6,10.25) {$\mathbf{N}_1$};
\node at (6.7,9.9) {\scriptsize{$2e$}};
\node at (6.7,9.25) {\scriptsize{$2e$-$\sum x_i$}};
\node (v14) at (7.6,9.3) {\scriptsize{2}};
\draw  (v8) edge (v14);
\draw  (v14) edge (v10);
\draw  (v9) edge (v15);
\draw  (v20) edge (v12);
\end{scope}
\draw  (v10) edge (v9);
\end{tikzpicture}
\ee
\bit
\item $f$ in $\mathbf{N}_{1}$ is glued to $f-x-y$ in $\mathbf{S}_i$ for $i=2,\cdots,m$.
\item $f-x_i,x_i$ in $\mathbf{N}_{0}$ are glued to $x,y$ in $\mathbf{S}_{2i-1}$ for $i=2,\cdots,\lceil m/2\rceil$.
\item $f-x_i,x_i$ in $\mathbf{N}_{0}$ are glued to $y,x$ in $\mathbf{S}_{2i}$ for $i=1,\cdots,\lceil m/2\rceil$.
\item $f$ in $\mathbf{N}_{1}$ is glued to $e+h-x-y$ in $\mathbf{S}_1$.
\item $f-x_1,x_1$ in $\mathbf{N}_{0}$ are glued to $f-x,f-y$ in $\mathbf{S}_{1}$.
\eit

The gluing rules for the $\su(2)^{(1)}$ delocalized flavor symmetry of
\be
\begin{tikzpicture} [scale=1.9]
\node (v1) at (-1.5,0.8) {2};
\node at (-1.5,1.1) {$\su(1)^{(1)}$};
\node (v3) at (-2.9,0.8) {2};
\node at (-2.9,1.1) {$\su(1)^{(1)}$};
\node (v2) at (-2.2,0.8) {$\cdots$};
\draw  (v2) edge (v3);
\draw  (v2) edge (v1);
\begin{scope}[shift={(0,0.45)}]
\node at (-2.65,-0.15) {$m$};
\draw (-4,0.15) .. controls (-4,0.1) and (-4,0.05) .. (-3.9,0.05);
\draw (-3.9,0.05) -- (-2.75,0.05);
\draw (-2.65,0) .. controls (-2.65,0.05) and (-2.7,0.05) .. (-2.75,0.05);
\draw (-2.65,0) .. controls (-2.65,0.05) and (-2.6,0.05) .. (-2.55,0.05);
\draw (-2.55,0.05) -- (-1.4,0.05);
\draw (-1.3,0.15) .. controls (-1.3,0.1) and (-1.3,0.05) .. (-1.4,0.05);
\end{scope}
\node (v4) at (-3.9,0.8) {2};
\node at (-3.8,1.1) {$\su(1)^{(1)}$};
\node (v5) at (-3.4,0.8) {\tiny{2}};
\draw [<-] (v4) edge (v5);
\draw  (v5) edge (v3);
\end{tikzpicture}
\ee
are described by
\be
\begin{tikzpicture} [scale=1.9]
\begin{scope}[rotate=90]
\begin{scope}[rotate=90, shift={(7.4,-5.6)}]
\node (v10) at (1.05,-2) {$\mathbf{3}^{1+1}_{0}$};
\node[rotate=0] (v9) at (0.1,-2) {$\cdots$};
\node (v8) at (2.7,-2) {$\mathbf{2}^{1+1}_{0}$};
\node (v7_1) at (4.4,-2) {$\mathbf{1}^{1+1}_{0}$};
\node at (1.5,-1.9) {\scriptsize{$f$-$x,x$}};
\node at (3.9,-1.9) {\scriptsize{$2f$-$y,y$}};
\node at (2.25,-1.9) {\scriptsize{$f$-$y,y$}};
\node at (3.15,-1.9) {\scriptsize{$f$-$x,x$}};
\node at (0.6,-1.9) {\scriptsize{$f$-$y,y$}};
\draw (v10) .. controls (0.45,-2.7) and (1.65,-2.7) .. (v10);
\node at (0.65,-2.3) {\scriptsize{$e$-$x$}};
\node at (1.45,-2.3) {\scriptsize{$e$-$y$}};
\node (v3_1) at (3.5,-2) {\scriptsize{2}};
\draw  (v8) edge (v3_1);
\draw  (v3_1) edge (v7_1);
\draw (v8) .. controls (2.1,-2.7) and (3.3,-2.7) .. (v8);
\node at (2.3,-2.3) {\scriptsize{$e$-$x$}};
\node at (3.1,-2.3) {\scriptsize{$e$-$y$}};
\draw (v7_1) .. controls (3.8,-2.7) and (5,-2.7) .. (v7_1);
\node at (4,-2.3) {\scriptsize{$e$-$x$}};
\node at (4.8,-2.3) {\scriptsize{$e$-$y$}};
\node (v15) at (-0.8,-2) {$\mathbf{m}^{1+1}_{0}$};
\draw (v15) .. controls (-1.4,-2.7) and (-0.2,-2.7) .. (v15);
\node at (-1.2,-2.3) {\scriptsize{$e$-$x$}};
\node at (-0.4,-2.3) {\scriptsize{$e$-$y$}};
\node at (-0.3,-1.9) {\scriptsize{$f$-$x,x$}};
\end{scope}
\node (v12) at (6.6,8.55) {$\mathbf{N}^{\lceil m/2\rceil}_0$};
\node (v20) at (6.6,10.25) {$\mathbf{N}_1$};
\node at (6.7,9.9) {\scriptsize{$2e$}};
\node at (6.7,9.25) {\scriptsize{$2e$-$\sum x_i$}};
\node (v14) at (7.6,9.3) {\scriptsize{2}};
\draw  (v8) edge (v14);
\draw  (v14) edge (v10);
\draw  (v9) edge (v15);
\draw  (v20) edge (v12);
\end{scope}
\draw  (v10) edge (v9);
\end{tikzpicture}
\ee
\bit
\item $f$ in $\mathbf{N}_{1}$ is glued to $f-x-y$ in $\mathbf{S}_i$ for $i=1,\cdots,m$.
\item $f-x_i,x_i$ in $\mathbf{N}_{0}$ are glued to $x,y$ in $\mathbf{S}_{2i-1}$ for $i=1,\cdots,\lceil m/2\rceil$.
\item $f-x_i,x_i$ in $\mathbf{N}_{0}$ are glued to $y,x$ in $\mathbf{S}_{2i}$ for $i=1,\cdots,\lceil m/2\rceil$.
\eit

The gluing rules for the $\su(2)^{(1)}$ delocalized flavor symmetry of
\be
\begin{tikzpicture} [scale=1.9]
\node (v1) at (-1.5,0.8) {2};
\node at (-1.5,1.1) {$\su(1)^{(1)}$};
\node (v3) at (-2.9,0.8) {2};
\node at (-2.9,1.1) {$\su(1)^{(1)}$};
\node (v2) at (-2.2,0.8) {$\cdots$};
\draw  (v2) edge (v3);
\draw  (v2) edge (v1);
\begin{scope}[shift={(0,0.45)}]
\node at (-2.65,-0.15) {$m$};
\draw (-4,0.15) .. controls (-4,0.1) and (-4,0.05) .. (-3.9,0.05);
\draw (-3.9,0.05) -- (-2.75,0.05);
\draw (-2.65,0) .. controls (-2.65,0.05) and (-2.7,0.05) .. (-2.75,0.05);
\draw (-2.65,0) .. controls (-2.65,0.05) and (-2.6,0.05) .. (-2.55,0.05);
\draw (-2.55,0.05) -- (-1.4,0.05);
\draw (-1.3,0.15) .. controls (-1.3,0.1) and (-1.3,0.05) .. (-1.4,0.05);
\end{scope}
\node (v4) at (-3.9,0.8) {2};
\node at (-3.8,1.1) {$\su(1)^{(1)}$};
\node (v5) at (-3.4,0.8) {\tiny{2}};
\draw [-] (v4) edge (v5);
\draw [->] (v5) edge (v3);
\end{tikzpicture}
\ee
are described by
\be
\begin{tikzpicture} [scale=1.9]
\begin{scope}[rotate=90]
\begin{scope}[rotate=90, shift={(7.4,-5.6)}]
\node (v10) at (1.05,-2) {$\mathbf{3}^{1+1}_{0}$};
\node[rotate=0] (v9) at (0.1,-2) {$\cdots$};
\node (v8) at (2.7,-2) {$\mathbf{2}^{1+1}_{0}$};
\node (v7_1) at (4.4,-2) {$\mathbf{1}^{1+1}_{0}$};
\node at (1.5,-1.9) {\scriptsize{$f$-$x,x$}};
\node at (3.95,-1.9) {\scriptsize{$f$-$y,y$}};
\node at (2.25,-1.9) {\scriptsize{$f$-$y,y$}};
\node at (3.2,-1.9) {\scriptsize{$2f$-$x,x$}};
\node at (0.6,-1.9) {\scriptsize{$f$-$y,y$}};
\draw (v10) .. controls (0.45,-2.7) and (1.65,-2.7) .. (v10);
\node at (0.65,-2.3) {\scriptsize{$e$-$x$}};
\node at (1.45,-2.3) {\scriptsize{$e$-$y$}};
\node (v3_1) at (3.6,-2) {\scriptsize{2}};
\draw  (v8) edge (v3_1);
\draw  (v3_1) edge (v7_1);
\draw (v8) .. controls (2.1,-2.7) and (3.3,-2.7) .. (v8);
\node at (2.3,-2.3) {\scriptsize{$e$-$x$}};
\node at (3.1,-2.3) {\scriptsize{$e$-$y$}};
\draw (v7_1) .. controls (3.8,-2.7) and (5,-2.7) .. (v7_1);
\node at (4,-2.3) {\scriptsize{$e$-$x$}};
\node at (4.8,-2.3) {\scriptsize{$e$-$y$}};
\node (v15) at (-0.8,-2) {$\mathbf{m}^{1+1}_{0}$};
\draw (v15) .. controls (-1.4,-2.7) and (-0.2,-2.7) .. (v15);
\node at (-1.2,-2.3) {\scriptsize{$e$-$x$}};
\node at (-0.4,-2.3) {\scriptsize{$e$-$y$}};
\node at (-0.3,-1.9) {\scriptsize{$f$-$x,x$}};
\end{scope}
\node (v12) at (6.6,8.55) {$\mathbf{N}^{\lceil m/2\rceil}_0$};
\node (v20) at (6.6,10.25) {$\mathbf{N}_1$};
\node at (6.7,9.9) {\scriptsize{$2e$}};
\node at (6.7,9.25) {\scriptsize{$2e$-$\sum x_i$}};
\node (v14) at (7.6,9.3) {\scriptsize{2}};
\draw  (v8) edge (v14);
\draw  (v14) edge (v10);
\draw  (v9) edge (v15);
\draw  (v20) edge (v12);
\end{scope}
\draw  (v10) edge (v9);
\end{tikzpicture}
\ee
\bit
\item $f$ in $\mathbf{N}_{1}$ is glued to $f-x-y$ in $\mathbf{S}_i$ for $i=1,\cdots,m$.
\item $f-x_i,x_i$ in $\mathbf{N}_{0}$ are glued to $x,y$ in $\mathbf{S}_{2i-1}$ for $i=1,\cdots,\lceil m/2\rceil$.
\item $f-x_i,x_i$ in $\mathbf{N}_{0}$ are glued to $y,x$ in $\mathbf{S}_{2i}$ for $i=1,\cdots,\lceil m/2\rceil$.
\eit

The gluing rules for the $\su(2)^{(1)}$ delocalized flavor symmetry of
\be
\begin{tikzpicture} [scale=1.9]
\node (v3) at (-2.9,0.8) {2};
\node at (-2.9,1.1) {$\su(1)^{(1)}$};
\node (v4) at (-3.9,0.8) {2};
\node at (-3.8,1.1) {$\su(1)^{(1)}$};
\node (v5) at (-3.4,0.8) {\tiny{3}};
\draw [-] (v4) edge (v5);
\draw [->] (v5) edge (v3);
\end{tikzpicture}
\ee
are described by
\be
\begin{tikzpicture} [scale=1.9]
\begin{scope}[rotate=90]
\begin{scope}[rotate=90, shift={(7.4,-5.6)}]
\node (v8) at (2.7,-2) {$\mathbf{2}^{1+1}_{0}$};
\node (v7_1) at (4.4,-2) {$\mathbf{1}^{1+1}_{0}$};
\node at (3.95,-1.9) {\scriptsize{$f$-$y,y$}};
\node at (3.2,-1.9) {\scriptsize{$3f$-$x,x$}};
\node (v3_1) at (3.6,-2) {\scriptsize{2}};
\draw  (v8) edge (v3_1);
\draw  (v3_1) edge (v7_1);
\draw (v8) .. controls (2.1,-2.7) and (3.3,-2.7) .. (v8);
\node at (2.3,-2.3) {\scriptsize{$e$-$x$}};
\node at (3.1,-2.3) {\scriptsize{$e$-$y$}};
\draw (v7_1) .. controls (3.8,-2.7) and (5,-2.7) .. (v7_1);
\node at (4,-2.3) {\scriptsize{$e$-$x$}};
\node at (4.8,-2.3) {\scriptsize{$e$-$y$}};
\end{scope}
\node (v12) at (6.8,10.3) {$\mathbf{N}^{1}_0$};
\node (v20) at (6.8,11.7) {$\mathbf{N}_1$};
\node at (6.9,11.35) {\scriptsize{$2e$}};
\node at (6.9,10.7) {\scriptsize{$2e$-$x$}};
\draw  (v20) edge (v12);
\end{scope}
\end{tikzpicture}
\ee
\bit
\item $f$ in $\mathbf{N}_{1}$ is glued to $f-x-y$ in $\mathbf{S}_i$ for $i=1,2$.
\item $f-x_1,x_1$ in $\mathbf{N}_{0}$ are glued to $x,y$ in $\mathbf{S}_{1}$.
\item $f-x_1,x_1$ in $\mathbf{N}_{0}$ are glued to $y,x$ in $\mathbf{S}_{2}$.
\eit

The gluing rules for the $\su(2)^{(1)}$ delocalized flavor symmetry of
\be
\begin{tikzpicture} [scale=1.9]
\node (v1) at (-2,0.8) {2};
\node at (-2,1.1) {$\su(1)^{(1)}$};
\node (v3) at (-2.9,0.8) {2};
\node at (-2.9,1.1) {$\su(1)^{(1)}$};
\node (v4) at (-3.9,0.8) {2};
\node at (-3.9,1.1) {$\su(1)^{(1)}$};
\node (v5) at (-3.4,0.8) {\tiny{2}};
\draw [-] (v4) edge (v5);
\draw [->] (v5) edge (v3);
\draw  (v3) edge (v1);
\node (v2) at (-4.85,0.8) {2};
\node at (-4.85,1.1) {$\su(1)^{(1)}$};
\draw  (v2) edge (v4);
\end{tikzpicture}
\ee
are described by
\be
\begin{tikzpicture} [scale=1.9]
\begin{scope}[rotate=90]
\begin{scope}[rotate=90, shift={(7.4,-5.6)}]
\node (v10) at (1.05,-2) {$\mathbf{3}^{1+1}_{0}$};
\node[rotate=0] (v9) at (0.2,-2) {\scriptsize{2}};
\node (v8) at (2.7,-2) {$\mathbf{2}^{1+1}_{0}$};
\node (v7_1) at (4.4,-2) {$\mathbf{1}^{1+1}_{0}$};
\node at (1.5,-1.9) {\scriptsize{$2f$-$x,x$}};
\node at (3.95,-1.9) {\scriptsize{$f$-$y,y$}};
\node at (2.25,-1.9) {\scriptsize{$f$-$y,y$}};
\node at (3.2,-1.9) {\scriptsize{$f$-$x,x$}};
\node at (0.6,-1.9) {\scriptsize{$f$-$y,y$}};
\draw (v10) .. controls (0.45,-2.7) and (1.65,-2.7) .. (v10);
\node at (0.65,-2.3) {\scriptsize{$e$-$x$}};
\node at (1.45,-2.3) {\scriptsize{$e$-$y$}};
\node (v3_1) at (3.6,-2) {\scriptsize{2}};
\draw  (v8) edge (v3_1);
\draw  (v3_1) edge (v7_1);
\draw (v8) .. controls (2.1,-2.7) and (3.3,-2.7) .. (v8);
\node at (2.3,-2.3) {\scriptsize{$e$-$x$}};
\node at (3.1,-2.3) {\scriptsize{$e$-$y$}};
\draw (v7_1) .. controls (3.8,-2.7) and (5,-2.7) .. (v7_1);
\node at (4,-2.3) {\scriptsize{$e$-$x$}};
\node at (4.8,-2.3) {\scriptsize{$e$-$y$}};
\node (v15) at (-0.7,-2) {$\mathbf{m}^{1+1}_{0}$};
\draw (v15) .. controls (-1.3,-2.7) and (-0.1,-2.7) .. (v15);
\node at (-1.1,-2.3) {\scriptsize{$e$-$x$}};
\node at (-0.3,-2.3) {\scriptsize{$e$-$y$}};
\node at (-0.2,-1.9) {\scriptsize{$f$-$x,x$}};
\end{scope}
\node (v12) at (6.6,8.55) {$\mathbf{N}^{2}_0$};
\node (v20) at (6.6,10.25) {$\mathbf{N}_1$};
\node at (6.7,9.9) {\scriptsize{$2e$}};
\node at (6.7,9.05) {\scriptsize{$2e$-$\sum x_i$}};
\node (v14) at (7.6,9.3) {\scriptsize{2}};
\draw  (v8) edge (v14);
\draw  (v14) edge (v10);
\draw  (v9) edge (v15);
\draw  (v20) edge (v12);
\end{scope}
\draw  (v10) edge (v9);
\end{tikzpicture}
\ee
\bit
\item $f$ in $\mathbf{N}_{1}$ is glued to $f-x-y$ in $\mathbf{S}_i$ for $i=1,\cdots,4$.
\item $f-x_i,x_i$ in $\mathbf{N}_{0}$ are glued to $x,y$ in $\mathbf{S}_{2i-1}$ for $i=1,2$.
\item $f-x_i,x_i$ in $\mathbf{N}_{0}$ are glued to $y,x$ in $\mathbf{S}_{2i}$ for $i=1,2$.
\eit

The gluing rules for the $(\su(2)^{(1)})^{\oplus 2}$ delocalized flavor symmetry of
\be
\begin{tikzpicture} [scale=1.9]
\node (v1) at (-1.5,0.8) {2};
\node at (-1.5,1.1) {$\su(2)^{(1)}$};
\node (v3) at (-2.9,0.8) {2};
\node at (-2.9,1.1) {$\su(2)^{(1)}$};
\node (v2) at (-2.2,0.8) {$\cdots$};
\draw  (v2) edge (v3);
\draw  (v2) edge (v1);
\begin{scope}[shift={(0,0.45)}]
\node at (-2.65,-0.15) {$m$};
\draw (-4,0.15) .. controls (-4,0.1) and (-4,0.05) .. (-3.9,0.05);
\draw (-3.9,0.05) -- (-2.75,0.05);
\draw (-2.65,0) .. controls (-2.65,0.05) and (-2.7,0.05) .. (-2.75,0.05);
\draw (-2.65,0) .. controls (-2.65,0.05) and (-2.6,0.05) .. (-2.55,0.05);
\draw (-2.55,0.05) -- (-1.4,0.05);
\draw (-1.3,0.15) .. controls (-1.3,0.1) and (-1.3,0.05) .. (-1.4,0.05);
\end{scope}
\node (v4) at (-3.9,0.8) {2};
\node at (-3.8,1.1) {$\su(2)^{(1)}$};
\draw  (v4) edge (v3);
\draw (v4) .. controls (-4.55,1.2) and (-4.55,0.4) .. (v4);
\end{tikzpicture}
\ee
for $m$ even are described by
\be
\begin{tikzpicture} [scale=1.9]
\node (v2) at (0.9,1.3) {$\mathbf{m}_{0}$};
\node (v3) at (3.1,1.3) {$\mathbf{(m-1)}_{0}^{2+2}$};
\node (v10) at (0.9,-0.8) {$\mathbf{(m+2)}_0^{2+2}$};
\node (v9) at (3.1,-0.8) {$\mathbf{(m+3)}_0$};
\node (v4) at (4.5,1.3) {$\cdots$};
\node (v8) at (4.2,-0.8) {$\cdots$};
\node (v5) at (5.3,1.3) {$\mathbf{2}_0$};
\node (v7) at (5.3,-0.8) {$\mathbf{2m}_{0}^{2+2}$};
\node (v6) at (7.3,1.3) {$\mathbf{1}_0^{2+2}$};
\draw  (v2) edge (v3);
\draw  (v3) edge (v4);
\draw  (v4) edge (v5);
\draw  (v5) edge (v6);
\draw  (v7) edge (v8);
\draw  (v8) edge (v9);
\draw  (v9) edge (v10);
\node at (1.3,1.4) {\scriptsize{$f$}};
\node at (2.3,1.4) {\scriptsize{$f$-$\sum y_i$}};
\node at (3.9,1.4) {\scriptsize{$f$-$\sum x_i$}};
\node at (5,1.4) {\scriptsize{$f$}};
\node at (5.6,1.4) {\scriptsize{$f$}};
\node at (5.9,-0.9) {\scriptsize{$f$-$\sum x_i$}};
\node at (4.7,-0.9) {\scriptsize{$f$-$\sum y_i$}};
\node at (1.7,-0.9) {\scriptsize{$f$-$\sum x_i$}};
\node at (1.1646,-0.3943) {\scriptsize{$x_i$}};
\node at (2.6,1) {\scriptsize{$y_i$}};
\node at (3.4,0.9) {\scriptsize{$x_i$}};
\node at (2.5,-0.9) {\scriptsize{$f$}};
\node at (0.75,0.8) {\scriptsize{$e$}};
\node at (3.7,-0.9) {\scriptsize{$f$}};
\node at (6.6,-0.9) {\scriptsize{$f$}};
\node at (4.8,-0.5) {\scriptsize{$y_i$}};
\node (v13) at (7.3,-0.8) {$\mathbf{(2m+1)}_{0}$};
\draw  (v7) edge (v13);
\node at (6.8,1.4) {\scriptsize{$f$-$\sum y_i$}};
\node at (5.6,-0.3) {\scriptsize{$x_i$}};
\node at (6.8,1) {\scriptsize{$y_i$}};
\node (v14) at (0.9,0.25) {\scriptsize{2}};
\draw  (v2) edge (v14);
\draw  (v10) edge (v14);
\node at (0.65,-0.15) {\scriptsize{$e$-$x_i$-$y_i$}};
\begin{scope}[shift={(2.2,-0.1)}]
\node at (0.75,-0.3) {\scriptsize{$e$}};
\end{scope}
\begin{scope}[shift={(2.2,0)}]
\node at (0.65,0.65) {\scriptsize{$e$-$x_i$-$y_i$}};
\end{scope}
\begin{scope}[shift={(2.2,0)}]
\node (v14) at (0.9,0.25) {\scriptsize{2}};
\end{scope}
\draw  (v3) edge (v14);
\draw  (v14) edge (v9);
\draw (v3) -- (3.9,0.6);
\draw (v7) -- (4.5,0);
\begin{scope}[shift={(4.4,0.1)}]
\node at (0.75,0.8) {\scriptsize{$e$}};
\end{scope}
\begin{scope}[shift={(4.4,0)}]
\node at (0.65,-0.15) {\scriptsize{$e$-$x_i$-$y_i$}};
\end{scope}
\begin{scope}[shift={(4.4,0)}]
\node (v14) at (0.9,0.25) {\scriptsize{2}};
\end{scope}
\draw  (v5) edge (v14);
\draw  (v14) edge (v7);
\begin{scope}[shift={(6.4,0)}]
\node (v14) at (0.9,0.25) {\scriptsize{2}};
\end{scope}
\draw  (v6) edge (v14);
\draw  (v14) edge (v13);
\begin{scope}[shift={(6.4,0)}]
\node at (0.65,0.65) {\scriptsize{$e$-$x_i$-$y_i$}};
\end{scope}
\begin{scope}[shift={(6.4,-0.1)}]
\node at (0.75,-0.3) {\scriptsize{$e$}};
\end{scope}
\node at (4.3,0.3) {$\cdots$};
\begin{scope}[shift={(1.8,0)}]
\node (v24) at (1,-1.6) {$\mathbf{M}^{(m+1)+(m+1)}_{0}$};
\node (v25) at (3.6,-1.6) {$\mathbf{M}_{1}$};
\node[rotate=0] at (2.1,-1.5) {\scriptsize{$2e$-$\sum x_i$}};
\node[rotate=0] at (3.2,-1.5) {\scriptsize{$2e$}};
\draw  (v24) edge (v25);
\end{scope}
\begin{scope}[shift={(-1.2,-1)}]
\node (v24) at (1.5,-1.6) {$\mathbf{P}^{2}_{0}$};
\node (v25) at (3.6,-1.6) {$\mathbf{P}_{1}$};
\node[rotate=0] at (2.1,-1.5) {\scriptsize{$2e$-$\sum x_i$}};
\node[rotate=0] at (3.2,-1.5) {\scriptsize{$2e$}};
\draw  (v24) edge (v25);
\end{scope}
\begin{scope}[shift={(3.6,-0.9)}]
\node (v24) at (1.5,-1.6) {$\mathbf{Q}^{2}_{0}$};
\node (v25) at (3.6,-1.6) {$\mathbf{Q}_{1}$};
\node[rotate=0] at (2.1,-1.5) {\scriptsize{$2e$-$\sum x_i$}};
\node[rotate=0] at (3.2,-1.5) {\scriptsize{$2e$}};
\draw  (v24) edge (v25);
\end{scope}
\node (v26) at (1.9941,0.2425) {\scriptsize{2}};
\draw  (v10) edge (v26);
\draw  (v26) edge (v3);
\node (v11) at (6.3526,0.3133) {\scriptsize{2}};
\draw  (v7) edge (v11);
\draw  (v11) edge (v6);
\end{tikzpicture}
\ee
\bit
\item $y_1,y_2$ in $\mathbf{S}_{m+2}$ are glued to $x_1,x_2$ in $\mathbf{P}_0$.
\item $f-y_1-y_2$ in $\mathbf{S}_{m+2}$ is glued to $f$ in $\mathbf{P}_1$.
\item $f$ in $\mathbf{S}_{m}$ is glued to $f-x_1-x_2$ in $\mathbf{P}_0$.
\item $x_1,x_2$ in $\mathbf{S}_{1}$ are glued to $x_1,x_2$ in $\mathbf{Q}_0$.
\item $f-x_1-x_2$ in $\mathbf{S}_{1}$ is glued to $f$ in $\mathbf{Q}_1$.
\item $f$ in $\mathbf{S}_{2m+1}$ is glued to $f-x_1-x_2$ in $\mathbf{Q}_0$.
\item $x_1-x_2,y_2-y_1$ in $\mathbf{S}_{m+2i}$ are glued to $f,f$ in $\mathbf{M}_1$ for $i=1,\cdots,\frac m2$.
\item $x_2-x_1,y_1-y_2$ in $\mathbf{S}_{m+1-2i}$ are glued to $f,f$ in $\mathbf{M}_1$ for $i=1,\cdots,\frac m2$.
\item $f-x_1,x_2,f-y_2,y_1$ in $\mathbf{S}_{m+2i}$ are glued to $f-x_{2i},y_{2i},x_{2i-1},y_{2i-1}$ in $\mathbf{M}_0$ for $i=1,\cdots,\frac m2$.
\item $f-x_2,x_1,f-y_1,y_2$ in $\mathbf{S}_{m+1-2i}$ are glued to $f-x_{2i+1},y_{2i+1},x_{2i},y_{2i}$ in $\mathbf{M}_0$ for $i=1,\cdots,\frac m2$.
\item $f,f$ in $\mathbf{S}_{m+2-2i}$ are glued to $x_{2i}-y_{2i},f-x_{2i-1}-y_{2i-1}$ in $\mathbf{M}_0$ for $i=1,\cdots,\frac m2$.
\item $f,f$ in $\mathbf{S}_{m+1+2i}$ are glued to $x_{2i+1}-y_{2i+1},f-x_{2i}-y_{2i}$ in $\mathbf{M}_0$ for $i=1,\cdots,\frac m2$.
\item $x_2-x_1$ in $\mathbf{P}_{0}$ is glued to $f$ in $\mathbf{M}_1$.
\item $f-x_2,x_1$ in $\mathbf{P}_{0}$ is glued to $f-x_1,y_1$ in $\mathbf{M}_0$.
\item $f$ in $\mathbf{P}_{1}$ is glued to $x_1-y_1$ in $\mathbf{M}_0$.
\item $x_2-x_1$ in $\mathbf{Q}_{0}$ is glued to $f$ in $\mathbf{M}_1$.
\item $f-x_2,x_1$ in $\mathbf{Q}_{0}$ is glued to $x_{m+1},y_{m+1}$ in $\mathbf{M}_0$.
\item $f$ in $\mathbf{Q}_{1}$ is glued to $f-x_{m+1}-y_{m+1}$ in $\mathbf{M}_0$.
\eit
Let us note, in order to avoid confusion, that there is no surface $\mathbf{S}_{m+1}$ in the above geometry. For $m$ odd, the gluing rules are described by
\be
\begin{tikzpicture} [scale=1.9]
\node (v2) at (0.9,1.3) {$\mathbf{m}_{0}$};
\node (v3) at (3.1,1.3) {$\mathbf{(m-1)}_{0}^{2+2}$};
\node (v10) at (0.9,-0.8) {$\mathbf{(m+2)}_0^{2+2}$};
\node (v9) at (3.1,-0.8) {$\mathbf{(m+3)}_0$};
\node (v4) at (4.5,1.3) {$\cdots$};
\node (v8) at (4,-0.8) {$\cdots$};
\node (v5) at (5.3,1.3) {$\mathbf{1}_0$};
\node (v7) at (5.3,-0.8) {$\mathbf{2m+1}_{0}^{2+2}$};
\draw  (v2) edge (v3);
\draw  (v3) edge (v4);
\draw  (v4) edge (v5);
\draw  (v7) edge (v8);
\draw  (v8) edge (v9);
\draw  (v9) edge (v10);
\node at (1.3,1.4) {\scriptsize{$f$}};
\node at (2.3,1.4) {\scriptsize{$f$-$\sum y_i$}};
\node at (3.9,1.4) {\scriptsize{$f$-$\sum x_i$}};
\node at (5,1.4) {\scriptsize{$f$}};
\node at (4.5,-0.9) {\scriptsize{$f$-$\sum y_i$}};
\node at (1.7,-0.9) {\scriptsize{$f$-$\sum x_i$}};
\node at (1.1646,-0.3943) {\scriptsize{$x_i$}};
\node at (2.6,1) {\scriptsize{$y_i$}};
\node at (3.4,0.9) {\scriptsize{$x_i$}};
\node at (2.5,-0.9) {\scriptsize{$f$}};
\node at (0.75,0.8) {\scriptsize{$e$}};
\node at (3.7,-0.9) {\scriptsize{$f$}};
\node at (4.8,-0.5) {\scriptsize{$y_i$}};
\node (v14) at (0.9,0.25) {\scriptsize{2}};
\draw  (v2) edge (v14);
\draw  (v10) edge (v14);
\node at (0.65,-0.15) {\scriptsize{$e$-$x_i$-$y_i$}};
\begin{scope}[shift={(2.2,-0.1)}]
\node at (0.75,-0.3) {\scriptsize{$e$}};
\end{scope}
\begin{scope}[shift={(2.2,0)}]
\node at (0.65,0.65) {\scriptsize{$e$-$x_i$-$y_i$}};
\end{scope}
\begin{scope}[shift={(2.2,0)}]
\node (v14) at (0.9,0.25) {\scriptsize{2}};
\end{scope}
\draw  (v3) edge (v14);
\draw  (v14) edge (v9);
\draw (v3) -- (3.9,0.6);
\draw (v7) -- (4.5,0);
\begin{scope}[shift={(4.4,0.1)}]
\node at (0.75,0.8) {\scriptsize{$e$}};
\end{scope}
\begin{scope}[shift={(4.4,0)}]
\node at (0.65,-0.15) {\scriptsize{$e$-$x_i$-$y_i$}};
\end{scope}
\begin{scope}[shift={(4.4,0)}]
\node (v14) at (0.9,0.25) {\scriptsize{2}};
\end{scope}
\draw  (v5) edge (v14);
\draw  (v14) edge (v7);
\node at (4.3,0.3) {$\cdots$};
\begin{scope}[shift={(0.2,-0.1)}]
\node (v24) at (1,-1.6) {$\mathbf{M}^{(m+1)+(m+1)}_{0}$};
\node (v25) at (3.6,-1.6) {$\mathbf{M}_{1}$};
\node[rotate=0] at (2.1,-1.5) {\scriptsize{$2e$-$\sum x_i$}};
\node[rotate=0] at (3.2,-1.5) {\scriptsize{$2e$}};
\draw  (v24) edge (v25);
\end{scope}
\begin{scope}[shift={(-1.6,-1)}]
\node (v24) at (1.5,-1.6) {$\mathbf{P}^{2}_{0}$};
\node (v25) at (3.6,-1.6) {$\mathbf{P}_{1}$};
\node[rotate=0] at (2.1,-1.5) {\scriptsize{$2e$-$\sum x_i$}};
\node[rotate=0] at (3.2,-1.5) {\scriptsize{$2e$}};
\draw  (v24) edge (v25);
\end{scope}
\begin{scope}[shift={(1.6,-1)}]
\node (v24) at (1.5,-1.6) {$\mathbf{Q}^{2}_{0}$};
\node (v25) at (3.6,-1.6) {$\mathbf{Q}_{1}$};
\node[rotate=0] at (2.1,-1.5) {\scriptsize{$2e$-$\sum x_i$}};
\node[rotate=0] at (3.2,-1.5) {\scriptsize{$2e$}};
\draw  (v24) edge (v25);
\end{scope}
\node (v26) at (1.9941,0.2425) {\scriptsize{2}};
\draw  (v10) edge (v26);
\draw  (v26) edge (v3);
\end{tikzpicture}
\ee
\bit
\item $y_1,y_2$ in $\mathbf{S}_{m+2}$ are glued to $x_1,x_2$ in $\mathbf{P}_0$.
\item $f-y_1-y_2$ in $\mathbf{S}_{m+2}$ is glued to $f$ in $\mathbf{P}_1$.
\item $f$ in $\mathbf{S}_{m}$ is glued to $f-x_1-x_2$ in $\mathbf{P}_0$.
\item $x_1,x_2$ in $\mathbf{S}_{2m+1}$ are glued to $x_1,x_2$ in $\mathbf{Q}_0$.
\item $f-x_1-x_2$ in $\mathbf{S}_{2m+1}$ is glued to $f$ in $\mathbf{Q}_1$.
\item $f$ in $\mathbf{S}_{1}$ is glued to $f-x_1-x_2$ in $\mathbf{Q}_0$.
\item $x_1-x_2,y_2-y_1$ in $\mathbf{S}_{m+2i}$ are glued to $f,f$ in $\mathbf{M}_1$ for $i=1,\cdots,\frac{m+1}2$.
\item $x_2-x_1,y_1-y_2$ in $\mathbf{S}_{m+1-2i}$ are glued to $f,f$ in $\mathbf{M}_1$ for $i=1,\cdots,\frac{m-1}2$.
\item $f-x_1,x_2,f-y_2,y_1$ in $\mathbf{S}_{m+2i}$ are glued to $f-x_{2i},y_{2i},x_{2i-1},y_{2i-1}$ in $\mathbf{M}_0$ for $i=1,\cdots,\frac{m+1}2$.
\item $f-x_2,x_1,f-y_1,y_2$ in $\mathbf{S}_{m+1-2i}$ are glued to $f-x_{2i+1},y_{2i+1},x_{2i},y_{2i}$ in $\mathbf{M}_0$ for $i=1,\cdots,\frac{m-1}2$.
\item $f,f$ in $\mathbf{S}_{m+2-2i}$ are glued to $x_{2i}-y_{2i},f-x_{2i-1}-y_{2i-1}$ in $\mathbf{M}_0$ for $i=1,\cdots,\frac{m+1}2$.
\item $f,f$ in $\mathbf{S}_{m+1+2i}$ are glued to $x_{2i+1}-y_{2i+1},f-x_{2i}-y_{2i}$ in $\mathbf{M}_0$ for $i=1,\cdots,\frac{m-1}2$.
\item $x_2-x_1$ in $\mathbf{P}_{0}$ is glued to $f$ in $\mathbf{M}_1$.
\item $f-x_2,x_1$ in $\mathbf{P}_{0}$ is glued to $f-x_1,y_1$ in $\mathbf{M}_0$.
\item $f$ in $\mathbf{P}_{1}$ is glued to $x_1-y_1$ in $\mathbf{M}_0$.
\item $x_1-x_2$ in $\mathbf{Q}_{0}$ is glued to $f$ in $\mathbf{M}_1$.
\item $f-x_1,x_2$ in $\mathbf{Q}_{0}$ is glued to $x_{m+1},y_{m+1}$ in $\mathbf{M}_0$.
\item $f$ in $\mathbf{Q}_{1}$ is glued to $f-x_{m+1}-y_{m+1}$ in $\mathbf{M}_0$.
\eit
Let us note, in order to avoid confusion, that there is no surface $\mathbf{S}_{m+1}$ in the above geometry.

\section{Flops and blowdowns}\label{FB}
In this section, we illustrate using a simple example how the flops and blowdowns associated to an RG flow from a $5d$ KK theory to a $5d$ SCFT transform the non-compact surfaces in the geometry associated to the $5d$ KK theory, which allows us not only to read the flavor symmetry of the $5d$ SCFT, but also describe how the non-abelian part of the flavor symmetry of the $5d$ SCFT is captured by non-compact surfaces in the geometry associated to the $5d$ SCFT. That is, we also obtain the data of how the non-compact surfaces associated to the flavor symmetry of the $5d$ SCFT are coupled to the compact part of the geometry. 

Many more examples of diverse kinds of RG flows are discussed in Part 2 \cite{Bhardwaj:2020avz} of this series of papers. 

The example we study is of the $5d$ KK theory
\be\label{sp1eg}
\begin{tikzpicture} [scale=1.9]
\node (v2) at (-3.5,0.8) {1};
\node at (-3.5,1.1) {$\sp(1)^{(1)}$};
\end{tikzpicture}
\ee
The $6d$ SCFT has a tensor branch description as a $6d$ $\cN=(1,0)$ gauge theory $\sp(1)+10\F$ which has an $\so(20)$ flavor symmetry, and hence we expect to be able to couple the compact part of the geometry (which was described in \cite{Bhardwaj:2018yhy,Bhardwaj:2018vuu,Bhardwaj:2019fzv}) to non-compact $\P^1$ surfaces comprising $\so(20)^{(1)}$, since there is no twist involved in the circle compactification. The gluing rules for 
\be
\begin{tikzpicture}
\node (w1) at (-0.5,0.9) {$\sp(1)^{(1)}$};
\begin{scope}[shift={(3,0)}]
\node (w2) at (-0.5,0.9) {$\so(20)^{(1)}$};
\end{scope}
\draw (w1)--(w2);
\end{tikzpicture}
\ee
were discussed in \cite{Bhardwaj:2018vuu,Bhardwaj:2019fzv} from which we find that the full geometry containing both compact and non-compact surfaces can be written as
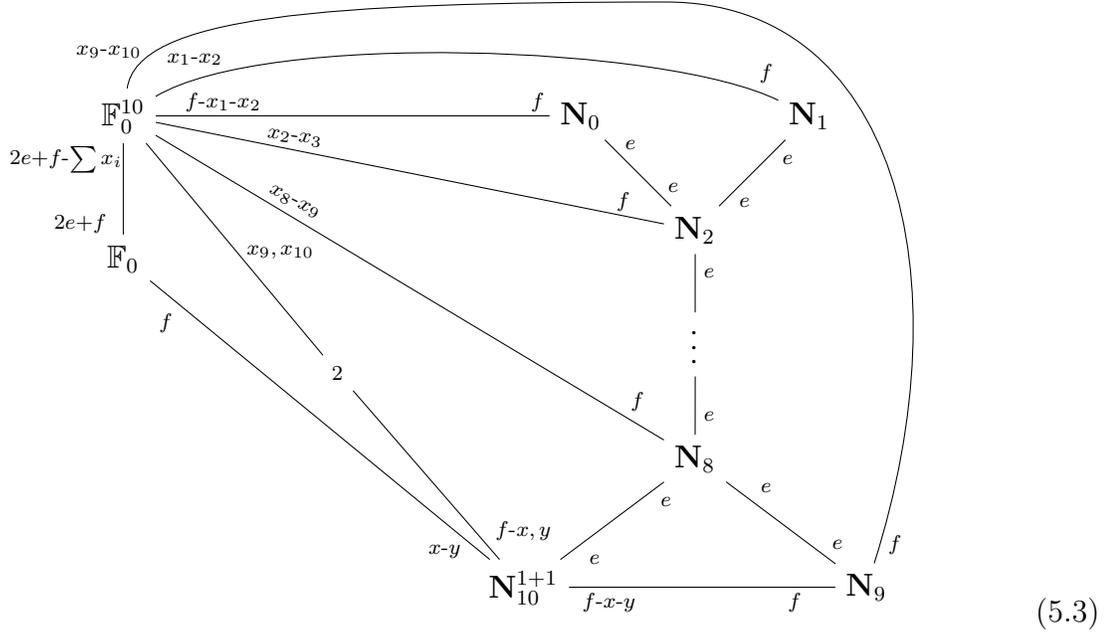
\be
\begin{tikzpicture} [scale=1.9]
\node (v1) at (7.6,13) {$\bF_0^{10}$};
\node (v10) at (7.6,12) {$\bF_{0}$};
\draw  (v10) edge (v1);
\node at (7.3,12.25) {\scriptsize{$2e$+$f$}};
\node at (7.2,12.7) {\scriptsize{$2e$+$f$-$\sum x_i$}};
\node (v2) at (10.8,13) {$\mathbf{N}_0$};
\node (v4) at (12.4,13) {$\mathbf{N}_1$};
\node (v3) at (11.6,12.2) {$\mathbf{N}_2$};
\node[rotate=90] (v6) at (11.6,11.4) {$\cdots$};
\node (v11) at (11.6,10.6) {$\mathbf{N}_{8}$};
\node (v12) at (10.4,9.7) {$\mathbf{N}^{1+1}_{10}$};
\node (v13) at (12.8,9.7) {$\mathbf{N}_{9}$};
\draw  (v2) edge (v3);
\draw  (v3) edge (v4);
\draw  (v3) edge (v6);
\draw  (v6) edge (v11);
\draw  (v11) edge (v12);
\draw  (v11) edge (v13);
\node at (11,9.6) {\scriptsize{$f$-$x$-$y$}};
\node at (12.3,9.6) {\scriptsize{$f$}};
\draw  (v1) edge (v2);
\draw  (v1) edge (v3);
\draw  (v1) edge (v11);
\node at (8.3,13.1) {\scriptsize{$f$-$x_1$-$x_2$}};
\node at (10.5,13.1) {\scriptsize{$f$}};
\node[rotate=-8] at (8.8,12.85) {\scriptsize{$x_2$-$x_3$}};
\node at (11.1,12.4) {\scriptsize{$f$}};
\node[rotate=-27] at (8.8,12.4) {\scriptsize{$x_{8}$-$x_{9}$}};
\node at (11.2,11) {\scriptsize{$f$}};
\node (v15) at (9.1,11.2) {\scriptsize{2}};
\draw  (v1) edge (v15);
\draw  (v15) edge (v12);
\node at (8.7,12.05) {\scriptsize{$x_{9},x_{10}$}};
\node at (10.4,10.1) {\scriptsize{$f$-$x,y$}};
\node at (7.9,11.55) {\scriptsize{$f$}};
\node at (9.85,9.95) {\scriptsize{$x$-$y$}};
\draw (v1) .. controls (8.6,13.6) and (11.4,13.5) .. (v4);
\node at (8.1,13.4) {\scriptsize{$x_1$-$x_2$}};
\node at (12.1,13.3) {\scriptsize{$f$}};
\draw (v1) .. controls (7.7,13.7) and (9,13.8) .. (11.4,13.8);
\draw (11.4,13.8) .. controls (13,13.8) and (13.5,11.9) .. (v13);
\node at (7.5,13.45) {\scriptsize{$x_{9}$-$x_{10}$}};
\node at (13,10) {\scriptsize{$f$}};
\node at (11.15,12.8) {\scriptsize{$e$}};
\node at (11.45,12.5) {\scriptsize{$e$}};
\node at (12.25,12.7) {\scriptsize{$e$}};
\node at (11.95,12.4) {\scriptsize{$e$}};
\node at (11.7,11.9) {\scriptsize{$e$}};
\node at (11.7,10.9) {\scriptsize{$e$}};
\node at (12.1,10.4) {\scriptsize{$e$}};
\node at (12.6,10) {\scriptsize{$e$}};
\node at (10.9,9.9) {\scriptsize{$e$}};
\node at (11.4,10.3) {\scriptsize{$e$}};
\draw  (v12) edge (v13);
\draw  (v10) edge (v12);
\end{tikzpicture}
\ee
where we display a Hirzebruch surface of degree $n$ and carrying $b$ blowups as $\bF_n^b$. Applying the isomorphism $\cS$ (which exchanges $e$ and $f$ in $\bF_0^b$) on the top compact surface, we can write the above geometry as
\be
\begin{tikzpicture} [scale=1.9]
\node (v1) at (7.6,13) {$\bF_0^{10}$};
\node (v10) at (7.6,12) {$\bF_{0}$};
\draw  (v10) edge (v1);
\node at (7.3,12.25) {\scriptsize{$2e$+$f$}};
\node at (7.2,12.7) {\scriptsize{$e$+$2f$-$\sum x_i$}};
\node (v2) at (10.8,13) {$\mathbf{N}_0$};
\node (v4) at (12.4,13) {$\mathbf{N}_1$};
\node (v3) at (11.6,12.2) {$\mathbf{N}_2$};
\node[rotate=90] (v6) at (11.6,11.4) {$\cdots$};
\node (v11) at (11.6,10.6) {$\mathbf{N}_{8}$};
\node (v12) at (10.4,9.7) {$\mathbf{N}^{1+1}_{10}$};
\node (v13) at (12.8,9.7) {$\mathbf{N}_{9}$};
\draw  (v2) edge (v3);
\draw  (v3) edge (v4);
\draw  (v3) edge (v6);
\draw  (v6) edge (v11);
\draw  (v11) edge (v12);
\draw  (v11) edge (v13);
\node at (11,9.6) {\scriptsize{$f$-$x$-$y$}};
\node at (12.3,9.6) {\scriptsize{$f$}};
\draw  (v1) edge (v2);
\draw  (v1) edge (v3);
\draw  (v1) edge (v11);
\node at (8.3,13.1) {\scriptsize{$e$-$x_1$-$x_2$}};
\node at (10.5,13.1) {\scriptsize{$f$}};
\node[rotate=-8] at (8.8,12.85) {\scriptsize{$x_2$-$x_3$}};
\node at (11.1,12.4) {\scriptsize{$f$}};
\node[rotate=-27] at (8.8,12.4) {\scriptsize{$x_{8}$-$x_{9}$}};
\node at (11.2,11) {\scriptsize{$f$}};
\node (v15) at (9.1,11.2) {\scriptsize{2}};
\draw  (v1) edge (v15);
\draw  (v15) edge (v12);
\node at (8.7,12.05) {\scriptsize{$x_{9},x_{10}$}};
\node at (10.4,10.1) {\scriptsize{$f$-$x,y$}};
\node at (7.9,11.55) {\scriptsize{$f$}};
\node at (9.85,9.95) {\scriptsize{$x$-$y$}};
\draw (v1) .. controls (8.6,13.6) and (11.4,13.5) .. (v4);
\node at (8.1,13.4) {\scriptsize{$x_1$-$x_2$}};
\node at (12.1,13.3) {\scriptsize{$f$}};
\draw (v1) .. controls (7.7,13.7) and (9,13.8) .. (11.4,13.8);
\draw (11.4,13.8) .. controls (13,13.8) and (13.5,11.9) .. (v13);
\node at (7.5,13.45) {\scriptsize{$x_{9}$-$x_{10}$}};
\node at (13,10) {\scriptsize{$f$}};
\node at (11.15,12.8) {\scriptsize{$e$}};
\node at (11.45,12.5) {\scriptsize{$e$}};
\node at (12.25,12.7) {\scriptsize{$e$}};
\node at (11.95,12.4) {\scriptsize{$e$}};
\node at (11.7,11.9) {\scriptsize{$e$}};
\node at (11.7,10.9) {\scriptsize{$e$}};
\node at (12.1,10.4) {\scriptsize{$e$}};
\node at (12.6,10) {\scriptsize{$e$}};
\node at (10.9,9.9) {\scriptsize{$e$}};
\node at (11.4,10.3) {\scriptsize{$e$}};
\draw  (v12) edge (v13);
\draw  (v10) edge (v12);
\end{tikzpicture}
\ee
Now from the compact surfaces we can read that the geometry also describes the $5d$ gauge theory $\sp(2)+10\F$ \cite{Bhardwaj:2019ngx}. Integrating out $\F$ from this theory leads to $5d$ SCFTs $\sp(2)+n\F$ with $n\le9$. We would like to determine the flavor symmetry for these $5d$ SCFTs. To integrate out one $\F$, we can flop the curve $f-x_1$ living in the top compact surface, which leads to the following geometry
\be
\begin{tikzpicture} [scale=1.9]
\node (v1) at (7.6,13) {$\bF_1^{9}$};
\node (v10) at (7.6,12) {$\bF_{0}$};
\draw  (v10) edge (v1);
\node at (7.3,12.25) {\scriptsize{$2e$+$f$}};
\node at (7.2,12.7) {\scriptsize{$h$+$f$-$\sum x_i$}};
\node (v2) at (10.8,13) {$\mathbf{N}_0$};
\node (v4) at (12.4,13) {$\mathbf{N}^1_1$};
\node (v3) at (11.6,12.2) {$\mathbf{N}_2$};
\node[rotate=90] (v6) at (11.6,11.4) {$\cdots$};
\node (v11) at (11.6,10.6) {$\mathbf{N}_{8}$};
\node (v12) at (10.4,9.7) {$\mathbf{N}^{1+1}_{10}$};
\node (v13) at (12.8,9.7) {$\mathbf{N}_{9}$};
\draw  (v2) edge (v3);
\draw  (v3) edge (v4);
\draw  (v3) edge (v6);
\draw  (v6) edge (v11);
\draw  (v11) edge (v12);
\draw  (v11) edge (v13);
\node at (11,9.6) {\scriptsize{$f$-$x$-$y$}};
\node at (12.3,9.6) {\scriptsize{$f$}};
\draw  (v1) edge (v2);
\draw  (v1) edge (v3);
\draw  (v1) edge (v11);
\node at (8.3,13.1) {\scriptsize{$e$-$x_2$}};
\node at (10.5,13.1) {\scriptsize{$f$}};
\node[rotate=-8] at (8.8,12.85) {\scriptsize{$x_2$-$x_3$}};
\node at (11.1,12.4) {\scriptsize{$f$}};
\node[rotate=-27] at (8.8,12.4) {\scriptsize{$x_{8}$-$x_{9}$}};
\node at (11.2,11) {\scriptsize{$f$}};
\node (v15) at (9.1,11.2) {\scriptsize{2}};
\draw  (v1) edge (v15);
\draw  (v15) edge (v12);
\node at (8.7,12.05) {\scriptsize{$x_{9},x_{10}$}};
\node at (10.4,10.1) {\scriptsize{$f$-$x,y$}};
\node at (7.9,11.55) {\scriptsize{$f$}};
\node at (9.85,9.95) {\scriptsize{$x$-$y$}};
\draw (v1) .. controls (8.6,13.6) and (11.4,13.5) .. (v4);
\node at (8.1,13.4) {\scriptsize{$f$-$x_2$}};
\node at (12.1,13.3) {\scriptsize{$f$-$x$}};
\draw (v1) .. controls (7.7,13.7) and (9,13.8) .. (11.4,13.8);
\draw (11.4,13.8) .. controls (13,13.8) and (13.5,11.9) .. (v13);
\node at (7.5,13.45) {\scriptsize{$x_{9}$-$x_{10}$}};
\node at (13,10) {\scriptsize{$f$}};
\node at (11.15,12.8) {\scriptsize{$e$}};
\node at (11.45,12.5) {\scriptsize{$e$}};
\node at (12.25,12.7) {\scriptsize{$e$}};
\node at (11.95,12.4) {\scriptsize{$e$}};
\node at (11.7,11.9) {\scriptsize{$e$}};
\node at (11.7,10.9) {\scriptsize{$e$}};
\node at (12.1,10.4) {\scriptsize{$e$}};
\node at (12.6,10) {\scriptsize{$e$}};
\node at (10.9,9.9) {\scriptsize{$e$}};
\node at (11.4,10.3) {\scriptsize{$e$}};
\draw  (v12) edge (v13);
\draw  (v10) edge (v12);
\end{tikzpicture}
\ee
where let us note, in order to avoid confusion, that we don't have any blowup labeled as $x_1$ living in the top compact surface anymore. The process of integrating out $\F$ is completed when we send the size of the blowup $x$ living in $\mathbf{N}_1$ to infinity. However, during this process, we would like to keep the size of the compact curve $f-x_2$ living in the top compact surface finite, since sending it to infinity will decompactify the top compact surface and hence change the theory away from the one we want to study. Since $f-x_2$ living in top compact surface $\bF_1^9$ is glued to $f-x$ living in $\mathbf{N}_1$, we must send the size of fiber $f$ of $\mathbf{N}_1$ to infinity. Thus, after integrating out $\F$, the surface $\mathbf{N}_1$ won't remain $\P^1$ fibered anymore, thus leaving us with the geometry
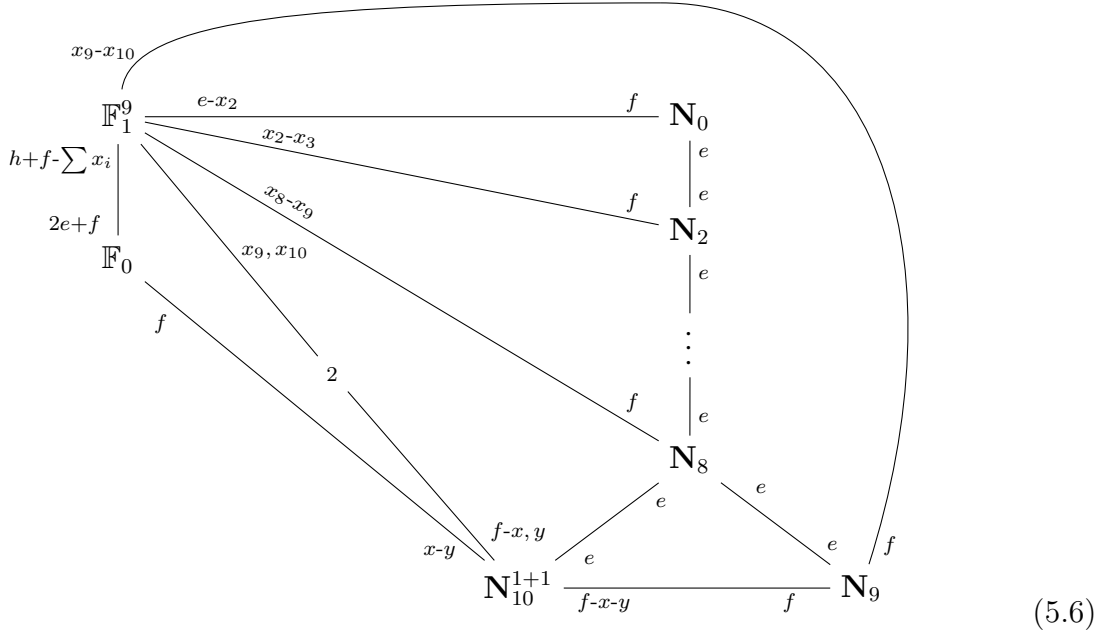
\be
\begin{tikzpicture} [scale=1.9]
\node (v1) at (7.6,13) {$\bF_1^{9}$};
\node (v10) at (7.6,12) {$\bF_{0}$};
\draw  (v10) edge (v1);
\node at (7.3,12.25) {\scriptsize{$2e$+$f$}};
\node at (7.2,12.7) {\scriptsize{$h$+$f$-$\sum x_i$}};
\node (v2) at (11.6,13) {$\mathbf{N}_0$};
\node (v3) at (11.6,12.2) {$\mathbf{N}_2$};
\node[rotate=90] (v6) at (11.6,11.4) {$\cdots$};
\node (v11) at (11.6,10.6) {$\mathbf{N}_{8}$};
\node (v12) at (10.4,9.7) {$\mathbf{N}^{1+1}_{10}$};
\node (v13) at (12.8,9.7) {$\mathbf{N}_{9}$};
\draw  (v2) edge (v3);
\draw  (v3) edge (v6);
\draw  (v6) edge (v11);
\draw  (v11) edge (v12);
\draw  (v11) edge (v13);
\node at (11,9.6) {\scriptsize{$f$-$x$-$y$}};
\node at (12.3,9.6) {\scriptsize{$f$}};
\draw  (v1) edge (v2);
\draw  (v1) edge (v3);
\draw  (v1) edge (v11);
\node at (8.3,13.1) {\scriptsize{$e$-$x_2$}};
\node at (11.2,13.1) {\scriptsize{$f$}};
\node[rotate=-8] at (8.8,12.85) {\scriptsize{$x_2$-$x_3$}};
\node at (11.2,12.4) {\scriptsize{$f$}};
\node[rotate=-27] at (8.8,12.4) {\scriptsize{$x_{8}$-$x_{9}$}};
\node at (11.2,11) {\scriptsize{$f$}};
\node (v15) at (9.1,11.2) {\scriptsize{2}};
\draw  (v1) edge (v15);
\draw  (v15) edge (v12);
\node at (8.7,12.05) {\scriptsize{$x_{9},x_{10}$}};
\node at (10.4,10.1) {\scriptsize{$f$-$x,y$}};
\node at (7.9,11.55) {\scriptsize{$f$}};
\node at (9.85,9.95) {\scriptsize{$x$-$y$}};
\draw (v1) .. controls (7.7,13.7) and (9,13.8) .. (11.4,13.8);
\draw (11.4,13.8) .. controls (13,13.8) and (13.5,11.9) .. (v13);
\node at (7.5,13.45) {\scriptsize{$x_{9}$-$x_{10}$}};
\node at (13,10) {\scriptsize{$f$}};
\node at (11.7,12.75) {\scriptsize{$e$}};
\node at (11.7,12.45) {\scriptsize{$e$}};
\node at (11.7,11.9) {\scriptsize{$e$}};
\node at (11.7,10.9) {\scriptsize{$e$}};
\node at (12.1,10.4) {\scriptsize{$e$}};
\node at (12.6,10) {\scriptsize{$e$}};
\node at (10.9,9.9) {\scriptsize{$e$}};
\node at (11.4,10.3) {\scriptsize{$e$}};
\draw  (v12) edge (v13);
\draw  (v10) edge (v12);
\end{tikzpicture}
\ee
from which we can read that the $5d$ SCFT admitting a mass deformation to $\sp(2)+9\F$ has $\so(20)$ flavor symmetry. This is done by computing 
\be
M_{ij}:=-f_i\cdot \mathbf{N}_j
\ee
(where $f_i$ labels the $\P^1$ fiber of $\mathbf{N}_i$) which gives rise to the Cartan matrix for $\so(20)$ finite Lie algebra. However, the classical flavor symmetry, that is the flavor symmetry of the $5d$ $\cN=1$ gauge theory $\sp(2)+9\F$ is only $\so(18)\oplus\u(1)$. Thus, we find that the flavor symmetry of the theory is enhanced from $\so(18)\oplus\u(1)$ to $\so(20)$ at the conformal point.

In a similar way, one can proceed to integrate out more $\F$ and determine the flavor symmetry for $5d$ SCFTs $\sp(2)+n\F$ for $n<9$. Let us instead apply $\cS$ on the bottom compact surface, which leads to the following geometry
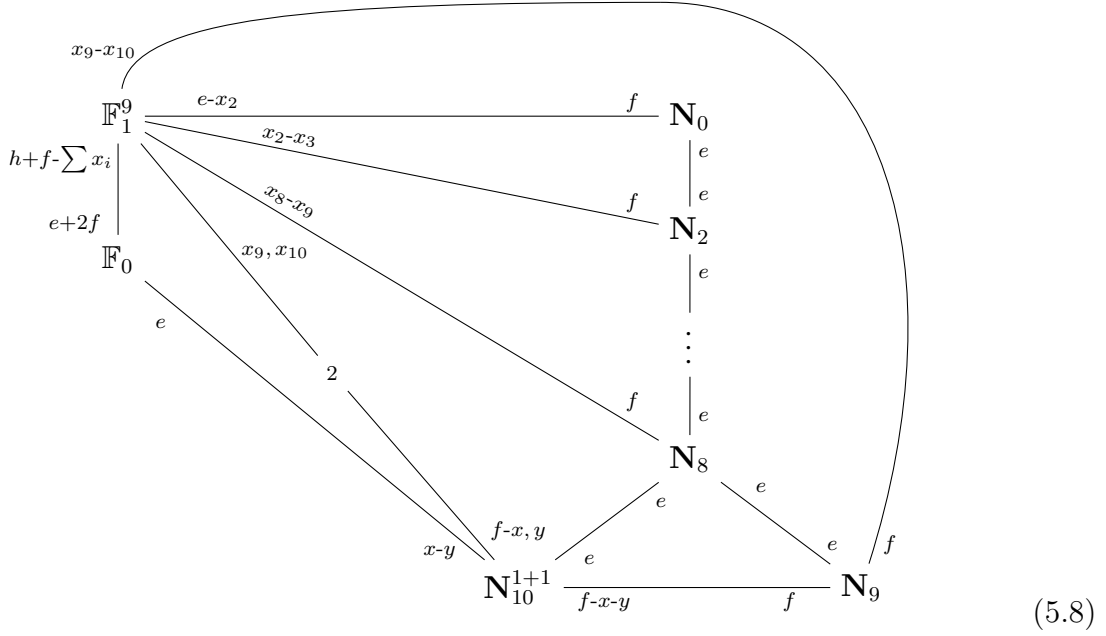
\be
\begin{tikzpicture} [scale=1.9]
\node (v1) at (7.6,13) {$\bF_1^{9}$};
\node (v10) at (7.6,12) {$\bF_{0}$};
\draw  (v10) edge (v1);
\node at (7.3,12.25) {\scriptsize{$e$+$2f$}};
\node at (7.2,12.7) {\scriptsize{$h$+$f$-$\sum x_i$}};
\node (v2) at (11.6,13) {$\mathbf{N}_0$};
\node (v3) at (11.6,12.2) {$\mathbf{N}_2$};
\node[rotate=90] (v6) at (11.6,11.4) {$\cdots$};
\node (v11) at (11.6,10.6) {$\mathbf{N}_{8}$};
\node (v12) at (10.4,9.7) {$\mathbf{N}^{1+1}_{10}$};
\node (v13) at (12.8,9.7) {$\mathbf{N}_{9}$};
\draw  (v2) edge (v3);
\draw  (v3) edge (v6);
\draw  (v6) edge (v11);
\draw  (v11) edge (v12);
\draw  (v11) edge (v13);
\node at (11,9.6) {\scriptsize{$f$-$x$-$y$}};
\node at (12.3,9.6) {\scriptsize{$f$}};
\draw  (v1) edge (v2);
\draw  (v1) edge (v3);
\draw  (v1) edge (v11);
\node at (8.3,13.1) {\scriptsize{$e$-$x_2$}};
\node at (11.2,13.1) {\scriptsize{$f$}};
\node[rotate=-8] at (8.8,12.85) {\scriptsize{$x_2$-$x_3$}};
\node at (11.2,12.4) {\scriptsize{$f$}};
\node[rotate=-27] at (8.8,12.4) {\scriptsize{$x_{8}$-$x_{9}$}};
\node at (11.2,11) {\scriptsize{$f$}};
\node (v15) at (9.1,11.2) {\scriptsize{2}};
\draw  (v1) edge (v15);
\draw  (v15) edge (v12);
\node at (8.7,12.05) {\scriptsize{$x_{9},x_{10}$}};
\node at (10.4,10.1) {\scriptsize{$f$-$x,y$}};
\node at (7.9,11.55) {\scriptsize{$e$}};
\node at (9.85,9.95) {\scriptsize{$x$-$y$}};
\draw (v1) .. controls (7.7,13.7) and (9,13.8) .. (11.4,13.8);
\draw (11.4,13.8) .. controls (13,13.8) and (13.5,11.9) .. (v13);
\node at (7.5,13.45) {\scriptsize{$x_{9}$-$x_{10}$}};
\node at (13,10) {\scriptsize{$f$}};
\node at (11.7,12.75) {\scriptsize{$e$}};
\node at (11.7,12.45) {\scriptsize{$e$}};
\node at (11.7,11.9) {\scriptsize{$e$}};
\node at (11.7,10.9) {\scriptsize{$e$}};
\node at (12.1,10.4) {\scriptsize{$e$}};
\node at (12.6,10) {\scriptsize{$e$}};
\node at (10.9,9.9) {\scriptsize{$e$}};
\node at (11.4,10.3) {\scriptsize{$e$}};
\draw  (v12) edge (v13);
\draw  (v10) edge (v12);
\end{tikzpicture}
\ee
Now the compact part describes the gauge theory $\su(3)_\half+9\F$ which only has a $\u(9)\oplus\u(1)$ classical flavor symmetry, but as we have seen above, the enhanced flavor symmetry for this theory is $\so(20)$.

Let us integrate out an $\F$ to obtain the theory $\su(3)_0+8\F$. This is done by first flopping $x_{10}$ living in the top compact surface, which leads to the geometry
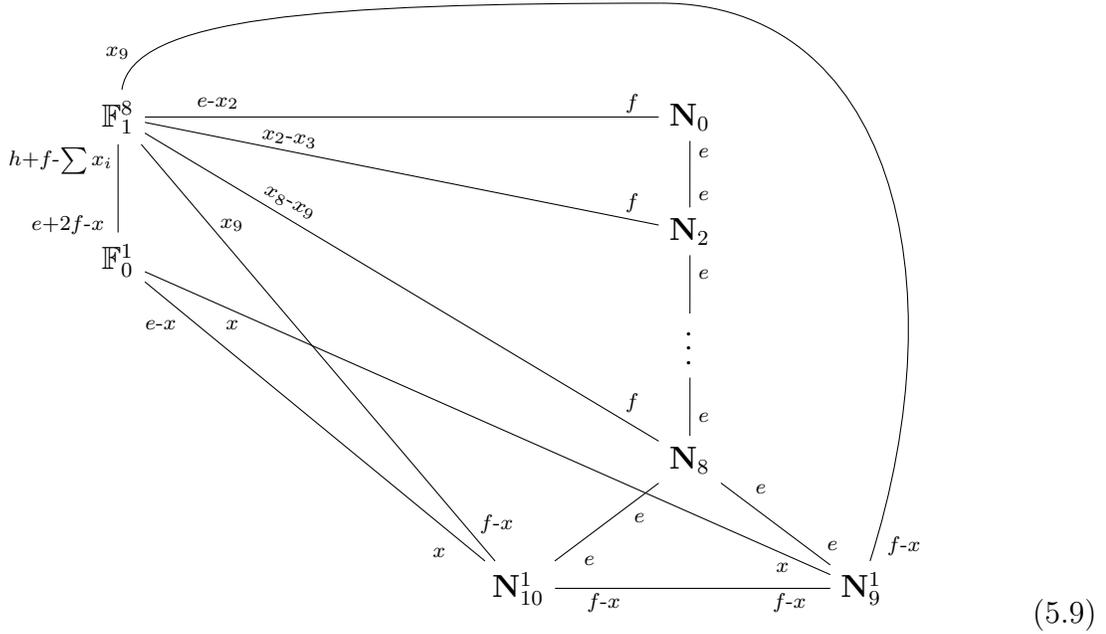
\be
\begin{tikzpicture} [scale=1.9]
\node (v1) at (7.6,13) {$\bF_1^{8}$};
\node (v10) at (7.6,12) {$\bF^1_{0}$};
\draw  (v10) edge (v1);
\node at (7.25,12.25) {\scriptsize{$e$+$2f$-$x$}};
\node at (7.2,12.7) {\scriptsize{$h$+$f$-$\sum x_i$}};
\node (v2) at (11.6,13) {$\mathbf{N}_0$};
\node (v3) at (11.6,12.2) {$\mathbf{N}_2$};
\node[rotate=90] (v6) at (11.6,11.4) {$\cdots$};
\node (v11) at (11.6,10.6) {$\mathbf{N}_{8}$};
\node (v12) at (10.4,9.7) {$\mathbf{N}^{1}_{10}$};
\node (v13) at (12.8,9.7) {$\mathbf{N}^1_{9}$};
\draw  (v2) edge (v3);
\draw  (v3) edge (v6);
\draw  (v6) edge (v11);
\draw  (v11) edge (v12);
\draw  (v11) edge (v13);
\node at (11,9.6) {\scriptsize{$f$-$x$}};
\node at (12.3,9.6) {\scriptsize{$f$-$x$}};
\draw  (v1) edge (v2);
\draw  (v1) edge (v3);
\draw  (v1) edge (v11);
\node at (8.3,13.1) {\scriptsize{$e$-$x_2$}};
\node at (11.2,13.1) {\scriptsize{$f$}};
\node[rotate=-8] at (8.8,12.85) {\scriptsize{$x_2$-$x_3$}};
\node at (11.2,12.4) {\scriptsize{$f$}};
\node[rotate=-27] at (8.8,12.4) {\scriptsize{$x_{8}$-$x_{9}$}};
\node at (11.2,11) {\scriptsize{$f$}};
\node at (8.4,12.25) {\scriptsize{$x_{9}$}};
\node at (10.25,10.15) {\scriptsize{$f$-$x$}};
\node at (7.9,11.55) {\scriptsize{$e$-$x$}};
\node at (9.85,9.95) {\scriptsize{$x$}};
\draw (v1) .. controls (7.7,13.7) and (9,13.8) .. (11.4,13.8);
\draw (11.4,13.8) .. controls (13,13.8) and (13.5,11.9) .. (v13);
\node at (7.6,13.45) {\scriptsize{$x_{9}$}};
\node at (13.1,10) {\scriptsize{$f$-$x$}};
\node at (11.7,12.75) {\scriptsize{$e$}};
\node at (11.7,12.45) {\scriptsize{$e$}};
\node at (11.7,11.9) {\scriptsize{$e$}};
\node at (11.7,10.9) {\scriptsize{$e$}};
\node at (12.1,10.4) {\scriptsize{$e$}};
\node at (12.6,10) {\scriptsize{$e$}};
\node at (10.9,9.9) {\scriptsize{$e$}};
\node at (11.25,10.2) {\scriptsize{$e$}};
\draw  (v12) edge (v13);
\draw  (v10) edge (v12);
\draw  (v1) edge (v12);
\draw  (v10) edge (v13);
\node at (8.4,11.55) {\scriptsize{$x$}};
\node at (12.25,9.85) {\scriptsize{$x$}};
\end{tikzpicture}
\ee
Now, we flop $f-x$ living in the bottom compact surface to obtain
\be
\begin{tikzpicture} [scale=1.9]
\node (v1) at (7.6,13) {$\bF_1^{8}$};
\node (v10) at (7.6,12) {$\bF_{1}$};
\draw  (v10) edge (v1);
\node at (7.35,12.25) {\scriptsize{$h$+$f$}};
\node at (7.2,12.7) {\scriptsize{$h$+$f$-$\sum x_i$}};
\node (v2) at (11.6,13) {$\mathbf{N}_0$};
\node (v3) at (11.6,12.2) {$\mathbf{N}_2$};
\node[rotate=90] (v6) at (11.6,11.4) {$\cdots$};
\node (v11) at (11.6,10.6) {$\mathbf{N}_{8}$};
\node (v12) at (10.4,9.7) {$\mathbf{N}^{1}_{10}$};
\node (v13) at (12.8,9.7) {$\mathbf{N}^2_{9}$};
\draw  (v2) edge (v3);
\draw  (v3) edge (v6);
\draw  (v6) edge (v11);
\draw  (v11) edge (v12);
\draw  (v11) edge (v13);
\node at (11,9.6) {\scriptsize{$f$-$x$}};
\node at (12.3,9.6) {\scriptsize{$f$-$x_1$}};
\draw  (v1) edge (v2);
\draw  (v1) edge (v3);
\draw  (v1) edge (v11);
\node at (8.3,13.1) {\scriptsize{$e$-$x_2$}};
\node at (11.2,13.1) {\scriptsize{$f$}};
\node[rotate=-8] at (8.8,12.85) {\scriptsize{$x_2$-$x_3$}};
\node at (11.2,12.4) {\scriptsize{$f$}};
\node[rotate=-27] at (8.8,12.4) {\scriptsize{$x_{8}$-$x_{9}$}};
\node at (11.2,11) {\scriptsize{$f$}};
\node at (8.4,12.25) {\scriptsize{$x_{9}$}};
\node at (10.25,10.15) {\scriptsize{$f$-$x$}};
\node at (7.85,11.65) {\scriptsize{$e$}};
\node at (9.85,9.95) {\scriptsize{$x$}};
\draw (v1) .. controls (7.7,13.7) and (9,13.8) .. (11.4,13.8);
\draw (11.4,13.8) .. controls (13,13.8) and (13.5,11.9) .. (v13);
\node at (7.6,13.45) {\scriptsize{$x_{9}$}};
\node at (13.1,10) {\scriptsize{$f$-$x_1$}};
\node at (11.7,12.75) {\scriptsize{$e$}};
\node at (11.7,12.45) {\scriptsize{$e$}};
\node at (11.7,11.9) {\scriptsize{$e$}};
\node at (11.7,10.9) {\scriptsize{$e$}};
\node at (12.1,10.4) {\scriptsize{$e$}};
\node at (12.6,10) {\scriptsize{$e$}};
\node at (10.9,9.9) {\scriptsize{$e$}};
\node at (11.25,10.2) {\scriptsize{$e$}};
\draw  (v12) edge (v13);
\draw  (v10) edge (v12);
\draw  (v1) edge (v12);
\draw  (v10) edge (v13);
\node at (8.4,11.55) {\scriptsize{$f$}};
\node at (12.15,9.85) {\scriptsize{$x_1$-$x_2$}};
\end{tikzpicture}
\ee
To completely integrate out the $\F$, we need to send the size of $x_2$ in $\mathbf{N}_9$ to infinity which forces the size of $x_1$ in $\mathbf{N}_9$ to go to infinity (in order to keep the size of $f$ in bottom compact surface finite), which in turn forces $f$ of $\mathbf{N}_9$ to go to infinite size (to keep the size of $x_9$ in top compact surface finite). Thus, we see that $\mathbf{N}_9$ decouples and we can write the resulting geometry as
\be
\begin{tikzpicture} [scale=1.9]
\node (v1) at (7.6,13) {$\bF_1^{8}$};
\node (v10) at (7.6,12) {$\bF_{1}$};
\draw  (v10) edge (v1);
\node at (7.35,12.25) {\scriptsize{$h$+$f$}};
\node at (7.2,12.7) {\scriptsize{$h$+$f$-$\sum x_i$}};
\node (v2) at (11.6,13) {$\mathbf{N}_0$};
\node (v3) at (11.6,12.2) {$\mathbf{N}_2$};
\node[rotate=90] (v6) at (11.6,11.4) {$\cdots$};
\node (v11) at (11.6,10.6) {$\mathbf{N}_{8}$};
\node (v12) at (10.4,9.7) {$\mathbf{N}^{1}_{10}$};
\draw  (v2) edge (v3);
\draw  (v3) edge (v6);
\draw  (v6) edge (v11);
\draw  (v11) edge (v12);
\draw  (v1) edge (v2);
\draw  (v1) edge (v3);
\draw  (v1) edge (v11);
\node at (8.3,13.1) {\scriptsize{$e$-$x_2$}};
\node at (11.2,13.1) {\scriptsize{$f$}};
\node[rotate=-8] at (8.8,12.85) {\scriptsize{$x_2$-$x_3$}};
\node at (11.2,12.4) {\scriptsize{$f$}};
\node[rotate=-27] at (8.8,12.4) {\scriptsize{$x_{8}$-$x_{9}$}};
\node at (11.2,11) {\scriptsize{$f$}};
\node at (8.4,12.25) {\scriptsize{$x_{9}$}};
\node at (10.25,10.15) {\scriptsize{$f$-$x$}};
\node at (7.85,11.65) {\scriptsize{$e$}};
\node at (9.85,9.95) {\scriptsize{$x$}};
\node at (11.7,12.75) {\scriptsize{$e$}};
\node at (11.7,12.45) {\scriptsize{$e$}};
\node at (11.7,11.9) {\scriptsize{$e$}};
\node at (11.7,10.9) {\scriptsize{$e$}};
\node at (10.9,9.9) {\scriptsize{$e$}};
\node at (11.25,10.2) {\scriptsize{$e$}};
\draw  (v10) edge (v12);
\draw  (v1) edge (v12);
\end{tikzpicture}
\ee
from which we can read that the $5d$ SCFT $\su(3)_0+8\F$ has an enhanced $\su(10)$ flavor symmetry.

Now let us integrate out two blowups through the top compact surface and two blowups through the bottom compact surface to obtain the theory $\su(3)_0+4\F$ with the associated geometry
\be
\begin{tikzpicture} [scale=1.9]
\node (v1) at (7.6,13) {$\bF_1^{4}$};
\node (v10) at (7.6,12) {$\bF_{1}$};
\draw  (v10) edge (v1);
\node at (7.45,12.25) {\scriptsize{$h$}};
\node at (7.3,12.7) {\scriptsize{$h$-$\sum x_i$}};
\node (v2) at (11.6,13) {$\mathbf{N}_4$};
\node (v3) at (11.6,12.2) {$\mathbf{N}_5$};
\node (v11) at (11.6,10.6) {$\mathbf{N}_{6}$};
\draw  (v2) edge (v3);
\draw  (v1) edge (v2);
\draw  (v1) edge (v3);
\draw  (v1) edge (v11);
\node at (8.3,13.1) {\scriptsize{$x_4$-$x_5$}};
\node at (11.2,13.1) {\scriptsize{$f$}};
\node[rotate=-8] at (8.8,12.85) {\scriptsize{$x_5$-$x_6$}};
\node at (11.2,12.4) {\scriptsize{$f$}};
\node[rotate=-27] at (8.8,12.4) {\scriptsize{$x_{6}$-$x_{7}$}};
\node at (11.2,11) {\scriptsize{$f$}};
\node at (11.7,12.75) {\scriptsize{$e$}};
\node at (11.7,12.45) {\scriptsize{$e$}};
\node at (11.7,11.9) {\scriptsize{$e$}};
\node at (11.7,10.9) {\scriptsize{$e$}};
\draw  (v3) edge (v11);
\end{tikzpicture}
\ee
for which the geometry tells us that the classical flavor symmetry $\u(4)\oplus\u(1)$ is not enhanced. Actually the above geometry only manifests the non-abelian $\su(4)$ part of the flavor symmetry but the $\u(1)$ factors can be deduced by comparing with the rank of the flavor symmetry of the gauge theory. 

Now we can integrate out the $e$ curves living in both the compact surfaces to obtain a non-Lagrangian phase with geometry
\be\label{NLG}
\begin{tikzpicture} [scale=1.9]
\node (v1) at (7.6,13) {$\dP^{4}$};
\node (v10) at (7.6,12) {$\dP$};
\draw  (v10) edge (v1);
\node at (7.45,12.25) {\scriptsize{$l$}};
\node at (7.3,12.7) {\scriptsize{$l$-$\sum x_i$}};
\node (v2) at (11.6,13) {$\mathbf{N}_4$};
\node (v3) at (11.6,12.2) {$\mathbf{N}_5$};
\node (v11) at (11.6,10.6) {$\mathbf{N}_{6}$};
\draw  (v2) edge (v3);
\draw  (v1) edge (v2);
\draw  (v1) edge (v3);
\draw  (v1) edge (v11);
\node at (8.3,13.1) {\scriptsize{$x_4$-$x_5$}};
\node at (11.2,13.1) {\scriptsize{$f$}};
\node[rotate=-8] at (8.8,12.85) {\scriptsize{$x_5$-$x_6$}};
\node at (11.2,12.4) {\scriptsize{$f$}};
\node[rotate=-27] at (8.8,12.4) {\scriptsize{$x_{6}$-$x_{7}$}};
\node at (11.2,11) {\scriptsize{$f$}};
\node at (11.7,12.75) {\scriptsize{$e$}};
\node at (11.7,12.45) {\scriptsize{$e$}};
\node at (11.7,11.9) {\scriptsize{$e$}};
\node at (11.7,10.9) {\scriptsize{$e$}};
\draw  (v3) edge (v11);
\end{tikzpicture}
\ee
where we denote a $\P^2$ with $n$ blowups as $\dP^n$ and a $\P^2$ without any blowups as $\dP$. According to the geometry, the flavor symmetry of the $5d$ SCFT corresponding to the above geometry is $\su(4)$. In fact, this $5d$ SCFT admits a $5d$ gauge theory description as $\su(2)\oplus\su(2)$ gauge theory with a bifundamental hyper and both theta angles being zero.

\section{Comparison with CFDs}\label{CFD}
In a series of recent papers \cite{Apruzzi:2019vpe,Apruzzi:2019opn,Apruzzi:2019enx} a neat way of encapsulating the flavor symmetry of $5d$ SCFTs was proposed. This was done in terms of a graph referred to as the Combined Fiber Diagram (CFD) of the $5d$ theory. The approach in those papers is similar to the one pursued here. That is, one starts from the CFD of a $5d$ KK theory parent of a $5d$ SCFT whose flavor symmetry is desired. Then, the RG flows to that $5d$ SCFT descend to operations on the CFD of the KK theory. In the end, one obtains the CFD for the desired $5d$ SCFT, from which the flavor symmetry of the $5d$ SCFT can be readily read.

However, the approach pursued there suffers from some limitations. The first limitation is that there is no known way of obtaining the CFD associated to an arbitrary $5d$ KK theory. The second limitation, which we discuss in this section, is that CFDs do not see all RG flows corresponding to integrating out BPS particles. If the BPS particle being integrated out descends from an M2 brane wrapping a gluing curve between two surfaces, then this RG flow is invisible in the CFD machinery. This invisibility is fine if integrating out this BPS particle decomposes the $5d$ theory into a product of two $5d$ theories of lower rank, since then the individual lower rank theories could be obtained inside the decoupling tree of a $5d$ KK theory of that lower rank. However, it is possible to preserve the rank of the $5d$ theory in such a flow\footnote{The possibility of this scenario was pointed out in Section 5 of \cite{Bhardwaj:2018yhy} where the $n=1$ version of the following example also appeared.}. An example of a $5d$ theory where such an RG flow is possible is provided by
\be
\begin{tikzpicture} [scale=1.9]
\node (v2) at (-3.5,0.8) {2};
\node at (-3.5,1.1) {$\su(n)^{(1)}$};
\end{tikzpicture}
\ee
which is obtained via an untwisted compactification of the $6d$ SCFT with tensor branch description $\su(n)+2n\F$.

Let us discuss the case of $n=2$ in the rest of this section. The $6d$ SCFT in this case has an $\so(7)$ flavor symmetry. Thus, we expect to be able to couple the compact part of the geometry (which was described in \cite{Bhardwaj:2018yhy,Bhardwaj:2018vuu,Bhardwaj:2019fzv}) to a collection of non-compact $\P^1$ fibered surfaces parametrizing $\so(7)^{(1)}$, the gluing rules for which were described in \cite{Bhardwaj:2018vuu,Bhardwaj:2019fzv}. The combined full geometry can be represented as
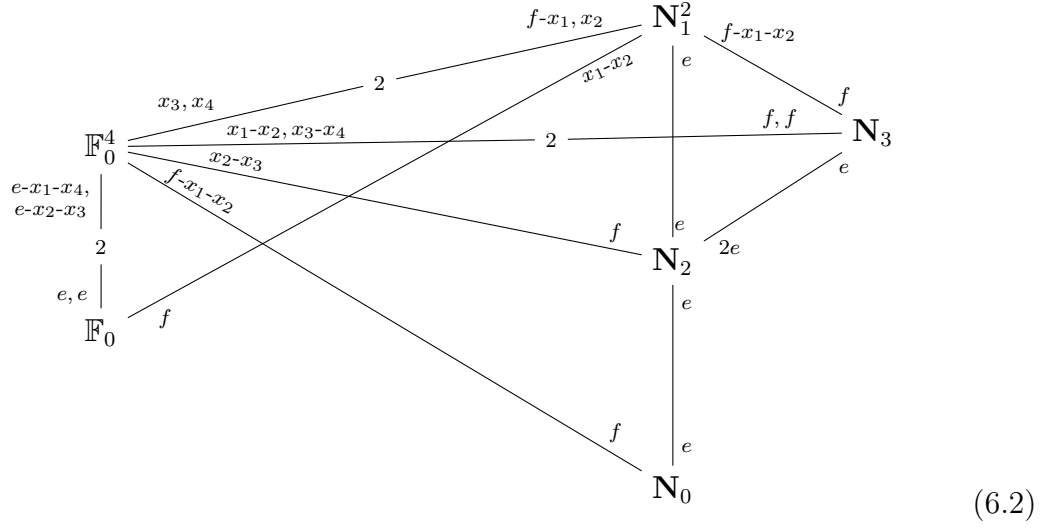
\be
\begin{tikzpicture} [scale=1.9]
\node (v1) at (7.6,13) {$\bF_0^{4}$};
\node (v10) at (7.6,11.7) {$\bF_0$};
\node at (7.4,11.95) {\scriptsize{$e,e$}};
\node at (7.25,12.7) {\scriptsize{$e$-$x_1$-$x_4,$}};
\node (v2) at (13,13.1) {$\mathbf{N}_3$};
\node (v3) at (11.6,12.2) {$\mathbf{N}_2$};
\node (v11) at (11.6,10.6) {$\mathbf{N}_{0}$};
\draw  (v2) edge (v3);
\draw  (v1) edge (v3);
\draw  (v1) edge (v11);
\node at (8.9,13.1) {\scriptsize{$x_1$-$x_2,x_3$-$x_4$}};
\node at (12.35,13.2) {\scriptsize{$f,f$}};
\node[rotate=-8] at (8.55,12.9) {\scriptsize{$x_2$-$x_3$}};
\node at (11.2,12.4) {\scriptsize{$f$}};
\node[rotate=-27] at (8.3,12.7) {\scriptsize{$f$-$x_{1}$-$x_{2}$}};
\node at (11.2,11) {\scriptsize{$f$}};
\node at (12.8,12.85) {\scriptsize{$e$}};
\node at (12,12.3) {\scriptsize{$2e$}};
\node at (11.7,11.9) {\scriptsize{$e$}};
\node at (11.7,10.9) {\scriptsize{$e$}};
\draw  (v3) edge (v11);
\node (v4) at (7.6,12.3) {\scriptsize{2}};
\draw  (v1) edge (v4);
\draw  (v4) edge (v10);
\node (v5) at (11.6,13.9) {$\mathbf{N}^2_1$};
\draw  (v10) edge (v5);
\node (v6) at (9.55,13.45) {\scriptsize{2}};
\draw  (v1) edge (v6);
\draw  (v6) edge (v5);
\node at (8.2,13.3) {\scriptsize{$x_3,x_4$}};
\node at (11.7,13.6) {\scriptsize{$e$}};
\node at (11.65,12.45) {\scriptsize{$e$}};
\node at (12.8,13.35) {\scriptsize{$f$}};
\node at (12.2,13.8) {\scriptsize{$f$-$x_1$-$x_2$}};
\node at (10.85,13.9) {\scriptsize{$f$-$x_1,x_2$}};
\node[rotate=25] at (11.15,13.55) {\scriptsize{$x_1$-$x_2$}};
\node at (8.05,11.8) {\scriptsize{$f$}};
\node at (7.25,12.55) {\scriptsize{$e$-$x_2$-$x_3$}};
\draw  (v5) edge (v3);
\node (v7) at (10.75,13.05) {\scriptsize{2}};
\draw  (v1) edge (v7);
\draw  (v7) edge (v2);
\draw  (v5) edge (v2);
\end{tikzpicture}
\ee
Flopping $x_4$ living in the top compact surface, we can write the above geometry as
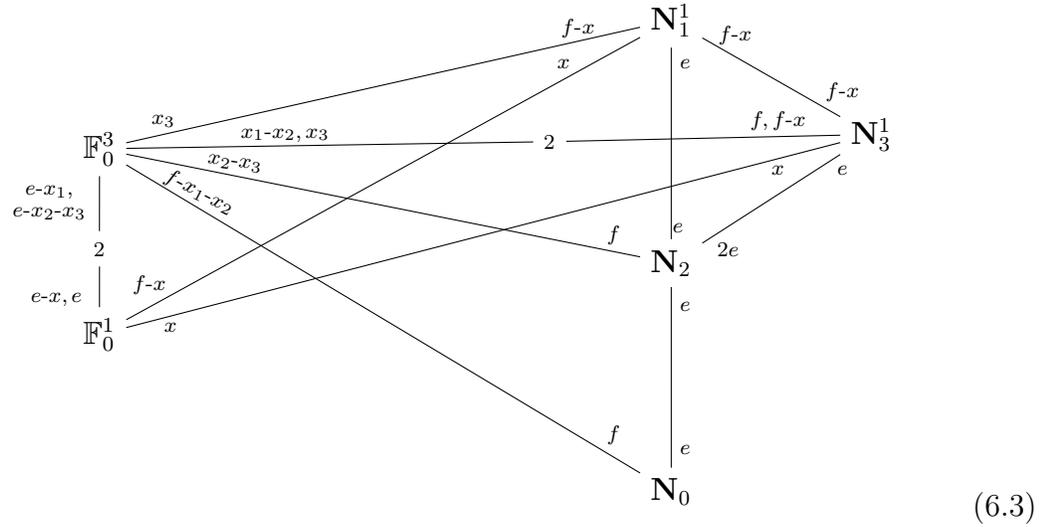
\be
\begin{tikzpicture} [scale=1.9]
\node (v1) at (7.6,13) {$\bF_0^{3}$};
\node (v10) at (7.6,11.7) {$\bF^1_0$};
\node at (7.3,11.95) {\scriptsize{$e$-$x,e$}};
\node at (7.25,12.7) {\scriptsize{$e$-$x_1,$}};
\node (v2) at (13,13.1) {$\mathbf{N}^1_3$};
\node (v3) at (11.6,12.2) {$\mathbf{N}_2$};
\node (v11) at (11.6,10.6) {$\mathbf{N}_{0}$};
\draw  (v2) edge (v3);
\draw  (v1) edge (v3);
\draw  (v1) edge (v11);
\node at (8.9,13.1) {\scriptsize{$x_1$-$x_2,x_3$}};
\node at (12.35,13.2) {\scriptsize{$f,f$-$x$}};
\node[rotate=-8] at (8.55,12.9) {\scriptsize{$x_2$-$x_3$}};
\node at (11.2,12.4) {\scriptsize{$f$}};
\node[rotate=-27] at (8.3,12.7) {\scriptsize{$f$-$x_{1}$-$x_{2}$}};
\node at (11.2,11) {\scriptsize{$f$}};
\node at (12.8,12.85) {\scriptsize{$e$}};
\node at (12,12.3) {\scriptsize{$2e$}};
\node at (11.7,11.9) {\scriptsize{$e$}};
\node at (11.7,10.9) {\scriptsize{$e$}};
\draw  (v3) edge (v11);
\node (v4) at (7.6,12.3) {\scriptsize{2}};
\draw  (v1) edge (v4);
\draw  (v4) edge (v10);
\node (v5) at (11.6,13.9) {$\mathbf{N}^1_1$};
\draw  (v10) edge (v5);
\node at (8.05,13.2) {\scriptsize{$x_3$}};
\node at (11.7,13.6) {\scriptsize{$e$}};
\node at (11.65,12.45) {\scriptsize{$e$}};
\node at (12.8,13.4) {\scriptsize{$f$-$x$}};
\node at (12.05,13.8) {\scriptsize{$f$-$x$}};
\node at (10.95,13.85) {\scriptsize{$f$-$x$}};
\node[rotate=0] at (10.85,13.6) {\scriptsize{$x$}};
\node at (7.95,12.05) {\scriptsize{$f$-$x$}};
\node at (7.25,12.55) {\scriptsize{$e$-$x_2$-$x_3$}};
\draw  (v5) edge (v3);
\node (v7) at (10.75,13.05) {\scriptsize{2}};
\draw  (v1) edge (v7);
\draw  (v7) edge (v2);
\draw  (v5) edge (v2);
\draw  (v1) edge (v5);
\draw  (v10) edge (v2);
\node[rotate=0] at (8.1,11.75) {\scriptsize{$x$}};
\node[rotate=0] at (12.35,12.85) {\scriptsize{$x$}};
\end{tikzpicture}
\ee
The above geometry is isomorphic to the following geometry
\be
\begin{tikzpicture} [scale=1.9]
\node (v1) at (7.6,13) {$\bF_1^{3}$};
\node (v10) at (7.6,11.7) {$\bF^1_1$};
\node at (7.3,11.95) {\scriptsize{$e,h$-$x$}};
\node at (7.3,12.7) {\scriptsize{$e,$}};
\node (v2) at (13,13.1) {$\mathbf{N}^1_3$};
\node (v3) at (11.6,12.2) {$\mathbf{N}_2$};
\node (v11) at (11.6,10.6) {$\mathbf{N}_{0}$};
\draw  (v2) edge (v3);
\draw  (v1) edge (v3);
\draw  (v1) edge (v11);
\node at (8.9,13.1) {\scriptsize{$f$-$x_1$-$x_2,x_3$}};
\node at (12.35,13.2) {\scriptsize{$f,f$-$x$}};
\node[rotate=-8] at (8.55,12.9) {\scriptsize{$x_2$-$x_3$}};
\node at (11.2,12.4) {\scriptsize{$f$}};
\node[rotate=-27] at (8.3,12.7) {\scriptsize{$x_{1}$-$x_{2}$}};
\node at (11.2,11) {\scriptsize{$f$}};
\node at (12.8,12.85) {\scriptsize{$e$}};
\node at (12,12.3) {\scriptsize{$2e$}};
\node at (11.7,11.9) {\scriptsize{$e$}};
\node at (11.7,10.9) {\scriptsize{$e$}};
\draw  (v3) edge (v11);
\node (v4) at (7.6,12.3) {\scriptsize{2}};
\draw  (v1) edge (v4);
\draw  (v4) edge (v10);
\node (v5) at (11.6,13.9) {$\mathbf{N}^1_1$};
\draw  (v10) edge (v5);
\node at (8.05,13.2) {\scriptsize{$x_3$}};
\node at (11.7,13.6) {\scriptsize{$e$}};
\node at (11.65,12.45) {\scriptsize{$e$}};
\node at (12.8,13.4) {\scriptsize{$f$-$x$}};
\node at (12.05,13.8) {\scriptsize{$f$-$x$}};
\node at (10.95,13.85) {\scriptsize{$f$-$x$}};
\node[rotate=0] at (10.85,13.6) {\scriptsize{$x$}};
\node at (7.95,12) {\scriptsize{$x$}};
\node at (7.15,12.55) {\scriptsize{$h$-$x_1$-$x_2$-$x_3$}};
\draw  (v5) edge (v3);
\node (v7) at (10.75,13.05) {\scriptsize{2}};
\draw  (v1) edge (v7);
\draw  (v7) edge (v2);
\draw  (v5) edge (v2);
\draw  (v1) edge (v5);
\draw  (v10) edge (v2);
\node[rotate=0] at (8.25,11.75) {\scriptsize{$f$-$x$}};
\node[rotate=0] at (12.35,12.85) {\scriptsize{$x$}};
\end{tikzpicture}
\ee
Now we notice that we can integrate out the $-1$ curve $e$ living in the bottom compact surface, which also integrates out $e$ living in top compact surface, since the two $e$ curves are glued to each other. Flopping this curve, we obtain
\be
\begin{tikzpicture} [scale=1.9]
\node (v1) at (7.6,13) {$\dP^{3}$};
\node (v10) at (7.6,11.7) {$\dP^1$};
\node at (7.45,12) {\scriptsize{$l$-$x$}};
\node (v2) at (13,13.1) {$\mathbf{N}^2_3$};
\node (v3) at (11.6,12.2) {$\mathbf{N}_2$};
\node (v11) at (11.6,10.6) {$\mathbf{N}_{0}$};
\draw  (v2) edge (v3);
\draw  (v1) edge (v3);
\draw  (v1) edge (v11);
\node at (8.9,13.1) {\scriptsize{$l$-$x_1$-$x_2,x_3$}};
\node at (12.25,13.2) {\scriptsize{$f$-$x_2,f$-$x_1$}};
\node[rotate=-8] at (8.55,12.9) {\scriptsize{$x_2$-$x_3$}};
\node at (11.2,12.4) {\scriptsize{$f$}};
\node[rotate=-27] at (8.3,12.7) {\scriptsize{$x_{1}$-$x_{2}$}};
\node at (11.2,11) {\scriptsize{$f$}};
\node at (12.8,12.85) {\scriptsize{$e$}};
\node at (12,12.3) {\scriptsize{$2e$}};
\node at (11.7,11.9) {\scriptsize{$e$}};
\node at (11.7,10.9) {\scriptsize{$e$}};
\draw  (v3) edge (v11);
\node (v5) at (11.6,13.9) {$\mathbf{N}^1_1$};
\draw  (v10) edge (v5);
\node at (8.05,13.2) {\scriptsize{$x_3$}};
\node at (11.7,13.6) {\scriptsize{$e$}};
\node at (11.65,12.45) {\scriptsize{$e$}};
\node at (12.8,13.4) {\scriptsize{$f$-$x$}};
\node at (12.05,13.8) {\scriptsize{$f$-$x$}};
\node at (10.95,13.85) {\scriptsize{$f$-$x$}};
\node[rotate=0] at (10.85,13.6) {\scriptsize{$x$}};
\node at (7.95,12) {\scriptsize{$x$}};
\node at (7.35,12.65) {\scriptsize{$l$-$\sum x_i$}};
\draw  (v5) edge (v3);
\node (v7) at (10.75,13.05) {\scriptsize{2}};
\draw  (v1) edge (v7);
\draw  (v7) edge (v2);
\draw  (v5) edge (v2);
\draw  (v1) edge (v5);
\draw  (v10) edge (v2);
\node[rotate=0] at (8.25,11.75) {\scriptsize{$l$-$x$}};
\node[rotate=20] at (12.35,12.85) {\scriptsize{$x_1$-$x_2$}};
\draw  (v1) edge (v10);
\end{tikzpicture}
\ee
Completing the integrating out process, i.e. sending $x_2$ living in $\mathbf{N}_3$ to infinite volume, gives rise to the geometry
\be
\begin{tikzpicture} [scale=1.9]
\node (v1) at (7.6,13) {$\dP^{3}$};
\node (v10) at (7.6,11.7) {$\dP^1$};
\node at (7.45,12) {\scriptsize{$l$-$x$}};
\node (v3) at (11.6,12.2) {$\mathbf{N}_2$};
\node (v11) at (11.6,10.6) {$\mathbf{N}_{0}$};
\draw  (v1) edge (v3);
\draw  (v1) edge (v11);
\node[rotate=-8] at (8.55,12.9) {\scriptsize{$x_2$-$x_3$}};
\node at (11.2,12.4) {\scriptsize{$f$}};
\node[rotate=-27] at (8.3,12.7) {\scriptsize{$x_{1}$-$x_{2}$}};
\node at (11.2,11) {\scriptsize{$f$}};
\node at (11.7,11.9) {\scriptsize{$e$}};
\node at (11.7,10.9) {\scriptsize{$e$}};
\draw  (v3) edge (v11);
\node (v5) at (11.6,13.9) {$\mathbf{N}^1_1$};
\draw  (v10) edge (v5);
\node at (8.05,13.2) {\scriptsize{$x_3$}};
\node at (11.7,13.6) {\scriptsize{$e$}};
\node at (11.65,12.45) {\scriptsize{$e$}};
\node at (10.95,13.85) {\scriptsize{$f$-$x$}};
\node[rotate=0] at (10.85,13.6) {\scriptsize{$x$}};
\node at (7.95,12) {\scriptsize{$x$}};
\node at (7.35,12.65) {\scriptsize{$l$-$\sum x_i$}};
\draw  (v5) edge (v3);
\draw  (v1) edge (v5);
\draw  (v1) edge (v10);
\end{tikzpicture}
\ee
Now flopping $x$ living in the bottom compact surface, we obtain
\be
\begin{tikzpicture} [scale=1.9]
\node (v1) at (7.6,13) {$\dP^{4}$};
\node (v10) at (7.6,11.7) {$\dP$};
\node at (7.45,12) {\scriptsize{$l$}};
\node (v3) at (11.6,12.2) {$\mathbf{N}_2$};
\node (v11) at (11.6,10.6) {$\mathbf{N}_{0}$};
\draw  (v1) edge (v3);
\draw  (v1) edge (v11);
\node[rotate=-8] at (8.55,12.9) {\scriptsize{$x_2$-$x_3$}};
\node at (11.2,12.4) {\scriptsize{$f$}};
\node[rotate=-27] at (8.3,12.7) {\scriptsize{$x_{1}$-$x_{2}$}};
\node at (11.2,11) {\scriptsize{$f$}};
\node at (11.7,11.9) {\scriptsize{$e$}};
\node at (11.7,10.9) {\scriptsize{$e$}};
\draw  (v3) edge (v11);
\node (v5) at (11.6,13.9) {$\mathbf{N}_1$};
\node at (8.05,13.25) {\scriptsize{$x_3$-$x_4$}};
\node at (11.7,13.6) {\scriptsize{$e$}};
\node at (11.65,12.45) {\scriptsize{$e$}};
\node at (11,13.9) {\scriptsize{$f$}};
\node at (7.35,12.65) {\scriptsize{$l$-$\sum x_i$}};
\draw  (v5) edge (v3);
\draw  (v1) edge (v5);
\draw  (v1) edge (v10);
\end{tikzpicture}
\ee
which we recognize as the geometry (\ref{NLG}) that we discussed at the end of the last section. There we had constructed this same theory by performing RG flows from a different $5d$ KK theory (\ref{sp1eg}) and all the RG flows were of the kind that are visible at the level of CFDs. Using both methods, we have arrived at the same prediction $\su(4)$ for the flavor symmetry of the $5d$ SCFT corresponding to the above geometry.

In conclusion, we see that the methods of this paper can be used to perform RG flows of the kind which are invisible to CFDs. Moreover, we have arrived at the same theory (\ref{NLG}) from two different $5d$ KK theories and both ways of arriving at this theory lead to the same prediction for the flavor symmetry of the corresponding $5d$ SCFT (\ref{NLG}). This demonstrates the robustness of our proposal for determining flavor symmetries of $5d$ SCFTs.

\section*{Acknowledgements}
The author thanks Patrick Jefferson, Hee-Cheol Kim, Sakura Schäfer-Nameki and Gabi Zafrir for discussions. The author is grateful to Sakura Schäfer-Nameki for providing comments on a draft version of this paper. This work is partly supported by ERC grants 682608 and 787185 under the European Union’s Horizon 2020 programme, and partly supported by NSF grant PHY-1719924. This work was partly completed at Perimeter Institute for Theoretical Physics. Research at Perimeter Institute is supported in part by the Government of Canada through the Department of Innovation, Science and Economic Development Canada and by the Province of Ontario through the Ministry of Colleges and Universities.

\bibliographystyle{ytphys}
\let\bbb\bibitem\def\bibitem{\itemsep4pt\bbb}
\bibliography{ref}

\end{document}